\documentclass[journal]{IEEEtran}
\usepackage{balance}
\usepackage{amsmath,amsfonts}
\usepackage{xfrac}
\usepackage{algorithmic}
\usepackage{algorithm}
\usepackage{enumitem}
\usepackage{xcolor}
\usepackage{array}
\usepackage{textcomp}
\usepackage{stfloats}
\usepackage[hyphens]{url}
\usepackage{hyperref}
\hypersetup{hidelinks}
\usepackage{verbatim}
\usepackage{graphicx}
\usepackage{cite}
\usepackage{multirow}
\usepackage{graphicx}
\usepackage{caption}
\usepackage{placeins}
\usepackage{subcaption}
\usepackage{booktabs}
\usepackage[export]{adjustbox}
\begin{document}

\title{Curbing the Ramifications of Authorship Abuse in Science}

\author{
	\IEEEauthorblockN{Md Somir Khan and Mehmet Engin Tozal}
	\\
	\IEEEauthorblockA{School of Computing and Informatics, University of Louisiana at Lafayette, Lafayette, Louisiana 70504, USA\\
		Email: md-somir.khan1@louisiana.edu and metozal@louisiana.edu
	}
}

\maketitle

\begin{abstract}
	Research performance is often measured using bibliometric indicators, such as publication count, total citations, and $h$-index.
	These metrics influence career advancements, salary adjustments, administrative opportunities, funding prospects, and professional recognition.
	However, the reliance on these metrics has also made them targets for manipulation, misuse, and abuse.
	One primary ethical concern is authorship abuse, which includes paid, ornamental, exploitative, cartel, and colonial authorships.
	These practices are prevalent because they artificially enhance multiple bibliometric indicators all at once.
	Our study confirms a significant rise in the mean and median number of authors per publication across multiple disciplines over the last 34 years.
	While it is important to identify the cases of authorship abuse, a thorough investigation of every paper proves impractical.
	In this study, we propose a credit allocation scheme based on the reciprocals of the Fibonacci numbers, designed to adjust credit for individual contributions while systematically reducing credit for potential authorship abuse.
	The proposed scheme aligns with rigorous authorship guidelines from scientific associations, which mandate significant contributions across most phases of a study, while accommodating more lenient guidelines from scientific publishers, which recognize authorship for minimal contributions.
	We recalibrate the traditional bibliometric indicators to emphasize author contribution rather than participation in publications. Additionally, we propose a new indicator, $L^{\prime}$-index, to quantify a researcher's labor contribution in their publications.
	Lastly, we advocate for simple contribution-based author ordering over complex conventions that create ambiguity and facilitate abuse.
	Our proposed credit allocation scheme mitigates the effects of authorship abuse and promotes a more ethical scientific ecosystem.
	
\end{abstract}


\section{Introduction\protect\footnote{This section is lengthier than usual to present a comprehensive background on the longstanding problem.}}
\IEEEPARstart{B}{ibliometric} indicators provide a quick assessment of a scientist's productivity and influence. Promotion, tenure, and funding of researchers, as well as their place in the social sphere of their communities, are highly dependent on their bibliometric indicators. Furthermore, scientists with superior metrics are granted a greater quantity of resources by both their institutions and government agencies. This affords them an enhanced ability to influence the trajectory of their respective fields and the scientific community, in general.

Initially, the total number of publications and citation counts were used to measure a researcher's performance. The total number of publications indicates a researcher's overall productivity, while the total citation count reflects their overall impact. However, both indicators have significant shortcomings. The total number of publications does not consider the individual contributions of authors, and a small number of highly cited papers can distort the total citation count. In 2005, J.E. Hirsch proposed the $h$-index, a single number combining publications and citation data~\cite{hirsch2005index}. It is described as having $h$ publications with a minimum of $h$ citations apiece. It has successfully addressed a few problems, such as representing researchers' impact in a single number and being robust to highly cited papers. $h$-index has become popular due to its adoption by popular citation databases like Scopus, Google Scholar, and Web of Science. It is widely used to evaluate the scientific impact of institutions, journals, and countries~\cite{hirsch2005index, Hirsch_2007, ROUSSEAU20072}. However, $h$-index has its own drawbacks, including its sensitivity to differences among scientific fields and researchers' career lengths as well as its disregard for multi-authorships~\cite{Bornmann_Daniel_2009}. 

Unfortunately, excessive dependency on bibliometric indicators for recruitment, promotion, salary, funding, institutional resources, social acceptance, and fame has made these indicators a target for gaming, misuse, and abuse. Particularly, authorship abuse has been rampant due to many bibliometric indicators relying on the number of publications and citations, both of which can be directly gamed by multi-authorships. Typical cases of authorship abuse~\cite{Harvey2018, CLAXTON200531} are listed as follows.

{
	\renewcommand{\arraystretch}{3.2} 
	\begin{table*}[h]
		\centering
		\caption{Bibliometric profiles of two researchers from Google Scholar. $P$ is the total number of publications. $C$ is the total number of citations. $h$ is the $h$-index.}
		\label{comparison_real_world}
		\begin{tabular}{|m{3cm}|l|l|l||l|l|l|m{3cm}|}
			\hline
			\multirow{4}{*}{%
				\centering
				\includegraphics[max width=\linewidth, max height=3cm]{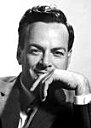}%
			} 
			& \textbf{Name:} Richard Feynman
			& $P$ & 157 & 1,775 & $P$ 
			& \textbf{Name:} Txxxxx Dxxxxxx 
			& \multirow{4}{*}{%
				\centering
				\includegraphics[max width=\linewidth, max height=3cm]{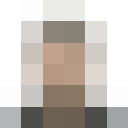}%
			} \\ \cline{2-7}
			& \multirow{2}{*}{\parbox{3cm}{\raggedright\textbf{Sub-field:} Physics}}
			& $C$ & 138,773 & 321,088 & $C$
			& \multirow{2}{*}{\parbox{3cm}{\raggedright\textbf{Sub-field:} Software Engineering, Computational Biology, High-Performance Computing (HPC), Physics}}
			& \\ \cline{3-6}
			& & $h$ & 62 & 239 & $h$ & & \\ \cline{2-7}
			& \multicolumn{3}{p{5cm}||}{}
			& \multicolumn{3}{p{5cm}|}{University of Sxxxxxxxxxx} 
			& \\ \hline
		\end{tabular}
	\end{table*}
}

\noindent \textbf{Paid Authorship:} Paying paper mills or independent researchers to be listed as authors, with fees typically based on the author rank in the byline. Although such publications include several authors, they often lack the novelty and/or scientific correctness needed to advance science. Hence, these publications not only boost the paying authors' bibliometric records but also contribute to the overall scientific pollution by generating excessive noise for other scientists to consume. Even if such a paper is of high quality, paying for publications is considered an unethical practice in science.

\noindent \textbf{Ornamental Authorship:} Honoring invitations to be included as an author in papers with hardly any meaningful contributions. 
The scientific community often overlooks the ornamental authorship. Such authors are typically placed toward the end in a paper's byline under the disguise of ``intellectual guidance'' and/or in the form of ``insignificant contributions''.
Invitations for ornamental authorships may have several motivations, such as including a well-known author to improve the acceptance chances and impact of a paper; including a colleague in exchange for a favor; or including a colleague due to their senior or administrative roles within an organization.
In contrast, ornamental authors are typically driven by a single motive: artificially boosting their bibliometric records.
Irrespective of the motivations involved, honoring authorship invitations without appropriate contributions is considered an unethical practice in science. 

\noindent \textbf{Exploitative Authorship:} Pressuring or manipulating a junior researcher for being listed as an author in their publications with hardly any substantial contributions. 
This abuse usually happens within an institution when a researcher in an administrative or senior position uses their authority and/or influence to coerce juniors into including their names in papers. 
Typically, the true authors of the paper appear at the beginning of the byline and the exploitative authors are listed toward the end.
While perpetrators often carry it out subtly in the form of ``advisory oversight'' or ``senior guidance'', it is still considered an unethical practice in science.

\noindent \textbf{Cartel Authorship:} Agreeing to add one another's names to their papers despite not directly contributing to each other's work. In most cases, the actual authors of a paper are positioned at the beginning of the byline, while other cartel members are placed toward the end, often ranked by seniority or hierarchy. This practice boosts the overall productivity of each researcher involved in the cartel. The perceived productivity tends to increase linearly with the size of the cartel. While the resulting publications may be of high quality, particularly when the cartel consists of the members of a prominent research laboratory, it is still considered an unethical practice in science.

\noindent \textbf{Colonial Authorship:} Forming asymmetric relationships with researchers from low- and middle-income countries to publish papers without substantial contributions, instead offering financial support for article processing charges and conference expenses.
This relationship typically involves a researcher from a developed country and a team of researchers from a developing country.
The former researcher uses their institution's financial capacity to fund the latter researchers' work in exchange for authorship.
Again, the actual authors of the paper are listed at the beginning of the byline, and the perpetrator appears at the end, masked as an ``international collaborator'' or ``senior collaborator''.
While these relationships may lead to high-quality publications, they are considered unethical due to the lack of meaningful contributions and hegemonic exploitation. 

\noindent \textbf{Passive Authorship:} Remaining passive or less involved in the key stages of research, including concept development, experimental design, empirical analyses, or manuscript drafting, but joining in for the revision phase before the paper submission. Passive authorship is a common form of ``professional authorship'' in academic settings. It is a business model, especially at larger laboratories, where graduate students are often supervised by their senior peers, post-docs or junior faculty rather than their advisors, who are also the directors of the respective research labs. The research lab director typically reviews the manuscript just before submission and is credited as the last author under the designation of ``overall coordination'' or ``senior oversight''. Passive authorship is not necessarily unethical, however passive authors should be credited appropriately for their coordination efforts.

The exploitation of multi-authorship practices without substantive scientific contributions has become increasingly prevalent across science.
To illustrate a notable example, Table~\ref{comparison_real_world} shows two researcher 
profiles drawn from Google Scholar.
The first is Richard Feynman, who held a terminal degree in Physics and was co-awarded 
the 1965 Nobel Prize for his original theoretical contributions 
to quantum electrodynamics. The second is a present-day research software engineer who does 
not hold a terminal degree.
The latter accrues $1{,}775$ publications and $321{,}088$ citations, against 
Feynman's $157$ publications\footnote{This figure also includes re-prints published after Richard Feynman's death in 1988.} and $138{,}773$ citations.
The larger totals come from participation in large-scale experimental 
collaborations rather than from scientific contributions.
Counting only participation makes no distinction between a contributing author 
and a supporting one.
Even the $h$-index, often regarded as more robust to inflation, places 
the present-day researcher at $239$ and Feynman at $62$.
Unfortunately, current bibliometrics consistently reward breadth of participation over depth of contribution.
	
Academic departments, journal editors, and scientific publishing stakeholders are well aware of the longstanding problem and have focused on advocating for policies to rectify the overt forms of authorship abuse.
Academic departments and research institutions have been putting more effort into scrutinizing CVs to verify the actual contributions of job applicants. Many departments mainly consider the publications in which the candidate or their students are the first authors when making decisions for tenure and promotions~\cite{LouisianaTenurePromotion}.

Scientific publishers, professional associations and journal editors, on the other hand develop guidelines to prevent multi-authorship abuse~\cite{ICMJE2025,ieee_authorship_guidelines,allen2014publishing,cope2024,cse_authorship,wame_authorship}.
According to the Institute of Electrical and Electronics Engineers (IEEE), all authors must fulfill three criteria: (i) making a significant intellectual contribution to the theoretical framework or experimental design and analysis; (ii) contributing explicitly to the drafting, revising, and reviewing of the work; and (iii) approving the final version~\cite{ieee_authorship_guidelines}.
Similarly, the authorship guideline established by the International Committee of Medical Journal Editors (ICMJE) mandates all authors meet four key criteria: (i) contribution to the conception, design, or analysis of the work; (ii) involvement in the intellectual content or drafting; (iii) approval of the final publication; and (iv) accountability for all aspects of the work~\cite{ICMJE2025} .
Likewise, the American Psychological Association's (APA) ethical principles of psychologists and code of conduct document addresses authorship credits in publications~\cite{APA2017}. According to the APA, psychologists can take authorship credit only (i) for work that they performed or (ii) for work that they substantially contributed; however psychologists cannot take authorship credit solely (iii) for their administrative roles or (iv) for minor contributions to research or writing.

Professional associations typically focus on determining who qualifies to be included as an author, while scientific publishers often consider which individuals should not be excluded from authorship.
As a result, many publishing businesses prefer a broader definition of authorship, such as the Contributor Roles Taxonomy (CRediT)~\cite{allen2014publishing}, which allows authorships for non-scientific contributions as well.
CRediT recommends that anyone involved in one of the 14 roles, including conceptualization, data curation, formal analysis, funding acquisition, investigation, methodology, project administration, resource management, software development, supervision, validation, visualization, writing, or revising, should be acknowledged as an author in a study.
Unfortunately, CRediT also enables multiple authorship claims with minimal effort on one single paper. Examples of trivial CRediT effort include but are not limited to ``making a critical commentary''; ``testing any code used in the work''; ``providing study materials''; ``supplying software''; ``preparing presentation''; ``maintaining research data''; as well as more vague forms of ``management of research activity''; ``formulation of overarching aims''; and ``leadership responsibility for research''~\cite{ElsevierCRediT}.

Unethical authorship practices, such as paid, ornamental, exploitative, cartel and colonial authorships, violate all authorship guidelines outlined above.
On the other hand, \emph{passive authorship} violates the IEEE, ICMJE and APA guidelines but adheres to CRediT, by definition.
In fact, CRediT can dismiss any instance of authorship abuse by allowing perpetrators to submit minimal contributions or assume vague roles in a study.
Passive authorship and similar conflicting practices can only be addressed by credit allocation schemes that distinguishes between more and less contributing authors.

Conversely, not all scientific disciplines follow a universally accepted standard authorship guideline. Some disciplines or research institutions appear to have their own operational definitions of authorship~\cite{Ioannidis2018}. For example, papers published by CERN are credited to all CERN members~\cite{CERN,dance2012authorship}. The increasing number of articles with hundreds or thousands of authors raises doubts about how actively each author abides by any standards~\cite{Bart2020}. In the past, authorship was typically related to scientific contributions, and non-scientific contributions were mentioned in acknowledgments. 
CERN does not differentiate between scientific and non-scientific contributions and is more aligned with CRediT authorship designations than the IEEE, ICMJE, or APA guidelines.
While CERN's practice is not considered authorship abuse per se, it adversely affects many disciplines by skewing their bibliometric indicators. 

Our analyses show that the number of coauthors has gradually increased over the years in every discipline. Because of the inflation in the number of coauthors, determining the actual productivity of an author has become complex. In a recent study by Ioannidis, authors who have published more than 60 papers in a full calendar year were defined as extremely productive authors, and the study suggests that there has been a significant increase in different scientific fields and countries since 2016~\cite{Ioannidis2023.11.23.568476}. In another study by the same author, 9000 hyper prolific authors who, during their careers, published 72 complete papers in at least one calendar year were reported. This means completing one full paper every five days, including weekends and holidays~\cite{Ioannidis2018}. A productivity rate of this type raises concerns about the actual performance of scientists and the amount of pollution/noise infused into science, in general. Surveys also report that hyper prolific authors often fail to meet the ICMJE criteria, despite the general reluctance of such authors taking the survey~\cite{Harvey2018, Ioannidis2018}.

To ensure transparency and accountability, some journals require authors to disclose their involvement and percentage of contributions as statements due to the rising number of authors credited in publications. However, the allocation of contribution percentage in multi-authored publications is highly subjective and prone to disagreement among the authors. Moreover, it is frequently influenced by seniority or laboratory politics~\cite{dance2012authorship, Brand2015BeyondAA}. The lack of a uniform policy for contributorship statements implemented by all publishers and the inherent social complexity hinder the integration of contributorship statements in bibliometric indicators.

On the other hand, it is easier to interpret bibliometric indicators derived from author rankings in a list of authors ordered by their contributions. Individual contributions often determine authorship order in most fields. In some fields, the last authorship slot is designated for the most senior or the author responsible for the overall coordination. However, the same last slot is frequently abused for ornamental, exploitative, cartel and colonial authorships as well as employed for passive authorship. 
It is important to differentiate between the passive authors responsible for the overall coordination and the second-most contributing authors of a publication.
The second-most contributing author is often a senior student, post-doc, or junior faculty member listed in the second position of the author byline, whereas the coordinating author's name generally appears at the end of the byline.
Another method is arranging the authors in alphabetical order. The alphabetical order of author names favors individuals with last name initials early in the alphabet~\cite{weber2018effects}.

The practice of ordering the authors by their contributions in the byline has been gaining traction among many scientists in academia and industry. Hence, a credit allocation scheme based on the order of authors in the byline can alleviate many problems related to multi-authorship abuse. Overall, this study also highlights the need for the scientific community to adopt a simple, contribution-based ordering of authors, in place of more complex authorship conventions that may facilitate abuse through ambiguity. There are two methods for counting author credits: \emph{whole counting}, where every author gets a whole credit of one, and \emph{fractional counting}, where every author gets a fraction of the whole credit.

The indefinite whole counting, which unfortunately is the current practice, fuels the authorship abuse by promoting scientists who network better over those who work better. Besides, it allows scientists in administrative positions to potentially distress the ones in subordinate positions. Moreover, it leads to an absurdity where the amount of output per publication \emph{indefinitely} increases by the number of authors. Different fractional credit allocation schemes have been proposed to fix the problem, including equalitarian fractional counting~\cite{price1981multiple}, harmonic counting~\cite{hodge1981publication, Hagen2008, Sekercioglu2008}, arithmetic counting~\cite{Hooydonk1997, Trueba2004}, geometric counting~\cite{Egghe2000} and others~\cite{Galam2011, Liu2012, Tol2011}.
The equalitarian fractional credit allocation scheme normalizes credit by the number of authors and gives equal credit to all authors. The other fractional counting methods allocate credit proportionally based on the author's rank in the byline.
The equalitarian fractional credit allocation scheme is unfair to the leading author(s) who contributed the most to a publication.
In contrast, proportional credit allocation schemes diminish rapidly for the first few authors, thus punishing the main contributors. More importantly, these credit allocation schemes often punish the most \emph{fundamental type} of scientific collaboration consisting of two authors, i.e., a student and their academic advisor or two peer scientists. 

{
	\renewcommand{\arraystretch}{1.2}
	\begin{table*}
		\centering
		\caption{Fibonacci numbers and their reciprocals for an example publication with 13 authors.}
		\label{tab_1:reciprocals_of_fibonacci_for_13_authors}
		\scalebox{0.9}{
			\begin{tabular}{|*{14}{c|}}
				\hline
				\textbf{Rank} & \textbf{1} & \textbf{2} & \textbf{3} & \textbf{4} & \textbf{5} & \textbf{6} & \textbf{7} & \textbf{8} & \textbf{9} & \textbf{10} & \textbf{11} & \textbf{12} & \textbf{13} \\
				\hline
				\hline
				\textbf{Fibonacci} & 1 & 1 & 2 & 3 & 5 & 8 & 13 & 21 & 34 & 55 & 89 & 144 & 233\\
				\hline
				\textbf{Reciprocals (Ratio)} & 1/1 & 1/1 & 1/2 & 1/3 & 1/5 & 1/8 & 1/13 & 1/21 & 1/34 & 1/55 & 1/89 & 1/144 & 1/233\\
				\hline
				\textbf{Reciprocals (Decimal)} & 1.0000 & 1.0000 & 0.5000 & 0.3333 & 0.2000 & 0.1250 & 0.0769 & 0.0476 & 0.0294 & 0.0182 & 0.0112 & 0.0069 & 0.0042 \\
				\hline
			\end{tabular}
		}
	\end{table*}
}

In this study, we propose a mathematically straightforward and computationally convenient approach for adjusting scientific indicators based on the reciprocals of the Fibonacci numbers, $F = [1, 1, 2, 3, 5, 8, 13, 21, \ldots]$.
Table~\ref{tab_1:reciprocals_of_fibonacci_for_13_authors} shows credits allocated by the reciprocals of the Fibonacci numbers for an example publication with 13 authors.
Demonstrated in the last row of the table, the first two authors leading the study receive whole credits of one, which is congruent with natural counting and its ordinary interpretation of productivity.
The rest of the authors supporting the study receive less credit but commensurate with their ranks in the byline.
The proposed scheme allocates more than one percent of a whole credit for each of the first eleven authors.
Besides, it allocates more than ten percent of a whole credit for each of the first six authors.
The proposed scheme inherently allocates whole credits for the authors of single-~and~dual-authorship papers.
Besides, the series consisting of the reciprocals of the Fibonacci numbers is convergent~\cite{OEIS_Fibonacci}.
That is, it stays consistent at approximately $3.35$, regardless of the number of authors involved in a publication, as shown in Equation~\ref{eq_sum_fib_constant}.

\begin{equation}
	\label{eq_sum_fib_constant}
	\psi = \sum_{i=1}^{\infty} \frac{1}{F_i} = \frac{1}{1} + \frac{1}{1} + \frac{1}{2} + \frac{1}{3}  + \frac{1}{5} + \ldots \approx 3.35
\end{equation}

The fundamental characteristics of the proposed credit allocation scheme based on the reciprocals of the Fibonacci numbers are as follows.
\begin{enumerate}[label=(\roman*), start=1]
	\item The proposed scheme converges to a constant value as the number of authors increases indefinitely.
	\item The proposed scheme consistently assigns the same credit for each author rank, irrespective of the number of authors.
	\item The proposed scheme allocates a whole credit to each author of the fundamental collaborations of two authors.
	\item The proposed scheme adheres to whole counting and its natural interpretation of productivity for the leading authors, i.e., ranked first and second in the byline.
	\item The proposed scheme does not decrease rapidly for the initial few supporting authors following the leading two.
	\item The proposed scheme exhibits a quick decrease after the initial several supporting authors following the leading two.
	\item The proposed scheme appropriately credits infinitely many authors, hence supporting operational authorship guidelines that recognize insignificant or non-scientific contributions, e.g., the CRediT and CERN.
	\item The proposed scheme appropriately credits the most contributing authors, hence supporting rigorous authorship guidelines that require significant contributions, e.g., the IEEE, ICMJE, and APA.
	\item The proposed scheme is straightforward and convenient, both mathematically and computationally.
\end{enumerate}

The approximations of the Fibonacci sequence and the golden ratio are reported to be observed in various aspects of the natural world, spanning from the spirals found in galaxies to the intricate organization of petals in flowers and even extending to the proportions exhibited by the human face. Independent of these claims, we are inspired by the utilization of the Fibonacci sequence in Computer Science, notably in the context of the Fibonacci coding employed in compression algorithms~\cite{FRAENKEL199631}, the Fibonacci search technique~\cite{Ferguson1960}, and the application of graphs as Fibonacci cubes to interconnect parallel and distributed systems~\cite{Apostolico1987}.

We used Google Scholar Profiles to collect a dataset of authors involving eight different fields, including Biology, Mathematics, Computer Science and Engineering, Electrical and Electronics Engineering, Economics, Marketing, Sociology, and Psychology, with three subfields from each.
However, the author dataset extends beyond the designated subfields, as the authors typically declare multiple subfields on their profile pages.
We are aware that the collected dataset may not represent the publication and authorship practices for every subfield in each discipline. 
However, our goal in this study is \emph{not} to analyze various disciplines' authorship and publication practices.
Instead, we utilize the dataset to illustrate the growing number of authors per publication across different fields, while also highlighting the effectiveness of the proposed Fibonacci-adjusted performance metrics against multi-authorship abuse.

Our analyses reveal that the average number of authors per publication has nearly doubled over the past 34 years. 
We observe a steady decline in the proportion of one-, two-, and three-author papers, which dropped from 64.44\% in 1991 to 34.35\% in 2024. In contrast, publications with multiple authors have consistently increased.
Notably, the share of publications with more than eight authors has risen from 5.08\% in 1991 to 21.18\% in 2024. 
This increase in authorship practices is reflected in the inflated values for the number of publications ($P$), citation counts ($C$), and Hirsch-indexes ($h$) in the Google Scholar Profile dataset across all collected fields.
To address these discrepancies, we introduce Fibonacci-adjusted total publications ($P^\prime$), Fibonacci-adjusted citation count ($C^\prime$), and Fibonacci-adjusted Hirsch index ($h^\prime$), which recalibrate $P$, $C$, and $h$ using the reciprocals of Fibonacci numbers.
Our analyses indicate that authors in our combined dataset typically exhibit a 25\% difference between $P$ and $P^\prime$.
This difference spikes to 74.26\% in the Biology dataset while decreasing to 12.74\% in the Mathematics dataset. We have identified three clusters of typical percentage differences: the first cluster consists solely of the Biology dataset, the second includes the Electrical and Electronics Engineering, Computer Science and Engineering, and Psychology datasets, and the third encompasses the remaining field datasets. We observe a similar trend in the percentage differences between $C$ and $C^\prime$. For the combined dataset, this difference is around $25\%$.
However, authors in the Biology, Electrical and Electronics Engineering, Computer Science and Engineering, and Psychology datasets show greater differences, ranging from $81.32\%$ to $27.20\%$, compared to the other datasets. We also see similar trends in the percentage differences between $h$ and $h^\prime$. 
The novel Labor-index ($L^\prime$-index) assumes a value between $0$ and $1$ and measures a researcher's labor contribution across their publications.
An $L^\prime$ near $1$ indicates the author contributes significantly. A low $L^\prime$ indicates that the author predominantly appears in supporting roles with less contributions.
The $L^\prime$-index fosters genuine collaborations while discouraging unethical authorship practices.
More importantly, it remains independent of publication practices across different disciplines.
We propose benchmark values of $0.61$ and $0.72$ for non-research-intensive and research-intensive institutions, respectively.
Our analysis shows that only $18.53\%$ of authors fall below $0.61$, indicating that a relatively small number of researchers predominantly take supporting roles.
However, this percentage is significantly higher in the Biology, Electrical and Electronics Engineering, and Computer Science and Engineering datasets compared to the Economics, Marketing, Sociology, Psychology, and Mathematics datasets. Some disciplines designate the last byline position for the most contributing author, a convention that also facilitates passive authorship. To assess the effect of this byline convention on the $L^\prime$-index, we conducted a separate experiment in which the last author of each publication was treated as the first author. Under this experiment, 70.16\% of authors in the Biology dataset remain below the upper benchmark, while 41.50\% remain below the lower benchmark. Details of this experiment are presented in Section~\ref{subsec:last_author}.

The rest of the paper is organized as follows.
In section~\ref{section:related_work} we present the related work.
In Section~\ref{section:datacoll} we outline the data collection and curation processes.
In Section~\ref{sec:motivation} we establish the motivation of this study.
In Section~\ref{sec:FibonacciIndicators} we introduce Fibonacci-adjusted bibliometric indicators.
In Section~\ref{section:empirical_evaluations}, we present the empirical evaluations.
Finally, we conclude our study in Section~\ref{section:conclusions}.

\section{Related Work}
\label{section:related_work}
Multi-authorship abuse is a well-known problem and many methods have been proposed in the literature to alleviate its ramifications. Many journals require contributorship statements to account for the contributions of the individual authors. These statements are intended to promote transparency, fairness, and accuracy in academic publishing. However, such statement templates vary between journals. In 2014, Allen et al. proposed the Contributor Roles Taxonomy (CRediT) and recommended its standardization in publishing~\cite{allen2014publishing, Brand2015BeyondAA}. CRediT is one of the most prevalent formats for disclosing author contributions. However, CRediT statements can be subjective and manipulated like the traditional authorship schemes~\cite{ilakovac2007reliability}.

Other researchers have proposed alternative credit allocation schemes based on CRediT. Yang et al.~\cite{YANG2022} proposed a model combining contribution statements and citation context to better quantify credit to researchers in the medical field. However, their method heavily relies on the availability and accuracy of detailed author statements, which may not always be consistent. Ding et al.~\cite{Ding2021} proposed a role-weighted credit system to allocate credit based on author roles. Yet, their approach does not differentiate between the varying aspects of the roles and give more credit to authors who had more roles. Yang et al.~\cite{Yang2017} proposed quantification of author contributions by calculating the ratio of the number of roles an author participated in to the total contributor roles listed in a publication. Corrêa Jr et al.~\cite{CORREAJR2017} allocated credits to co-authors based on the ratio of the number of each co-author’s contribution categories to the total number of the contribution categories presented in a publication. These methods rely on authorship statements. Yet, authorship statements are not standardized across all disciplines and not supported by all publishers~\cite{Vasilevsky2021, SHLOBIN2022}.

Unlike the contributorship statements the author byline is consistently present in every publication. Several methods have been proposed to allocate credit in multi-authorship publications based on the author byline. Price proposed fractional counting to allocate credit in multi-authorship publications~\cite{price1981multiple}. Each author receives equal credit regardless of their contribution to the work, with the publication and citations distributed among all authors. This scheme unfairly marginalizes the contributions of authors who have contributed more than the others. Hodge and Greenberg~\cite{hodge1981publication} proposed harmonic counting as a credit allocation method that addresses the limitations of equalitarian fractional counting. It employs the harmonic series and normalization to allocate credit in a progressively diminishing manner. Unfortunately, harmonic counting  is not naturally interpretable and does not support the fundamental collaborations between a student and their advisor or between two peer scientists. Hooydonk~\cite{Hooydonk1997} proposed arithmetic counting for credit allocation. In this approach credit allocation for authors is determined by subtracting their position in the group from the total number of authors and dividing this difference by the sum of an arithmetic series ranging from one to the total number of authors in the group. Egghe et al. suggested geometric counting, which involves using geometric series to assign credit to authors~\cite{Egghe2000}. However, these methods quickly reduce the credits of the first few authors in the byline. Besides, their interpretation may vary depending on the number of authors involved.

Researchers have also proposed many h-index variants to address multi-authorship abuse. Hirsch’s $\bar{h}$-index factors in coauthors’ $h$-indexes but can disadvantage junior collaborators and requires a recursive computation~\cite{hirsch2010index}. Batista et al. introduced the complementary h-index \cite{batista2006possible}, while Wan et al. proposed the pure h-index \cite{wan2007pure}. Both metrics normalizes $h$-index to adjust for large collaborative groups. However, they remain biased towards researchers whose $h$-core is high because of large collaborations. Adapted pure $h$-index~\cite{chai2008adapted} and fractional approaches by Egghe~\cite{egghe2008mathematical} and Schreiber~\cite{schreiber2008modification, schreiber2009fractionalized} adjust ranks or citation counts by the number of authors. Hagen’s harmonic h-index~\cite{Hagen2008} and Zhang’s weighted h-index~\cite{zhang2009proposal} alter citation credit distribution, often prioritizing first and last authors. Galam’s gh-index~\cite{Galam2011} tailors credits heterogeneously while ensuring total credit sums to one. Other methods, like Richard’s Pareto weighting~\cite{Tol2011} and the $hc$-index by Bhattacharya et al.~\cite{bhattacharya2022impact}, rely on authors’ citation histories and coauthors’ impacts, requiring access to extensive bibliometric data.

In this study, we propose a mathematically straightforward and computationally convenient approach based on the reciprocals of Fibonacci numbers to determine authorship credit for publications. Our method does not rely on bibliometric metadata or author contribution statements; instead, it is based on the authors' ranks in the byline. 
This approach is easy to understand. 
It does not grow indefinitely as the number of authors grow, yet it consistently assigns the same credit for the same rank. 
It encourages fundamental collaborations and avoids unfairness to the leading authors and the first few supporting authors, who relatively contribute more.
Nevertheless, it assigns proper credit to the supporting authors who contribute less.
Lastly, it is congruent with whole counting for the leading authors, which supports a natural interpretation of productivity.

\section{Author Data Collection and Curation}
\label{section:datacoll}
In this section, we first outline the author data collection process and then present the data preprocessing steps.

\subsection{Author Data Collection}
We utilized Google Scholar to compile our dataset. Unlike other citation databases such as Scopus and Web of Science (WoS), Google Scholar contains a broader range of bibliometric data that is publicly available and it is widely used among scholars. We summarize our data collection process as follows.

{
	\renewcommand{\arraystretch}{1.2} 
	\begin{table*}
		\centering
		\caption{Scientific fields used in this study and their sub-fields.}
		\label{tab_2:fields_subfields}
		\begin{tabular}{|l|p{10cm}|}
			\hline
			\textbf{Field} & \textbf{Sub-fields} \\ \hline
			\hline
			BIOL (Biology) & Genetics, Microbiology, Molecular Biology \\ \hline
			MATH (Mathematics) & Algebra, Number Theory, Combinatorics \\ \hline
			CSCE (Computer Science and Engineering) & Cyber Security, Computer Networks, Software Engineering\\ \hline
			EENG (Electrical and Electronics Engineering) & Power Electronics, Signal Processing, Microelectronics \\ \hline
			ECON (Economics) & Econometrics, Microeconomics, Macroeconomics \\ \hline
			MARK (Marketing) & Consumer Behavior, Brand Management, Marketing Strategy\\ \hline
			SOCI (Sociology) & Demography, Criminology, Economic Sociology \\ \hline
			PSYC (Psychology) & Cognitive Psychology, Developmental Psychology, Social Psychology\\ \hline
		\end{tabular}
	\end{table*}
}

\begin{enumerate}
	\item First, we selected eight fields of academic disciplines and three sub-fields from each field to encompass a broader range of data. 
	The set of fields consists of Biology, Mathematics, Computer Science and Engineering, Electrical and Electronics Engineering, Economics, Marketing, Sociology, and Psychology.
	We present these fields and their sub-fields in Table~\ref{tab_2:fields_subfields}. 
	
	We also avoided cross-disciplinary or popular subfields such as Artificial Intelligence (AI), Machine Learning (ML), or Biostatistics (BS) because the list of scientists contributing to these fields may be controversial. For example, many clinical medicine researchers declare their research field as AI, and many physicists associated with CERN declare their research field as ML. Besides, such authors from Medicine and Physics skew other discipline statistics as they publish disproportionate amounts of papers. 
	
	\item We used the sub-fields as search parameters to obtain lists of author profiles. Google Scholar ranks author profiles in each sub-field based on their citation counts, with ten author profiles on each page. We randomly selected one author from each page, covering 200 pages of each sub-field. For sub-fields which has less than 200 pages available, we randomly chose multiple authors from each page to ensure 200 author profiles are collected on average.
	
	It is important to note that the sub-fields are self-declared by authors as their research fields and authors typically declare multiple research fields on their Google Scholar pages. Therefore, variation in the collected data extends beyond the designated three subfields for each discipline. 
	
	\item We gathered each author's public data, including publication titles, author names, publication types, publication dates, and citation counts for each publication. 
\end{enumerate}

We ended up collecting 4,785 author profiles from at least 24 different sub-fields. Overall, 578,210 publication records were stored in an SQLite database.
Our data collection started on 19 November, 2024 and ended on 17 December, 2024 to avoid overloading Google services. 

\subsection{Author Data Curation}
The author data obtained from Google Scholar can be noisy. One of the limitations of Google Scholar is its tendency to under-report the number of authors for larger author lists, typically displaying around first 150 authors. In addition, an author's name might not appear in his/her publication or a variation of the name might appear. 
Moreover, some collaborations, such as the ATLAS and CMS collaborations at CERN, are listed as the sole authors in publication records and claimed by thousands of authors. Sometimes Google Scholar also displays incorrect publication years. To address these concerns, we preprocessed the collected data as follows.

{
	\renewcommand{\arraystretch}{1.2} 
	\begin{table*}
		\centering
		\caption{A summary of the final dataset before and after preprocessing.}
		\label{tab_3:data summary by field}
		\begin{tabular}{|l|ll|ll|}
			\hline
			\multirow{2}{*}{} & \multicolumn{2}{l|}{\textbf{Number of Author Profiles}}                                                                                                & \multicolumn{2}{l|}{\textbf{Number of Publications}}                                                                                                 \\ \cline{2-5} 
			& \multicolumn{1}{l|}{\begin{tabular}[c]{@{}l@{}}\textbf{Before}\end{tabular}} & \begin{tabular}[c]{@{}l@{}}\textbf{After}\end{tabular} & \multicolumn{1}{l|}{\begin{tabular}[c]{@{}l@{}}\textbf{Before}\end{tabular}} & \begin{tabular}[c]{@{}l@{}}\textbf{After}\end{tabular} \\ \hline \hline
			\textbf{BIOL (Biology)}                & \multicolumn{1}{l|}{600}                                                       & 600                                                         & \multicolumn{1}{l|}{169362}                                                   & 108465                                                      \\ \hline
			\textbf{MATH (Mathematics)}                 & \multicolumn{1}{l|}{590}                                                       & 586                                                         & \multicolumn{1}{l|}{29862}                                                   & 20614                                                      \\ \hline
			\textbf{CSCE (Computer Science and Engineering)}        & \multicolumn{1}{l|}{600}                                                       & 595                                                         & \multicolumn{1}{l|}{94487}                                                   & 61125                                                       \\ \hline
			\textbf{EENG (Electrical and Electronics Engineering)}               & \multicolumn{1}{l|}{595}                                                       & 560                                                         & \multicolumn{1}{l|}{111218}                                                    & 71876                                                       \\ \hline
			\textbf{ECON (Economics)}  & \multicolumn{1}{l|}{602}                                                       & 592                                                         & \multicolumn{1}{l|}{54837}                                                   & 34801                                                       \\ \hline
			\textbf{MARK (Marketing)}               & \multicolumn{1}{l|}{604}                                                       & 589                                                        & \multicolumn{1}{l|}{29899}                                                    & 21710                                                       \\ \hline
			\textbf{SOCI (Sociology)}                 & \multicolumn{1}{l|}{595}                                                       & 584                                                         & \multicolumn{1}{l|}{41797}                                                    & 28387                                                       \\ \hline
			\textbf{PSYC (Psychology)}  & \multicolumn{1}{l|}{599}                                                       & 594                                                         & \multicolumn{1}{l|}{46748}                                                    & 30031                                                       \\ \hline
		\end{tabular}
	\end{table*}
}

\begin{enumerate}
	\item We first removed authors who had either non-functional profile URLs or non-existent publications.
	\item Then, we employed fuzzy~logic~\cite{fuzzywuzzy} to tokenize and validate author names and also to determine authors' positions in the author byline.
	\item Further, we removed publication records that lacked publication dates or had unclear publication years or missing author names.
	\item Finally, we removed all publication records that contain keyword ``Patent'' as their publication type.
\end{enumerate}

To ensure that each sub-field consecutively has at least five publications per year, we analyzed the preprocessed publication records. Our findings revealed that all sub-fields consistently had at least five publications per year starting from 1991. Consequently, we selected 1991 as the starting point for our study. We excluded the data before 1991 and focused on the period from 1991 to 2024. Specifically, we removed all publication records before 1991 and after 2024 as well as the authors who don't have records in the same period. Ultimately, this resulted in 4,700 authors and 377,009 publication records.

While our author dataset may not reflect the extensive publication and authorship practices in these eight distinct disciplines, our focus in this study is \emph{not} on analyzing such practices. Instead, we employ the author dataset to illustrate the rising number of authors per publication across many disciplines and highlight the effectiveness of the proposed Fibonacci-adjusted performance metrics against multi-authorship abuse.
Therefore, we use the field names to only refer to the collected authors of the subfields and their extensions when authors declare multiple subfields on their profile pages.

Lastly, we summarize the number of author profiles and the number of publications before and after preprocessing in Table~\ref{tab_3:data summary by field}.

\section{Motivation}
\label{sec:motivation}

\begin{figure*}
	\centering
	\begin{subfigure}[t]{0.48\textwidth}
		\centering
		\includegraphics[height=0.24\textheight]{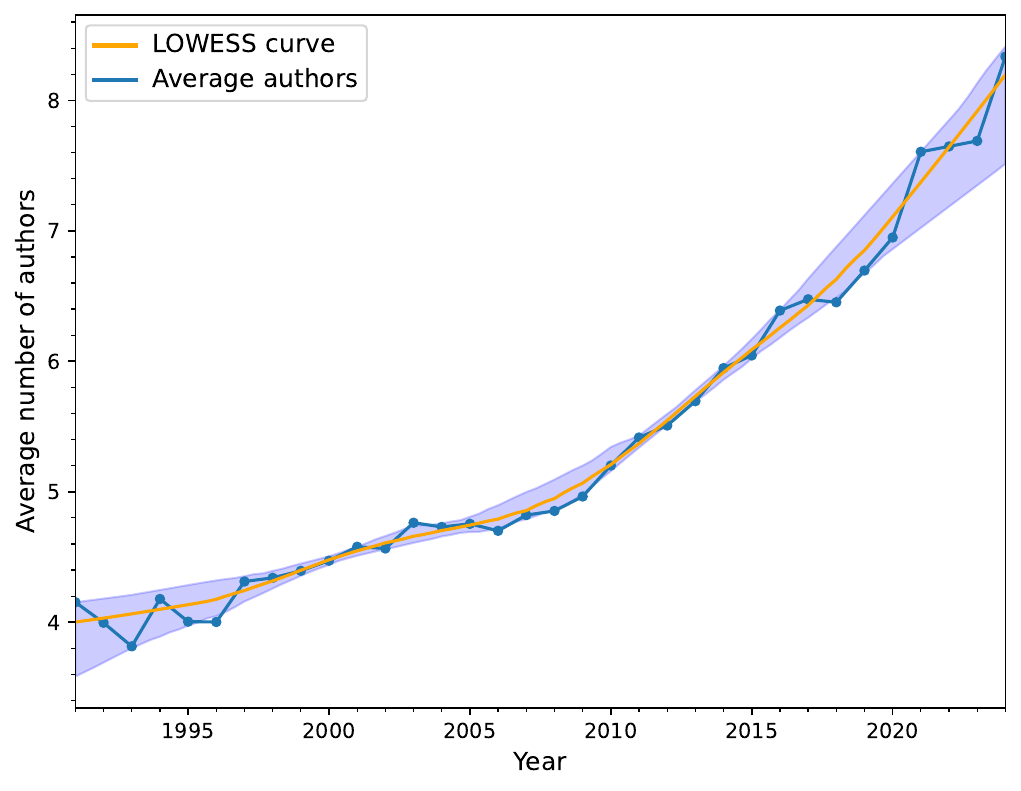}
		\caption{Average number of authors in publications from 1991 to 2024.}
		\label{fig:average_author_per_year}
	\end{subfigure}
	\hfill
	\begin{subfigure}[t]{0.48\textwidth}
		\centering
		\includegraphics[height=0.24\textheight]{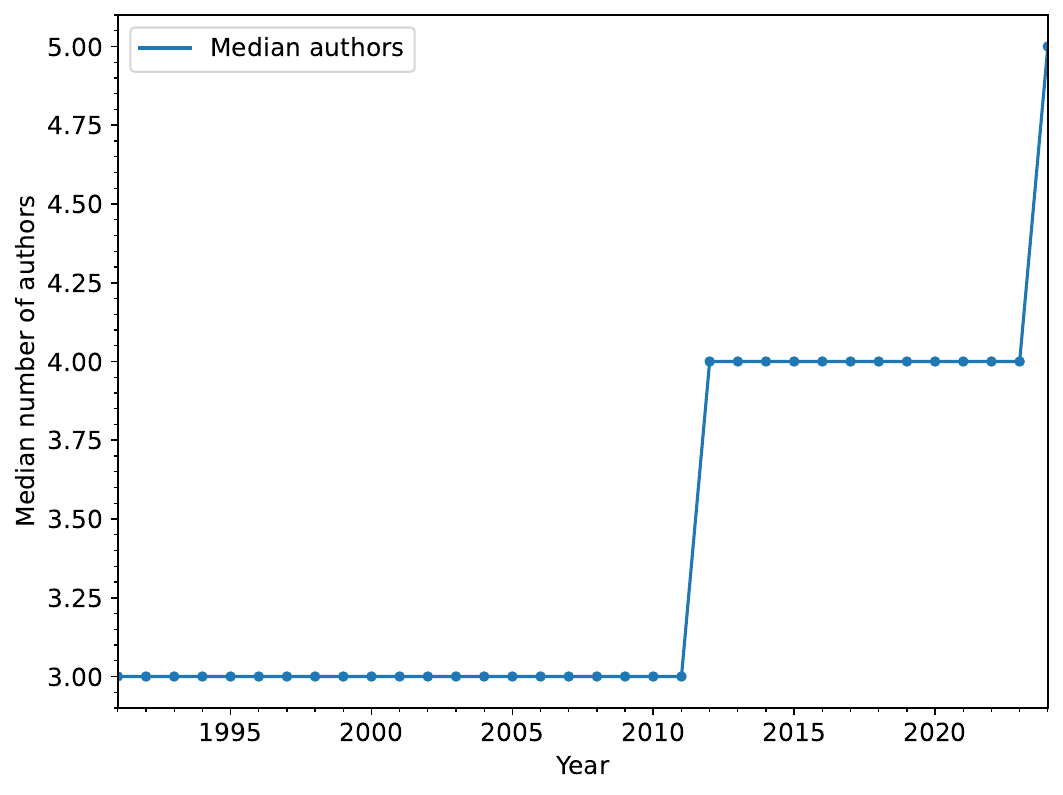}
		\caption{Median number of authors in publications from 1991 to 2024.}
		\label{fig:median_author_per_year}
	\end{subfigure}
	\hfill
	
	\begin{subfigure}[t]{0.48\textwidth}
		\centering
		\includegraphics[height=0.24\textheight]{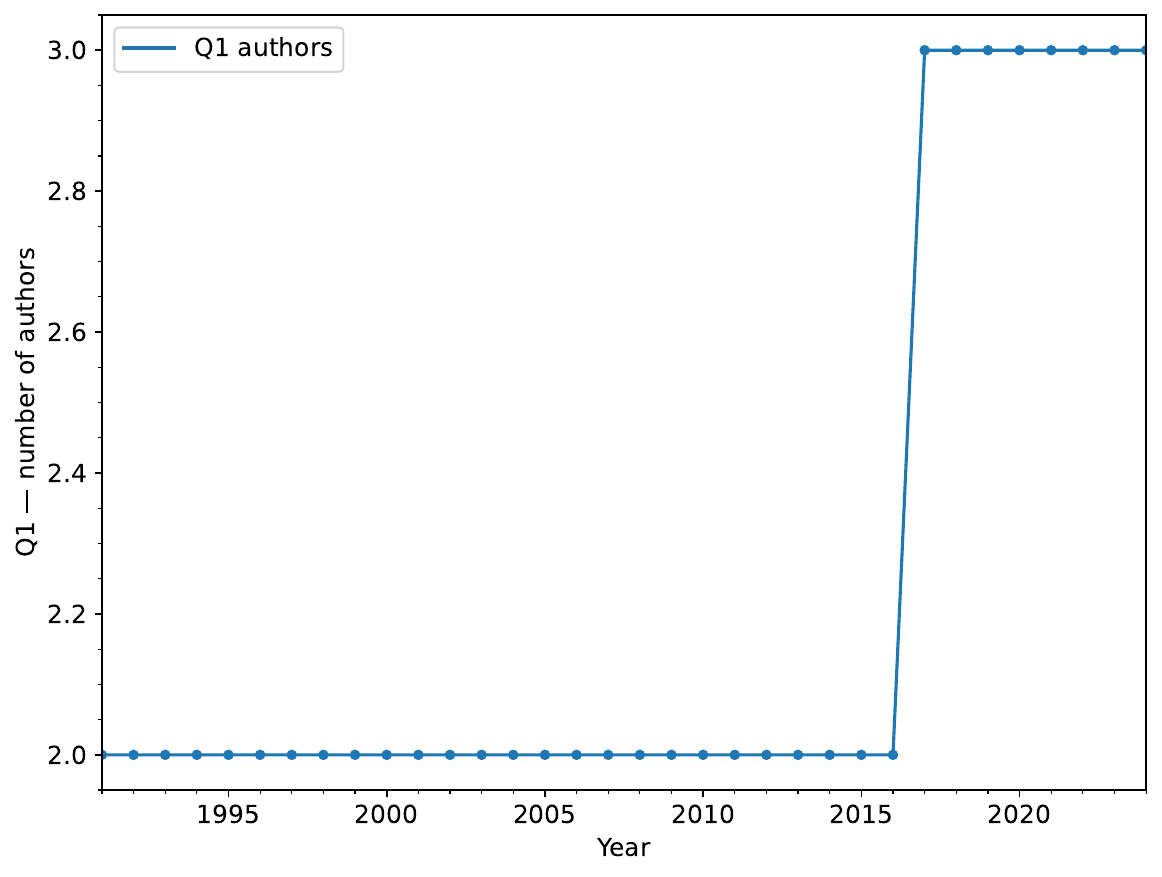}
		\caption{Q1 number of authors in publications from 1991 to 2024.}
		\label{fig:q1_author_per_year}
	\end{subfigure}
	\hfill
	\begin{subfigure}[t]{0.48\textwidth}
		\centering
		\includegraphics[height=0.24\textheight]{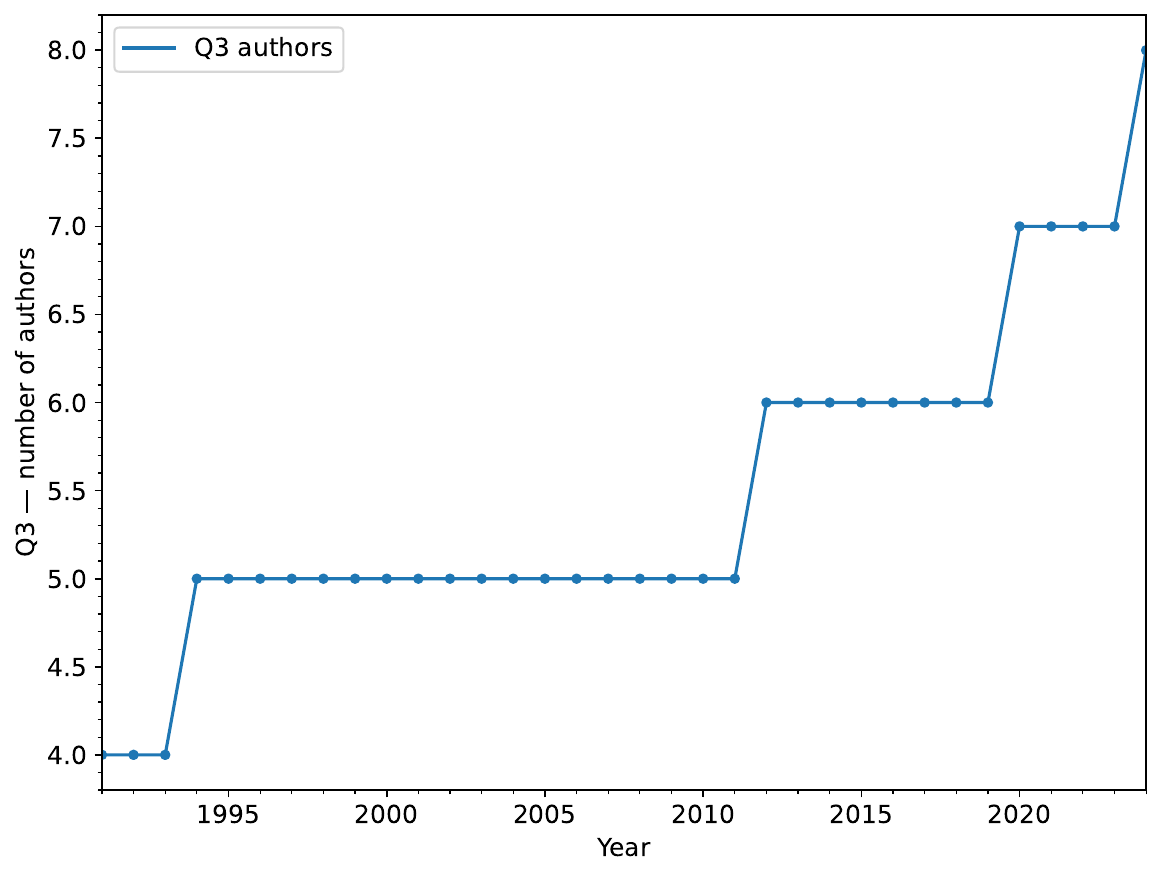}
		\caption{Q3 number of authors in publications from 1991 to 2024.}
		\label{fig:q3_author_per_year}
	\end{subfigure}
	
	\caption{Mean, median,  Q1 and Q2 statistics of the number of authors in publications from 1991 to 2024.}
	\label{fig:typical_number_of_authors}
\end{figure*}

In our comprehensive investigation into the trend of the typical number of authors in publications, we analyze a substantial 34-year period from 1991 to 2024. Figure~\ref{fig:typical_number_of_authors} shows the descriptive statistics of the number of authors per year, including the mean, median (Q2), first quartile (Q1) and third quartile (Q3) of the combined fields. The average number of authors per year has nearly doubled from four authors per publication to around eight as depicted in Figure~\ref{fig:average_author_per_year}. 

Figures~\ref{fig:median_author_per_year},~\ref{fig:q1_author_per_year},~and~\ref{fig:q3_author_per_year} demonstrate that the median (Q2), first quartile (Q1) and third quartile (Q3) statistics also show an upward trend over the years. The median number of authors has increased from three authors to five. Similarly, the Q3 statistic has risen from four to almost eight authors and the Q1 statistic has increased from two to three authors. Due to space limitations we present the mean, median, Q3 and Q1 statistics of the number of authors between 1991 and 2024 for our field-specific datasets in Appendices~\ref{avg_trend_field},~\ref{median_trend_field},~\ref{Q3_trend_field},~and~\ref{Q1_trend_field}, respectively.

\begin{figure}
	\includegraphics[width=0.48\textwidth,height=0.25\textheight]{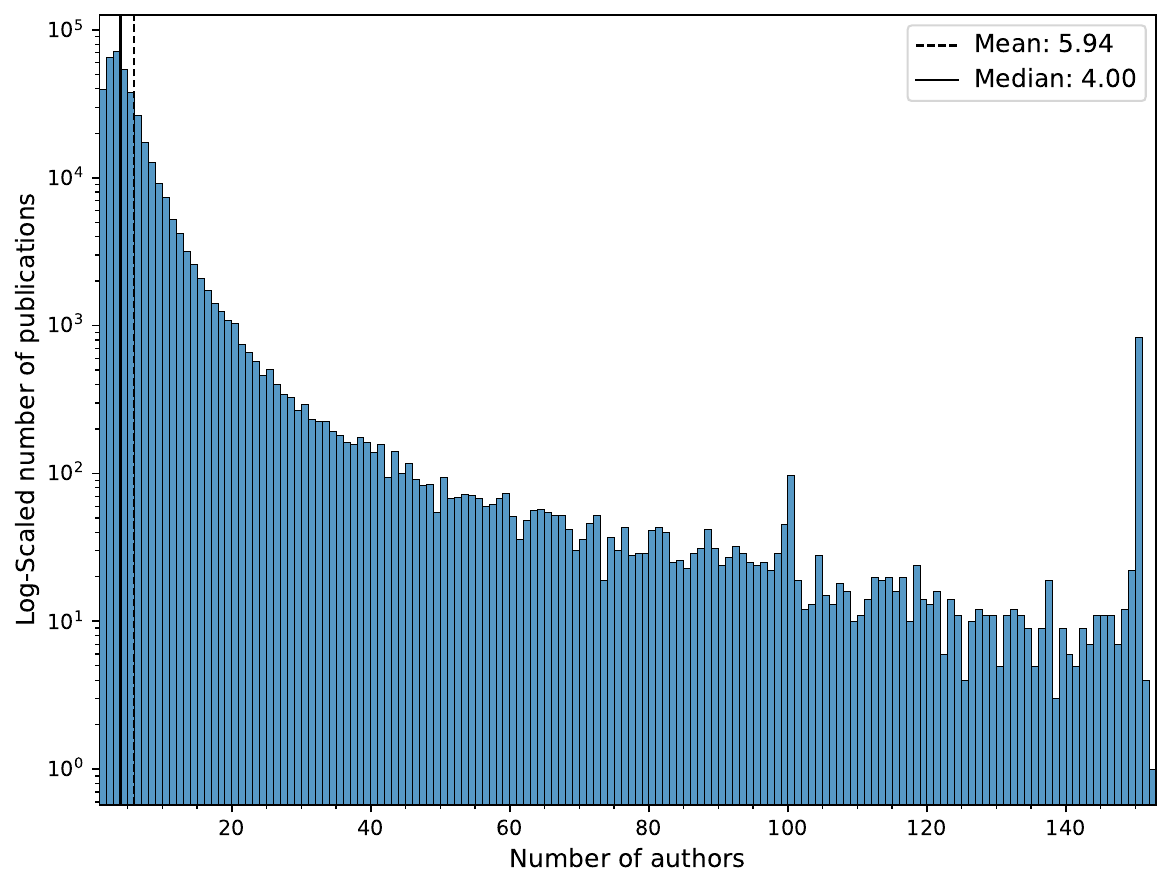}
	\caption{The distribution of the number of authors in publications from 1991 to 2024.}
	\label{fig:histogram_of_publications_vs_number_of_authors}
\end{figure}

\begin{figure}
	\includegraphics[width=0.48\textwidth,height=0.25\textheight]{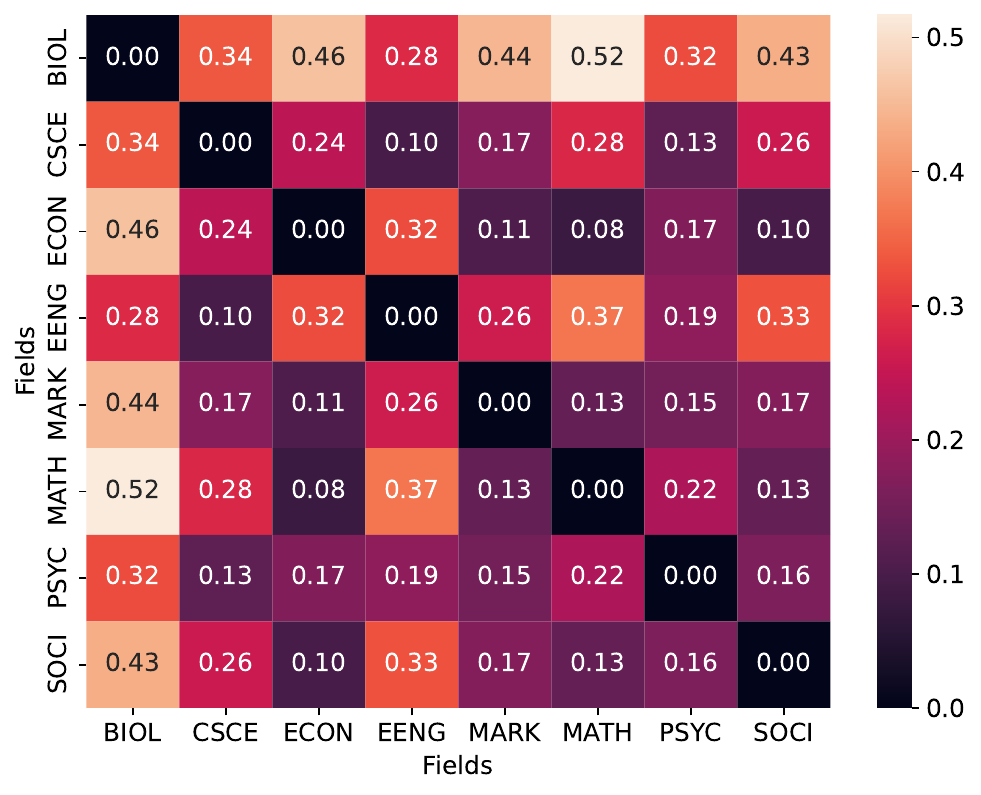}
	\caption{Hellinger distance matrix of the author count distributions of the field datasets.}
	\label{fig:hellinger_distance_matrix}
\end{figure}

Next, we investigate the authorship practices appearing in the combined and field-specific datasets.
Figure~\ref{fig:histogram_of_publications_vs_number_of_authors} shows the distribution of the number of authors in our combined author dataset. The distribution exhibits a right tail with a mean value of 5.94 and a median value of 4.0. 
We analyze the distributions of the number of authors further for each field dataset to uncover the differences in authorship trends. The Hellinger Distance matrix, presented in Figure~\ref{fig:hellinger_distance_matrix}, illustrates the distances between the author count distributions of the individual field datasets. Hellinger Distance measures how closely or distinctly two probability distributions compare \cite{Cieslak2008}. It takes on a value between 0 and 1, where a smaller value indicates greater similarity between the distributions.
Our findings show that the author count distributions exhibit a right tail for the field-specific datasets as well.
However the distribution of the authors in our Biology dataset has a heavier right tail compared to other fields, with a mean of 10.2 and  a median of 6 as shown in Appendix~\ref{dist_pub_each_field}. 
The Biology dataset also exhibits a higher dissimilarity compared to the other field datasets. 
The authors in the Electrical and Electronics Engineering, Computer Science and Engineering, and
Psychology datasets present similar authorship practices, compared to the rest. 
On the other hand, the Economics, Marketing, Sociology and Mathematics datasets exhibit comparatively lighter right tails with lower mean and median author counts. 
These four field datasets also cluster similarly in terms of their authorship practices.
Due to space limitations we show the distributions of the number of authors for each field dataset in Appendix~\ref{dist_pub_each_field}.

\begin{figure*}
	\centering
	\includegraphics[width=\linewidth, height=0.25\textheight]{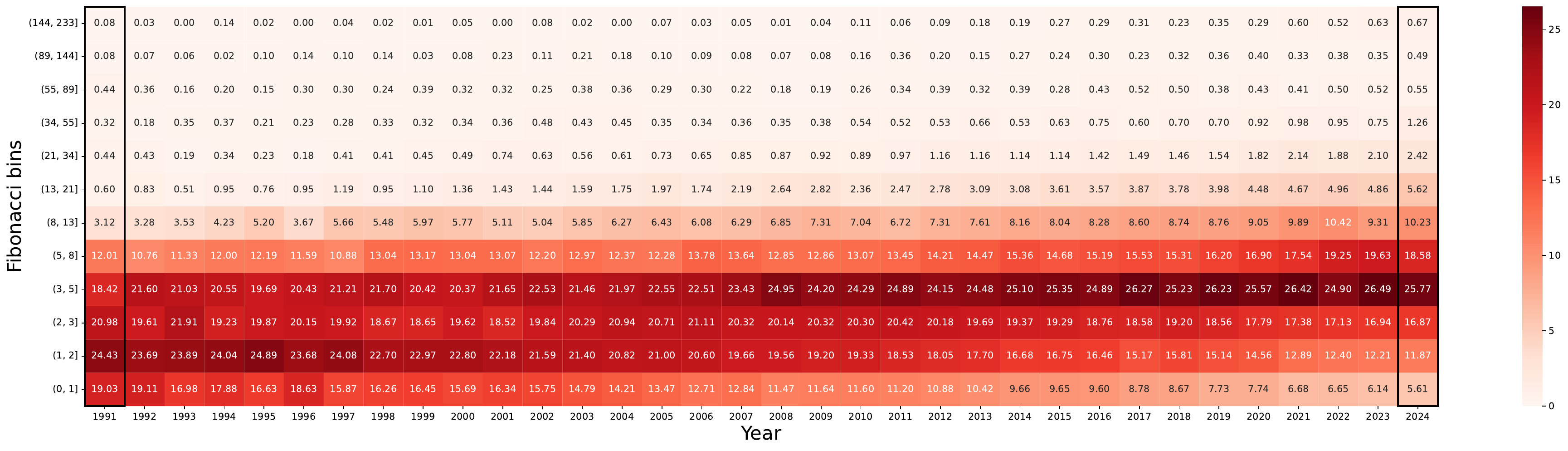}
	\caption{Percentages of the Fibonacci binned author counts for the publications between 1991 and 2024.}
	\label{fig:Diff_1991_2024}
\end{figure*}
{
	\renewcommand{\arraystretch}{1.5} 
	\begin{table*}
		\centering
		\caption{Percentages of the Fibonacci binned author counts for the publications in 1991 and 2024.}
		\label{tab_4:summaryDiff_1991_2024}
		\scalebox{1}{
			\begin{tabular}{|*{13}{c|}}
				\hline
				\textbf{Fibonacci Intervals} & \textbf{(0,1]} & \textbf{(1,2]} & \textbf{(2,3]} & \textbf{(3,5]} & \textbf{(5,8]} & \textbf{(8,13]} & \textbf{(13,21]} & \textbf{(21,34]} & \textbf{(34,55]} & \textbf{(55,89]} & \textbf{(89,144]} & \textbf{(144,233]}\\
				\hline
				\hline
				\textbf{1991} & 19.03 & 24.43 & 20.98 & 18.42 & 12.01 & 3.12 & 0.60 & 0.44 & 0.32 & 0.44 & 0.08 & 0.08\\
				\hline
				\textbf{2024} & 5.61 & 11.87 & 16.87 & 25.77 & 18.58 & 10.23 & 5.62 & 2.42 & 1.26 & 0.55 & 0.49 & 0.67\\
				\hline
			\end{tabular}
		}
	\end{table*}
}

Lastly, we investigate the change in authorship practices from 1991 to 2024.
Figure~\ref{fig:Diff_1991_2024} depicts the percentage of publications per year, grouping the number of authors into intervals of Fibonacci numbers up to 233, i.e., Fibonacci binning for the combined dataset. The figure demonstrates a noticeable trend of a decrease in the percentage of publications with one, two, and three authors and an increase in publications with four or more authors. Single-author publications began to decline gradually after approximately 1994. The percentage of single-author publications has declined from around 19.03\% in 1991 to 5.61\% in 2024. Two-author publications remained high until around 1997, after which there was a gradual decrease. Specifically, two-author publications decreased from around 24.43\% in 1991 to 11.87\% in 2024. Publications with three authors also show a decline, with the percentage of such publications decreasing from 20.98\% in 1991 to 16.87\% in 2024. On the other hand, publications with four and five authors exhibit an increasing trend, gradually rising since approximately 2000. It has increased from around 18.42\% in 1991 to 25.77\% in 2024. The percentage of publications with six to eight authors show an increase from 12.01\% in 1991 to 18.58\% in 2024. Similarly, publications with nine to thirteen authors also show an increasing trend approximately from 1997. Specifically, the percentage of such publications has increased from 3.12\% in 1991 to 10.23\% in 2024. Publications with 14 to 233 authors show an overall increase, with a notable increase in publications with 14 to 21 authors and publications with 145 to 233 authors. Notably, publications with 14 to 21 authors has increased from around 0.60\% to 5.62\%, while those with 145 to 233 authors has increased from around 0.08\% to 0.67\% between 1991 and 2024.

In Table~\ref{tab_4:summaryDiff_1991_2024}, we summarize Figure~\ref{fig:Diff_1991_2024} by highlighting the percentages of multi-author publications at the beginning, 1991, and in the end, 2024, of our study period using Fibonacci binning. Please note that Google Scholar typically displays around the first 150 authors. Hence, the increase in publications with 145 to 233 authors reflects the number of publications with 145 or more authors.
It is important to note that while specific figures may vary by discipline, we observe a consistent trend of an increasing number of authors per publication across all field datasets over the years.
The percentages of Fibonacci binned publications per year for the field-specific datasets are shown in Appendix~\ref{appendix:percent_pub_interval}.

\begin{figure*}
	\centering
	\begin{subfigure}[t]{0.48\textwidth}
		\centering
		\includegraphics[width=\linewidth,height=0.25\textheight]{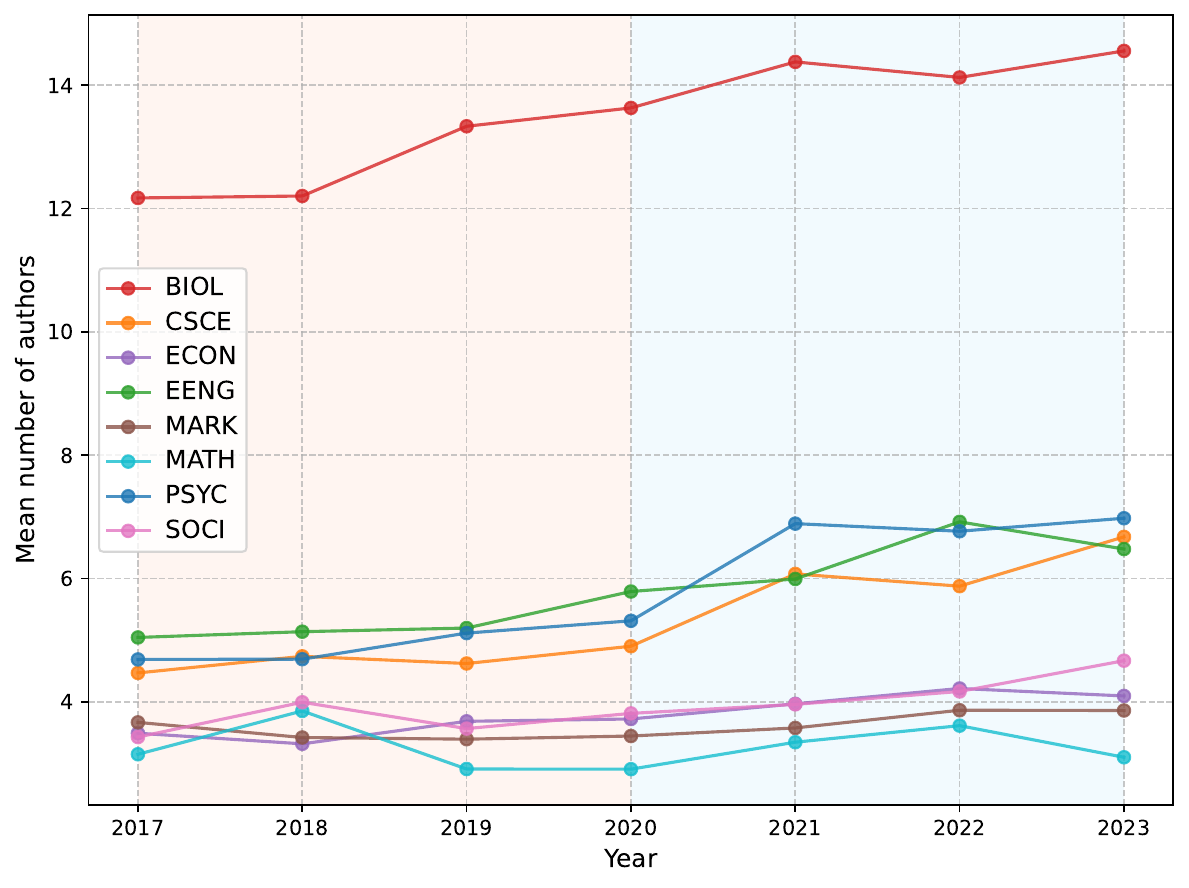}
		\caption{Mean number of authors per year in each field.}
		\label{fig:n_mean_author_field}
	\end{subfigure}
	\hfill
	\begin{subfigure}[t]{0.48\textwidth}
		\centering
		\includegraphics[width=\linewidth,height=0.25\textheight]{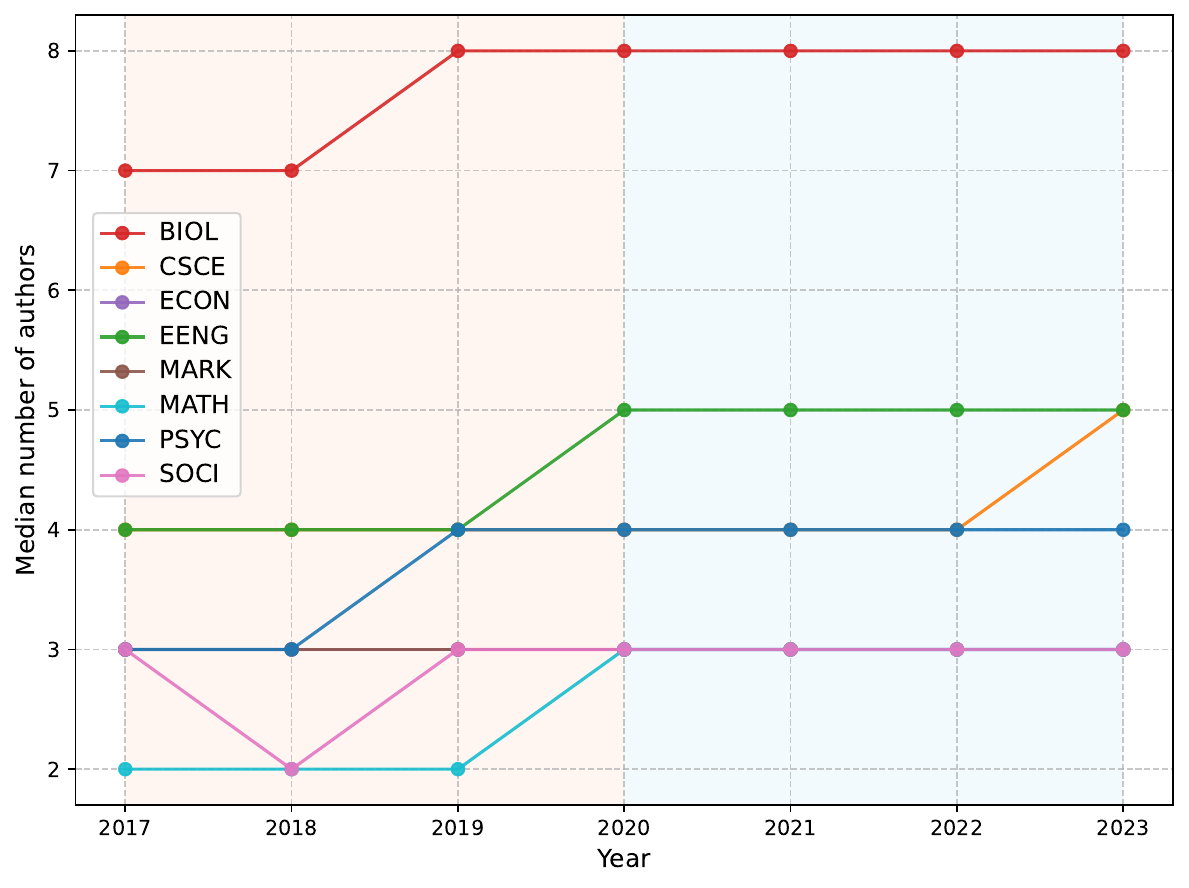}
		\caption{Median number of authors per year in each field.}
		\label{fig:n_median_author_field}
	\end{subfigure}
	\hfill
	
	\caption{Mean and median number of authors in publications before and after 2020 closures.}
	\label{fig:n_authors_2017_2023}
\end{figure*}

Proponents of long lists of authorships claim that modern science requires large collaborating groups by nature.
This is particularly true for experimental scientists who are involved in different aspects of a research project, including on-field data collection, experimental design, execution of the experiment at a laboratory, replication of the experiment across multiple laboratories,  collection of the experimental results, and their evaluations.
While refuting or proving these claims warrant their own studies, we visually looked into collaboration practices before and after the onset of COVID-19.
In 2020, a range of restrictive measures, including lockdowns, stay-at-home orders, movement restrictions, school and workplace closures, and remote work mandates, were implemented broadly.
Figure~\ref{fig:n_authors_2017_2023} illustrates the mean and median number of authors per publication, comparing three years prior to 2020 and three years following it. 
The figures show no significant changes in collaboration patterns before and after these closures.
Furthermore, the figures show that the trend in collaborations across the collected discipline datasets mostly continued to rise or remained stable despite the closures of universities, institutions, laboratories, and restrictions to field travels.
Therefore, the significant differences among the field datasets may also be attributed to \emph{authorship practices} along with (or rather than) \emph{collaboration practices}. 

\section{Fibonacci-adjusted Performance Indicators}
\label{sec:FibonacciIndicators}

Due to this increasing authorship trend, it has been reported that the traditional metrics such as the $h$-index no longer fully correlate with the scientific output of researchers \cite{Koltun2021}. A fractional index like the $h$-frac, which normalizes citation counts, provides a better correlation with researchers' scientific output \cite{Koltun2021,egghe2008mathematical}. However, simply normalizing citation counts by the number of authors does not adequately credit the authors who made significant contributions compared to the ones who made less and to the others in between. Our proposed method adjusts or counts publications and citations based on author rankings in the byline. It gives whole credits to the leading contributors and adjusts credits for the supporting contributors without placing any limit on the number of authors. Our approach encourages collaborations, while discouraging the manipulation of bibliometric indexes through multi-authorship abuse.

In the following, we introduce alternative indexes adjusted by the reciprocals of the Fibonacci numbers to evaluate scientific performance based on contribution rather than participation. Specifically, we present alternative indexes for counting the amount of publications, the amount of citations, and $h$-index.
In addition, we present the $L^\prime$-index to quantify a researcher's labor contribution across their publications.

\subsection{Total Publications by Fibonacci-adjusted Contributions}
Counting publications by participation is the traditional technique to quantify scientific productivity. Counting by participation gives a whole credit to each author of a publication without considering their appropriate contributions.
In this study, we propose a fractional counting technique based on the reciprocals of Fibonacci numbers referred as \emph{total publications by Fibonacci-adjusted contributions}, $P^{\prime}$-index.
Ranking authors in the byline of a publication with respect to their contributions has been steadily growing in academia and widely adopted in industry.
Coherently, $P^{\prime}$-index adjusts author credits according to their ranks in the byline of their publications. 
Equation~\ref{eq_authorship_credit} presents the $P^{\prime}$-index for an author.

\begin{equation}
	\label{eq_authorship_credit}
	P'= \sum_{k=1}^{P} \frac{1}{F_{R(k)}}
\end{equation}

\noindent where $P^{\prime}$ is the author's total publications by contribution; $P$ is their number of participated publications; $R(k)$ is the author's rank (position) in the $k$-th publication; and $F_{\boldsymbol{.}}$ is the Fibonacci number corresponding to the author's rank in the $k$-th publication.
$P^{\prime}$-index adds up the Fibonacci-adjusted credits of an author, whereas $P$-index adds up whole credits for participation.

\begin{figure}
	\includegraphics[width=0.48\textwidth,height=0.25\textheight]{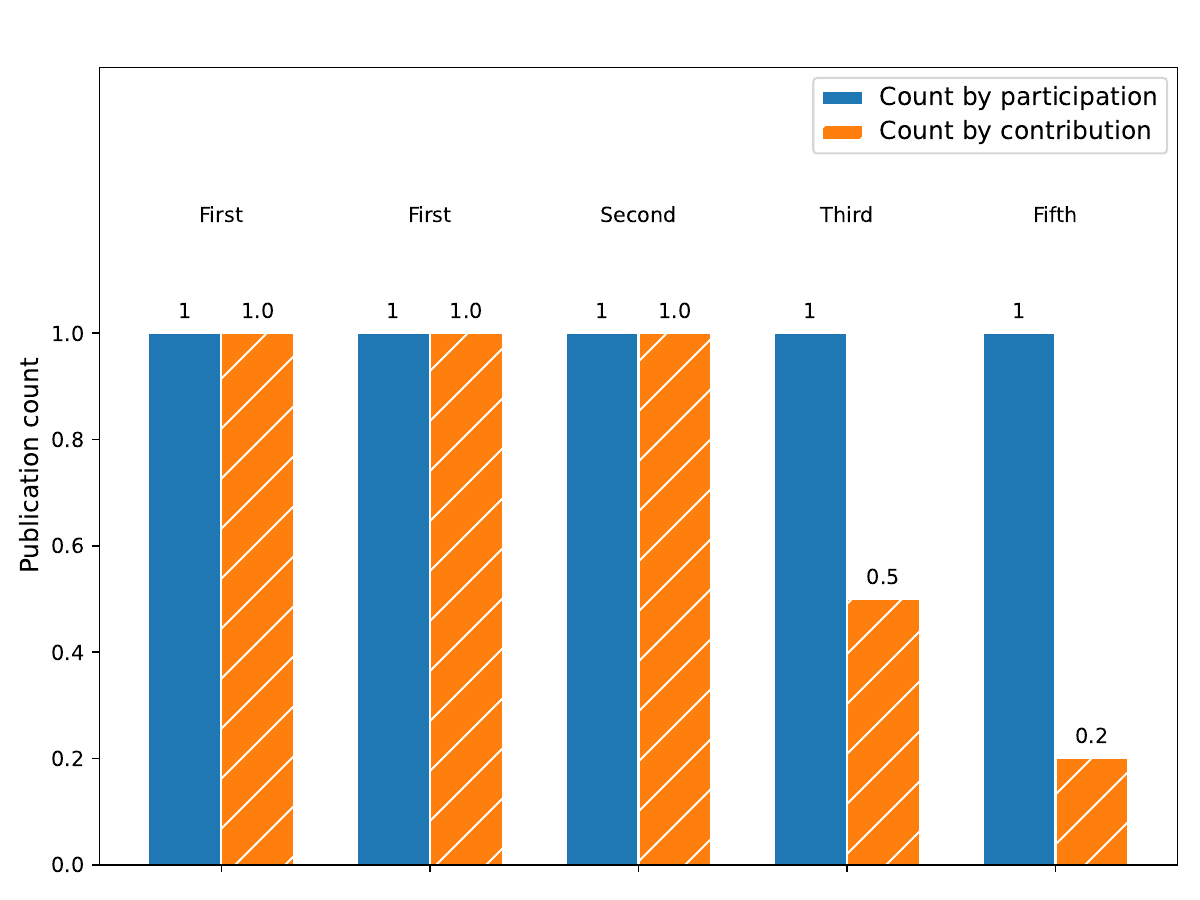}
	\caption{Total publications by participation and by contribution for the fictional author with credentials in Table~\ref{tab_5:example_author_bib}.}
	\label{fig:publication_count_by_contribution}
\end{figure}

\begin{table}
	\centering
	\caption{ Example credentials of a fictional author John Doe with five publications.}
	\label{tab_5:example_author_bib}
	\scalebox{1}{
		\begin{tabular}{|*{11}{c|}}
			\hline
			\textbf{Paper} & \textbf{\#1} & \textbf{\#2} & \textbf{\#3} & \textbf{\#4} & \textbf{\#5}\\
			\hline
			\hline
			\textbf{Author's Position (Rank)} & $1^{st}$ & $1^{st}$ & $2^{nd}$ & $3^{rd}$ & $5^{th}$\\
			\hline
			\textbf{Citation Count}        & 1 & 1 & 4 & 9 & 25\\
			\hline
		\end{tabular}
	}
\end{table}

Figure~\ref{fig:publication_count_by_contribution} demonstrates total publications by participation and by contributions for an example author, John Doe, with five papers.
Table~\ref{tab_5:example_author_bib} lists the papers of the fictional author along with his position in each paper and the number of citations each paper received.
The total publications by participation is 5.0 because the author receives a full credit regardless of the amount of his contributions in each paper, as shown in Figure~\ref{fig:publication_count_by_contribution}.
In the same figure, the author receives full credits for his Fibonacci-adjusted contributions based on his leading roles in the first three papers.
He receives half of a full credit for his third position in the fourth paper and one-fifth of a whole credit for his fifth position in the fifth paper.
As a result, the example author's total publications by adjusted contributions is $3.7$, compared to total publications by participation, $5.0$.

\subsection{Fibonacci-adjusted Labor Index}

The total publications by Fibonacci-adjusted contributions quantify the cumulative contributed labor of a researcher. However, cumulative measures do not indicate how much labor a scientist contributes when participating in publications. Therefore, we define the Fibonacci-adjusted labor index, denoted by $L^{\prime}$, as the contributed labor normalized by participation.

\begin{equation}
	\label{eq_Lprime_index}
	\begin{split}
		L^{\prime} &=\frac{P'}{P} \\
		&= \frac{1}{P} \sum_{k=1}^{P} \frac{1}{F_{R(k)}}
	\end{split}
\end{equation}

\noindent where $P^{\prime}$ denotes the total publications by Fibonacci-adjusted contributions and $P$ denotes the total publications by participation; $R(k)$ is the author's rank (position) in the $k$-th publication; and $F_{\boldsymbol{.}}$ is the Fibonacci number corresponding to the author's rank in the $k$-th publication.

The Fibonacci-adjusted labor index, $L^{\prime} $ introduced in Equation~\ref{eq_Lprime_index}, takes values in the interval of $(0,1]$. Larger values indicate that an author contributes more labor when participating in publications. Smaller values indicate that an author predominantly assumes supporting roles in their publications. Thus, $L^{\prime}$ measures the amount of Fibonacci-adjusted contributed labor in collaborations.

Equation~\ref{eq_Lprime_index} is equivalent to an author's expected Fibonacci-adjusted credit.
To derive reference values for the $L^{\prime}$-index, we examine the intrinsic center of gravity of the Fibonacci credit scheme itself.
Let $\mathcal{F} = \{\frac{1}{F_{1}}, \frac{1}{F_{2}}, \frac{1}{F_{3}}, \frac{1}{F_{4}}, \frac{1}{F_{5}}, \frac{1}{F_{6}}, \ldots\}$ be the well-ordered set of all possible Fibonacci-adjusted credit allocations where $\mathcal{F}_{i}=\frac{1}{F_{i}}$.
Let $\mathbf{X}$ be a random variable on domain $\mathcal{F}$ denoting the probability of the received credit on a paper such that
\begin{equation}
	\label{eq_prob_credit}
	Pr(\mathbf{X} = \mathcal{F}_i) = \frac{\mathcal{F}_i}{\psi}
\end{equation}

\noindent where $\mathcal{F}_i$ is the $i^{th}$ element in $\mathcal{F}$, and $\psi \approx 3.35$ is the sum of reciprocals of the Fibonacci series presented in Equation~\ref{eq_sum_fib_constant}.
The expected value for the random variable $\mathbf{X}$ is presented in Equation~\ref{eq_prob_credit}.
\begin{equation}
	\label{eq_expected_credit}
	E[\mathbf{X}] = \sum_{i=1}^{\infty} \frac{\mathcal{F}_i}{\psi } \mathcal{F}_i = \frac{1}{\psi} \sum_{i=1}^{\infty} \mathcal{F}_i^2
\end{equation}

\noindent where the summation term is convergent, as it is bounded above by the famous Basel problem plus one.
However, we were unable to find a closed-form solution for the summation term.
Computationally, the first 10,946 terms approximately sum up to $\approx 2.426 \leq (\pi^2/6)+1$.
Therefore, the expected value in Equation~\ref{eq_expected_credit} is approximately $0.72$, which may serve as a threshold value.

A more beautiful but less sound threshold value is derived from Binet's formula. The reciprocals of the Fibonacci numbers do not decay by a fixed amount, i.e., linearly, instead it approximately decays by a multiplicative factor, i.e., geometrically.  Let $\mathcal{F} = \{\frac{1}{F_{1}}, \frac{1}{F_{2}}, \frac{1}{F_{3}}, \frac{1}{F_{4}}, \frac{1}{F_{5}}, \frac{1}{F_{6}}, \ldots\}$ be the well-ordered set of all possible Fibonacci-adjusted credit allocations where $\mathcal{F}_{i}=\frac{1}{F_{i}}$. 
Equation

\begin{equation}
	\label{eq_binet_recip_fibonacci}
	\mathcal{F}_{i}=\frac{\sqrt{5}}{\varphi^{i}-\vartheta^{i}}
\end{equation}

\noindent where $\varphi=\frac{1+\sqrt5}{2}$ and $\vartheta=\frac{1-\sqrt5}{2}$.
Since $|\vartheta^{i}| < 1$ for very large $i$, $\mathcal{F}_{i}$ can be simplified to
\begin{equation}
	\label{eq_binet_recip_fibonacci_simplified}
	\mathcal{F}_{i}=\frac{\sqrt{5}}{\varphi^{i}}
\end{equation}
The asymptotic behavior of successive elements of $\mathcal{F}$, i.e., $\mathcal{F}_{i} / \mathcal{F}_{i-1}$ is shown in Equation~\ref{eq_binet_recip_fibonacci_limit}.
\begin{equation}
	\label{eq_binet_recip_fibonacci_limit}
	\lim_{i\to\infty} \frac{\mathcal{F}_{i}}{\mathcal{F}_{i-1}} = \varphi^{-1}
\end{equation}
\noindent where $\varphi^{-1} \approx 0.61$. That is, the reciprocals of Fibonacci space asymptotically decays by a constant factor of approximately $0.61$.
Hence, we adopt $0.61$ as the geometric threshold of the Fibonacci-adjusted labor index, $L^{\prime}$.
Note that $\varphi^{-1}$ is the reciprocal of the famous golden ratio.

We refer to $0.72$ as the \emph{upper benchmark} and $0.61$ as the \emph{lower benchmark} 
of the $L^{\prime}$-index. This lower benchmark is quite generous, allowing a researcher 
who publishes in a leading (first or second) position in every three publications to also serve as the 
third coauthor for up to $10.63$ publications or the infinity-th coauthor for up to 
$1.91$ publications without falling below the $0.61$ threshold. However, $0.72$ is a more viable 
benchmark value for institutions that invest in and expect higher research productivity. 
A benchmark value of $0.72$ allows a researcher who takes a leading role 
in every three publications to also serve as the third coauthor for up to $3.81$ 
publications or the infinity-th coauthor for up to $1.16$ publications without falling 
below $0.72$. We present the detailed breakdown at each byline position in 
Appendix~\ref{t_index_benchmark_table_appendix}.

The $L^{\prime}$-index is independent of variations in publication practices across different disciplines, allowing it to be applied to any field with a consistent benchmark value.
Besides, maintaining an $L^{\prime}$-index of $0.61$ or $0.72$ is quite reasonable, especially in the first twenty years of a scientific career.
In fact, the first twenty years of a scientific career often involve all significant events such as obtaining a doctoral degree, securing a job, achieving tenure and/or promotion, receiving salary increases, and taking on editorial or administrative roles.
Many of these milestones are, to some extent, evaluated based on scientific performance, which is prone to multi-authorship abuse via traditional indicators.
Saying that, looking into only the $L^{\prime}$-index is insufficient to assess a researcher's overall productivity like any other descriptive statistic.
It is essential to consider additional statistics, such as $P^{\prime}$-index, $C^{\prime}$-index, and $h^{\prime}$-index, to obtain an overall picture of scientific performance.
Our fictional author, John Doe, with credentials presented in Table~\ref{tab_5:example_author_bib} has an $L^{\prime}$-index of $0.74$, which is above $0.72$.

\begin{figure}
	\centering
	\includegraphics[width=0.48\textwidth,height=0.25\textheight]{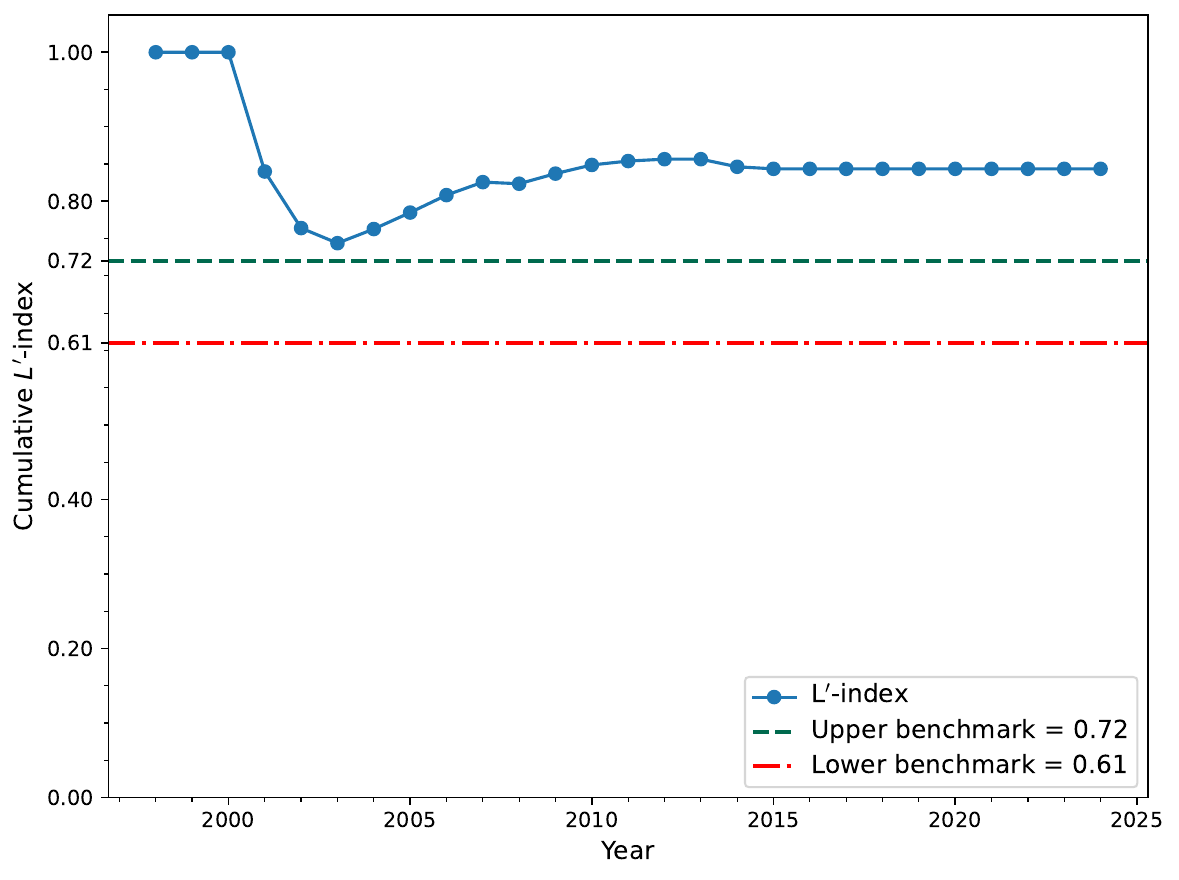}
	\caption{A prominent scientist's cumulative $L^\prime$-index over the years.}
	\label{fig:t_prime_a_prominent_scientist}
\end{figure}

Lastly,  we present the Google Scholar-derived cumulative $L^{\prime}$-index of a prominent scientist, who is also known to frequently collaborate with researchers in diverse fields in Figure~\ref{fig:t_prime_a_prominent_scientist}.
The scientist's $L^{\prime}$-index starts at $1.0$, followed by a quick decline, and then stabilizing around $0.84$.

\subsection{Total Citations by Fibonacci-adjusted Contributions}
In the traditional citation counting all participating authors get the same amount of citations per publication. We propose \emph{total citations by Fibonacci-adjusted contributions},  $C^{\prime}$-index, where each author is credited according to their position in the byline of a publication as shown in Equation~\ref{eq_citation_credit}.
\begin{equation}
	\label{eq_citation_credit}
	C^{\prime} = \sum_{k=1}^{P} \frac{C_k}{F_{R(k)}}
\end{equation}

\noindent where $C^{\prime}$ is an author's total citations by adjusted contributions; $P$ is their number of participated publications; $C_{k}$ is the total citations of the $k$-th publication; $R(k)$ is the author's rank (position) in the $k$-th publication; and $F_{\boldsymbol{.}}$ is the Fibonacci number corresponding to the author's rank in the $k$-th publication.

\begin{figure}
	\includegraphics[width=0.48\textwidth,height=0.25\textheight]{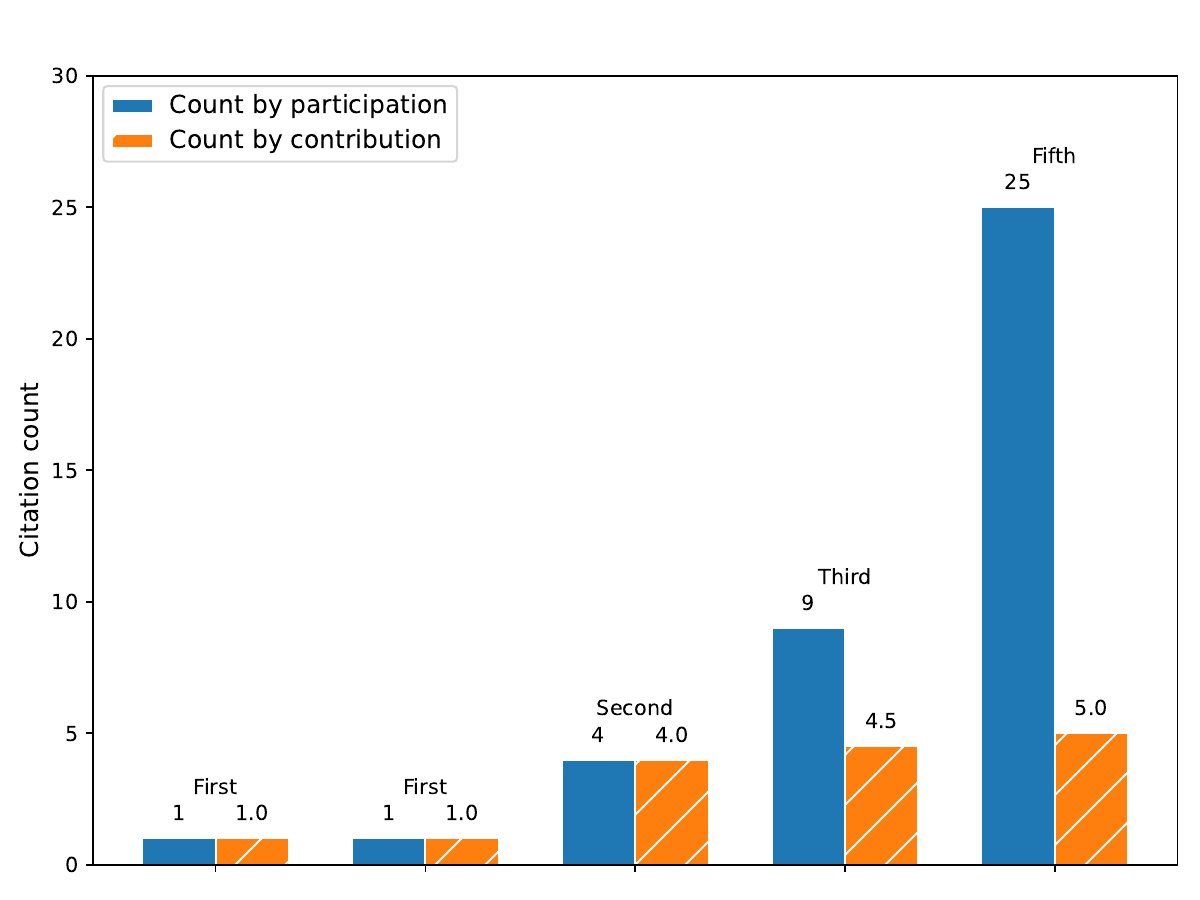}
	\caption{Total citations by participation and by contribution for the fictional author with credentials in Table~\ref{tab_5:example_author_bib}.}
	\label{fig:citation_count_by_contribution}
\end{figure}

Figure~\ref{fig:citation_count_by_contribution} demonstrates total citations by participation and by contributions for the example author, John Doe, with five papers.
Table~\ref{tab_5:example_author_bib} lists the papers of the fictional author along with his position in each paper and the number of citations each paper received.
Total citations by participation is $40.0$ because the author receives full citation credit per paper irrespective of his actual contributions, as shown in Figure~\ref{fig:citation_count_by_contribution}.
In fact more than half of John Doe's citations come from a paper in which he was the fifth author.
In the same figure, the author receives full citation credits for his contributions based on his leading roles in the first three papers.
However, he receives half of the full citation credit of $9$ for his third position in the fourth paper, i.e., $4.5$ citation credits.
Put in other words, the total citations of this publication is adjusted by $\sfrac{1}{F_3}=\sfrac{1}{2}$, which corresponds to the reciprocal of the Fibonacci number reflecting the author's position in the byline.
Similarly, he receives one-fifth of the full citation credit of $25$ for his fifth position in the fifth paper, i.e., $5$ citation credits.
As a result, the example author's total citations by adjusted contributions is $15.5$, compared to total citations by participation, $40.0$.

\subsection{Fibonacci-adjusted $h$-index}
The $h$-index of a scientist is defined as having $h$ publications which has at least $h$ citations each.
It is computed over full citation and publication credits for each paper that a scientist participates in.
Since $h$-index does not reflect authors' individual contributions in each paper, Google Scholar is full of authors with ludicrous $h$-indexes.
We propose, $h^{\prime}$-index, a \emph{Fibonacci-adjusted $h$-index} based on an author's Fibonacci-adjusted publication and citation credits.
Analogous to $h$-index, the $h^{\prime}$-index of a researcher is defined as having $h^{\prime}$ amount of publications by contribution which has at least $h^{\prime}$ amount of citations by contribution.

\begin{figure*}
	\centering
	\begin{subfigure}[t]{0.48\textwidth}
		\centering
		\includegraphics[width=\linewidth,height=0.25\textheight]{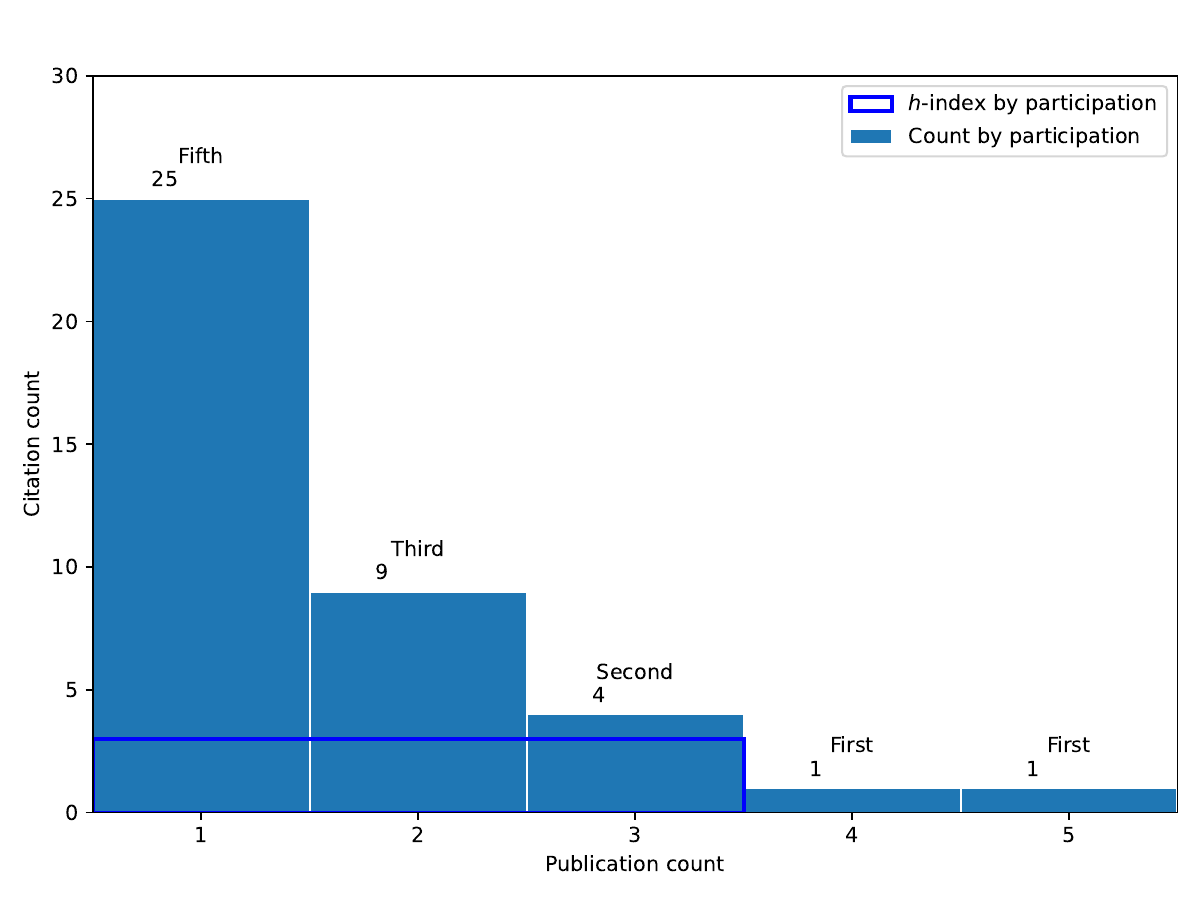}
		\caption{$h$-index by participation.}
		\label{fig:h-index_count}
	\end{subfigure}
	\hfill
	\begin{subfigure}[t]{0.48\textwidth}
		\centering
		\includegraphics[width=\linewidth,height=0.25\textheight]{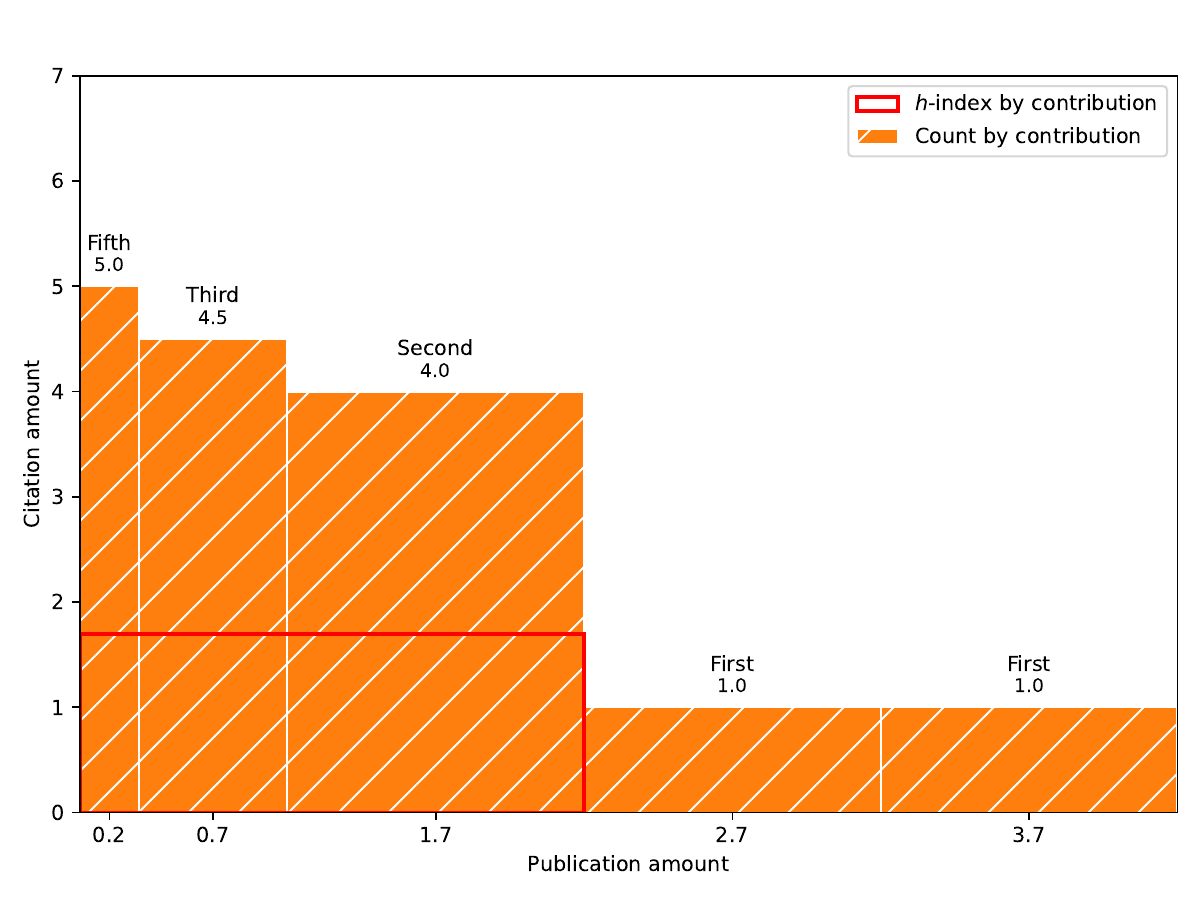}
		\caption{$h^{\prime}$-index by contributions.}
		\label{fig:h-index_count_by_contribution}
	\end{subfigure}
	\hfill
	
	\caption{$h$-index by participation and $h^{\prime}$-index by contributions for the fictional author with credentials in Table~\ref{tab_5:example_author_bib}.}
	\label{fig:h_index_by_contribution}
\end{figure*}

Figure~\ref{fig:h_index_by_contribution} demonstrates $h$-index by participation and $h^{\prime}$-index by contribution for the example author, John Doe, with five papers.
Table~\ref{tab_5:example_author_bib} lists the papers of the fictional author along with his position in each paper and the number of citations each paper received.
The traditional $h$-index is calculated by sorting the publications with respect to their citation counts and iterating through the sorted publications until the total count of the iterated publications is less than the citations of the next publication as shown in Figure~\ref{fig:h-index_count}.
The blue rectangle in the figure indicates that the threshold value is achieved at $3.0$.
That is the author has 3 publications with at  least 3 citations.
An analogous algorithm for $h^{\prime}$-index sorts the publications with respect to their Fibonacci-adjusted citation amounts and iterates through the sorted publications until the total Fibonacci-adjusted amount of the iterated publications is less than the Fibonacci-adjusted citation amount of the next publication as shown in Figure~\ref{fig:h-index_count_by_contribution}.
The red rectangle in the figure indicates that the threshold value is achieved at $1.7$.
That is, the author has 1.7 publications with at  least 1.7 citations by Fibonacci-adjusted contributions.
Note that while the traditional $h$-index is always an integer, the Fibonacci-adjusted $h^{\prime}$-index is a real value.

\section{Empirical Evaluations}
\label{section:empirical_evaluations}
In this section, we present the empirical analysis of the proposed Fibonacci-adjusted performance indicators. 
While our dataset may not reflect the extensive publication and authorship practices in each field, our focus in this study is \emph{not} on analyzing such practices.
Instead, we employ the author dataset to demonstrate the effectiveness of the Fibonacci-adjusted performance metrics against multi-authorship abuse. 
A thorough investigation of publication norms in different academic fields requires further examintaions.
In the following we analyze publication counts, citation counts, and $h$-indexes by participation and by contribution as well as the $L^{\prime}$-indexes.

In this section, we use percentage difference to compare the traditional indicators against the Fibonacci-adjusted indicators. Percentage difference compares the absolute difference of two indicator values over their means instead of a starting point, as shown in Equation~\ref{eq:percentageDifference}.

\begin{equation}
	\label{eq:percentageDifference}
	diff(x_{i}, x_{j})=2\frac{|x_{i}-x_{j}|}{x_{i}+x_{j}}100
\end{equation}

\subsection{Publications by Contribution}
\begin{figure}
	\includegraphics[width=0.48\textwidth,height=0.25\textheight]{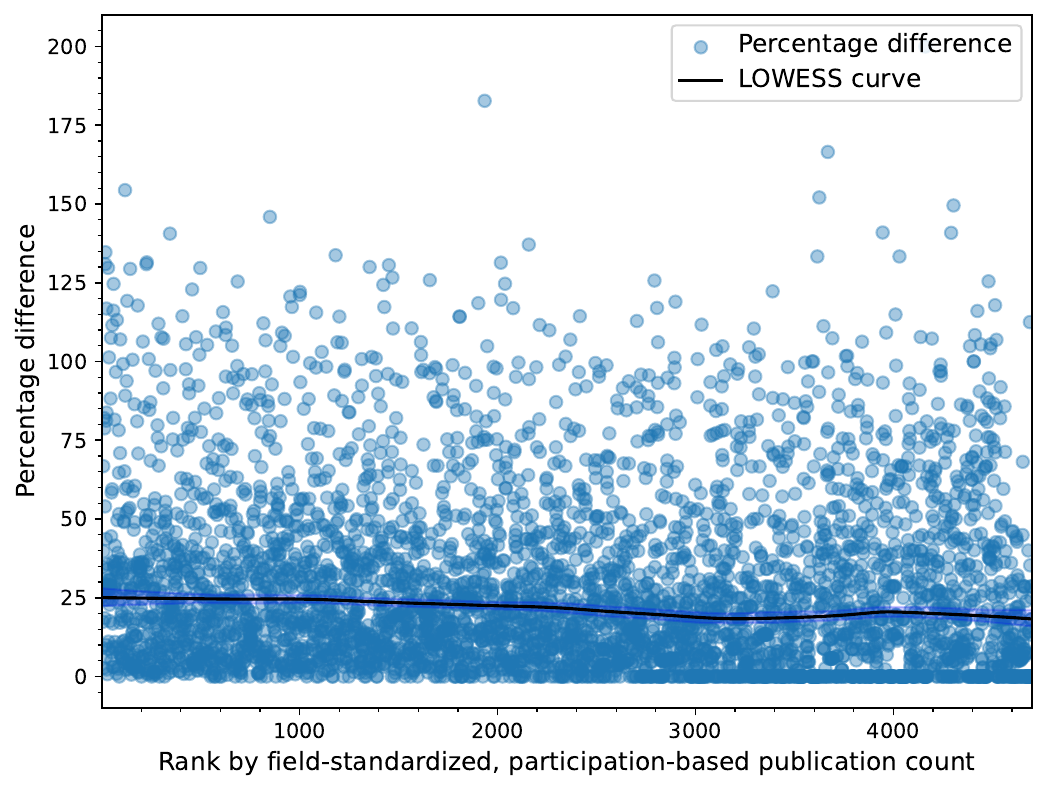}
	\caption{Percentage difference between publications by participation and by contribution ranked by participation.}
	\label{fig:percentage_difference_pub_count_participation_rank.}
\end{figure}

Figure~\ref{fig:percentage_difference_pub_count_participation_rank.} illustrates the percentage differences between publication counts based on participation and Fibonacci-adjusted contributions, with author rankings determined by field-standardized, participation-based publication counts in descending order. The typical percentage difference remains relatively stable at around 25\% across all ranks, suggesting that the majority of authors in our combined dataset experience a percentage difference of approximately 25\%. However, the wide spread of percentage differences above 25\% reflects the presence of authors whose publication counts are significantly affected by the proposed adjustment for contributions. This indicates that some authors contribute less to their publications despite having high participation counts.

In our study, we thoroughly examined the author datasets of all collected fields. Authors in the Biology dataset have the highest typical percentage difference of around 74.26\% in publication counts by participation and by contribution among other fields. The typical percentage difference for authors in Electrical and Electronics Engineering is 36.26\%, in Computer Science and Engineering is 29.14\% and in Psychology is 23.89\%. While these fields often require collaborative research, the trend in author hyperinflation is undeniable. These fields also generally receive more funding for research, hence set a competitive environment open to various types of abuse. Authors in the remaining fields show a typical percentage difference of less than 20\%. Sociology has 15.60\%, Economics has 14.96\%, Marketing has 14.48\%, and Mathematics has 12.74\%. The low typical percentage differences among the authors of these fields does not rule out the possibility of authorship abuse. In fact, we observe authors with high percentage differences, up to about 130\% in their publications by contributions and by participation. The comparatively low average number of authors in publications and lower percentage differences in these field datasets suggest less magnified collaborations and may also indicate stricter norms or fewer opportunities for unethical authorship practices. The percentage differences between publications by participation and by contribution for the collected authors of individual fields are shown in Appendix~\ref{percentage_difference_pub_appendix}, which propels the discussions below along with Appendix~\ref{appendix:percent_pub_interval} .

Specifically, in the Biology dataset, the average number of authors increases from around five to sixteen in the last 34 years. Publications with one to five authors have significantly decreased. Notably, single-author publications have decreased from around 7.8\% to 0.93\%, two-author publications have decreased from around 15\% to 3.24\%,  and three-author publications have decreased from  20.90\% to 4.22\%. Publications with four and five authors have decreased from around 26.5\% to 12.71\%. Publications with six to eight authors remain relatively stable, around 22\% each year. On the other hand, publications with nine to twenty-one authors have exhibited a noteworthy increase in recent years. Specifically, publications with nine to thirteen authors have increased from approximately 5.60\% to 23.65\%. Similarly, publications with 14 to 21 authors have increased from around 0.89\% to 16.76\%. Publications with more than 21 authors also show increases in recent years. Correspondingly, the typical percentage of difference in publication counts by participation and by contribution is the highest in the Biology dataset among all fields in our study. It starts around 80\% for higher-ranked authors and gradually decreases to around 60\% for the lower-ranked authors.

In the Electrical and Electronics Engineering dataset, the average number of authors per publication has increased from around three to seven. Publications with one to three authors show a steep decline. Single-author publications have decreased from around 11.69\% to 1.81\%. Likewise, two-author publications have decreased from around 30.27\% to 6.73\%. Additionally, three-author publications have decreased from around 27.29\% to 15.49\%. In contrast, publications with more than four authors have increased in the past decade. Specifically, publications with four and five authors have increased from around 20.18\% to 35\%. Publications with six to eight authors have increased from 8.25\% to 27.70\%, while those with nine to thirteen authors have increased from around 1.83\% to 8.79\%. Publications with more than 13 authors have increased from around 0.5\% to 5\%. In consequence, the typical percentage difference between publication counts by participation and by contribution in Electrical and Electronics Engineering dataset starts around 45\% for the higher-ranked authors and decreases to around 10\% for the lower-ranked authors. The sharp decline demonstrates that the percentage difference is more prominent among higher-prolific authors rather than lower-prolific.

In the Computer Science and Engineering dataset, the average number of authors per year has increased from around four to six between 1991 and 2024. Publications with one and two authors show significant declines. Single-author publications have decreased from approximately 18\% to 3\%, while two-author publications have dropped from around 30\% to 9\%. Additionally, three-author publications have decreased from around 25\% to 17\%. In contrast, publications with four to thirteen authors have shown relatively higher increases in recent years. Among them, publications with four and five authors have increased from around 20\% to 36\%. Similarly, publications with six to eight authors have increased from around 7\% to  25\%, and publications with nine to thirteen authors have increased from around 2\% to 6\%. Publications with more than 13 authors also show increase, though at a smaller scale. In consequence, the typical percentage difference starts around 35\% for the higher-ranked authors and gradually decreases to around 20\% for the lower-ranked authors. Similar to the Biology and Electrical and Electronics Engineering datasets, we observe that the percentage difference is more striking among higher-prolific authors rather than lower-prolific.

In our Psychology dataset, the average number of authors has increased from four to seven. Single-author publications have decreased from around 31.78\% to 6.82\%, and two-author publications have decreased from around 33.77\% to 14.10\%. Publications with three authors remain steady at around 17-20\% each year. Conversely, publications involving four to thirteen authors are on the rise. Among them, publications with four and five authors have increased from around 10\% to 29\%. Publications with six to eight authors have increased from approximately 2.6\% to 16\%. Publications with nine to thirteen authors have increased from around 1\% to 9\%. Publications with more than 13 authors have increased from around 2\% to 5\%. Analogously, the typical percentage difference shows a downward trend starting at about 30\% for the higher-ranked authors and gradually decreasing to around 20\% for the lower-ranked authors. This trend is similar to the overall trend in our combined dataset. Nevertheless, we observe that the Psychology dataset exhibits a resemblance to the previous three datasets in terms of the greater percentage differences among the higher-prolific authors.

In the Sociology dataset, the average number of authors per publication has increased from around two to four authors. There is a noticeable shift from publications with one to two authors to publications with three to thirteen authors. Single-author publications have decreased significantly, dropping from around 40.21\% to 15\%. Two-author publications have decreased from around 24.33\% to 19.48\%. In contrast, publications with three authors show a relatively higher increase from around 17\% to 21.22\%. Publications with four and five authors have increased from around 13.22\% to 24.79\%. Publications with six to eight authors have increased from 3.17\% to 11.60\%. Publications with nine to thirteen authors have increased from around 1\% to 5\%. Publications with more than 13 authors also show increase in recent years. In contrast, the typical percentage difference begins at approximately 10\% and remains relatively stable between 10\% and 15\% for ranks up to 500 before decreasing to 0\% for the remaining authors. Compared to previous datasets, the Sociology dataset exhibits a more stable percentage difference trend among higher-prolific and lower-prolific authors.

In the Economics dataset, the average number of authors per publication has increased from around three to four authors. The number of authors in publications shows a shift from one to two authors to publications with three to eight authors, leading to a significant decrease in single-author publications, which decreases from around 43.77\% to around 10.19\%. Two-author publications also show a decline from around 32.25\% to 22.41\%. However, 22\% is still very high compared to other fields. Meanwhile, the three-author publication shows an increase from around 11.46\% to 27.91\%. Publications with four and five authors increase from around 5\% to 24.65\%. Additionally, publications with six to eight authors increase from around 1\% to 7\%.  In recent years, publications with nine to thirteen authors have increased from around 0.7\% to 2.5\%. Publications with more than 13 authors also show increase in recent years. The typical percentage difference starts around 15\% remains consistent for ranks up to 500, and decreases to 0\% for the remaining lower-ranked authors. While the typical percentage difference in the Economics dataset starts higher than the Sociology dataset, we observe that the higher-prolific and lower-prolific authors exhibit a similar behavior.

The Marketing dataset demonstrates a consistent publication practice with three authors. The average number of authors has slightly increased but remains around three. Publications with one and two authors show a decline. Single-author publications have decreased from around 25\% to 9\%, and two-author publications have decreased from around 31\% to 19\%. Three-author publications are consistently around 20-30\% for most of the years. In contrast, publications with four to eight authors show an increase. Specifically, publications with four and five authors have increased from around 10\% to 28\%, and publications with six to eight authors have increased from around 2\% to 7\%. Publications with nine to thirteen authors have increased from around 0\% to 1.5\%. Publications with more than 13 authors also show smaller increase in recent years. On the other hand, the typical percentage difference begins around 20\% for the higher-ranked authors and decreases to approximately 15\% for authors ranked between 300 and 400. It decreases to 0\% for the remaining authors. Comparably, the marketing dataset does not demonstrate a sharp increase in the number of authors per paper. However, the dataset has some outlying authors whose contributions and participation exhibit larger differences.

In the Mathematics dataset, the average number of authors has risen from two to three. Single-author publications have seen a notable decrease from approximately 32.32\% to 14.55\%. Two-author publications have dropped from around 38.38\% to 27.22\%. Meanwhile, publications with three to thirteen authors have experienced a significant increase, with three-author publications rising from about 19.19\% to 26.96\% and publications with four to five authors increasing from around 9.09\% to 22.68\%. Publications with six to eight authors have grown from roughly 1\% to 6\%. Publications with nine to thirteen authors have increased from 0\% to about 2\%. Additionally, publications with more than thirteen authors show a small increase in recent years. Analogously, the typical percentage difference starts at around 15\% and remains relatively stable across the standardized ranks. The typical percentage difference in publication count within this field is the lowest in our combined dataset.

Lastly, we demonstrate the effects of publications by participation and by Fibonacci-adjusted contributions of two authors sharing the same research sub-field with similar publication counts by participation in Table~\ref{comparison_pub}.
While the left-hand side author has published $131$ papers and the right-hand side author published $135$ paper, the right-hand side author's publication count by Fibonacci-adjusted contributions is significantly lower, $44.43$, compared to the left-hand side author, $122.53$.
That is, the author on the right-hand side took more supporting roles down in the byline than the author on the left-hand side.

{
	\renewcommand{\arraystretch}{3.32} 
	\begin{table*}
		\centering
		\caption{Comparison of bibliometrics of two authors with similar publication counts by participation.}
		\label{comparison_pub}
		\begin{tabular}{|m{3cm}|l|l|l||l|l|l|m{3cm}|}
			\hline
			\multirow{3}{*}{%
				\centering
				\includegraphics[max width=\linewidth, max height=3cm]{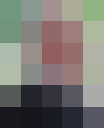}%
			} 
			& \textbf{Name:} Dxx Xxxxxxxxx 
			& P & 92 & 95 & P 
			& \textbf{Name:} Mx. Xxxxxx Xxxxx 
			& \multirow{3}{*}{%
				\centering
				\includegraphics[max width=\linewidth, max height=3cm]{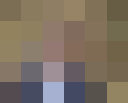}%
			} \\ \cline{2-7}
			& \parbox{3cm}{
				\raggedright
				\textbf{Field:} Computer Science and Engineering\\
				\textbf{Sub-field:} Cyber Security\\
			} 
			& P$^\prime$ & 82.39 & 55.91 & P$^\prime$ 
			& \parbox{3cm}{
				\raggedright
				\textbf{Field:} Computer Science and Engineering\\
				\textbf{Sub-field:} Cyber Security\\
			} 
			& \\ \cline{2-7}
			& \multicolumn{3}{p{5cm}||}{University of Dxxxxxx Xxxxx}
			& \multicolumn{3}{p{5cm}|}{Dxxxxx University} 
			& \\ \hline
		\end{tabular}
	\end{table*}
}

\subsection{$L^{\prime}$-index}

\begin{figure}
	\includegraphics[width=0.48\textwidth,height=0.25\textheight]{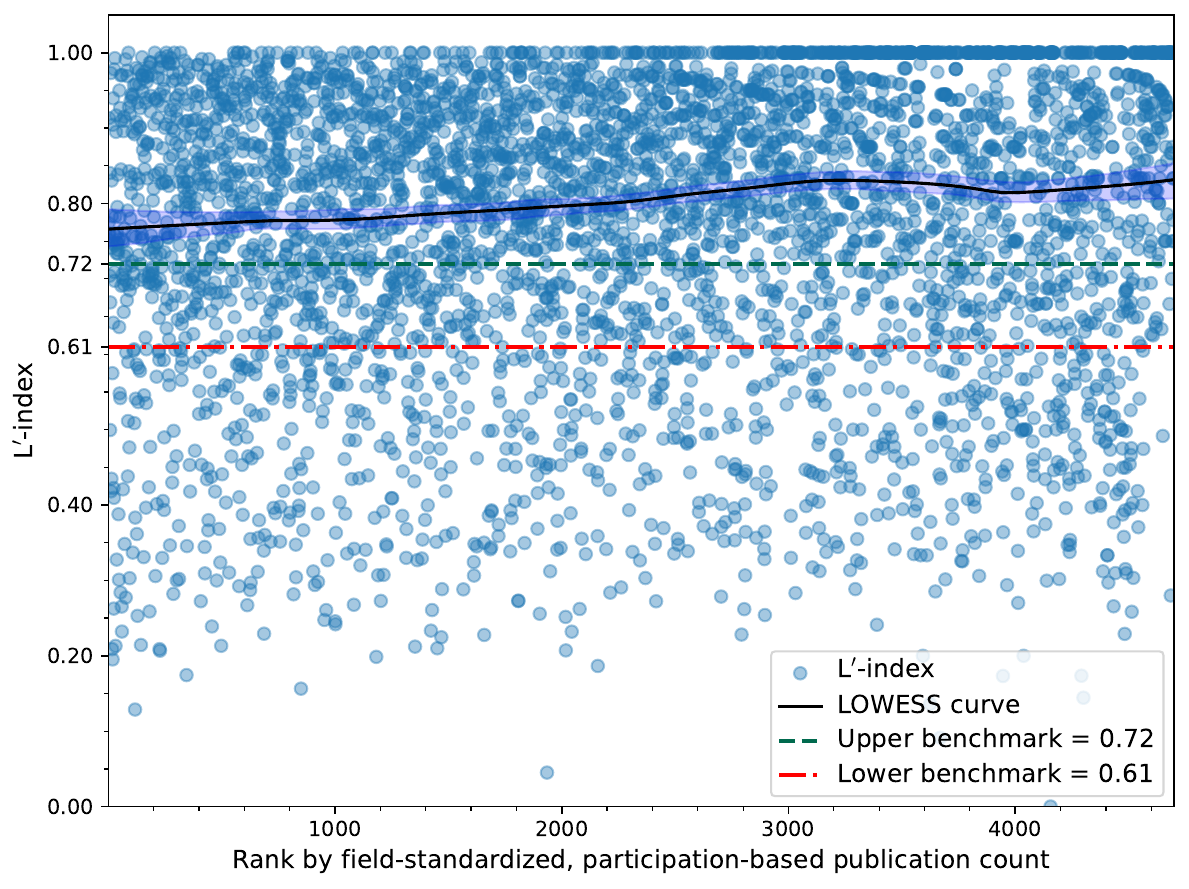}
	\caption{$L^{\prime}$-index of authors by standardized publication rank.}
	\label{fig:t_index_participation_rank}
\end{figure}

The distribution of the $L^\prime$-index of authors within our combined dataset is illustrated in Figure~\ref{fig:t_index_participation_rank}, where the x-axis represents the descending ranks of authors based on field-standardized, participation-based publication counts. The LOWESS curve reveals a slight upward trend in the $L^\prime$-index, starting at approximately 0.77 and stabilizing around 0.8 as the standardized participation rank increases. This suggests that most authors in our combined dataset have $L^\prime$-index values exceeding the upper and lower benchmarks of 0.72 and 0.61, respectively, indicating that they predominantly take leading roles in their publications. The wide spread of data points below the lower benchmark reflects those authors who more frequently take on supporting roles, a pattern observed across all ranks.
Notably, the data points situated below the lower benchmark and closer to higher ranks represent higher-prolific authors with $L^\prime$-index values below $0.61$. Around 18.53\% of authors in our combined dataset fall below the lower benchmark value of $0.61$. This suggests that these authors either predominantly take supporting roles or may be involved in potential cases of authorship abuse.
We further analyzed the $L^\prime$-index of authors in each field within our combined dataset and presented the figures in Appendix~\ref{t_index_by_publication_participation}, which drive the following discussions.

Authors in our Biology dataset have the lowest typical $L^\prime$-index at 0.47. Electrical and Electronics Engineering, and Computer Science Engineering datasets exhibit typical  $L^\prime$-indexes of 0.71 and 0.76, respectively. Authors in the Psychology dataset has a typical $L^\prime$-index of 0.78. Authors in the remaining field datasets, Sociology, Economics, Marketing, and Mathematics, have a typical $L^\prime$-index of around 0.87.

In our Biology dataset the typical $L^\prime$-index indicates an upward trend starting at 0.4 and gradually increasing towards 0.6 for the lower-ranked authors. The trend remains below both the upper and lower benchmark values. Some authors have an $L^\prime$-index above the lower benchmark of 0.61. The majority of authors in this field fall below it. In our dataset almost 93.5\% of authors in Biology are below the upper benchmark and 80.6\% of authors are below the lower benchmark. This trend suggests that authors in the Biology dataset are more likely to engage supporting roles rather than leading roles in a large number of papers, however.

In our Electrical and Electronics Engineering dataset, the typical $L^\prime$-index displays a clear upward trend, starting at approximately 0.64. It passes the upper benchmark value around rank 300, steadily rising to about 1.0 for the lower-ranked authors. In this dataset, roughly 50\% of the authors fall below the upper benchmark value, while 26.6\% are below the lower benchmark value. This trend signifies a gradual shift from supporting to leading roles as the rank decreases. The small group of higher-ranked authors with a low $L^\prime$-index may indicate that these contributors concentrate solely on supporting roles, despite their very high publication counts.

In our Computer Science and Engineering dataset, the typical $L^\prime$-index displays a gradual upward trend, beginning at approximately 0.7, crossing the upper benchmark value around rank 150, and gradually increasing to about 0.9. In this dataset, almost 36\% of the authors are below the upper benchmark, and around 14\% of the authors are below the lower benchmark. The small subset of higher-prolific authors near or below the benchmark indicates that these authors mostly take supporting roles compared to the others.

In our Psychology dataset, the typical $L^\prime$-index shows a slow upward trend, starting at around 0.75 and gradually increasing to around 0.85 as rank decreases. Most authors in this field demonstrate an $L^\prime$-index above both of the benchmark values, which may suggest stricter norms in authorship practices or that researchers predominantly take leading roles in their publications. The percentage of authors below the lower benchmark value is about 11\% in this dataset.

The typical $L^\prime$-index for the Marketing, Economics, Sociology, and Mathematics datasets reveal broadly similar trends, with field-specific nuances. In all four field datasets, the typical $L^\prime$-index values begin at relatively high levels, ranging from approximately 0.85 in Marketing to around 0.9 in Sociology and remain stable across most standardized ranks. This stability indicates that the majority of authors consistently take leading roles in their publications. Furthermore, a sharp rise in $L^\prime$-index values is observed for lower-ranked authors in Marketing, Economics, and Sociology, whereas Mathematics exhibits a consistent trend without such a steep increase. The proportion of authors falling below the lower benchmark of 0.61 is relatively low among the collected authors of these fields, with Mathematics showing the lowest proportion at $2.2\%$ and Marketing, Economics and Sociology slightly higher at 3\%, 4.3\% and 5\%, respectively. The overall pattern highlights a predominance of leading authorships across all four field datasets.

Lastly, we emphasize the effects of publications and $L^\prime$-index reflecting the leading and supporting roles of two authors sharing the same research sub-field with identical publication counts by participation in Table~\ref{comparison_t_index}.
While both authors have published $74$ papers, the right-hand side author's $L^\prime$-index is significantly lower, $0.48$, compared to the left-hand side author, $0.95$.
That is, the author on the right-hand side took more supporting roles down in the byline than the author on the left-hand side.

{
	\renewcommand{\arraystretch}{3.32} 
	\begin{table*}[htbp]
		\centering
		\caption{Comparison of $L^\prime$-index of two authors with identical publication counts by participation.}
		\label{comparison_t_index}
		\begin{tabular}{|m{3cm}|l|l|l||l|l|l|m{3cm}|}
			\hline
			\multirow{3}{*}{%
				\centering
				\includegraphics[max width=\linewidth, max height=3cm]{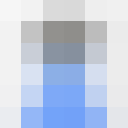}
			}
			& \textbf{Name:} Bxxxxxx Xxxxx 
			& P & 74 & 74 & P 
			& \textbf{Name:} Jxxxx Xxxxxxx
			& \multirow{3}{*}{%
				\centering
				\includegraphics[max width=\linewidth, max height=3cm]{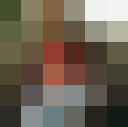}
			} \\ \cline{2-7}
			& \parbox{3cm}{%
				\raggedright
				\textbf{Field:} Psychology\\
				\textbf{Sub-field:} Developmental Psychology\\
			}  
			& $L^\prime$ & 0.95 & 0.48 & $L^\prime$  
			& \parbox{3cm}{%
				\raggedright
				\textbf{Field:} Psychology\\
				\textbf{Sub-field:} Developmental Psychology\\
			}  
			& \\ \cline{2-7}
			& \multicolumn{3}{p{5cm}||}{University of Wxxxxx}
			& \multicolumn{3}{p{5cm}|}{Sxxxxxxxx University of Xxxxxxxxxx} 
			& \\ \hline
		\end{tabular}
	\end{table*}
}

\subsection{Effect of Last-Author Convention on the
	$L^\prime$-Index}
\label{subsec:last_author}

Some disciplines designate the last byline position for the most contributing author. Unfortunately, this convention is also frequently abused through passive authorship by supervising or coordinating researchers. Because the $L^\prime$-index assigns Fibonacci-adjusted credit according to byline rank, this convention reduces the measured credit of a genuinely contributing last author. To assess the effect of this practice, we conducted a separate experiment in which the last author of each publication was moved to the beginning of the byline and assigned the corresponding full Fibonacci credit.

Under the modified convention, the percentage of authors in the Biology dataset falling below the lower benchmark decreases from 80.60\% to 41.50\%, while the percentage below the upper benchmark decreases from 93.50\% to 70.16\%. The typical $L^\prime$-index in the Biology dataset rises from 0.47 to 0.63, exceeding the lower benchmark of 0.61 but remaining below the upper benchmark of 0.72. Under this experiment, 70.16\% of authors in the Biology dataset remain below the upper benchmark and 41.50\% remain below the lower benchmark, indicating that the high prevalence of low $L^\prime$-index values observed in this dataset reflects a genuine pattern of frequent supporting authorship roles rather than an artifact of the byline convention.

For the Electrical and Electronics Engineering dataset, the percentage of authors below the lower benchmark drops from 26.61\% to 8.39\%, the highest among the remaining datasets. Most other datasets fall below 3\% under the lower benchmark, and the typical $L^\prime$-index exceeds both benchmark values in all datasets except the Biology dataset.

\subsection{Citations by Contribution}

\begin{figure}
	\includegraphics[width=0.48\textwidth,height=0.25\textheight]{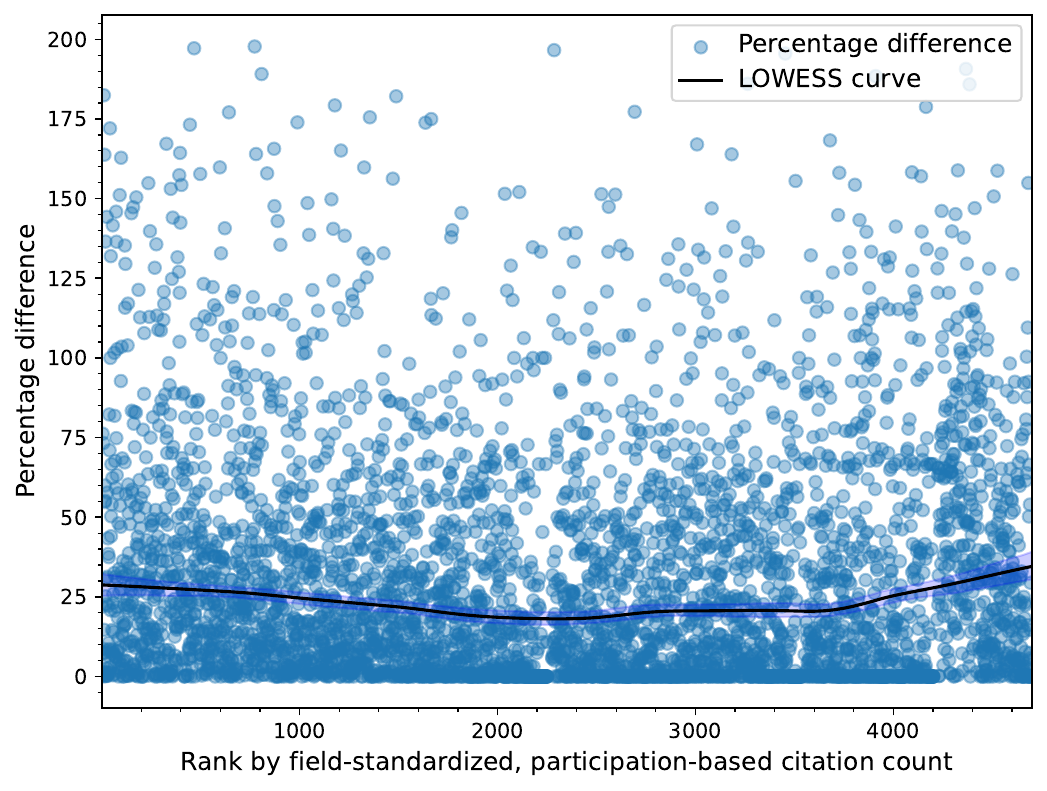}
	\caption{Percentage difference between citations by participation and by contribution ranked by participation.}
	\label{fig:percentage_difference_citation_count_participation_rank}
\end{figure} 

Figure~\ref{fig:percentage_difference_citation_count_participation_rank} depicts the percentage differences between citation counts based on participation and Fibonacci-adjusted contributions, with rankings determined by field-standardized, participation-based citation counts in descending order. Similar to publication counts, the typical percentage difference in citation counts starts around 30\% and remains relatively stable, around 25\% to 30\% across all ranks. However, we also observe a wide spread of percentage differences exceeding 30\%. This indicates the presence of authors whose citation counts are significantly affected by the proposed citations by Fibonacci-adjusted contributions. 

We further analyzed the citation distribution and percentage differences in the combined and individual field datasets, which are shown in Appendix~\ref{citation_distribution_per_publication}. In the combined dataset, the median and mean number of citations per publication are 4.0 and 40.29, respectively. The large difference between the median and mean values suggests a long-right tail of highly cited publications. Among the collected field datasets, Biology has the highest median of 6.0 and a mean of 63.66, indicating that a publication in Biology typically receives six citations, and the high mean value suggests a skewed tail of highly cited publications, which is a natural trend. The Biology dataset is followed by Electrical and Electronics Engineering, Mathematics, and Computer Science and Engineering datasets, showing high medians of 5.0, 5.0, and 4.0, respectively, but comparably moderate means of 28.24, 23.48, and 24.57, which indicate relatively light-tailed distributions. Psychology, Marketing, and Economics datasets show low medians of 3.0, 3.0, and 2.0 yet demonstrate relatively high means of 40.66, 45.27, and 35.36, suggesting that while most publications receive fewer citations, smaller subsets of publications have significantly higher citations. The Sociology dataset shows the lowest median of 2.0 but a relatively high mean of 29.41. Similar to the variations in authorship practices across different field datasets, the distributions of citations also varies from one collected field dataset to another.

Among all fields, the Biology dataset has the highest typical percentage difference of around 81.32\% in citation counts by participation and by contribution. This dataset is followed by Electrical and Electronics Engineering at 36.30\%, Computer Science and Engineering at 31.88\%, Psychology at 27.20\%, Sociology at 19.21\%, Marketing at 17.44\%, Economics at 17.36\%, and Mathematics at 14.75\%. The trend shows that achieving high numbers of citations via multi-authorship is more common in the Biology dataset compared to other field datasets. The percentage differences between citations by participation and by contribution for each field dataset are shown in Appendix~\ref{percentage_difference_citation_appendix}. Additionally, we showed the ECDF plot illustrating the cumulative distributions of percentage differences in citation counts for each field in Appendix~\ref{percentage_difference_citation_ecdf_appendix}, which together drive the following discussions.

In the Biology dataset, the typical percentage difference starts around 110\% and decreases to around 60\% as the rank increases. This indicates that the authors in the Biology dataset typically have a higher percentage difference compared to those in other fields. The ECDF plot shows that approximately 34\% of the authors in this field exhibit a percentage difference of more than 100. The relatively higher percentage difference across all ranks in this field dataset cannot be attributed solely to collaboration practices as discussed at the end of Section~\ref{sec:motivation}.

In the Electrical and Electronics Engineering dataset, the typical percentage difference starts at around 60\% and sharply decreases to around 35\% for authors ranked 1 to 100. It remains relatively stable at around 35\% for authors ranked 100 to 400, then gradually decreases to 0\% for the remaining authors. The sharp decrease for the higher-prolific authors suggests that the proposed citations by contribution affects prolific authors the most. The ECDF plot in Appendix~\ref{percentage_difference_citation_ecdf_appendix} shows that approximately 6\% of the authors in this field dataset have a percentage difference of 89\% or higher.

In our Computer Science and Engineering dataset, the typical percentage difference begins at around 40\% and slowly drops to around 25\% as the ranks increase. The typical percentage difference follows a similar trend to the combined dataset. The ECDF plot shows that around 4\% of the authors in this dataset have a percentage difference of 89\% or higher.

In our Psychology dataset, the typical percentage difference also follows a similar trend with the combined dataset, beginning at around 25\% and remaining around 25\% to 30\% for authors ranked between 1 and 400 and gradually dropping to 5\% for the remaining authors. The ECDF plot shows that around 5\% of the authors in this dataset exhibit a percentage difference of 89\% or higher.

In Sociology, Marketing, and Economics datasets, the typical percentage difference starts around 10\% and remains stable at around 10\% for the top 400 ranked authors, gradually declining to 0\% for the authors ranked lower. This suggests that most authors in these field datasets experience minimal deviation in citation counts based on participation and contribution. The ECDF plot in Appendix~\ref{percentage_difference_citation_ecdf_appendix} shows that a small portion, between 2\% and 3\%, of the authors exhibit percentage differences of 89\% or higher, indicating cases of disproportionate citation credit allocations. While the Mathematics dataset shows a similar stability, it has a relatively higher typical percentage difference of approximately 20\%. However, only 1\% of authors exhibit percentage differences of 89\% or higher, indicating a more proportional citation distribution. Overall, the authors in these field datasets demonstrate relatively low variability in citation counts by contributions. Nonetheless, the existence of a small subset of authors with extreme citation discrepancies does not rule out the possibility of authorship abuse on smaller scales.

Lastly, we demonstrate the effects of citation counts by participation and by Fibonacci-adjusted contributions of two authors sharing the same research sub-field with similar citation counts by participation in Table~\ref{comparison_citation}.
While both authors have received around $10,330$ citations, with left-hand side author receiving $10,352$ and the right-hand-side author receiving $10,331$, the right-hand side author's citation count by Fibonacci-adjusted contributions is significantly lower, $989.24$, compared to the left-hand side author, $8,679.45$.
That is, the author on the right-hand side took more supporting roles down in the byline than the author on the left-hand side in their highly cited papers.

{
	\renewcommand{\arraystretch}{3.32}
	\begin{table*}[htbp]
		\centering
		\caption{Comparison of bibliometrics of two authors with similar citation counts by participation.}
		\label{comparison_citation}
		\begin{tabular}{|m{3cm}|l|l|l||l|l|l|m{3cm}|}
			\hline
			\multirow{3}{*}{%
				\centering
				\includegraphics[max width=\linewidth, max height=3cm]{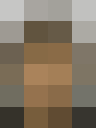}
			} 
			& \textbf{Name:} Axxxxx Xxxxx 
			& C & $10,352$ & $10,331$ & C 
			& \textbf{Name:} Wxxxxxx Xxxxx 
			& \multirow{3}{*}{%
				\centering
				\includegraphics[max width=\linewidth, max height=3cm]{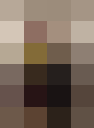}
			} \\ \cline{2-7}
			& \parbox{3cm}{%
				\textbf{Field:} Biology\\
				\textbf{Sub-field:} Genetics\\
			}
			& C$^\prime$ & $8,679.45$ & $989.24$ & C$^\prime$ 
			& \parbox{3cm}{%
				\textbf{Field:} Biology\\
				\textbf{Sub-field:} Genetics\\
			} 
			& \\ \cline{2-7}
			& \multicolumn{3}{p{5cm}||}{University of Xxxxx}
			& \multicolumn{3}{p{5cm}|}{University of Xxxxx Xxxxxxxx}
			& \\ \hline
		\end{tabular}
	\end{table*}
}

\subsection{$h$-index by Contribution}

\begin{figure}
	\includegraphics[width=0.48\textwidth,height=0.25\textheight]{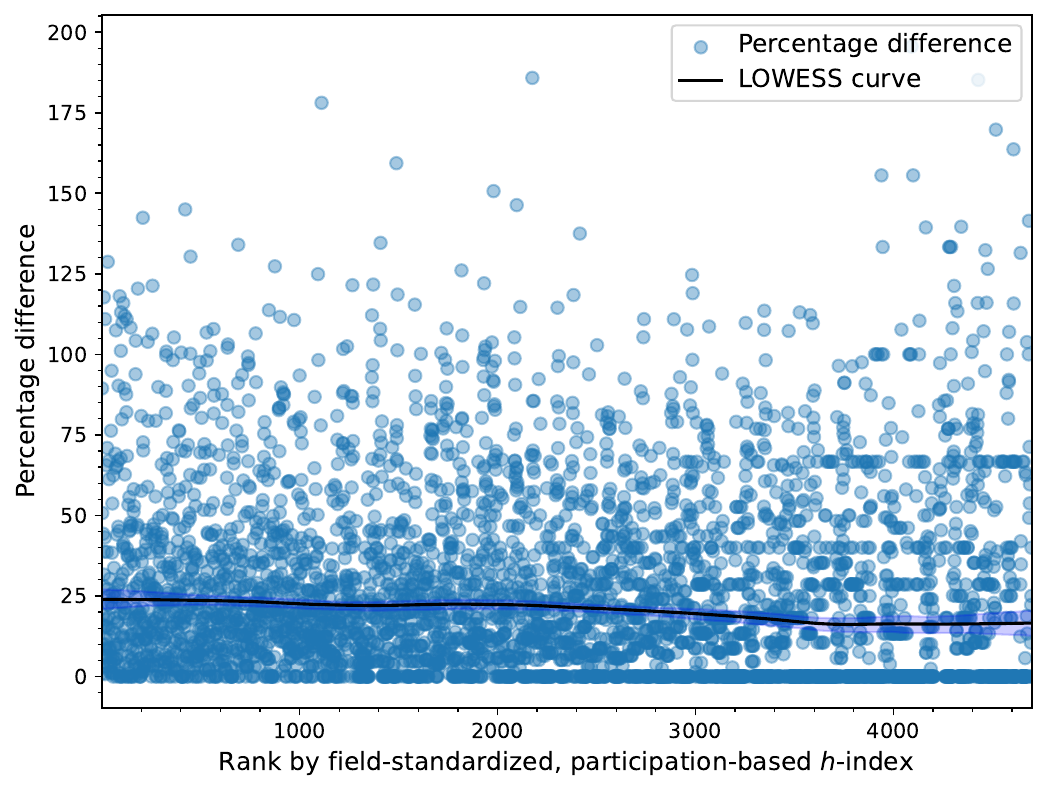}
	\caption{Percentage difference between $h$-index by participation and contribution on rank by participation.}
	\label{fig:percentage_difference_h_index_participation_rank}
\end{figure} 

Figure~\ref{fig:percentage_difference_h_index_participation_rank} depicts the percentage differences between $h$-index calculated by participation and by Fibonacci-adjusted contributions, with rankings determined by field-standardized, participation-based $h$-index in descending order. Similar to publication and citation counts, the percentage difference is roughly around 25\% across all ranks. Suggesting that the majority of authors in our combined dataset experience a percentage difference of approximately 25\% in their $h$-indexes. Nonetheless, the wide spread of percentage differences above 25\% reflects the presence of authors whose publication counts are significantly affected by our proposed $h$-index by Fibonacci-adjusted contributions.

We analyzed the distribution of the $h$-index by participation in our combined dataset and in field-specific datasets, depicted in Appendix~\ref{h_index_distribution_by_participation}. Across all authors in the combined dataset, the $h$-index distribution has a median of 11 and a mean of 16.83. This indicates that while a typical author has an $h$-index of 11, a subset of authors with relatively high $h$-index values raise the mean. Further examining the field distribution of the $h$-index by participation reveals that the Biology dataset stand out with the highest median of 35 and a mean of 39.98, suggesting that authors in our Biology dataset tend to have exceptionally high $h$-indexes driven by higher citations and publication counts. The Electrical and Electronics Engineering dataset, with a median of 19 and a mean of 20.71, and Computer Science and Engineering dataset, with a median of 15 and a mean of 18.41, both rank above the overall median and mean, though still noticeably below the Biology dataset as reflected by thier citation and publication counts. In contrast, the Psychology dataset has a median of 9 and a mean of 12.52, while the Economics dataset exhibits a median of 10 and a mean of 12.60, indicating relatively similar $h$-index statistics. The Marketing dataset has a median of 7 and a mean of 9.40, reflecting lower $h$-index statistics compared to the overall trend. Similarly, the Sociology dataset, with a median of 7 and a mean of 10.36, and the Mathematics dataset, with a median of 8 and a mean of 10.34, demonstrate relatively lower trends. These contrasts highlight variations in researchers' impact across disciplines and the publication and citation practices, as well as field-specific norms in our collected datasets.

Among the fields, the Biology dataset has the highest typical percentage difference of  67.19\% in $h$-index by participation and by contribution. This dataset is followed by Electrical and Electronics Engineering at 30.96\%, Computer Science and Engineering at 25.57\%, Psychology at 22.68\%, Sociology at 15.63\%, Marketing at 15.23\%, Economics at 14.47\%, and Mathematics at 12.48\%. This decline in the percentage differences of authors' $h$-indexes are correlated to the authorship practices in each field in our collected datasets. The percentage difference between $h$-index by participation and by Fibonacci-adjusted contribution for each field dataset is shown in Appendix~\ref{percentage_difference_h_index_appendix}. Additionally, we show the ECDF plot illustrating the cumulative distributions of percentage differences in $h$-index in Appendix~\ref{percentage_difference_h_index_ecdf_appendix}, which together drive the following discussions.

In our Biology dataset, the typical percentage difference shows a clear downward trend. It starts around 95\% and gradually decreases to around 50\% as the rank increases. This trend differs substantially from that observed in our combined dataset. The ECDF plot shows that approximately 20\% of the authors in this field exhibit a percentage difference of 89\% or higher, suggesting that a significant proportion of researchers have a high discrepancy between participation and contribution. This pattern suggests that the traditional bibliometric indicators may overestimate the actual contributions of some higher-prolific authors.

In our Electrical and Electronics Engineering dataset, the typical percentage difference shows a downward trend. It starts at around 50\% and decreases to around 30\% for authors ranked between 1 and 200. The trend stabilizes around 30\% for authors ranked between 200 and 400 and sharply decreases to 0\% for the remaining authors, indicating lower deviations among those with lower ranks. The ECDF plot in Appendix~\ref{percentage_difference_h_index_ecdf_appendix} shows that approximately 4\% of the authors in this dataset exhibit a percentage difference of 89\% or higher. Although this trend is closer to our combined dataset and lower than what is observed in the Biology dataset, the higher percentage differences exhibited by higher-prolific authors ranked between 1 and 200 suggests that the traditional bibliometrics remains biased toward these prolific authors.

In our Computer Science and Engineering dataset, the typical percentage difference begins at around 35\% and drops to around 25\% for authors ranked between 1 and 100. It remains around 25\% for the authors ranked between 200 and 400 and decreases to around 15\% for the remaining authors. The ECDF plot in Appendix~\ref{percentage_difference_h_index_ecdf_appendix} shows that approximately 1\% of the authors in this dataset exhibit a percentage difference of 89\% or higher. This trend aligns closer to the trend observed in the combined dataset, yet it still demonstrates that the higher-prolific authors take advantage of the traditional bibliometric indicators.

In our Psychology dataset, the typical percentage difference starts at around 25\% and remains consistent for authors ranked between 1 and 200. It gradually decreases to below 20\% for the remaining authors. The ECDF plot shows that approximately 2\% of the collected authors in this field exhibit a percentage difference of 89\% or higher. This trend is similar to the trend observed in the combined dataset. The lower percentage differences of 25\% or below indicates that most of the collected authors in this field are relatively less affected by the proposed $h^{\prime}$-index.

In Sociology, Economics, Marketing, and Mathematics datasets, the typical percentage difference starts around 10\% and remains relatively consistent for authors ranked between 1 and 400. It gradually decreases to 0\%, except in the Mathematics, where it stays around 10\% across all ranks. This consistently lower percentage differences across all authors suggests the prevalence of smaller collaborations as well as stricter authorship norms or fewer opportunities for unethical authorship practices.

Lastly, we demonstrate the effects of $h$-index by participation and by Fibonacci-adjusted contributions of two authors sharing the same research sub-field with the same $h$-index by participation values in Table~\ref{comparison_h_index}.
While both authors have an $h$-index of $43$, the right-hand side author's  $h$-index by Fibonacci-adjusted contributions is significantly lower, $18.47$, compared to the left-hand side author, $34.33$.
That is, the author on the right-hand side took more supporting roles down in the byline than the author on the left-hand side in their highly cited papers.

{
	\renewcommand{\arraystretch}{3.32} 
	\begin{table*}
		\centering
		\caption{Comparison of bibliometrics of two authors with identical $h$-indexes by participation.}
		\label{comparison_h_index}
		\begin{tabular}{|m{3cm}|l|l|l||l|l|l|m{3cm}|}
			\hline
			\multirow{3}{*}{%
				\centering
				\includegraphics[max width=\linewidth, max height=3cm]{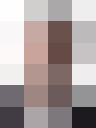}
			}
			& \textbf{Name:} Axxxx Xxxxx 
			& $h$ & 43 & 43 & $h$ 
			& \textbf{Name:} Sxxxxxx Xx 
			& \multirow{3}{*}{%
				\centering
				\includegraphics[max width=\linewidth, max height=3cm]{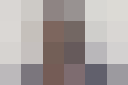}
			} \\ \cline{2-7}
			& \parbox{3cm}{%
				\textbf{Field:} Economics\\
				\textbf{Sub-field:} Econometrics\\
			}  
			& $h^\prime$ & 34.33 & 18.47 & $h^\prime$ 
			& \parbox{3cm}{%
				\textbf{Field:} Economics\\
				\textbf{Sub-field:} Econometrics\\
			}  
			& \\ \cline{2-7}
			& \multicolumn{3}{p{5cm}||}{Xxxx University of Xxxxxxxxx}
			& \multicolumn{3}{p{5cm}|}{University of Sxxxxx}
			& \\ \hline
		\end{tabular}
	\end{table*}
}

\section{Conclusions}
\label{section:conclusions}

In this study, we analyzed Google Scholar's bibliometric data of 4,700 authors from eight distinct fields and their 377,009 publication records between 1991 and 2024.
In the last 34 years, the average number of authors per publication has increased from four to eight. In particular, the one-, two-, and three-author publications, significantly dropped from 64.44\% in 1991 to 34.35\% in 2024.
In contrast, publications with multiple authors have consistently increased.
Notably, the share of publications with more than eight authors has risen from 5.08\% in 1991 to 21.18\% in 2024.
We believe that a significant part of this increase is related to authorship practices along with (or  rather than) collaboration practices.

A thorough investigation of every paper for authorship abuse is not feasable.
Hence, we proposed a credit allocation scheme based on the reciprocals of Fibonacci numbers, in this study. 
The proposed credit allocation scheme adjusts credit for individual contributions while systematically reducing credit for potential authorship abuse.
The proposed scheme supports both strict and lenient authorship guidelines, including the IEEE, ICMJE, and APA, which require major contributions, as well as the CRediT and CERN, which accept even minimal or no direct contributions for authorship.
We introduced three new indicators, namely Fibonacci-adjusted total publications ($P'$), Fibonacci-adjusted citation count ($C'$), and Fibonacci-adjusted Hirsch-index ($h'$).
These metrics recalibrate the original measures of total publications ($P$), citation count ($C$), and Hirsch-index ($h$) by using the reciprocals of Fibonacci numbers.
In addition, we introduced the $L^{\prime}$-index to quantify a scientist's labor contribution across their publications along with two benchmark values.
The $L^{\prime}$ inherently supports collaborations while curbing deceptive authorship practices.

Our analyses show that the authors in the combined dataset typically exhibit a 25\% percentage difference between $P$ and $P'$. This percentage difference increases to 74.26\% in the Biology dataset, while it decreases to 12.74\% in the Mathematics dataset. Specifically, the percentage difference is 36.26\% for the Electrical and Electronics Engineering dataset, 29.14\% for the Computer Science and Engineering dataset, 23.89\% for the Psychology dataset, 15.60\% for the Sociology dataset, 14.96\% for the Economics dataset, 14.48\% for the Marketing dataset, and 12.74\% for the Mathematics dataset.

Our analyses of $L^{\prime}$-index show that only $18.53\%$ of authors fall below $0.61$, indicating that a relatively small number of researchers primarily take on supporting roles. In contrast, the Biology dataset shows that $80.6\%$ of authors fall below this threshold. Notably, significant differences emerge across all collected field datasets: $26.61\%$ in Electrical and Electronics Engineering, $14.45\%$ in Computer Science and Engineering, $10.94\%$ in Psychology, $5.14\%$ in Sociology, $4.39\%$ in Economics, $3.06\%$ in Marketing, and $2.22\%$ in Mathematics.

The typical percentage differences between $C$ and $C^\prime$ for the combined dataset is around $25\%$.
In contrast, authors from the Biology, Electrical and Electronics Engineering, Computer Science and Engineering, and Psychology datasets exhibit larger differences, ranging from $81.32\%$ to $27.20\%$, compared to the other field-specific datasets, which is between $19.21\%$ and $14.75\%$.

The percentage differences between $h$ and $h^{\prime}$ is typically around $25\%$ for the combined dataset. However, in specific fields the difference is considerably higher, with $67.19\%$ in Biology, $30.96\%$ in Electrical and Electronics Engineering, $25.57\%$ in Computer Science and Engineering, and $22.68\%$ in Psychology datasets, while for the remaining field datasets the differences range from $15.63\%$ to $12.48\%$.

Lastly, authorship abuse is an ethical problem; hence, the proposed solution aims to curb its ramifications rather than eliminate it. The proposed bibliometric indicators limit the impact of deceptive authorship practices and promote fairer recognition of scholars based on contribution rather than participation.

\section*{Acknowledgement}
Mehmet Engin Tozal conceived the original concept for the study. Md Somir Khan collected the dataset and designed, implemented, and executed all experiments. Both authors analyzed the results and contributed to the preparation of the manuscript. Both authors reviewed and approved the final manuscript.

\section*{Acknowledgement}
This is the first paper in a series. The next study will use the notations $P^{\prime\prime}$, $C^{\prime\prime}$, $h^{\prime\prime}$, and $L^{\prime\prime}$ to further adjust the proposed bibliometric indicators based on the quality of the publication venues. Papers in low-quality, open access, mega, and predatory journals and conferences, which lack or have high variance in their standards, may have some value. Yet, they should not receive the same credit as those in high-quality venues prioritizing scientific novelty and advancement. 

In fact, low-quality publication venues are a more serious problem in science. They not only skew the bibliometrics indicators in favor of their authors but also cause scientific pollution by generating too much noise lacking originality, innovation, and scientific rigor. Consequently, this slows down the advancement of science by making it harder for researchers, especially junior ones, to navigate through an overwhelming volume of meaningless publications.

\bibliographystyle{IEEEtran}
\balance
\bibliography{references}

\appendices
\onecolumn

\section{The average number of authors per year in the combined and field datasets}

\label{avg_trend_field}
\begin{figure*}[!htbp]
	\centering
	\begin{subfigure}[t]{0.32\textwidth}
		\centering
		\includegraphics[width=\linewidth,keepaspectratio]{plots/avg_trend.pdf}
		\caption{Combined}
	\end{subfigure}
	\hfill
	\begin{subfigure}[t]{0.32\textwidth}
		\centering
		\includegraphics[width=\linewidth,keepaspectratio]{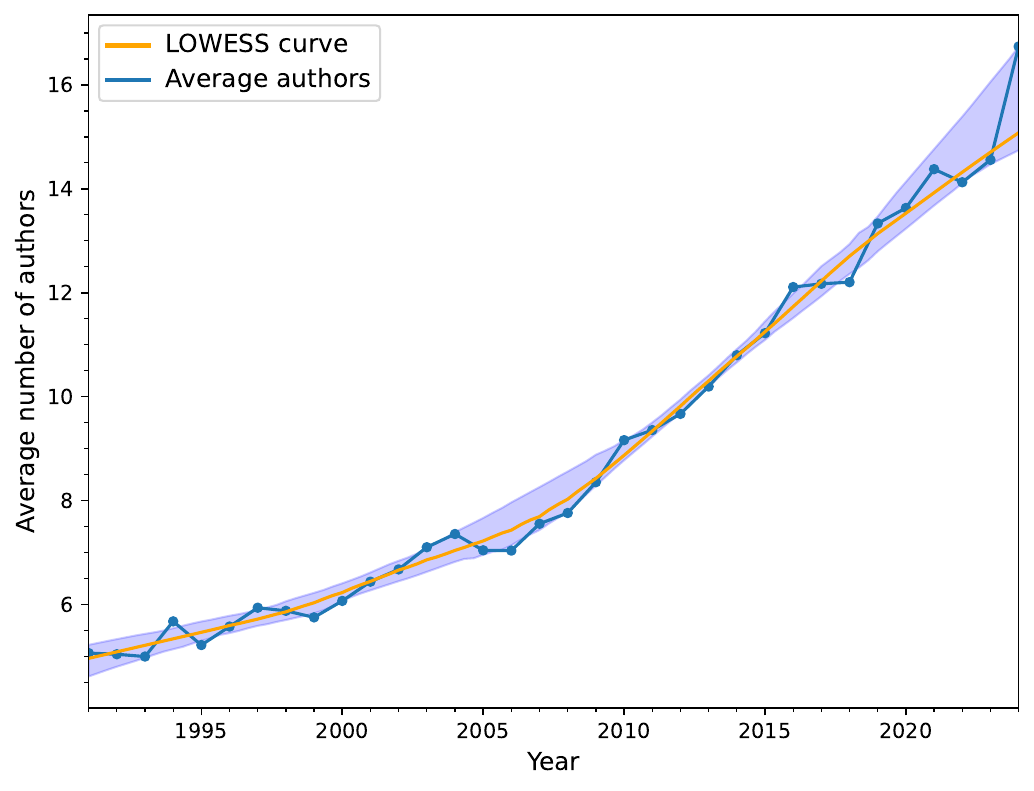}
		\caption{Biology}
	\end{subfigure}
	\hfill
	\begin{subfigure}[t]{0.32\textwidth}
		\centering
		\includegraphics[width=\linewidth,keepaspectratio]{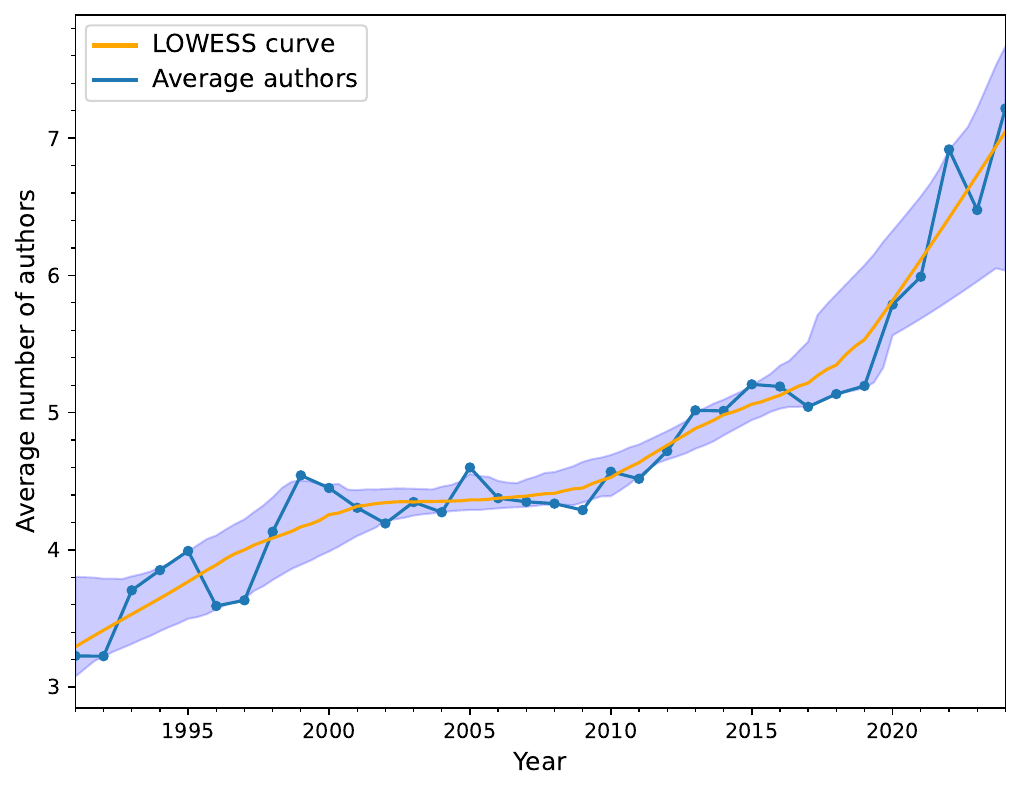}
		\caption{Electrical and Electronics Engineering}
	\end{subfigure}
	
	\begin{subfigure}[t]{0.32\textwidth}
		\centering
		\includegraphics[width=\linewidth,keepaspectratio]{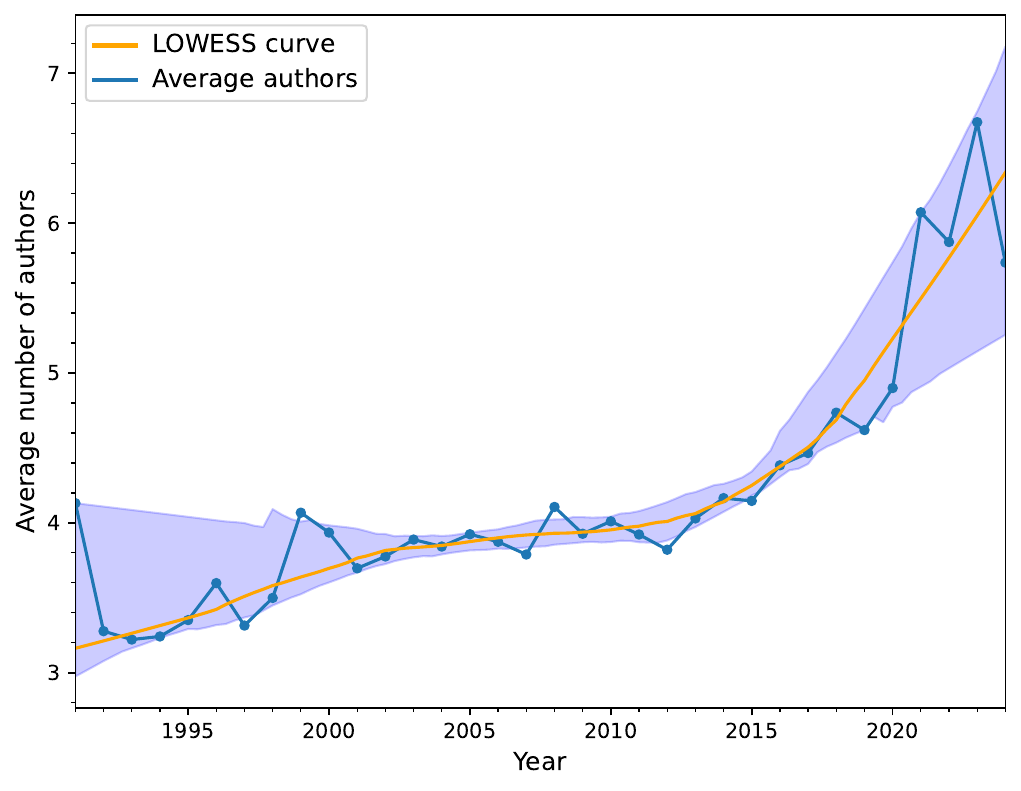}
		\caption{Computer Science and Engineering}
	\end{subfigure}
	\hfill
	\begin{subfigure}[t]{0.32\textwidth}
		\centering
		\includegraphics[width=\linewidth,keepaspectratio]{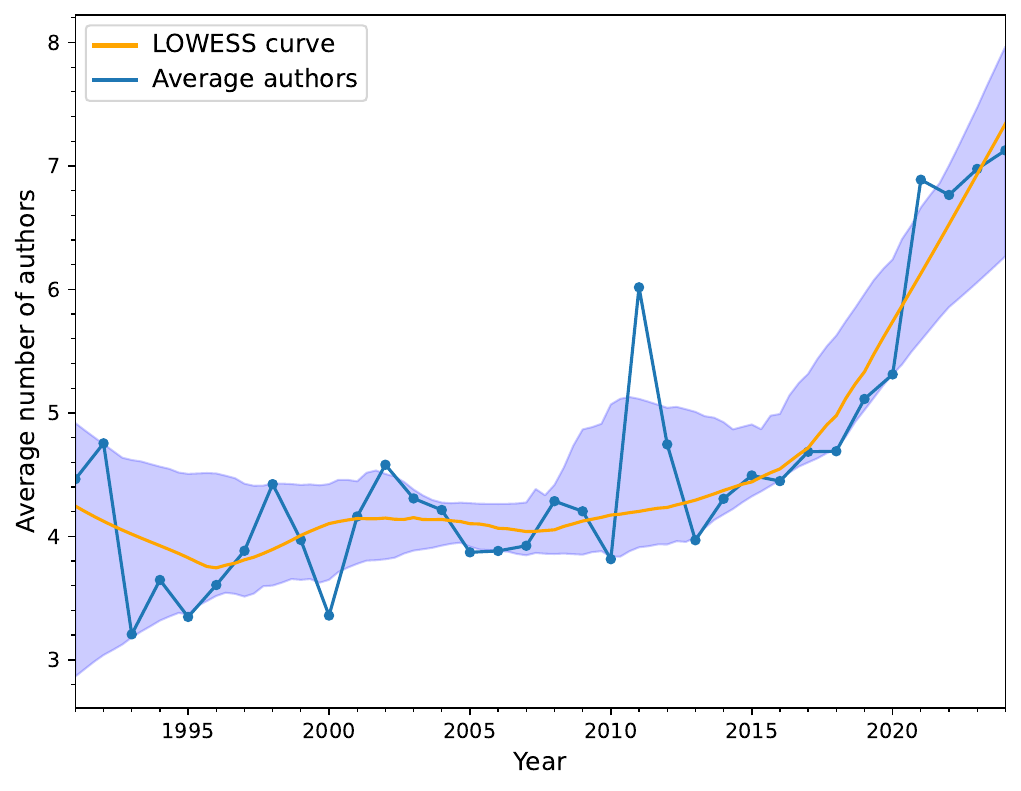}
		\caption{Psychology}
	\end{subfigure}
	\hfill
	\begin{subfigure}[t]{0.32\textwidth}
		\centering
		\includegraphics[width=\linewidth,keepaspectratio]{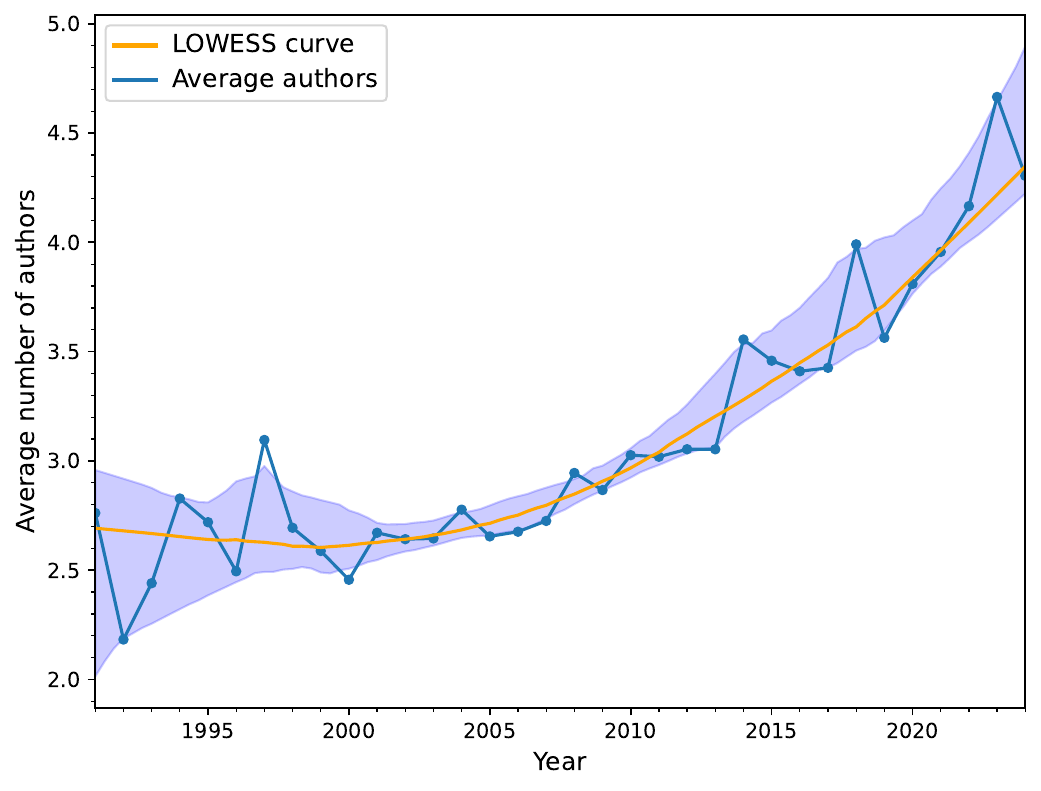}
		\caption{Sociology}
	\end{subfigure}
	
	\begin{subfigure}[t]{0.32\textwidth}
		\centering
		\includegraphics[width=\linewidth,keepaspectratio]{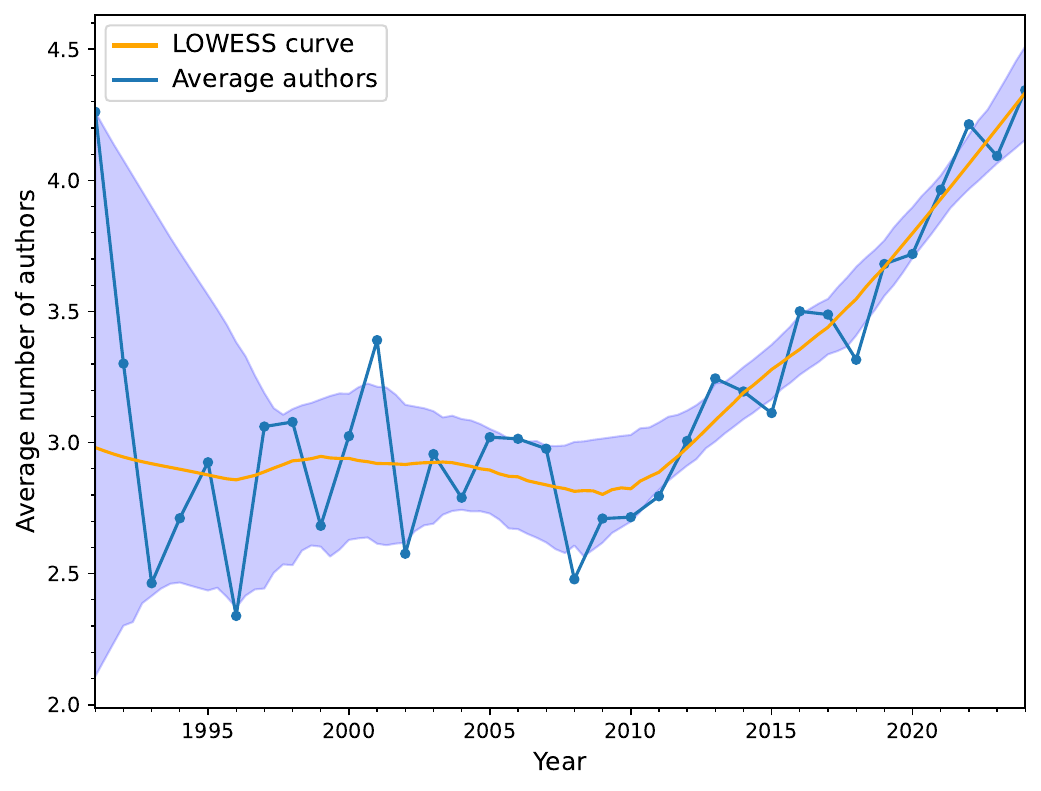}
		\caption{Economics}
	\end{subfigure}
	\hfill
	\begin{subfigure}[t]{0.32\textwidth}
		\centering
		\includegraphics[width=\linewidth,keepaspectratio]{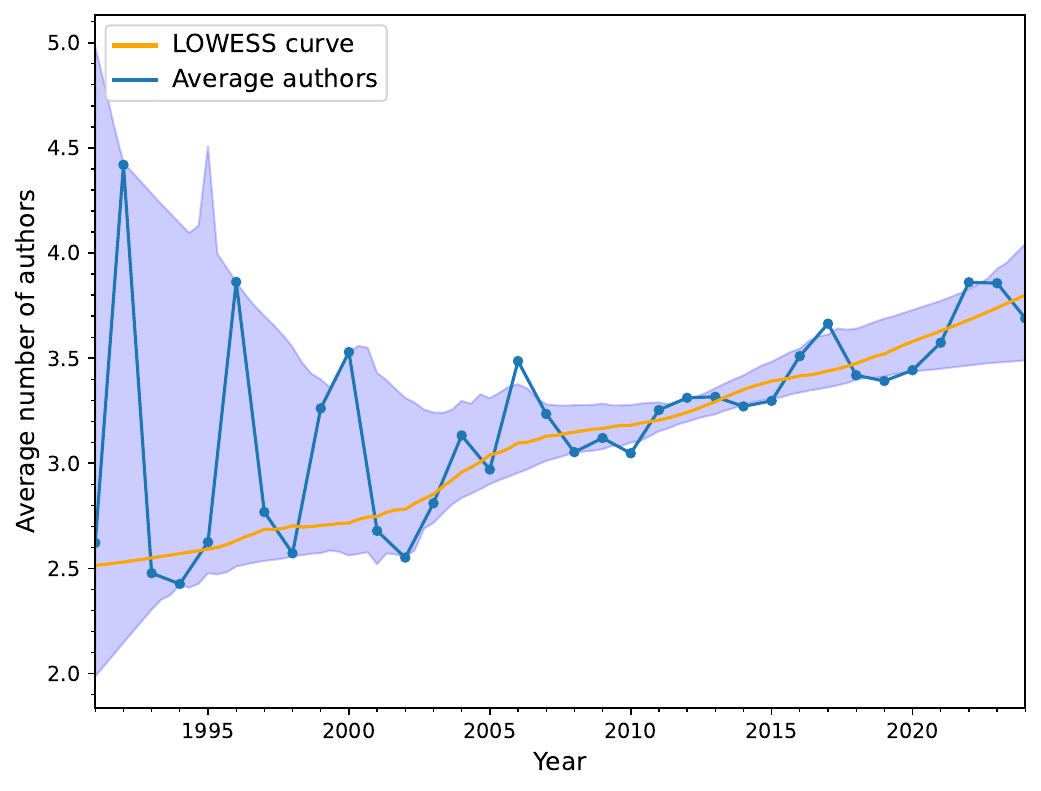}
		\caption{Marketing}
	\end{subfigure}
	\hfill
	\begin{subfigure}[t]{0.32\textwidth}
		\centering
		\includegraphics[width=\linewidth,keepaspectratio]{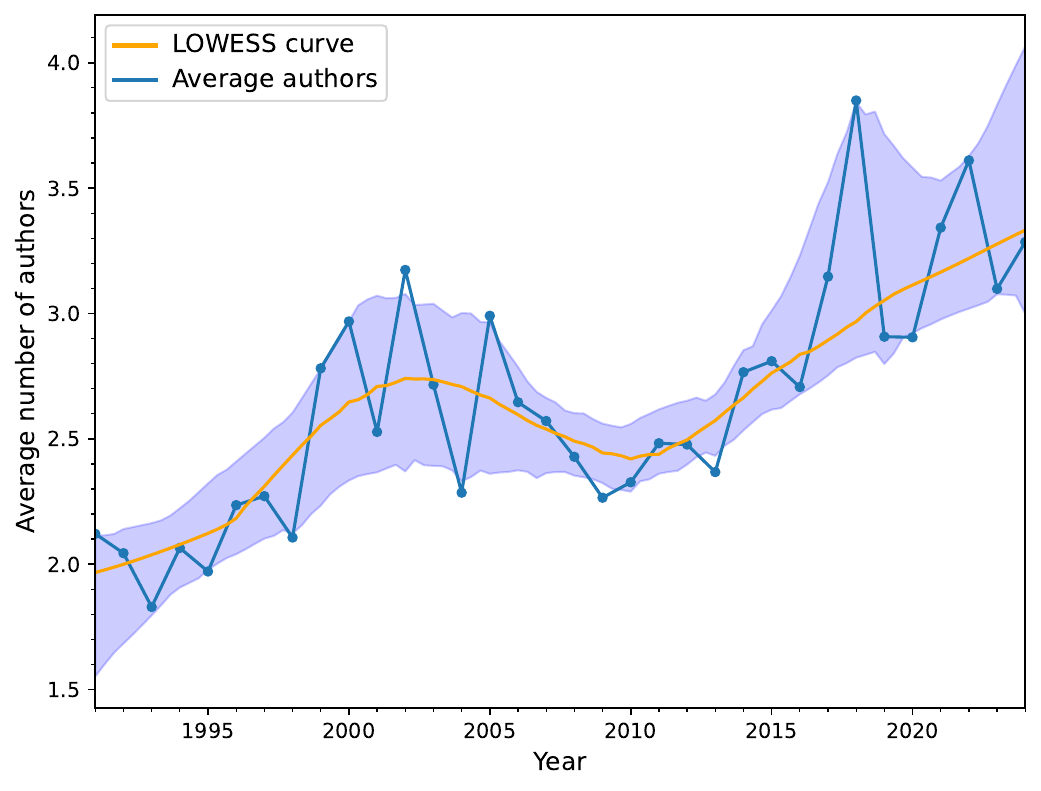}
		\caption{Mathematics}
	\end{subfigure}
	
	\caption{The average number of authors per year in the combined and field datasets.}
	\label{fig:avg_field_trend}
\end{figure*}
\FloatBarrier
\clearpage

\section{The median number of authors per year in the combined and field datasets}

\label{median_trend_field}
\begin{figure*}[!htbp]
	\centering
	\begin{subfigure}[t]{0.32\textwidth}
		\centering
		\includegraphics[width=\linewidth,keepaspectratio]{plots/median_authors_per_year.pdf}
		\caption{Combined}
	\end{subfigure}
	\hfill
	\begin{subfigure}[t]{0.32\textwidth}
		\centering
		\includegraphics[width=\linewidth,keepaspectratio]{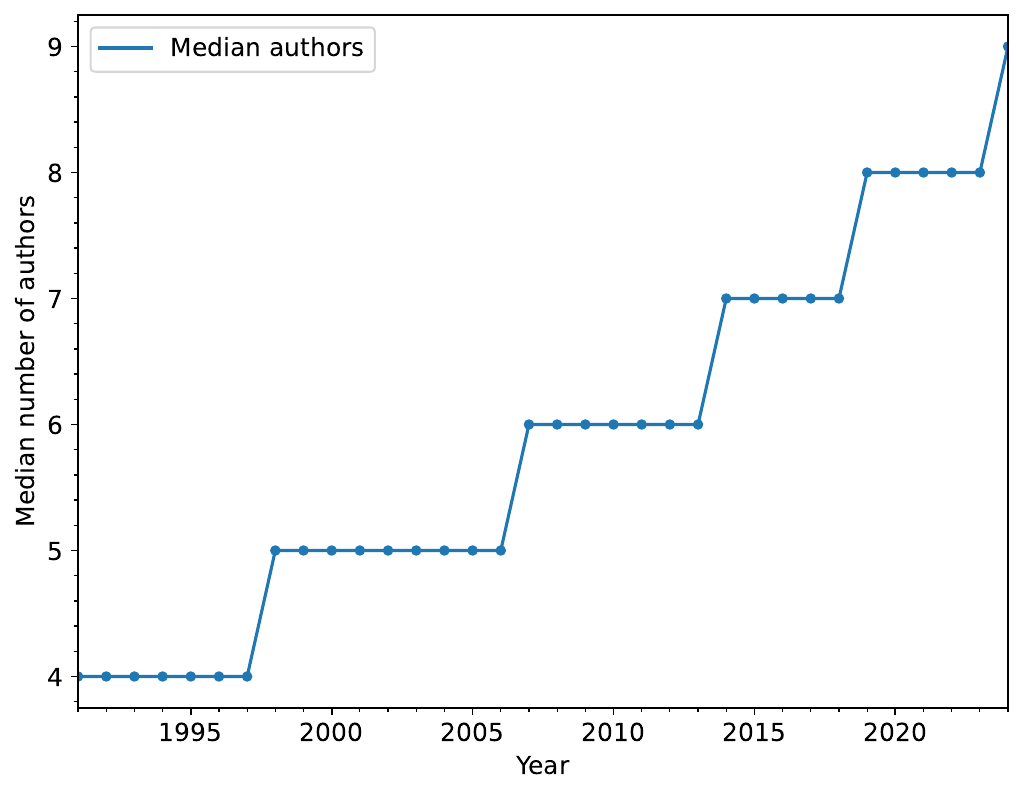}
		\caption{Biology}
	\end{subfigure}
	\hfill
	\begin{subfigure}[t]{0.32\textwidth}
		\centering
		\includegraphics[width=\linewidth,keepaspectratio]{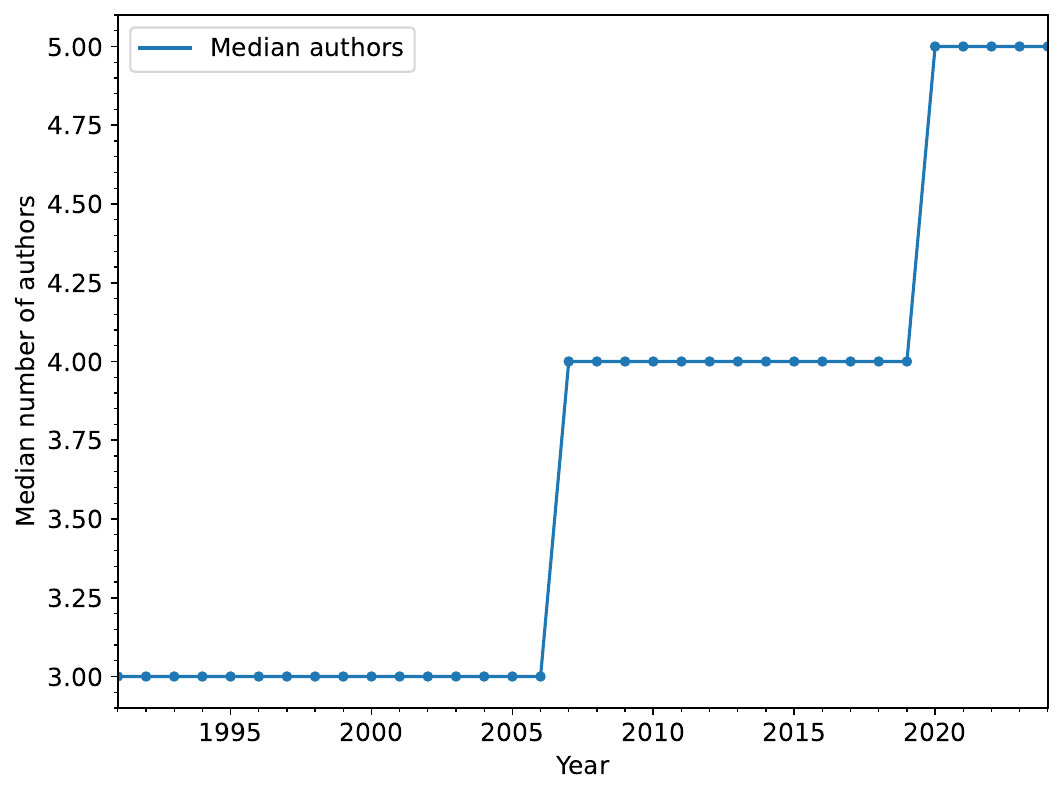}
		\caption{Electrical and Electronics Engineering}
	\end{subfigure}
	
	\begin{subfigure}[t]{0.32\textwidth}
		\centering
		\includegraphics[width=\linewidth,keepaspectratio]{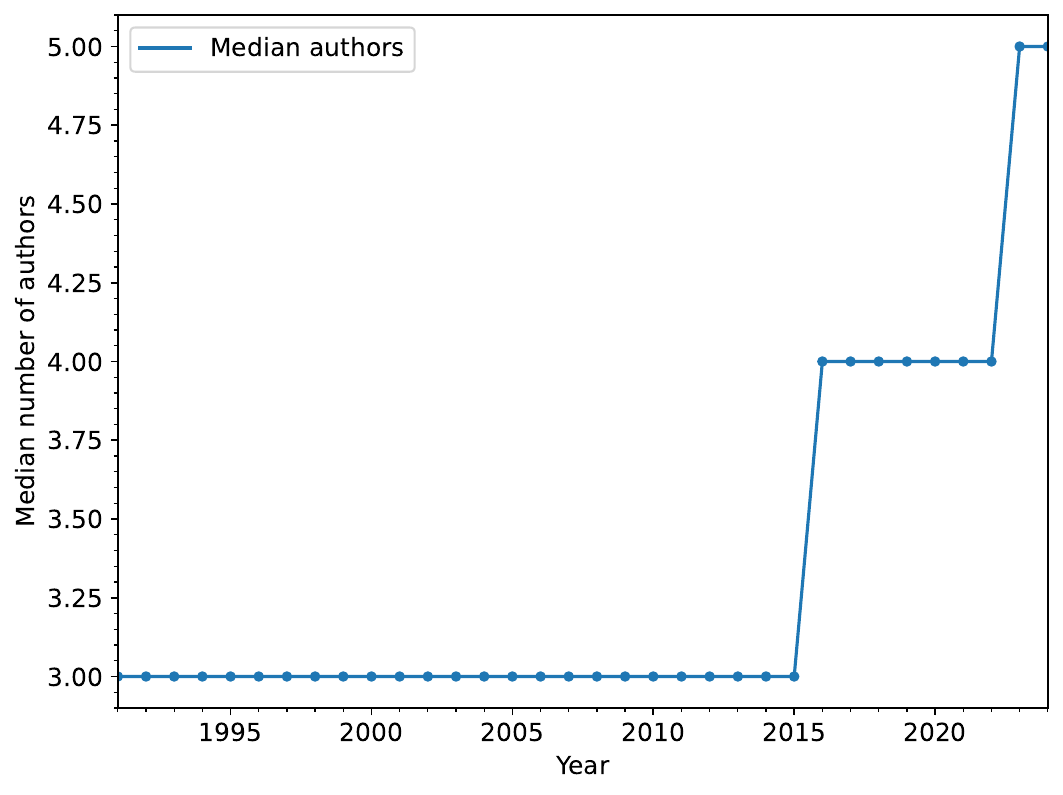}
		\caption{Computer Science and Engineering}
	\end{subfigure}
	\hfill
	\begin{subfigure}[t]{0.32\textwidth}
		\centering
		\includegraphics[width=\linewidth,keepaspectratio]{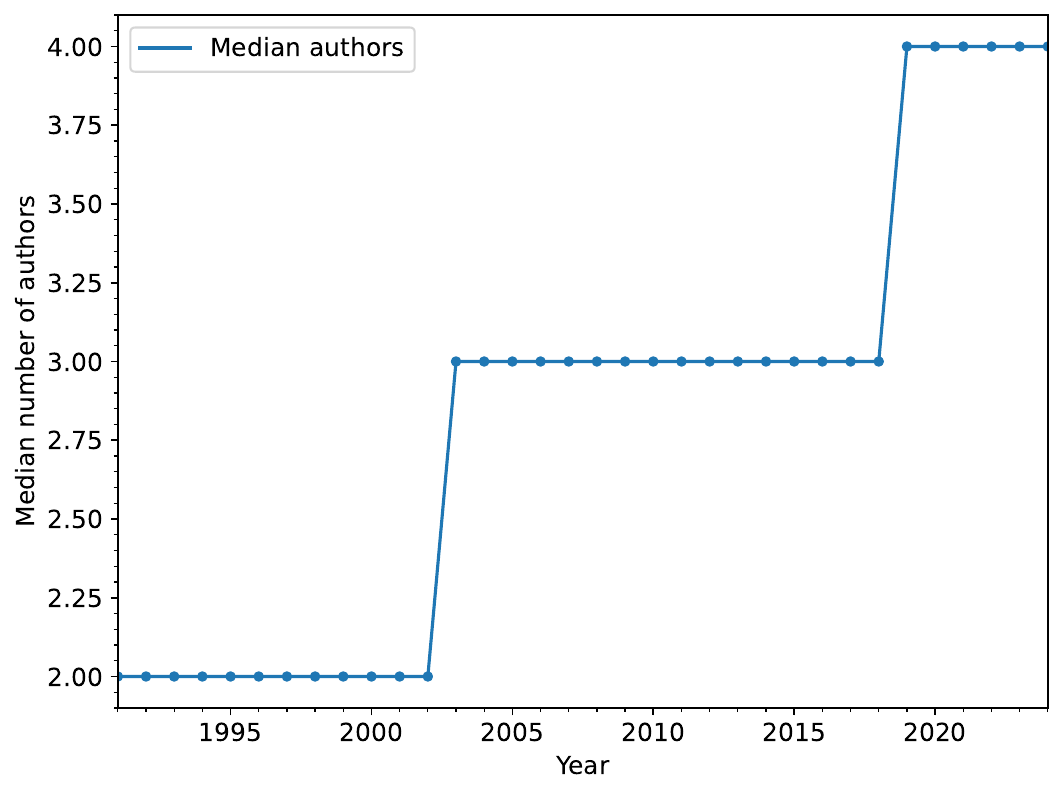}
		\caption{Psychology}
	\end{subfigure}
	\hfill
	\begin{subfigure}[t]{0.32\textwidth}
		\centering
		\includegraphics[width=\linewidth,keepaspectratio]{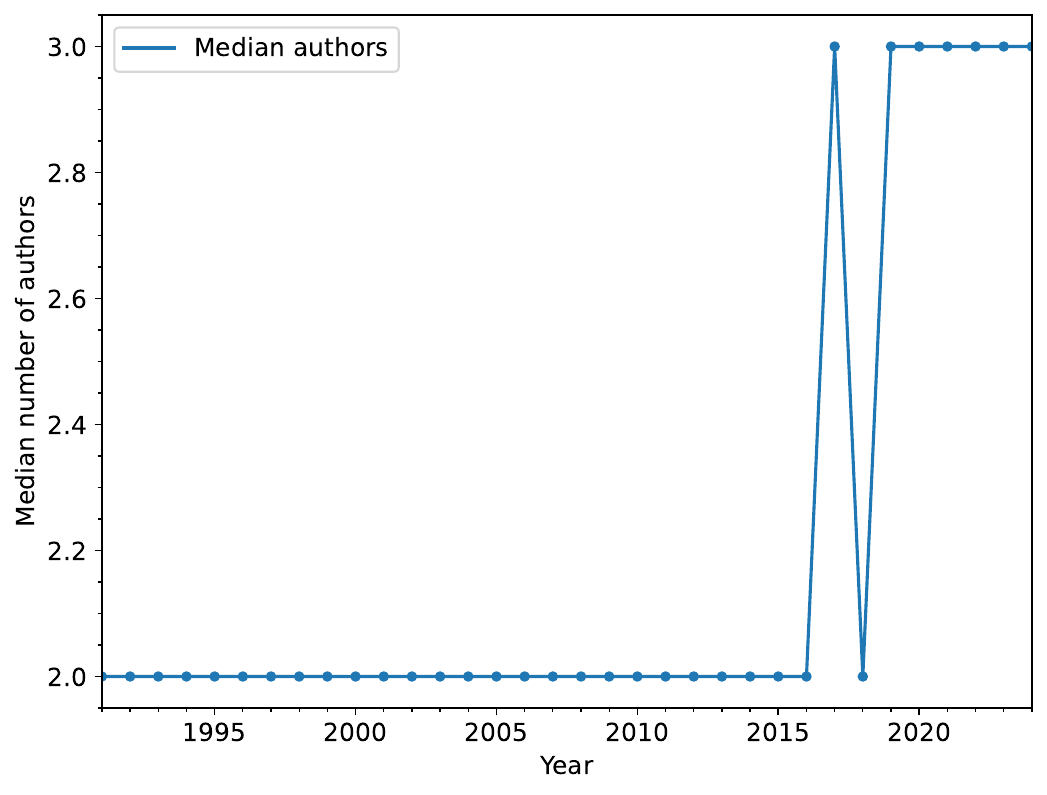}
		\caption{Sociology}
	\end{subfigure}
	
	\begin{subfigure}[t]{0.32\textwidth}
		\centering
		\includegraphics[width=\linewidth,keepaspectratio]{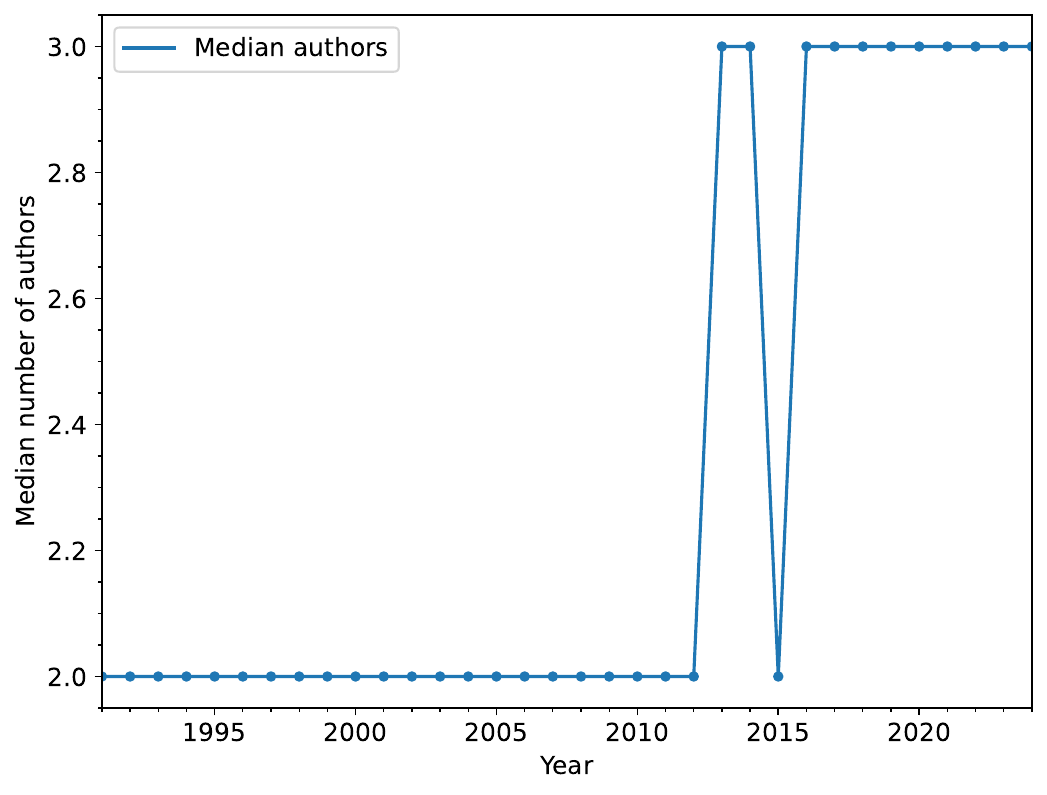}
		\caption{Economics}
	\end{subfigure}
	\hfill
	\begin{subfigure}[t]{0.32\textwidth}
		\centering
		\includegraphics[width=\linewidth,keepaspectratio]{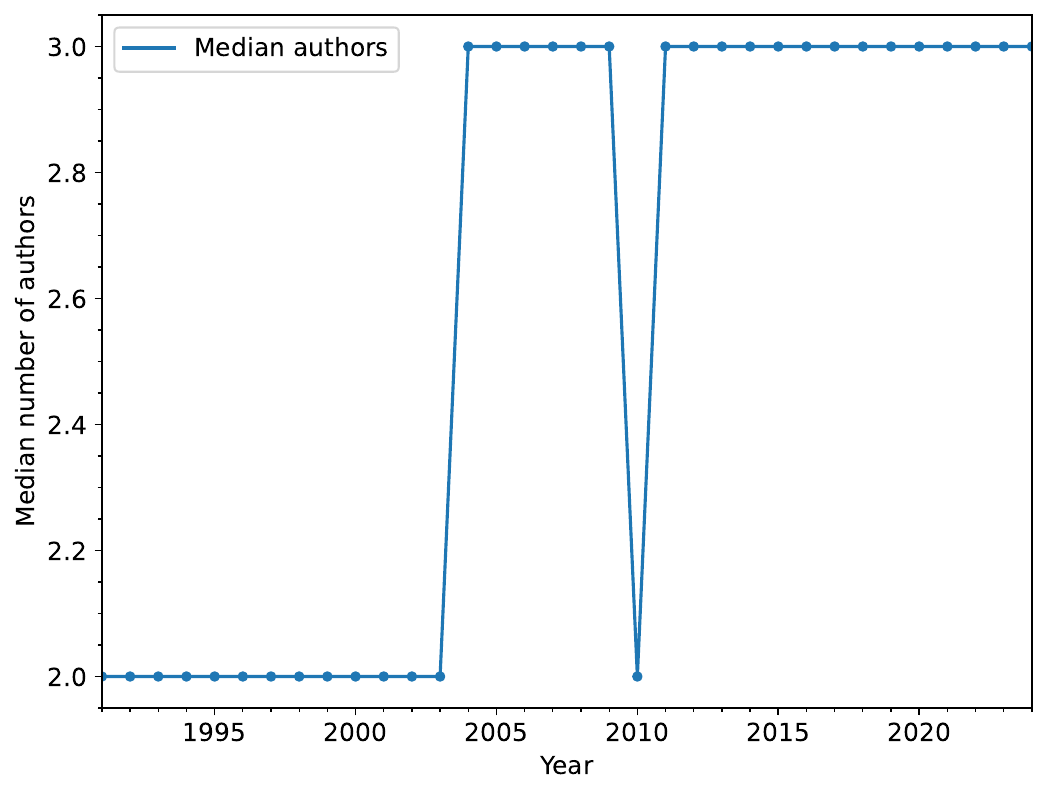}
		\caption{Marketing}
	\end{subfigure}
	\hfill
	\begin{subfigure}[t]{0.32\textwidth}
		\centering
		\includegraphics[width=\linewidth,keepaspectratio]{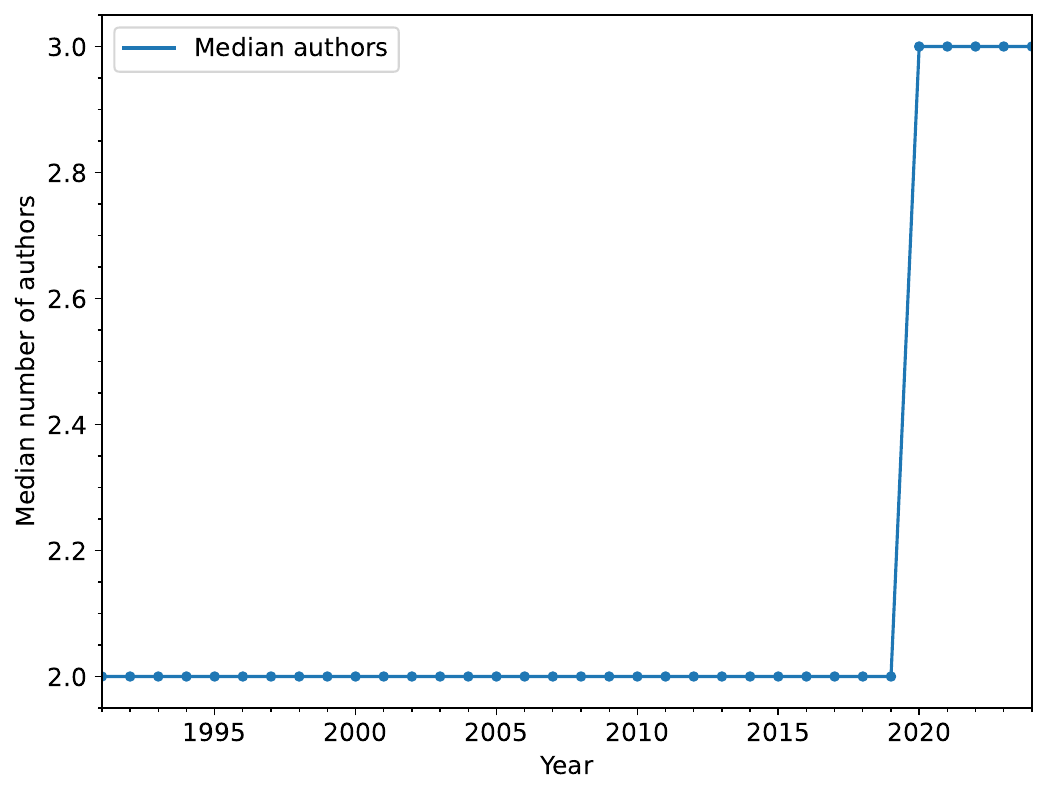}
		\caption{Mathematics}
	\end{subfigure}
	
	\caption{The median number of authors per year in the combined and field datasets.}
	\label{fig:median_field_trend}
\end{figure*}
\FloatBarrier
\clearpage

\section{The Q3 statistic number of authors per year in the combined and field datasets}

\label{Q3_trend_field}
\begin{figure*}[!htbp]
	\centering
	\begin{subfigure}[t]{0.32\textwidth}
		\centering
		\includegraphics[width=\linewidth,keepaspectratio]{plots/third_quantile_authors_per_year.pdf}
		\caption{Combined}
	\end{subfigure}
	\hfill
	\begin{subfigure}[t]{0.32\textwidth}
		\centering
		\includegraphics[width=\linewidth,keepaspectratio]{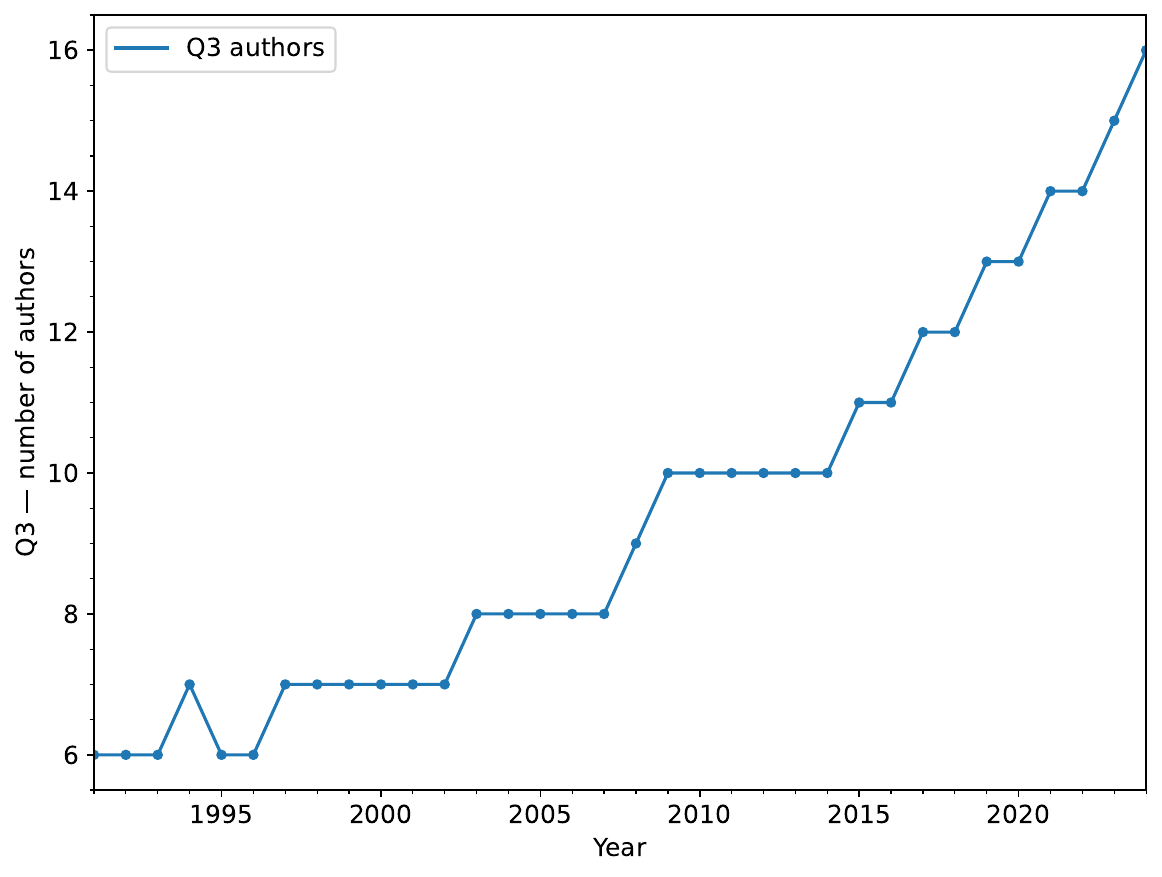}
		\caption{Biology}
	\end{subfigure}
	\hfill
	\begin{subfigure}[t]{0.32\textwidth}
		\centering
		\includegraphics[width=\linewidth,keepaspectratio]{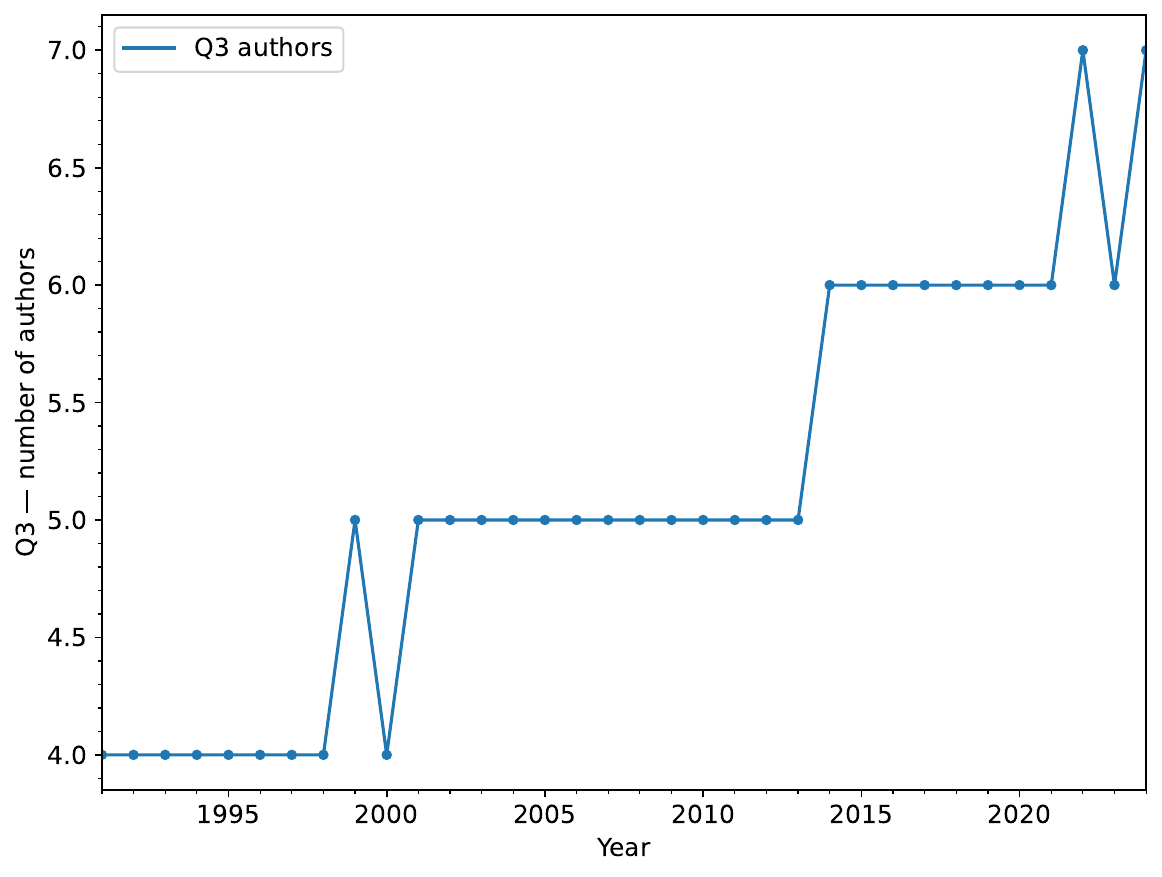}
		\caption{Electrical and Electronics Engineering}
	\end{subfigure}
	
	\begin{subfigure}[t]{0.32\textwidth}
		\centering
		\includegraphics[width=\linewidth,keepaspectratio]{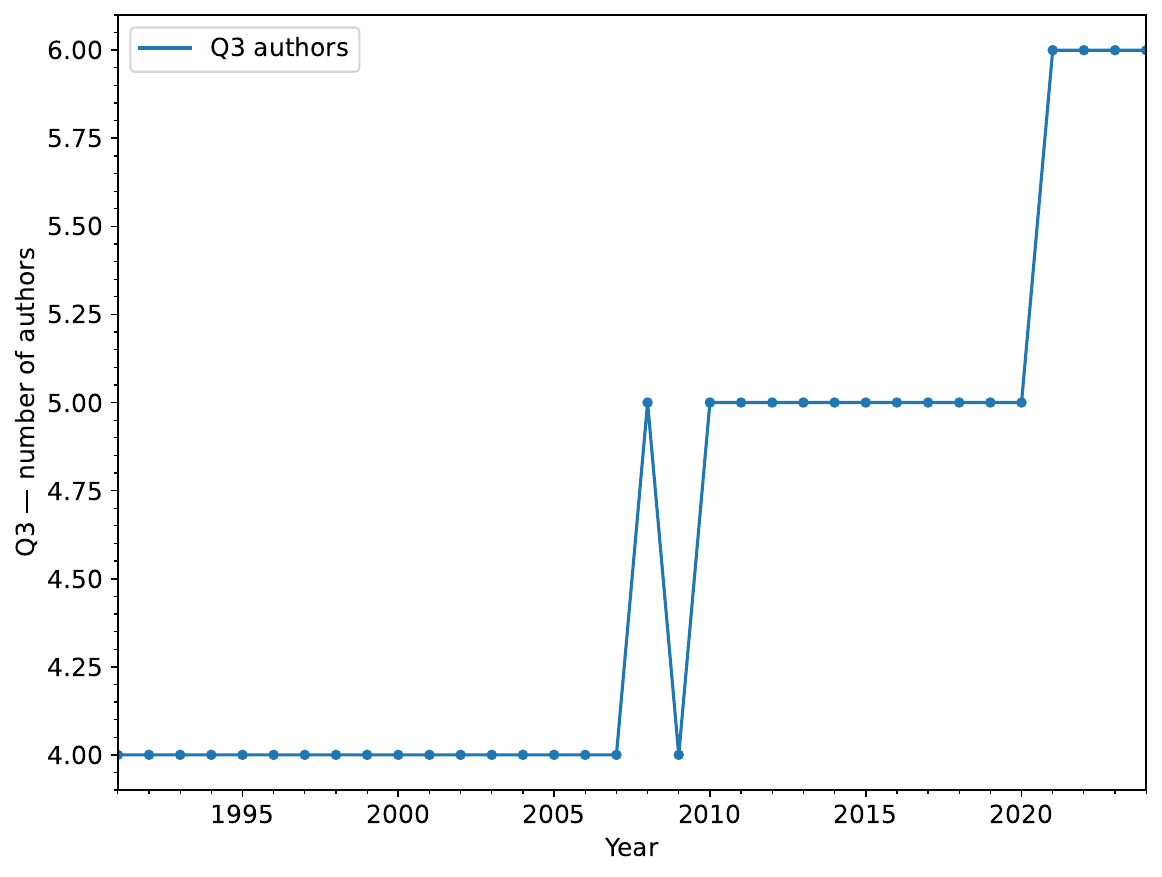}
		\caption{Computer Science and Engineering}
	\end{subfigure}
	\hfill
	\begin{subfigure}[t]{0.32\textwidth}
		\centering
		\includegraphics[width=\linewidth,keepaspectratio]{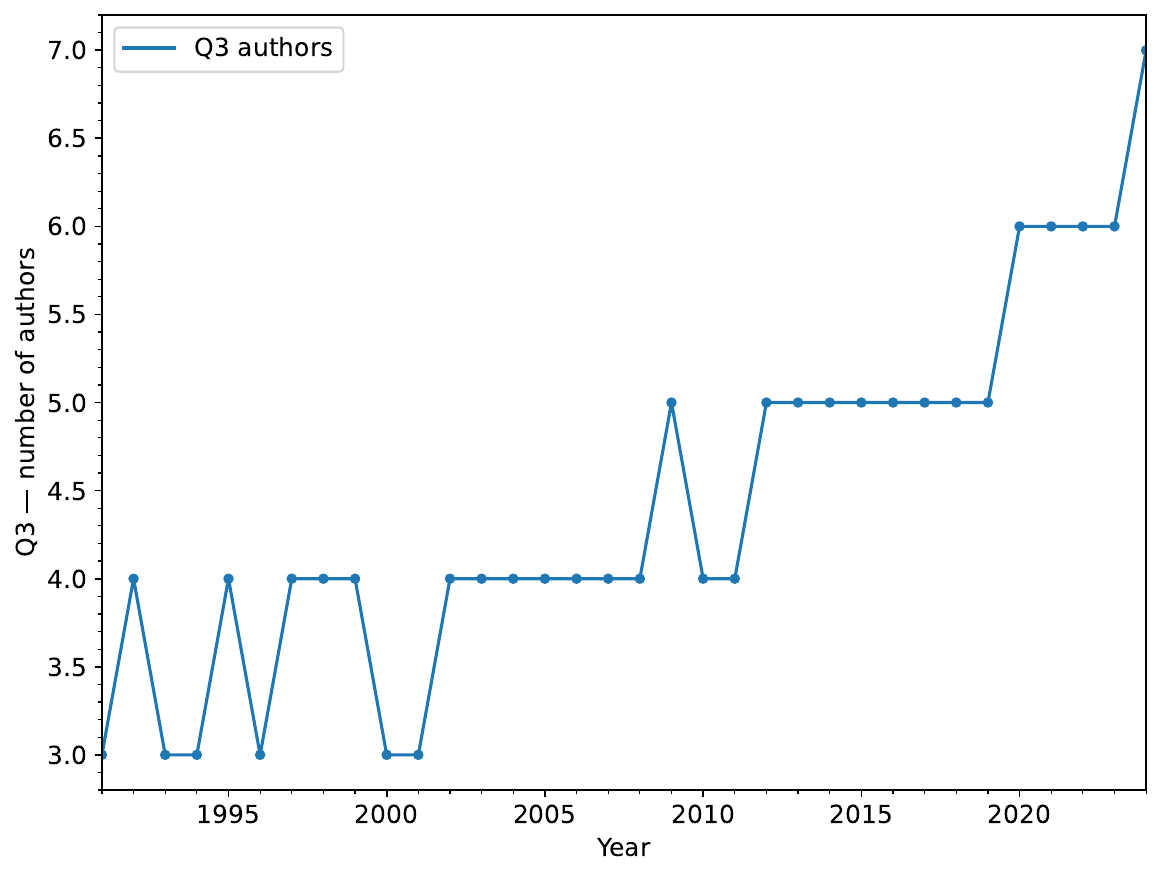}
		\caption{Psychology}
	\end{subfigure}
	\hfill
	\begin{subfigure}[t]{0.32\textwidth}
		\centering
		\includegraphics[width=\linewidth,keepaspectratio]{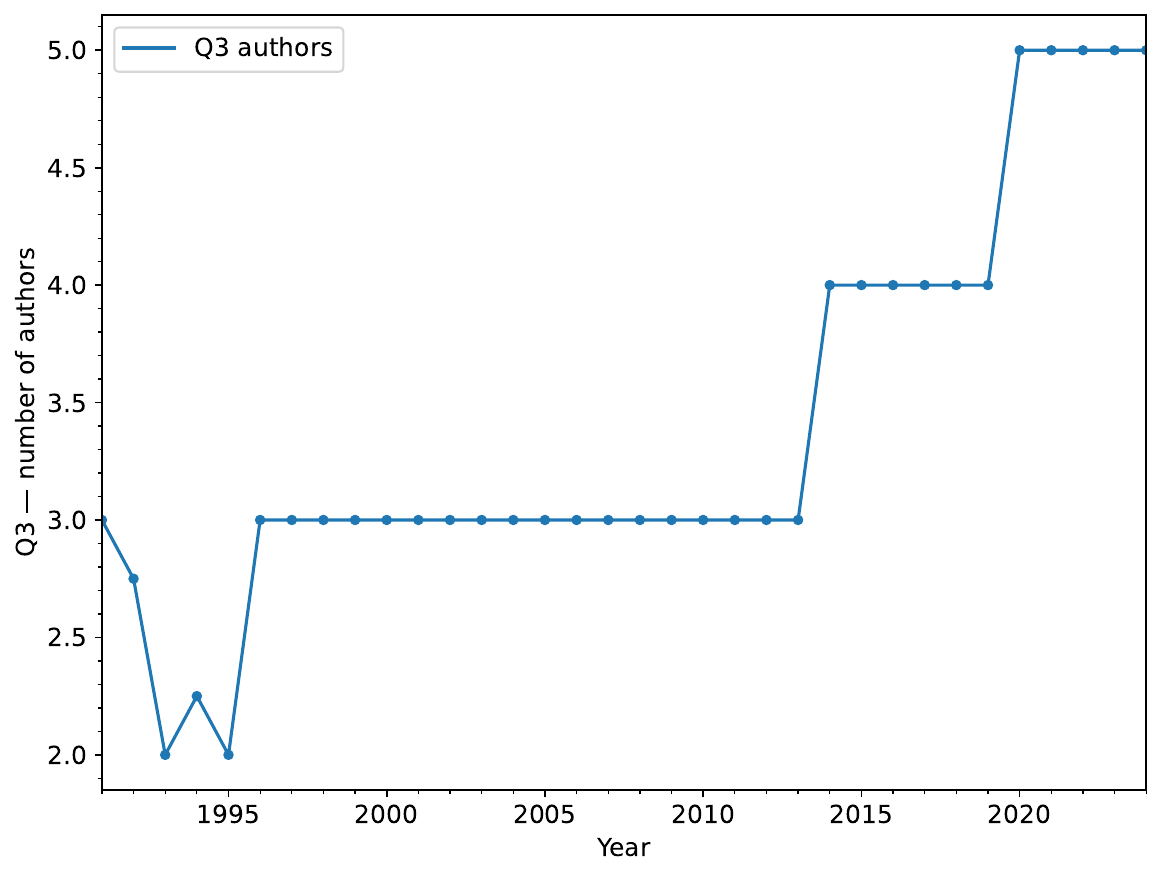}
		\caption{Sociology}
	\end{subfigure}
	
	\begin{subfigure}[t]{0.32\textwidth}
		\centering
		\includegraphics[width=\linewidth,keepaspectratio]{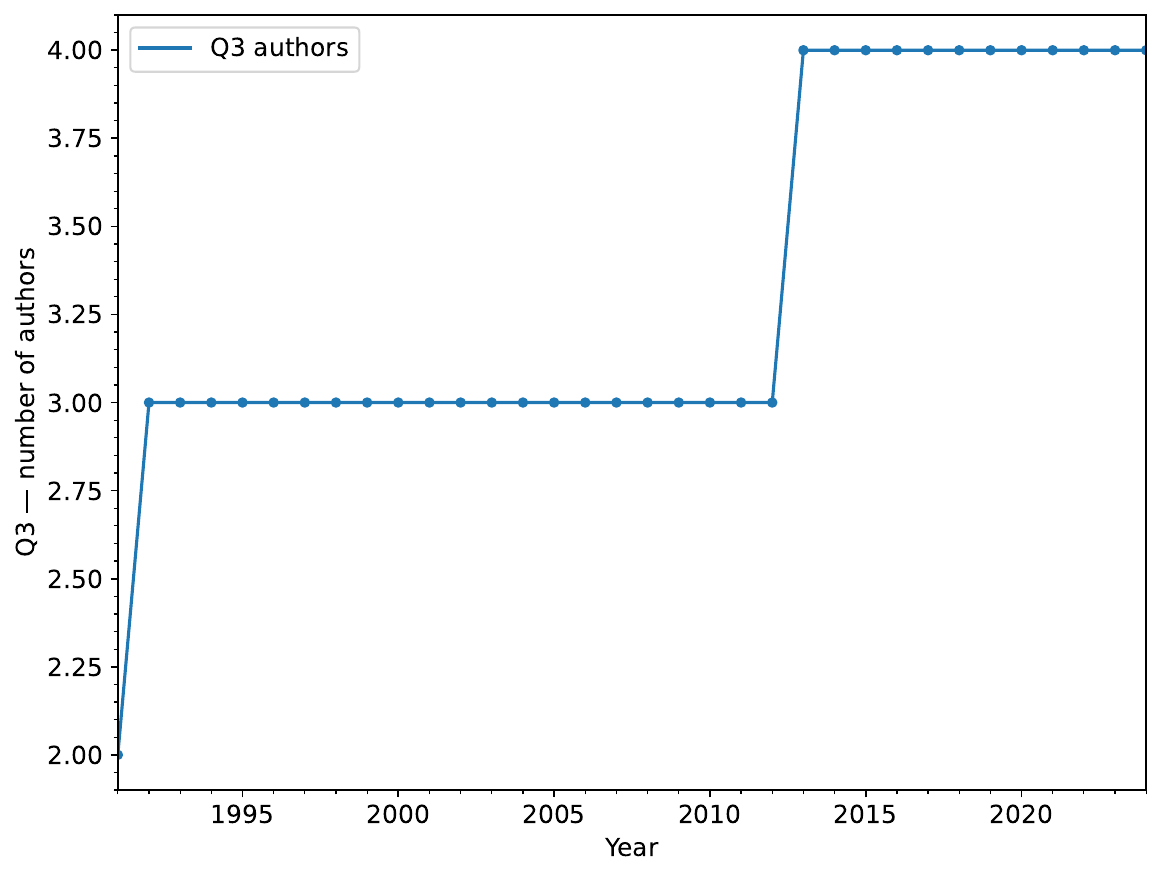}
		\caption{Economics}
	\end{subfigure}
	\hfill
	\begin{subfigure}[t]{0.32\textwidth}
		\centering
		\includegraphics[width=\linewidth,keepaspectratio]{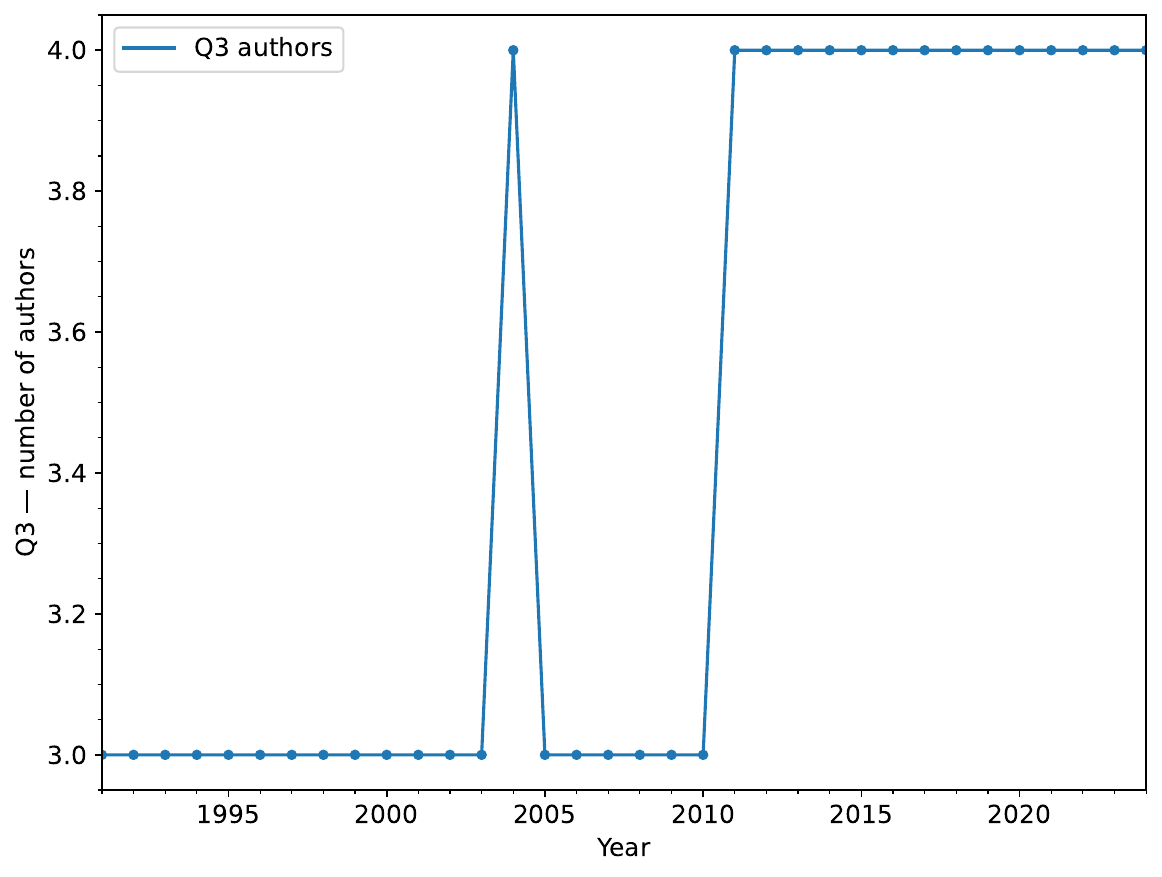}
		\caption{Marketing}
	\end{subfigure}
	\hfill
	\begin{subfigure}[t]{0.32\textwidth}
		\centering
		\includegraphics[width=\linewidth,keepaspectratio]{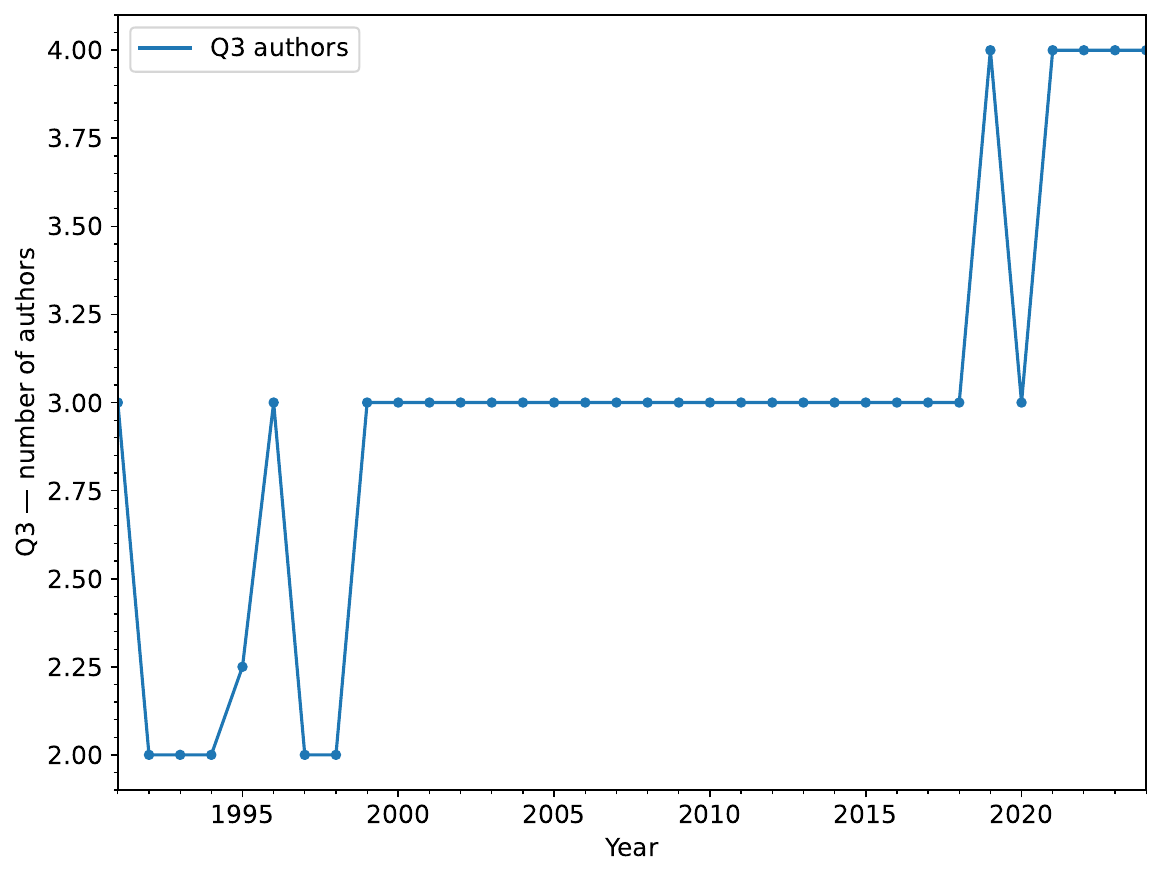}
		\caption{Mathematics}
	\end{subfigure}
	
	\caption{The Q3 number of authors per year in the combined and field datasets.}
	\label{fig:Q3_field_trend}
\end{figure*}
\clearpage
\FloatBarrier

\section{The Q1 statistic number of authors per year in the combined and field datasets}

\label{Q1_trend_field}
\begin{figure*}[!htbp]
	\centering
	\begin{subfigure}[t]{0.32\textwidth}
		\centering
		\includegraphics[width=\linewidth,keepaspectratio]{plots/first_quantile_authors_per_year.pdf}
		\caption{Combined}
	\end{subfigure}
	\hfill
	\begin{subfigure}[t]{0.32\textwidth}
		\centering
		\includegraphics[width=\linewidth,keepaspectratio]{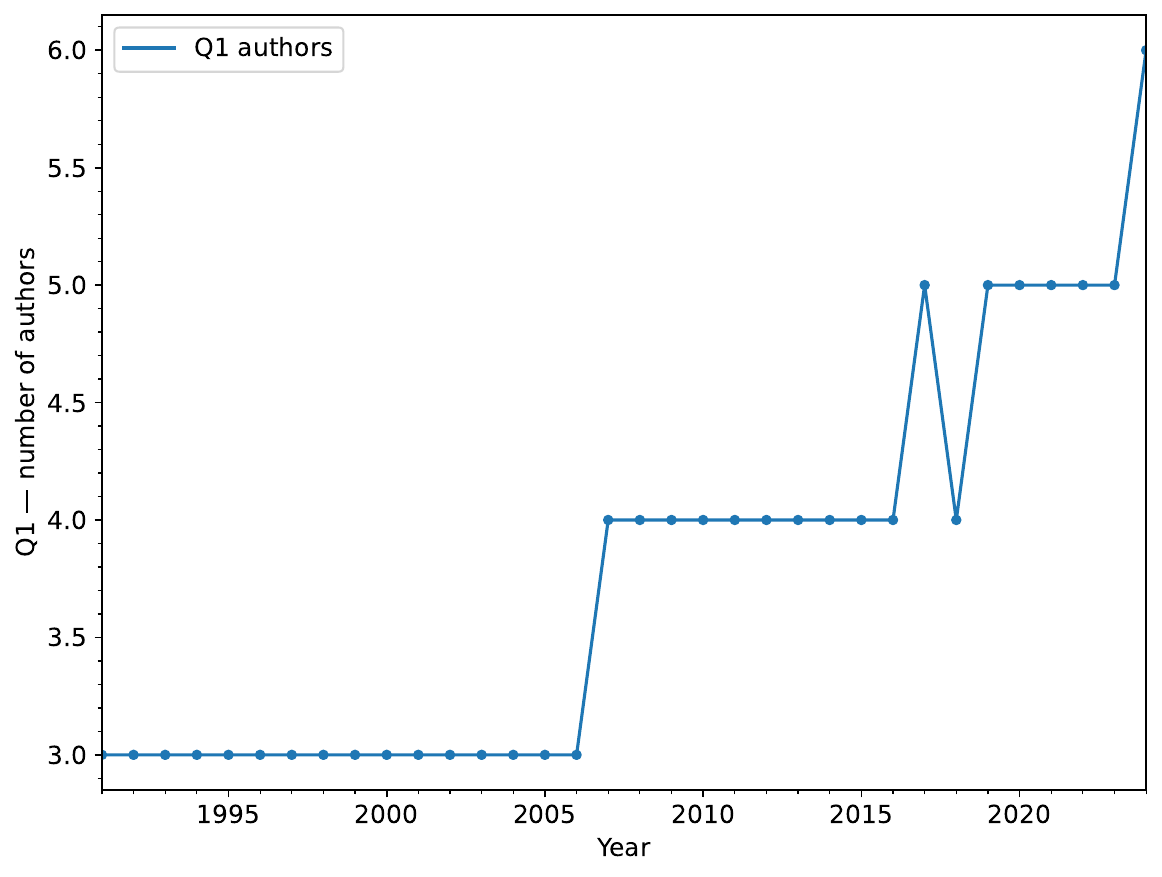}
		\caption{Biology}
	\end{subfigure}
	\hfill
	\begin{subfigure}[t]{0.32\textwidth}
		\centering
		\includegraphics[width=\linewidth,keepaspectratio]{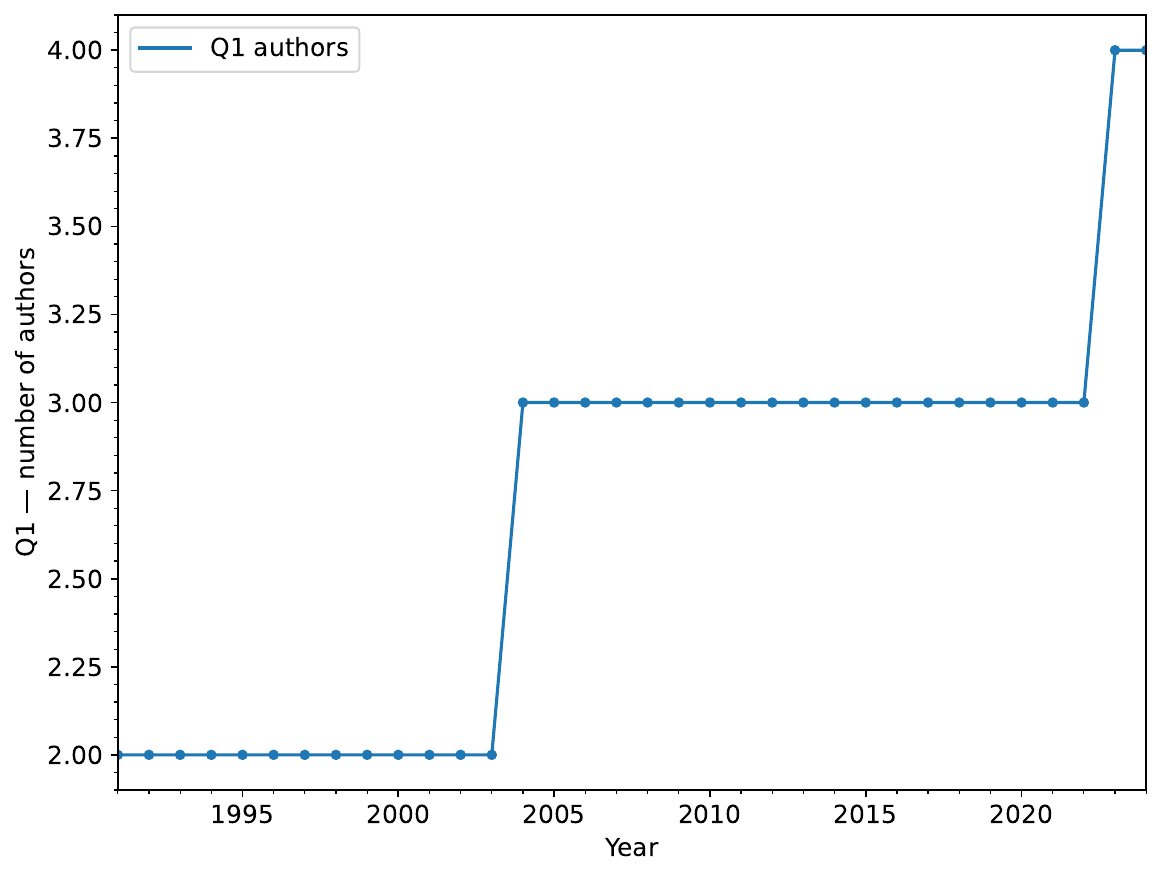}
		\caption{Electrical and Electronics Engineering}
	\end{subfigure}
	
	\begin{subfigure}[t]{0.32\textwidth}
		\centering
		\includegraphics[width=\linewidth,keepaspectratio]{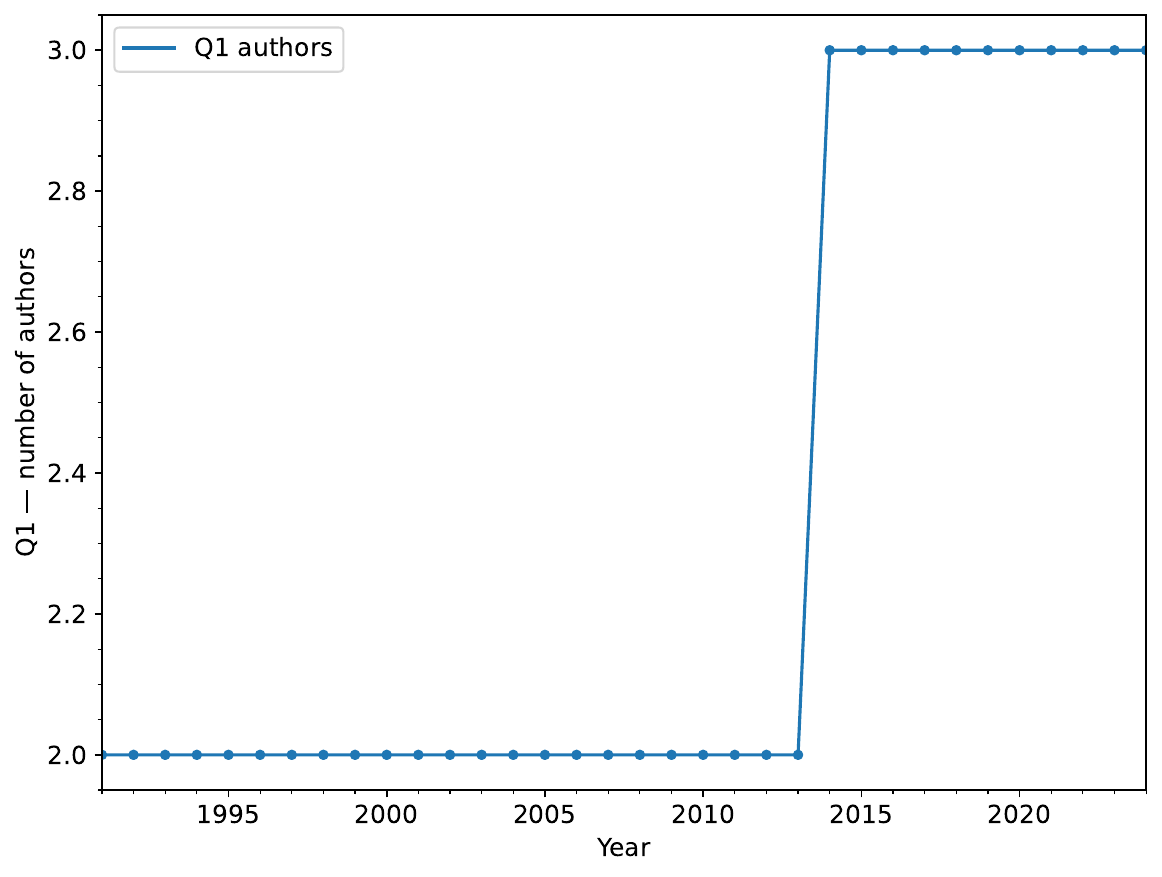}
		\caption{Computer Science and Engineering}
	\end{subfigure}
	\hfill
	\begin{subfigure}[t]{0.32\textwidth}
		\centering
		\includegraphics[width=\linewidth,keepaspectratio]{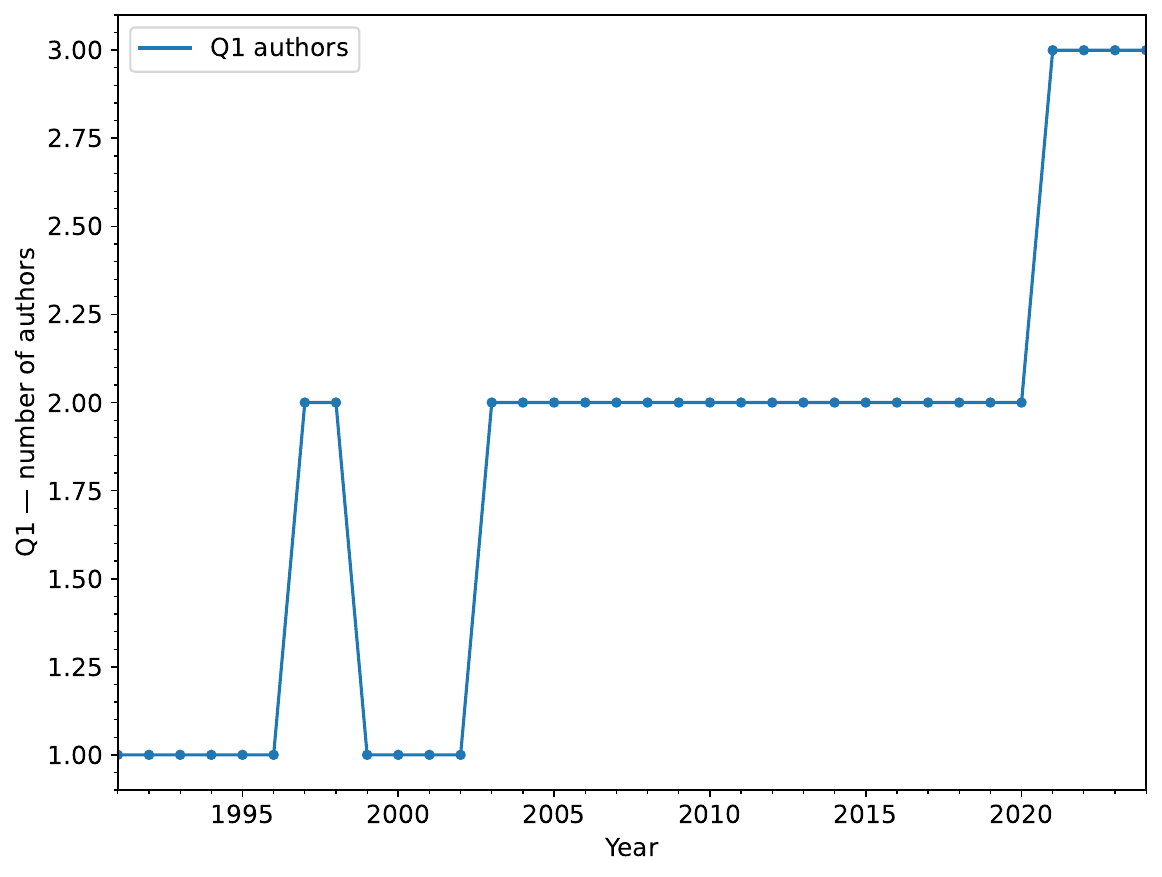}
		\caption{Psychology}
	\end{subfigure}
	\hfill
	\begin{subfigure}[t]{0.32\textwidth}
		\centering
		\includegraphics[width=\linewidth,keepaspectratio]{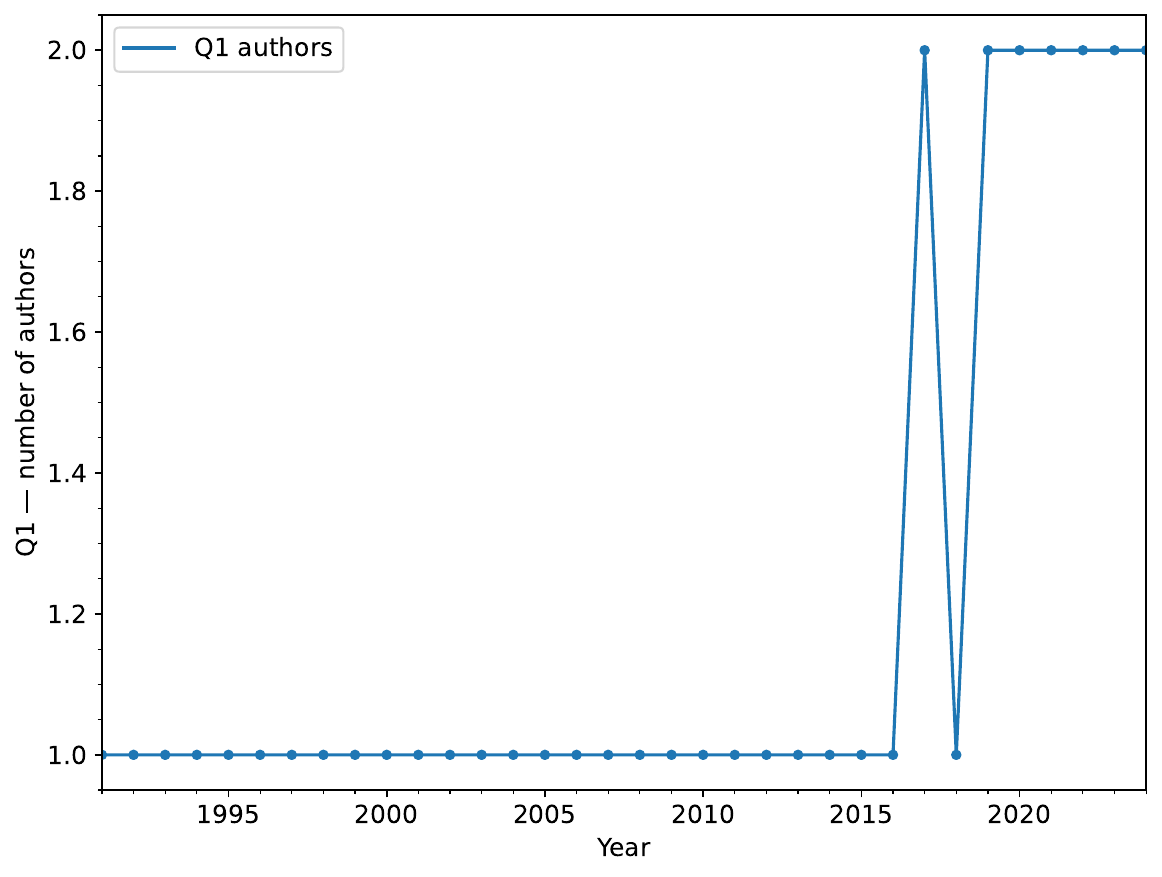}
		\caption{Sociology}
	\end{subfigure}
	
	\begin{subfigure}[t]{0.32\textwidth}
		\centering
		\includegraphics[width=\linewidth,keepaspectratio]{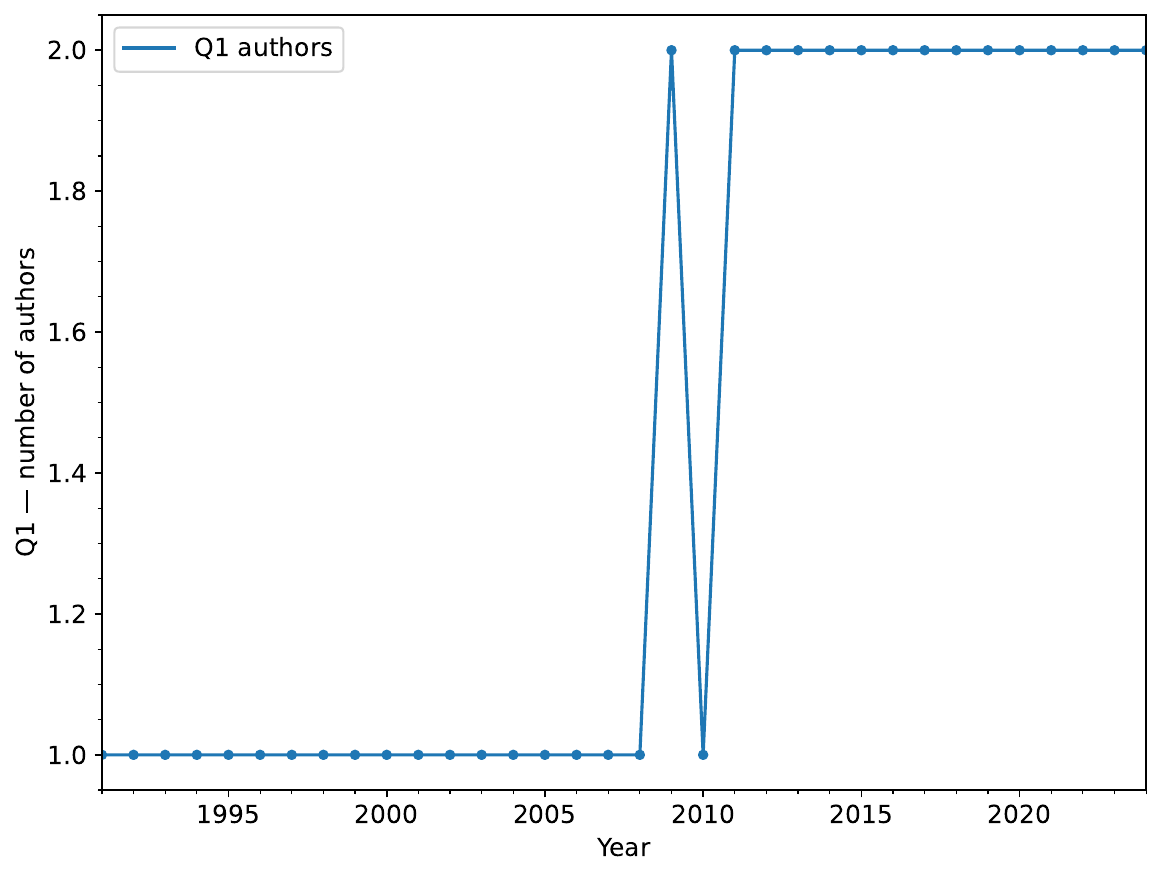}
		\caption{Economics}
	\end{subfigure}
	\hfill
	\begin{subfigure}[t]{0.32\textwidth}
		\centering
		\includegraphics[width=\linewidth,keepaspectratio]{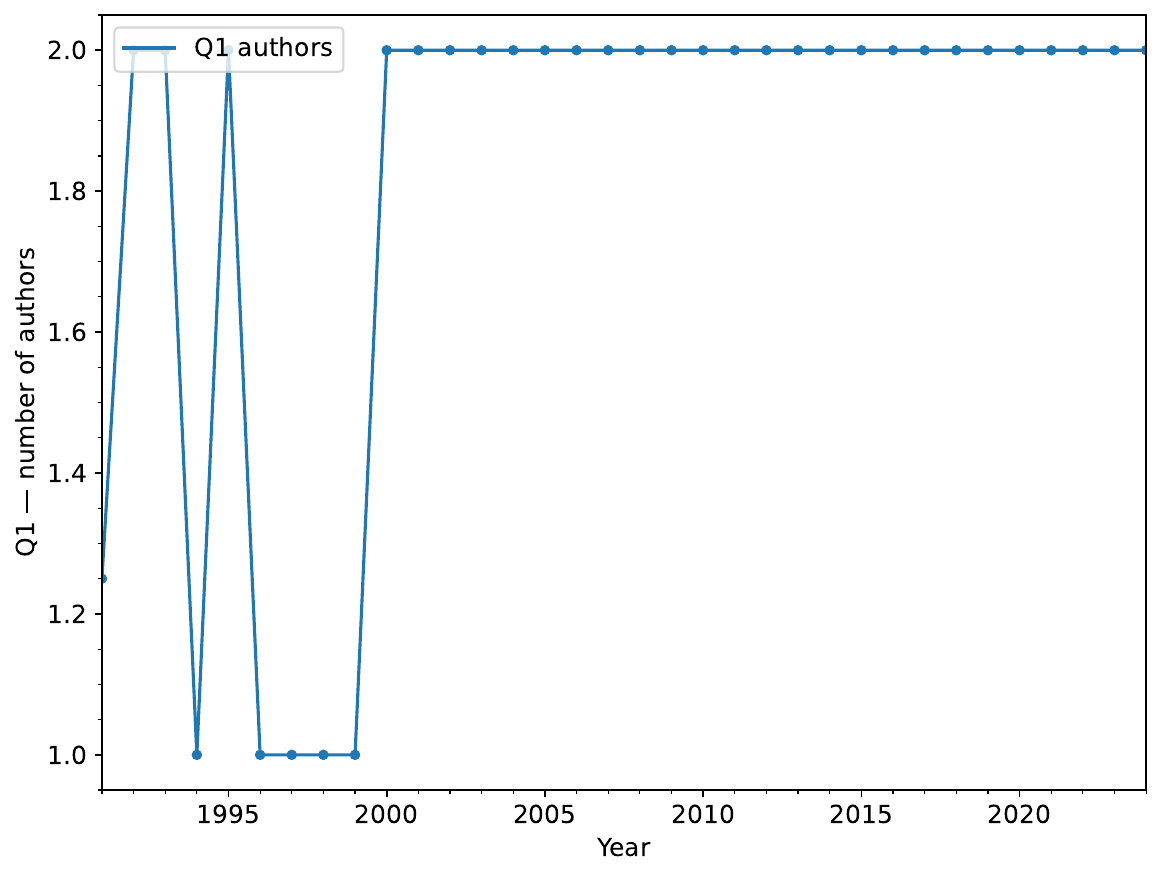}
		\caption{Marketing}
	\end{subfigure}
	\hfill
	\begin{subfigure}[t]{0.32\textwidth}
		\centering
		\includegraphics[width=\linewidth,keepaspectratio]{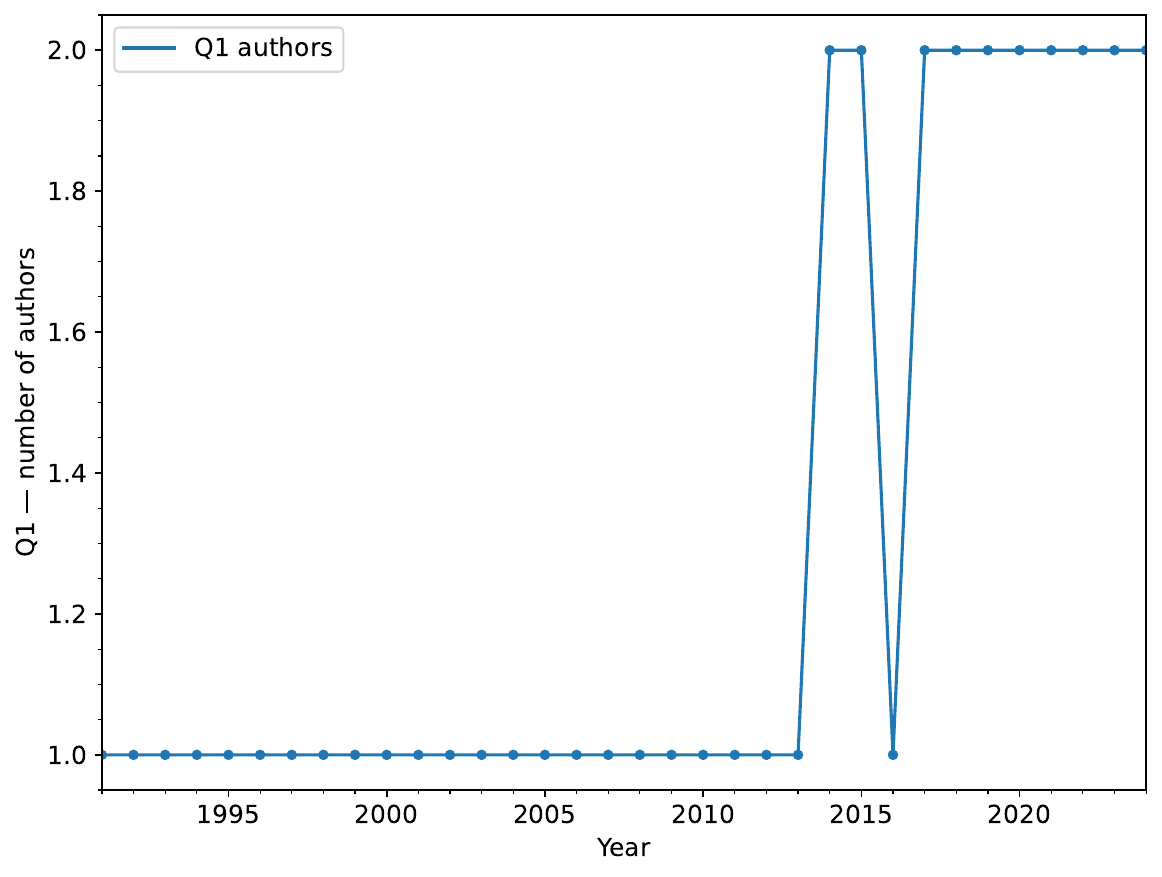}
		\caption{Mathematics}
	\end{subfigure}
	
	\caption{The Q1 number of authors per year in the combined and field datasets.}
	\label{fig:Q1_field_trend}
\end{figure*}
\FloatBarrier
\clearpage

\section{Distribution of Number of Authors in the Combined and Field Datasets}

\label{dist_pub_each_field}
\begin{figure*}[!htbp]
	\centering
	\begin{subfigure}[t]{0.32\textwidth}
		\centering
		\includegraphics[width=\linewidth,keepaspectratio]{plots/dist_y_log.pdf}
		\caption{Combined}
	\end{subfigure}
	\hfill
	\begin{subfigure}[t]{0.32\textwidth}
		\centering
		\includegraphics[width=\linewidth,keepaspectratio]{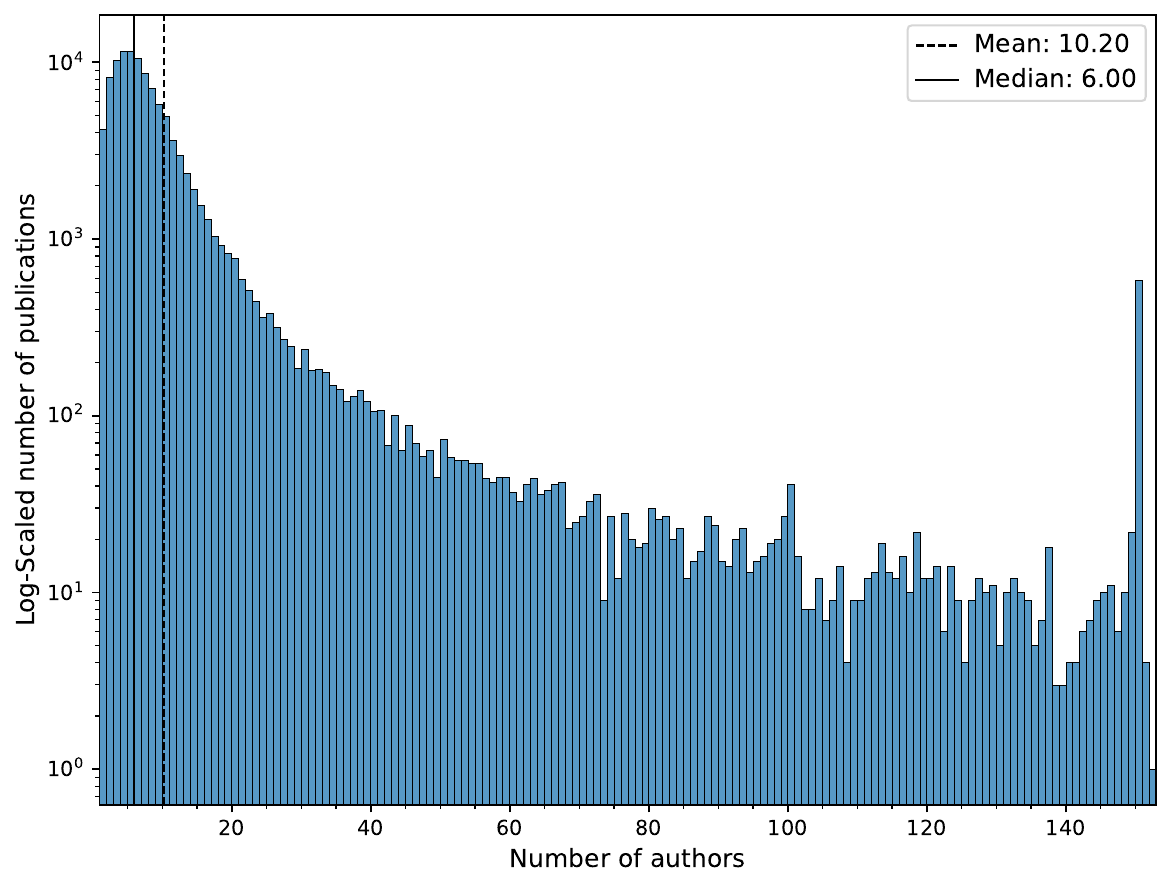}
		\caption{Biology}
	\end{subfigure}
	\hfill
	\begin{subfigure}[t]{0.32\textwidth}
		\centering
		\includegraphics[width=\linewidth,keepaspectratio]{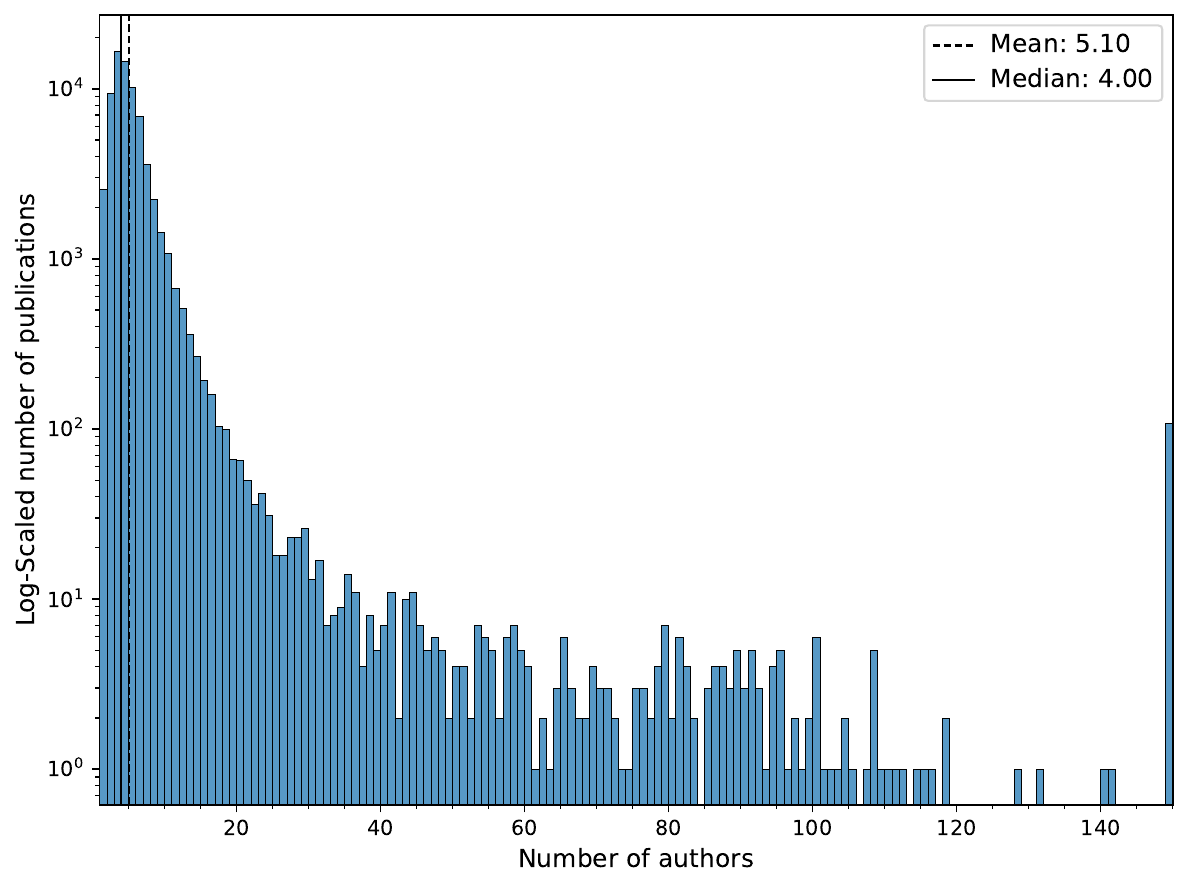}
		\caption{Electrical and Electronics Engineering}
	\end{subfigure}
	
	\begin{subfigure}[t]{0.32\textwidth}
		\centering
		\includegraphics[width=\linewidth,keepaspectratio]{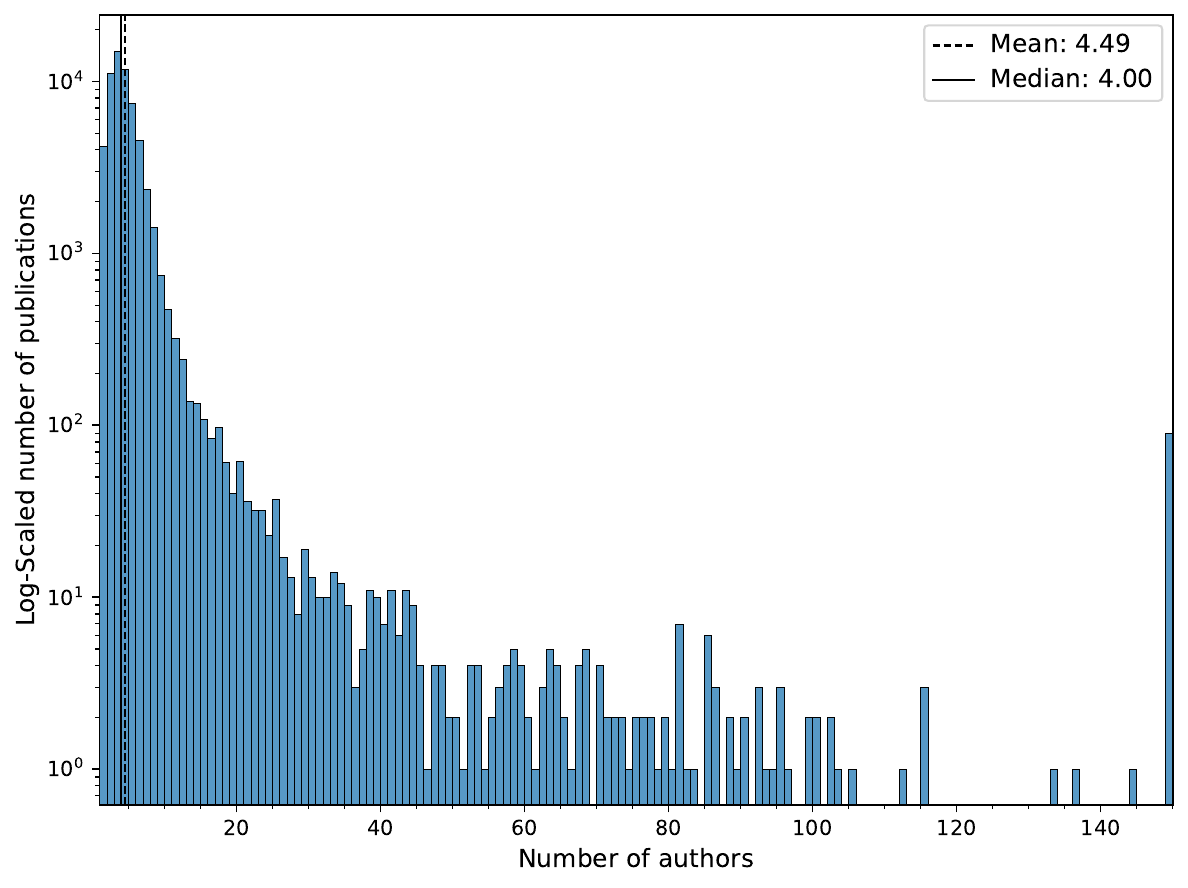}
		\caption{Computer Science and Engineering}
	\end{subfigure}
	\hfill
	\begin{subfigure}[t]{0.32\textwidth}
		\centering
		\includegraphics[width=\linewidth,keepaspectratio]{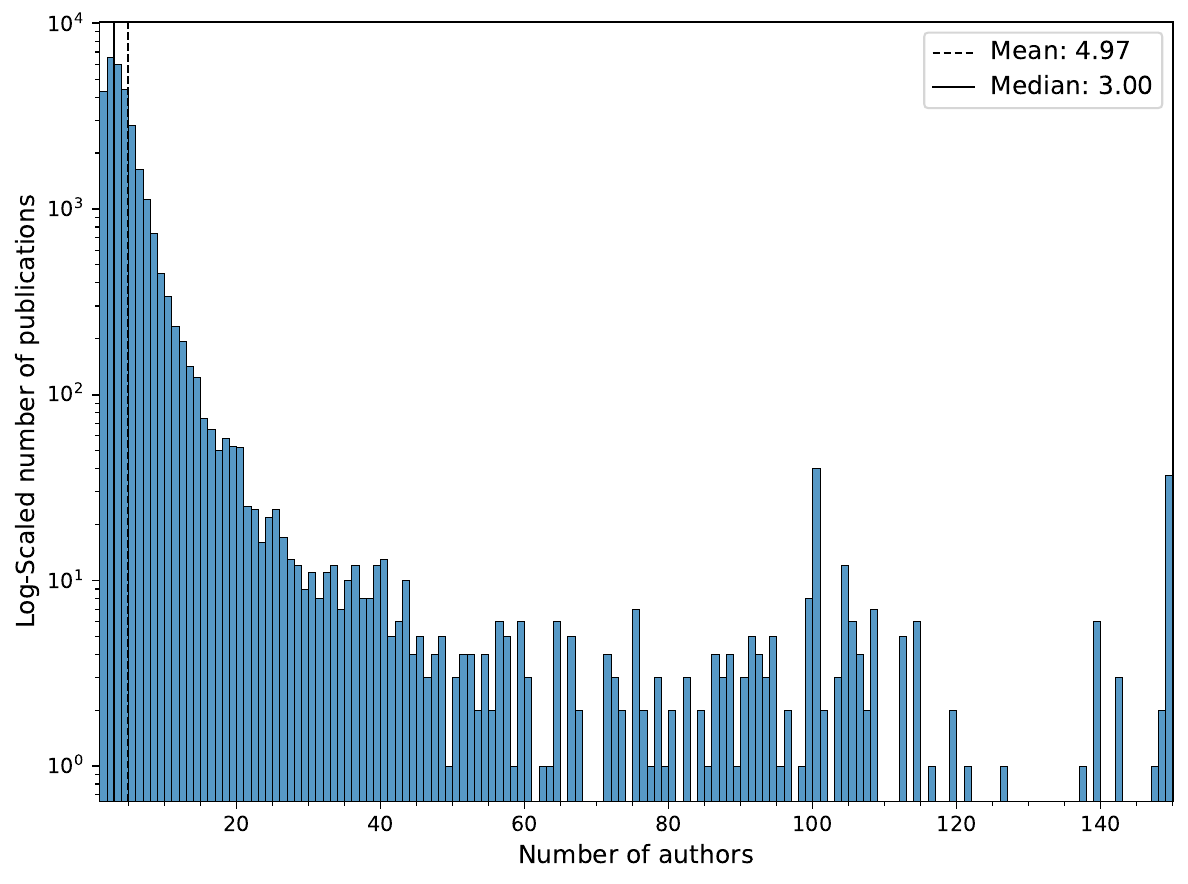}
		\caption{Psychology}
	\end{subfigure}
	\hfill
	\begin{subfigure}[t]{0.32\textwidth}
		\centering
		\includegraphics[width=\linewidth,keepaspectratio]{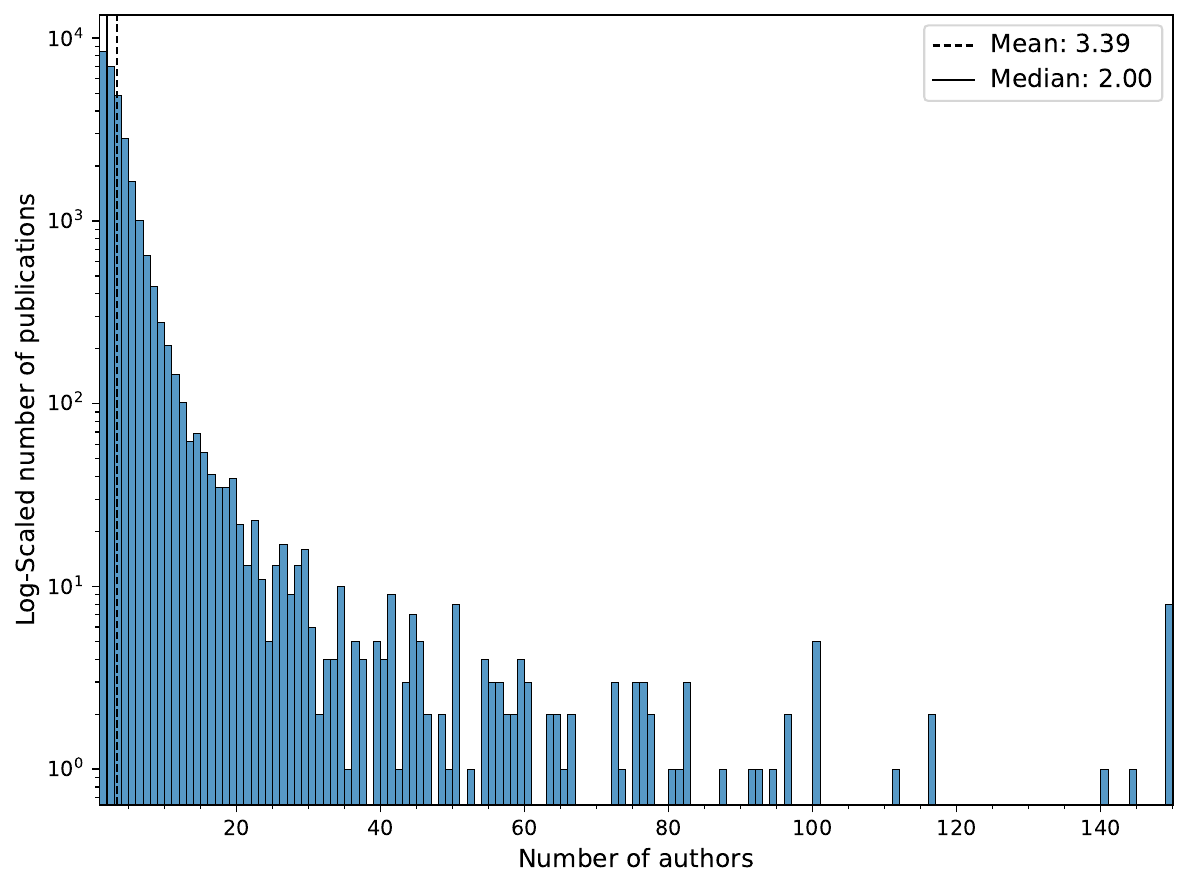}
		\caption{Sociology}
	\end{subfigure}
	
	\begin{subfigure}[t]{0.32\textwidth}
		\centering
		\includegraphics[width=\linewidth,keepaspectratio]{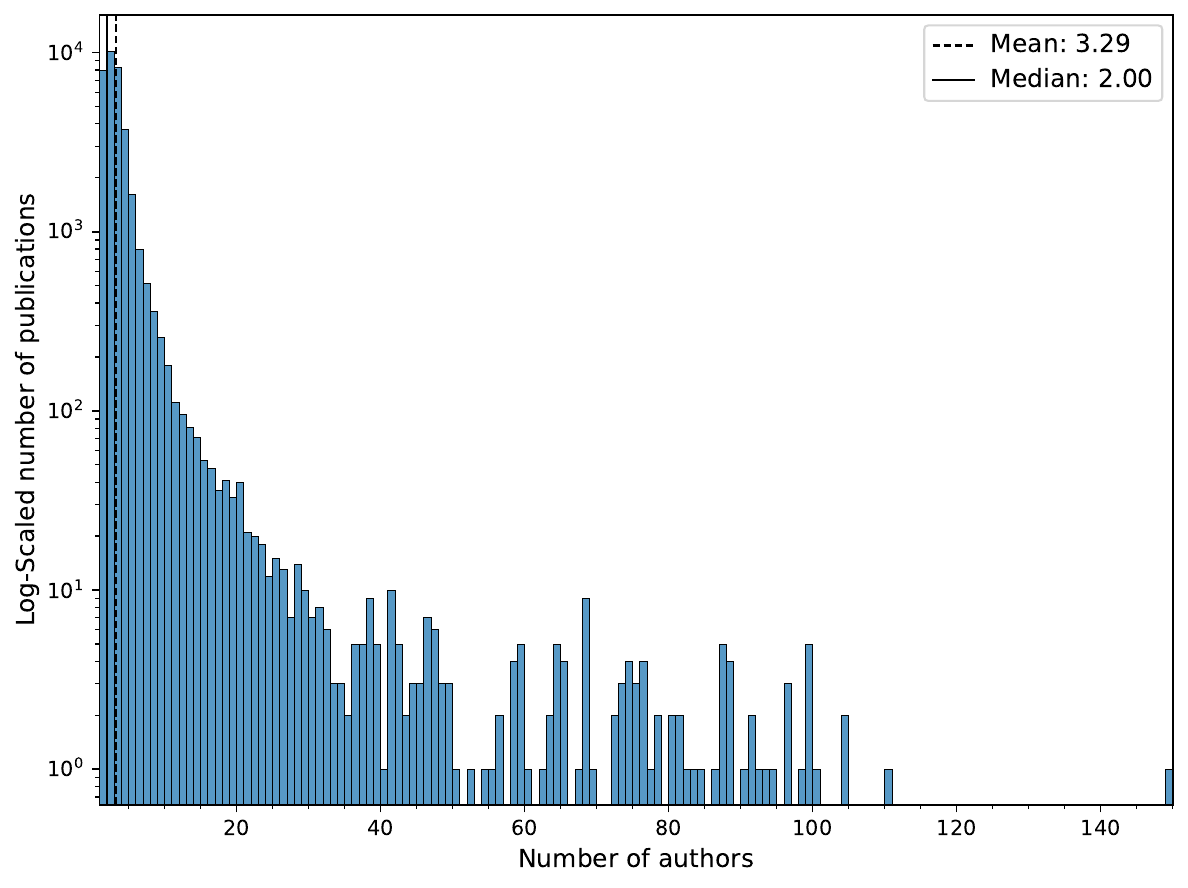}
		\caption{Economics}
	\end{subfigure}
	\hfill
	\begin{subfigure}[t]{0.32\textwidth}
		\centering
		\includegraphics[width=\linewidth,keepaspectratio]{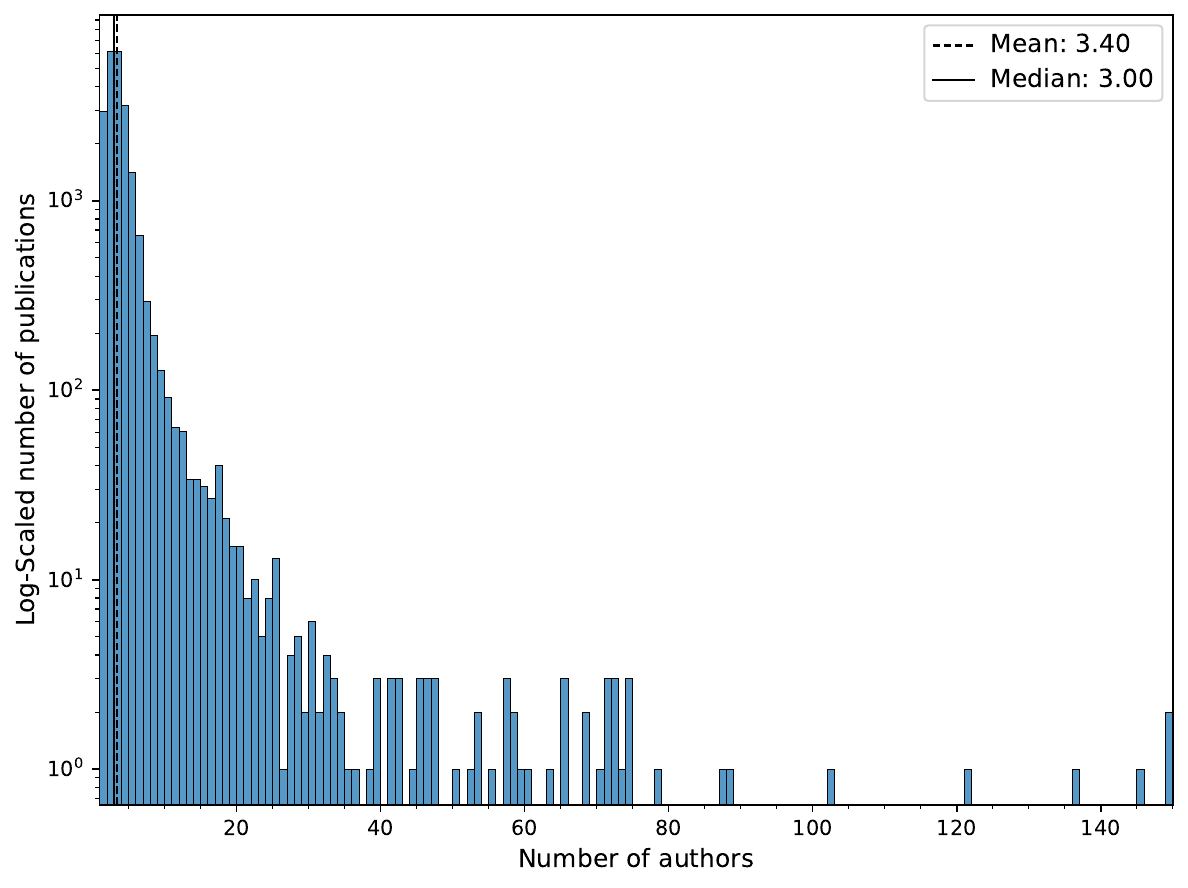}
		\caption{Marketing}
	\end{subfigure}
	\hfill
	\begin{subfigure}[t]{0.32\textwidth}
		\centering
		\includegraphics[width=\linewidth,keepaspectratio]{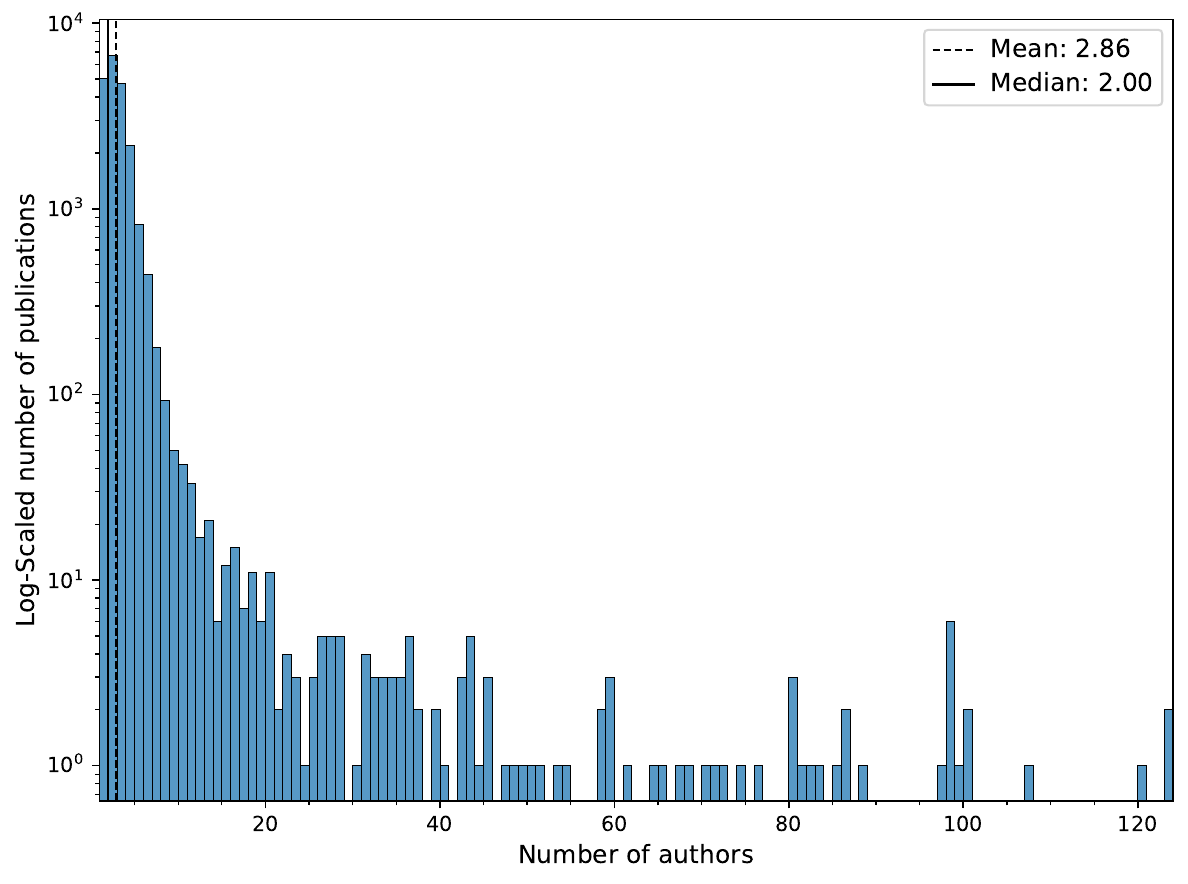}
		\caption{Mathematics}
	\end{subfigure}
	
	\caption{Log transformed distribution of number of authors in the combined and field datasets.}
	\label{fig:dist_pub_field}
\end{figure*}
\clearpage
\FloatBarrier

\section{Percentage of Publications Based on Fibinacci-binned Number of Authors Per Year}
\label{appendix:percent_pub_interval}
\begin{figure*}[!htbp]
	\centering
	\begin{subfigure}[t]{\textwidth}
		\centering
		\includegraphics[width=\linewidth,keepaspectratio]{plots/heatmap.pdf}
		\caption{Combined}
	\end{subfigure}
	\hfill
	\begin{subfigure}[t]{\textwidth}
		\centering
		\includegraphics[width=\linewidth,keepaspectratio]{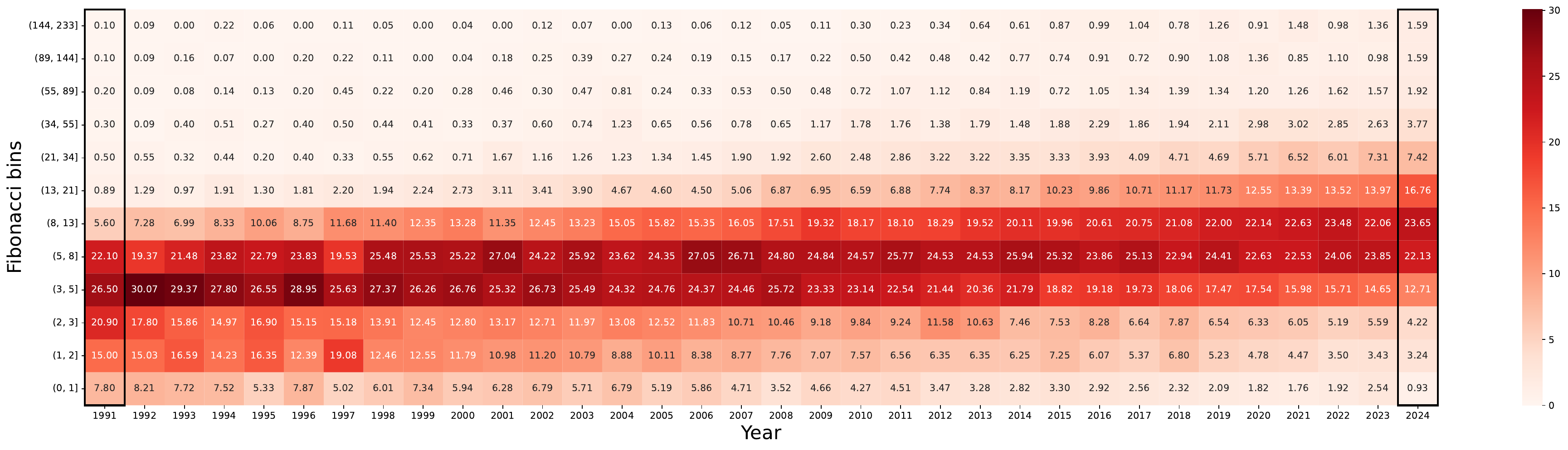}
		\caption{Biology}
	\end{subfigure}
	\hfill
	
	\begin{subfigure}[t]{\textwidth}
		\centering
		\includegraphics[width=\linewidth,keepaspectratio]{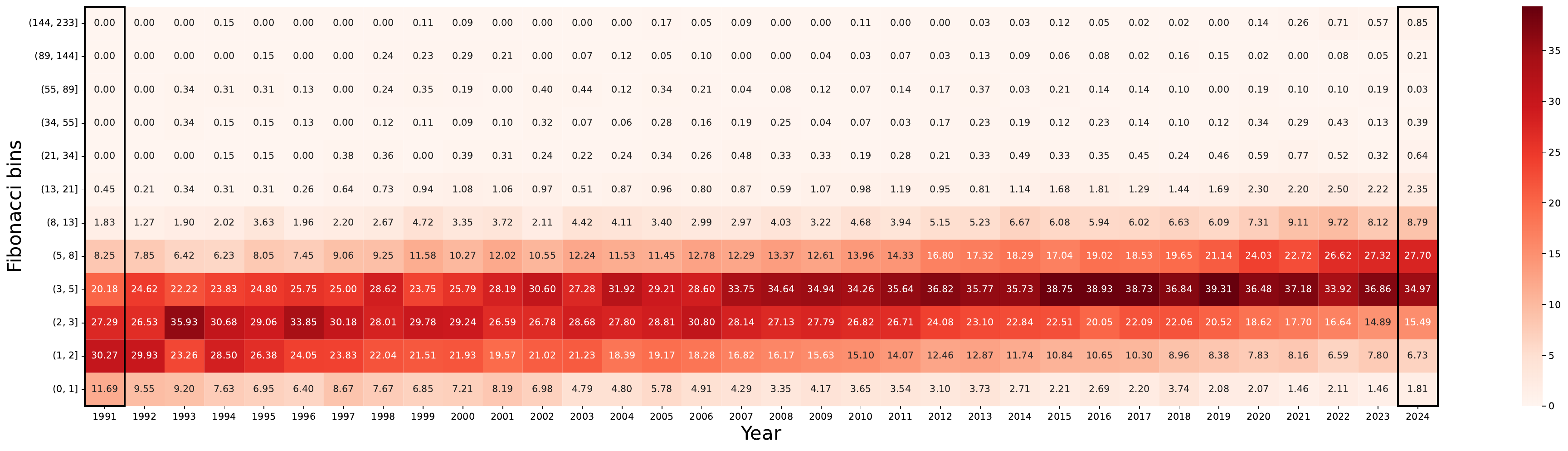}
		\caption{Electrical and Electronics Engineering}
	\end{subfigure}
	\hfill
	
	\caption{Percentage of publications based on Fibinacci-binned number of authors per year in the Combined, Biology, and Electrical and Electronics Engineering datasets.}
	\label{fig:percent_pub_interval_1}
\end{figure*}

\begin{figure*}[!htbp]
	\centering
	\begin{subfigure}[t]{\textwidth}
		\centering
		\includegraphics[width=\linewidth,keepaspectratio]{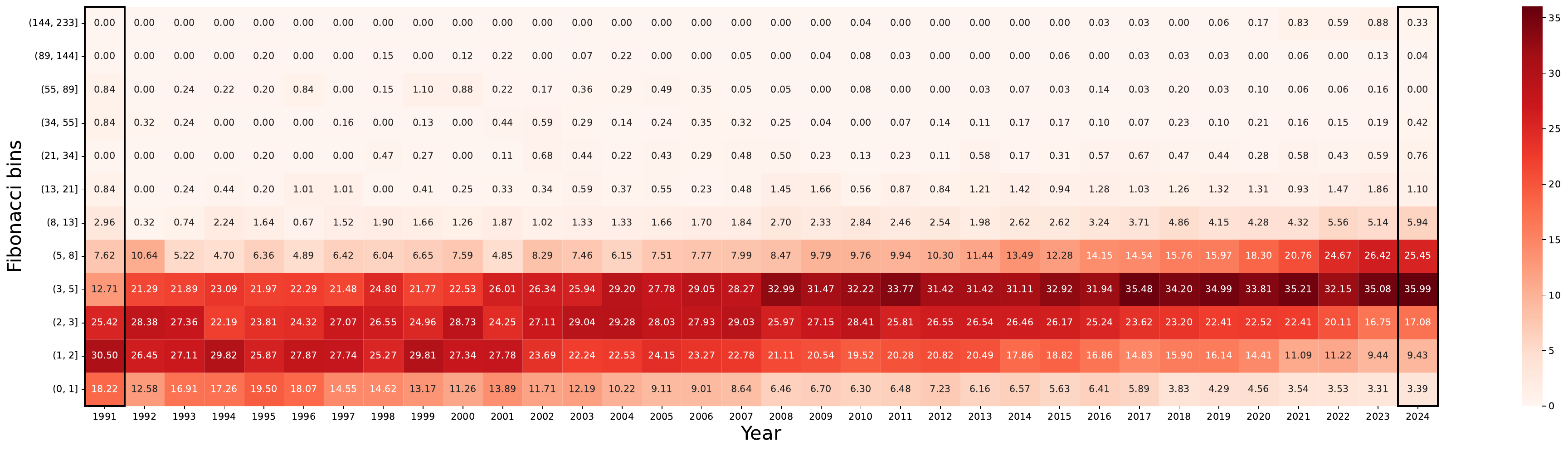}
		\caption{Computer Science and Engineering}
	\end{subfigure}
	\hfill
	\begin{subfigure}[t]{\textwidth}
		\centering
		\includegraphics[width=\linewidth,keepaspectratio]{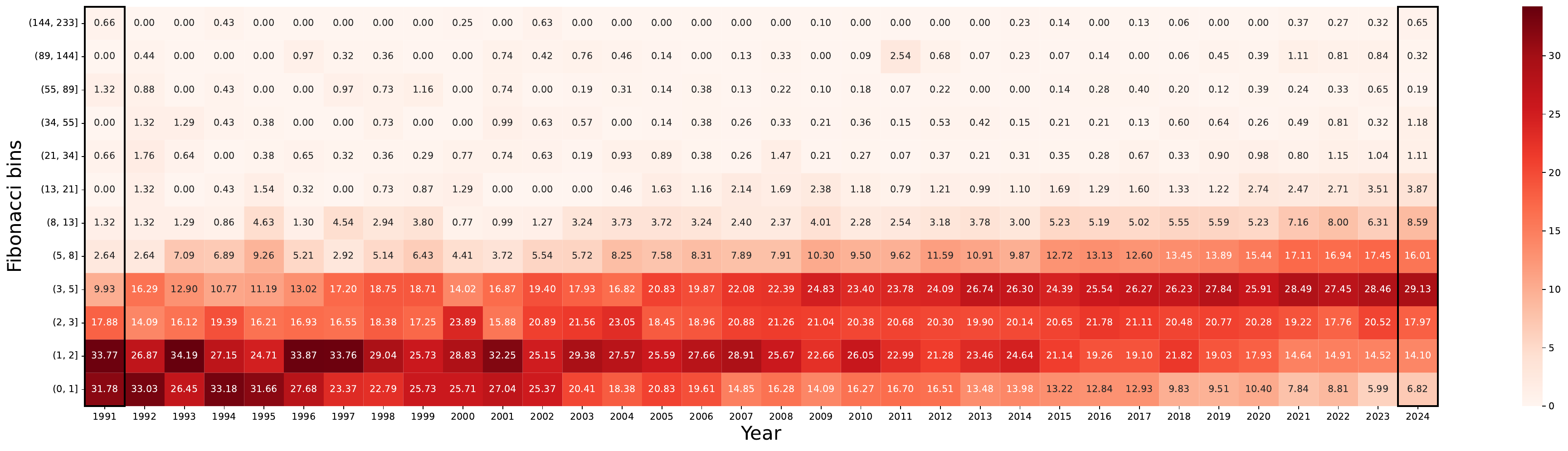}
		\caption{Psychology}
	\end{subfigure}
	\hfill
	\begin{subfigure}[t]{\textwidth}
		\centering
		\includegraphics[width=\linewidth,keepaspectratio]{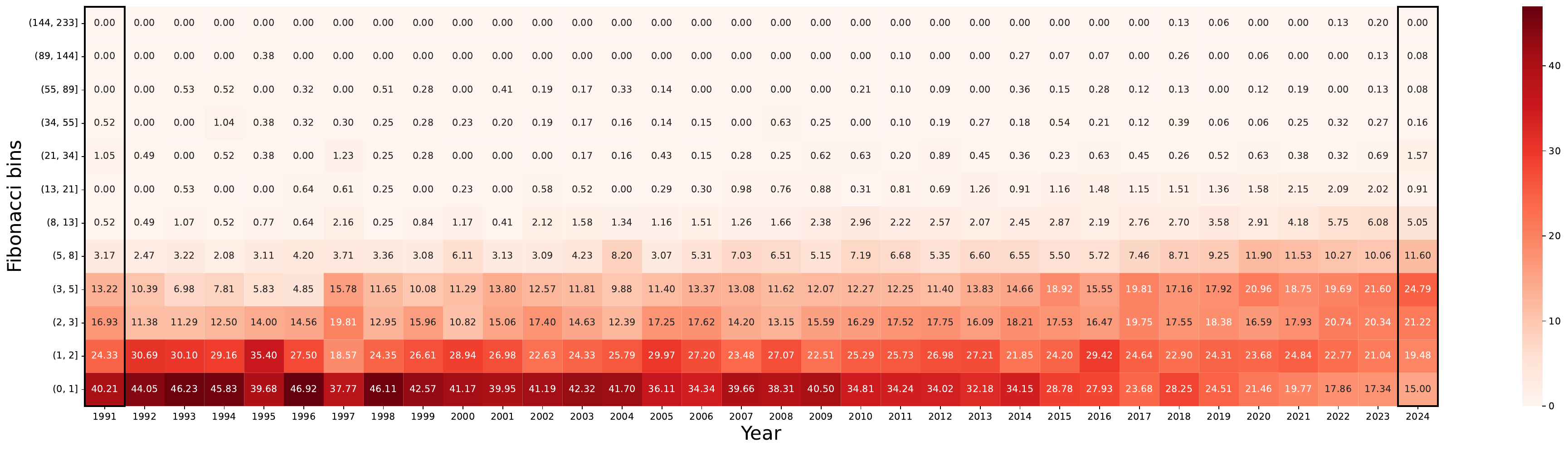}
		\caption{Sociology}
	\end{subfigure}

	\caption{Percentage of publications based on Fibinacci-binned number of authors per year in the Computer Science and Engineering, Psychology, and Sociology datasets.}
	\label{fig:percent_pub_interval_2}
\end{figure*}

\begin{figure*}[!htbp]
	\centering
	\begin{subfigure}[t]{\textwidth}
		\centering
		\includegraphics[width=\linewidth,keepaspectratio]{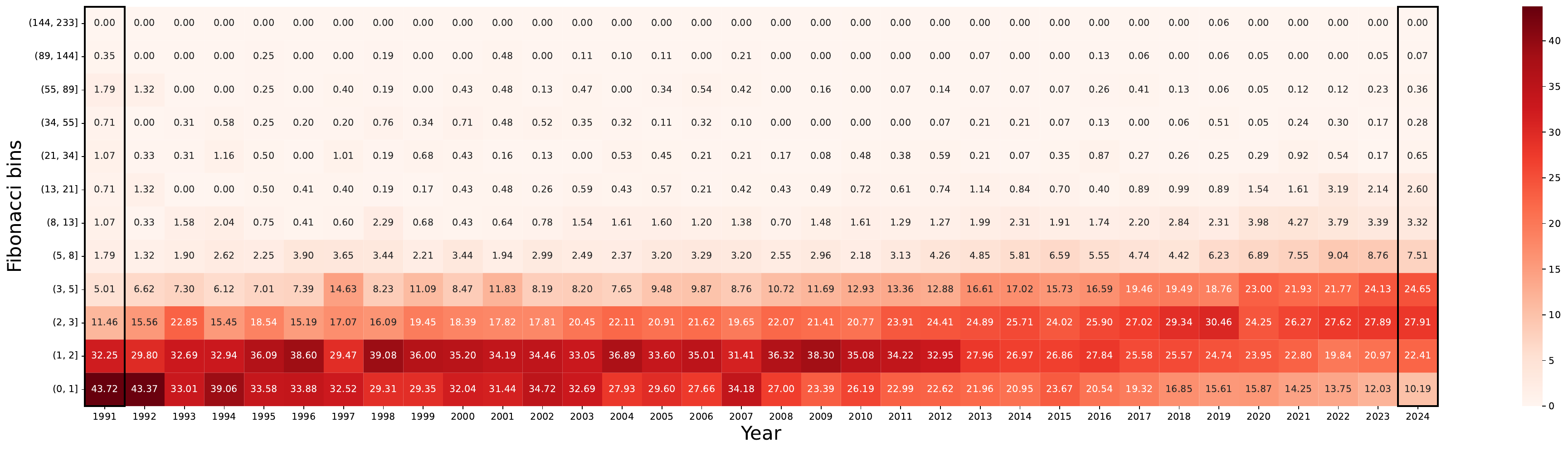}
		\caption{Economics}
	\end{subfigure}
	\hfill
	\begin{subfigure}[t]{\textwidth}
		\centering
		\includegraphics[width=\linewidth,keepaspectratio]{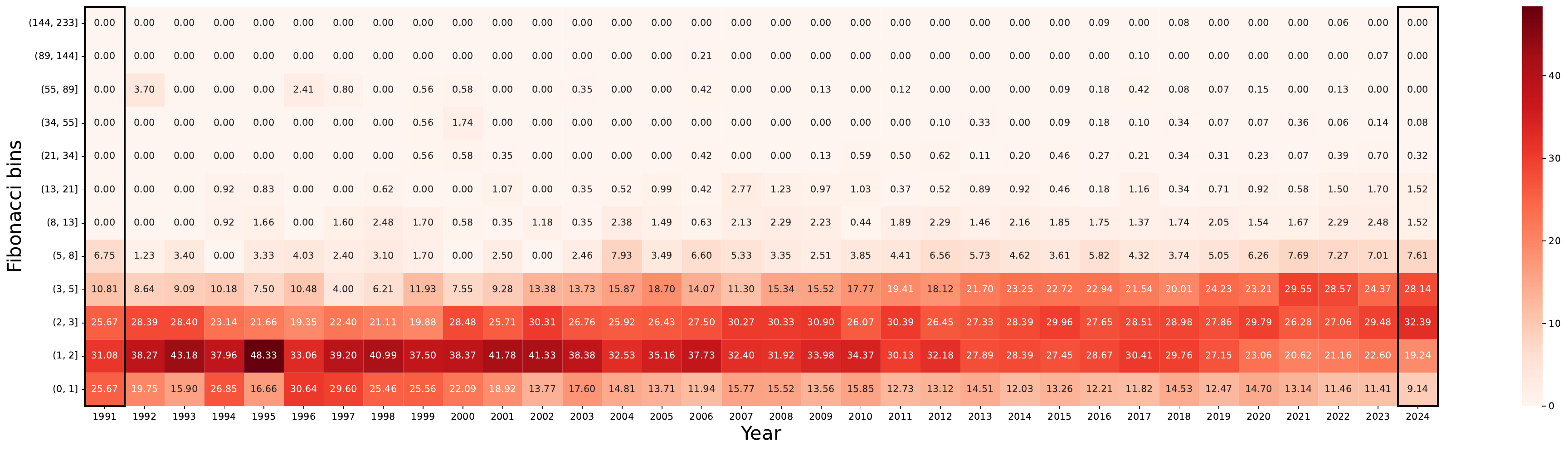}
		\caption{Marketing}
	\end{subfigure}
	\hfill
	
	\begin{subfigure}[t]{\textwidth}
		\centering
		\includegraphics[width=\linewidth,keepaspectratio]{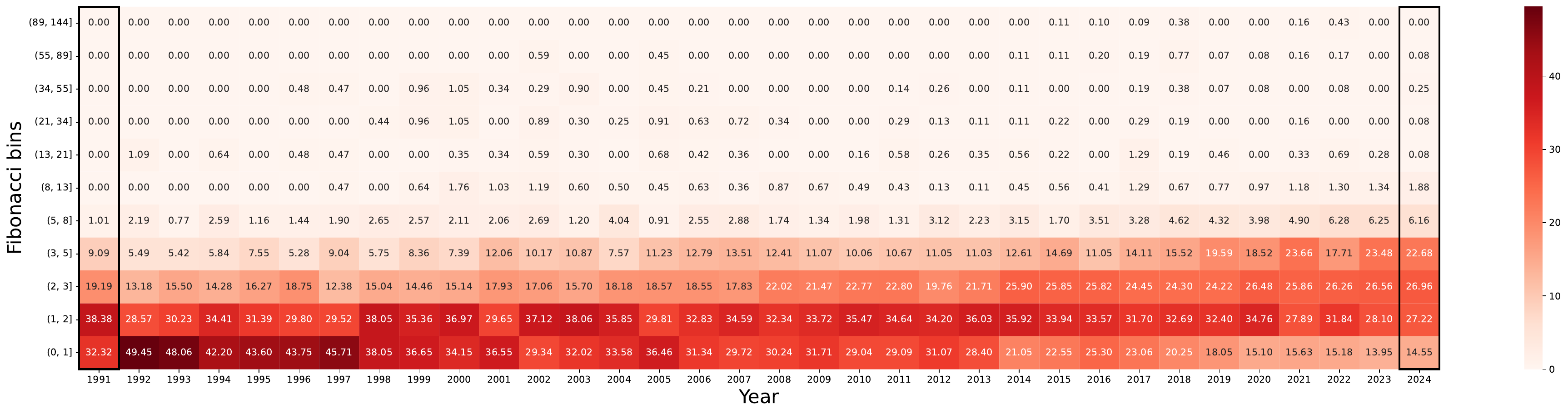}
		\caption{Mathematics}
	\end{subfigure}
	
	\caption{Percentage of publications based on Fibinacci-binned number of authors per year in the Economics, Marketing, and Mathematics datasets.}
	\label{fig:percent_pub_interval_3}
\end{figure*}
\FloatBarrier
\clearpage

\section{Supporting-role publications that can accompany every three full-credit publications}
\label{t_index_benchmark_table_appendix}
\begin{table}[htbp]
	\centering
	\begin{tabular}{cccc}
		\toprule
		\textbf{Author's Position} & \textbf{Fibonacci Number} & \textbf{Upper Benchmark (0.72)} & \textbf{Lower Benchmark (0.61)} \\
		\midrule
		1  & 1      &       &       \\
		2  & 1      &       &       \\
		3  & 2      & 3.81  & 10.63  \\
		4  & 3      & 2.17  & 4.22  \\
		5  & 5      & 1.61  & 2.85  \\
		6  & 8      & 1.41  & 2.41  \\
		7  & 13     & 1.30  & 2.19  \\
		8  & 21     & 1.24  & 2.08  \\
		9  & 34     & 1.21  & 2.01  \\
		10 & 55     & 1.19  & 1.97  \\
		11 & 89     & 1.18  & 1.95  \\
		12 & 144    & 1.17  & 1.94  \\
		13 & 233    & 1.17  & 1.93  \\
		14 & 377    & 1.17  & 1.92  \\
		15 & 610    & 1.16  & 1.92  \\
		16 & 987    & 1.16  & 1.92  \\
		17 & 1597   & 1.16  & 1.92  \\
		18 & 2584   & 1.16  & 1.91  \\
		19 & 4181   & 1.16  & 1.91  \\
		20 & 6765   & 1.16  & 1.91  \\
		21 & 10946  & 1.16  & 1.91  \\
		22 & 17711  & 1.16  & 1.91  \\
		23 & 28657  & 1.16  & 1.91  \\
		24 & 46368  & 1.16  & 1.91  \\
		25 & 75025  & 1.16  & 1.91  \\
		26 & 121393 & 1.16  & 1.91  \\
		27 & 196418 & 1.16  & 1.91  \\
		28 & 317811 & 1.16  & 1.91  \\
		29 & 514229 & 1.16  & 1.91  \\
		30 & 832040 & 1.16  & 1.91  \\
		31 & 1346269 & 1.16  & 1.91  \\
		32 & 2178309 & 1.16  & 1.91  \\
		33 & 3524578 & 1.16  & 1.91  \\
		34 & 5702887 & 1.16  & 1.91  \\
		\bottomrule
	\end{tabular}
	\caption{The number of supporting-role publications that can accompany every three full-credit publications to maintain $L^\prime$-index benchmark values.}
	\label{tab:t_index_benchmark}
\end{table}
\FloatBarrier
\clearpage

\section{Percentage Difference in Publications by Participation and by Contribution}

\label{percentage_difference_pub_appendix}
\begin{figure*}[!htbp]
	\centering
	\begin{subfigure}[t]{0.32\textwidth}
		\centering
		\includegraphics[width=\linewidth,keepaspectratio]{plots/percent_diff_hc.pdf}
		\caption{Combined}
	\end{subfigure}
	\hfill
	\begin{subfigure}[t]{0.32\textwidth}
		\centering
		\includegraphics[width=\linewidth,keepaspectratio]{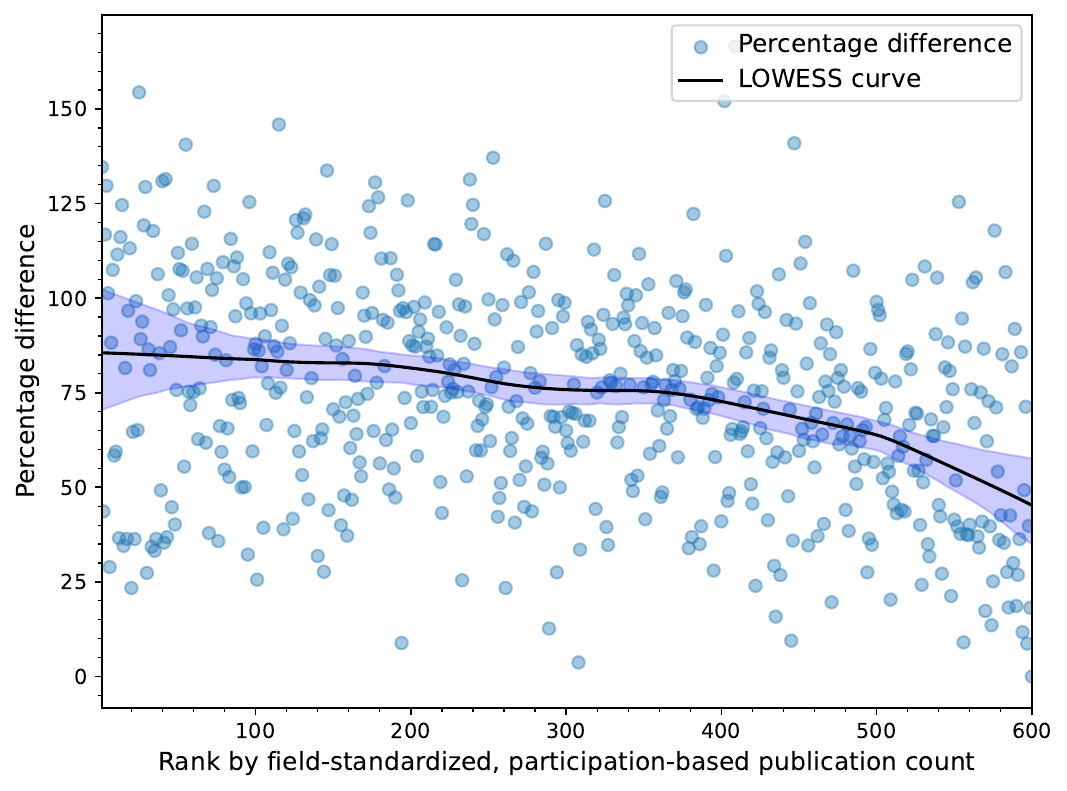}
		\caption{Biology}
	\end{subfigure}
	\hfill
	\begin{subfigure}[t]{0.32\textwidth}
		\centering
		\includegraphics[width=\linewidth,keepaspectratio]{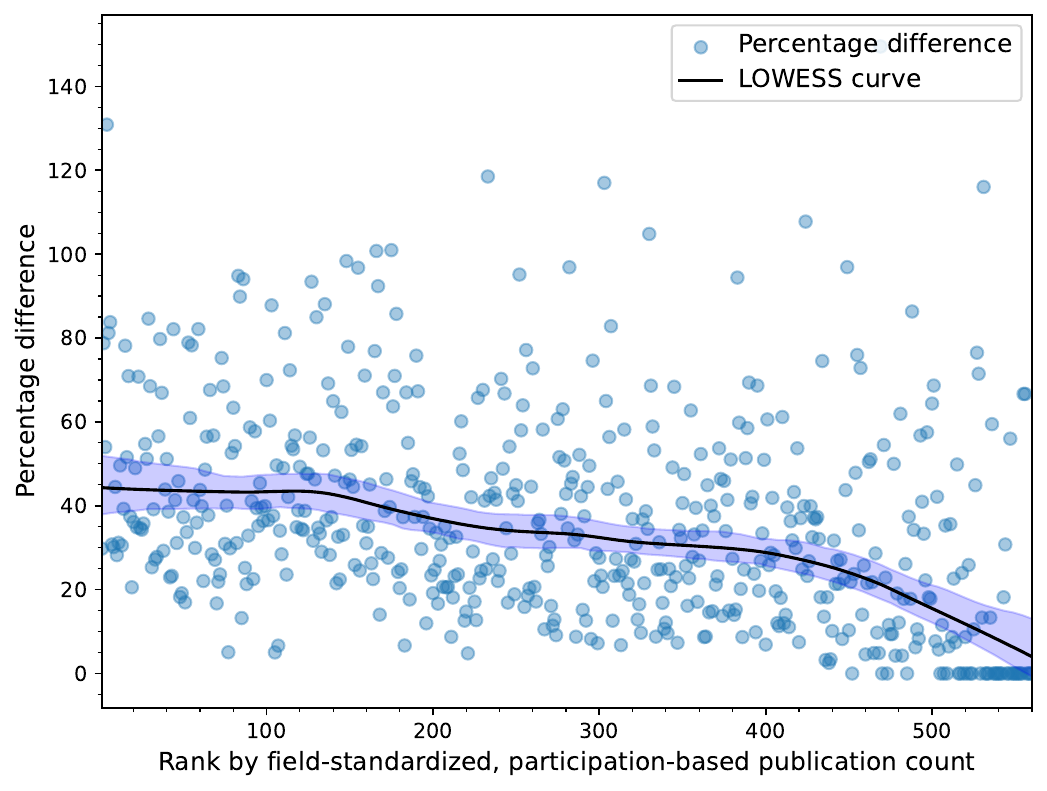}
		\caption{Electrical and Electronics Engineering}
	\end{subfigure}
	
	
	\begin{subfigure}[t]{0.32\textwidth}
		\centering
		\includegraphics[width=\linewidth,keepaspectratio]{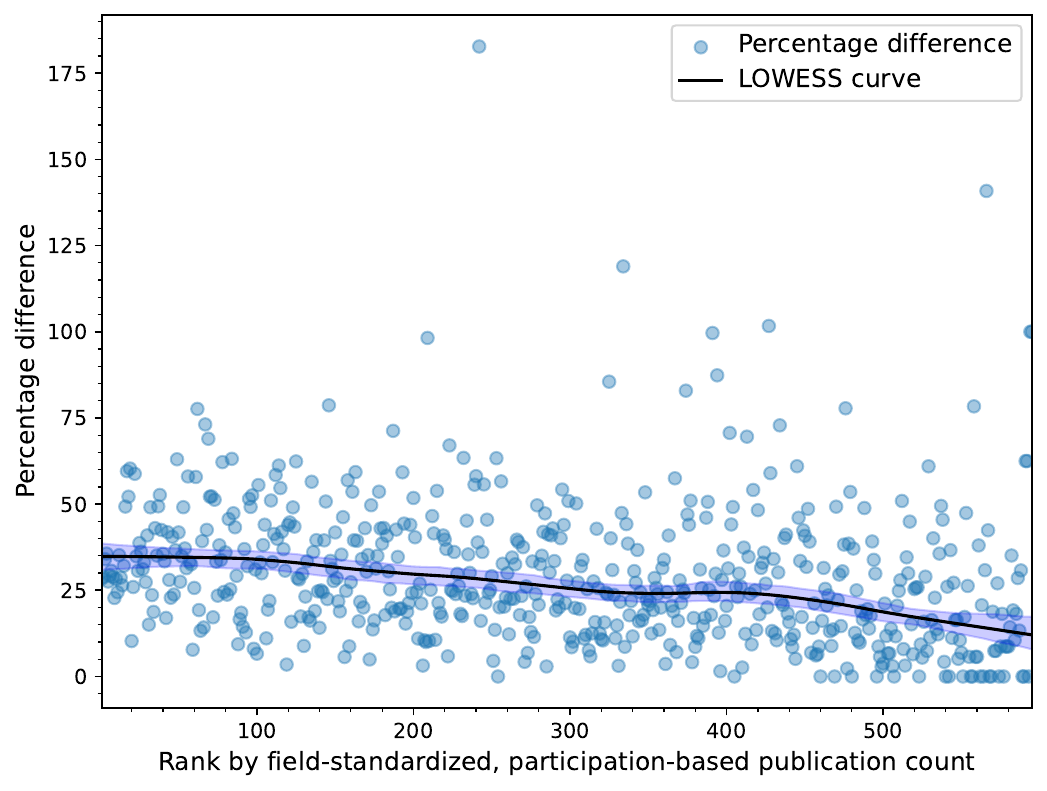}
		\caption{Computer Science and Engineering}
	\end{subfigure}
	\hfill
	\begin{subfigure}[t]{0.32\textwidth}
		\centering
		\includegraphics[width=\linewidth,keepaspectratio]{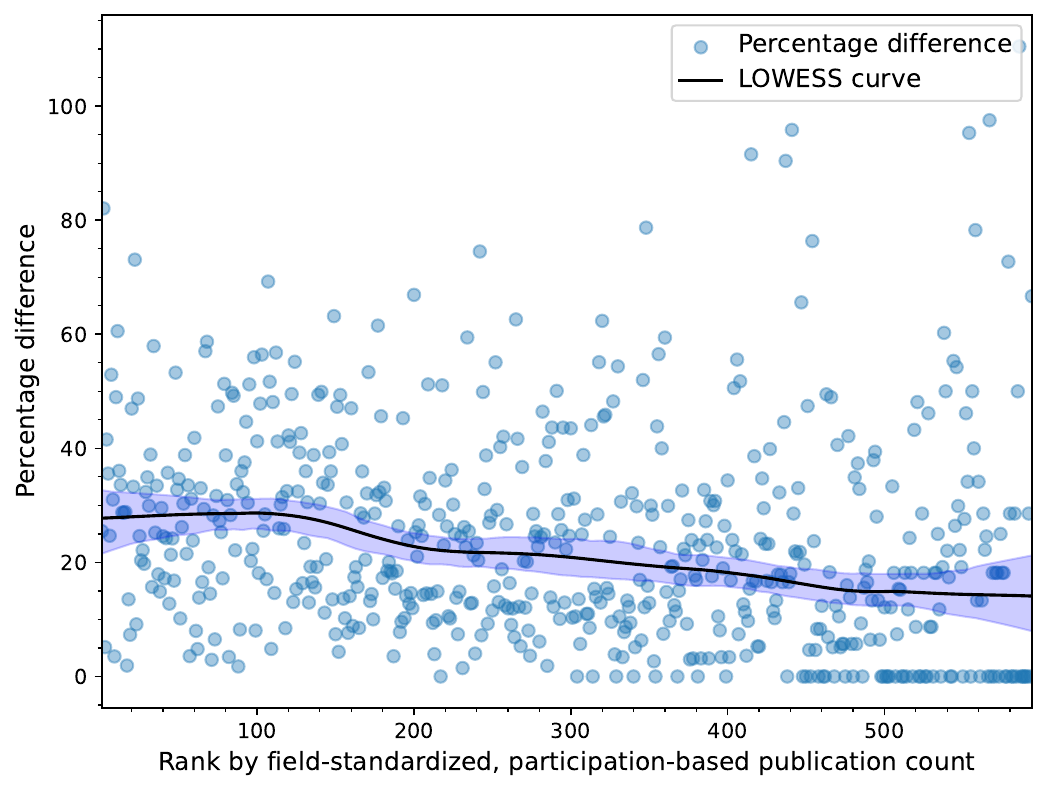}
		\caption{Psychology}
	\end{subfigure}
	\hfill
	\begin{subfigure}[t]{0.32\textwidth}
		\centering
		\includegraphics[width=\linewidth,keepaspectratio]{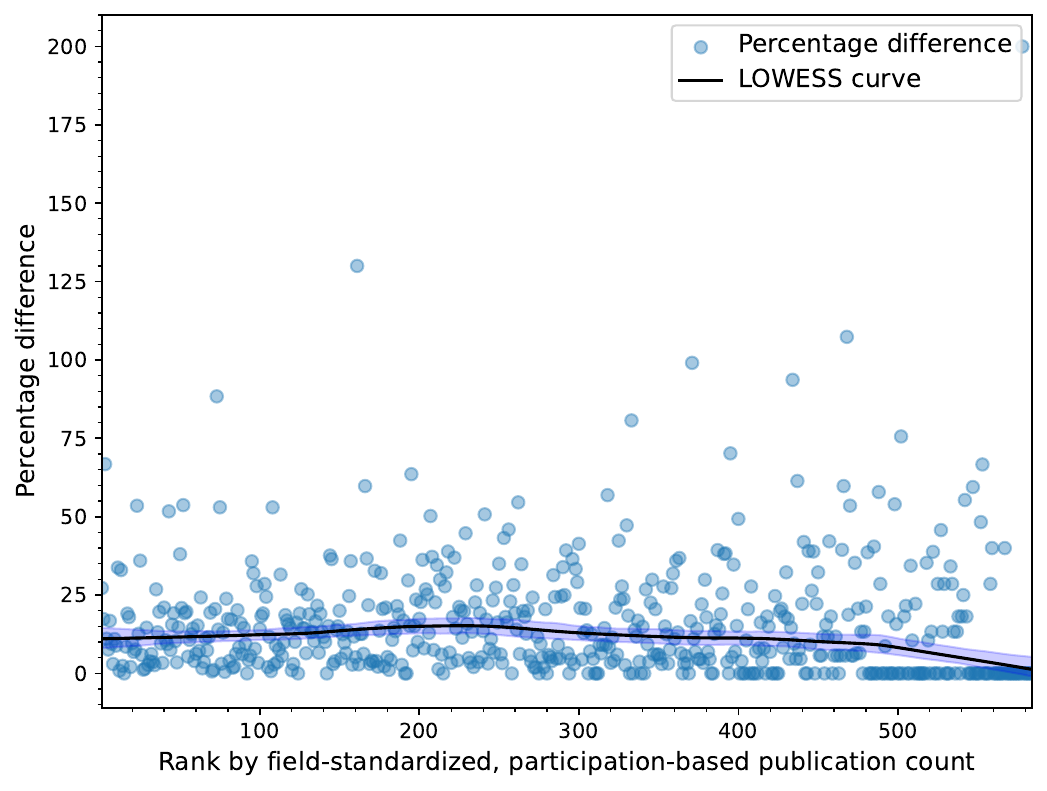}
		\caption{Sociology}
	\end{subfigure}
	
	\begin{subfigure}[t]{0.32\textwidth}
		\centering
		\includegraphics[width=\linewidth,keepaspectratio]{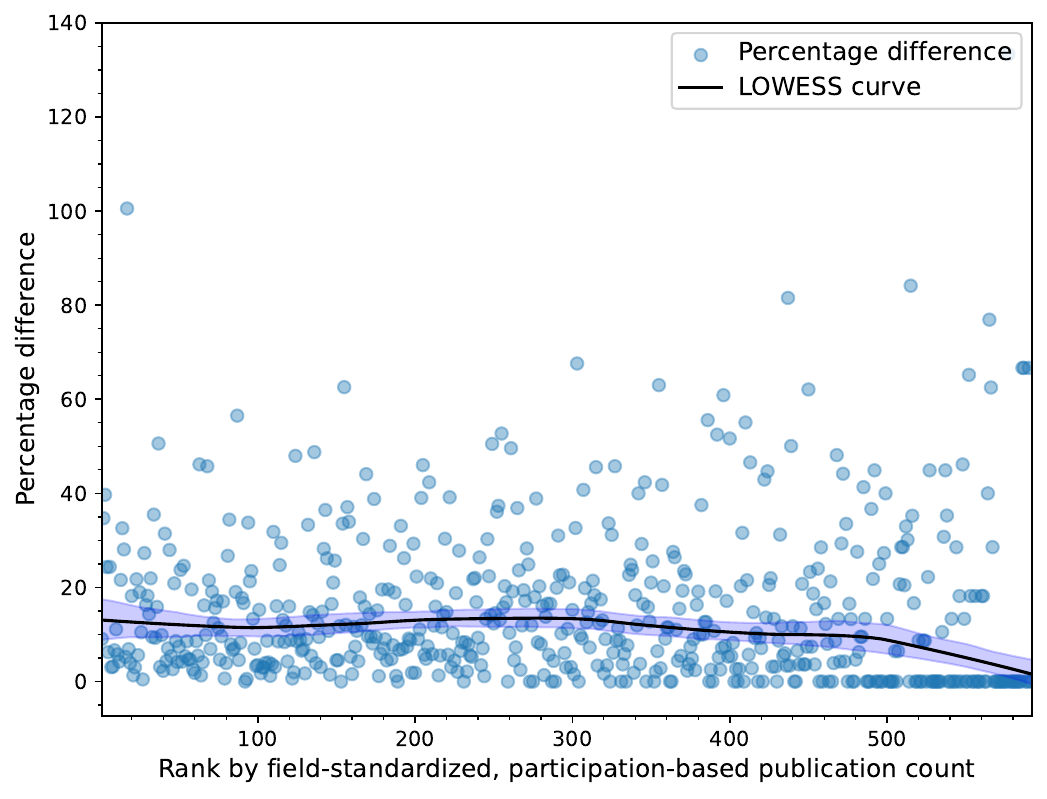}
		\caption{Economics}
	\end{subfigure}
	\hfill
	\begin{subfigure}[t]{0.32\textwidth}
		\centering
		\includegraphics[width=\linewidth,keepaspectratio]{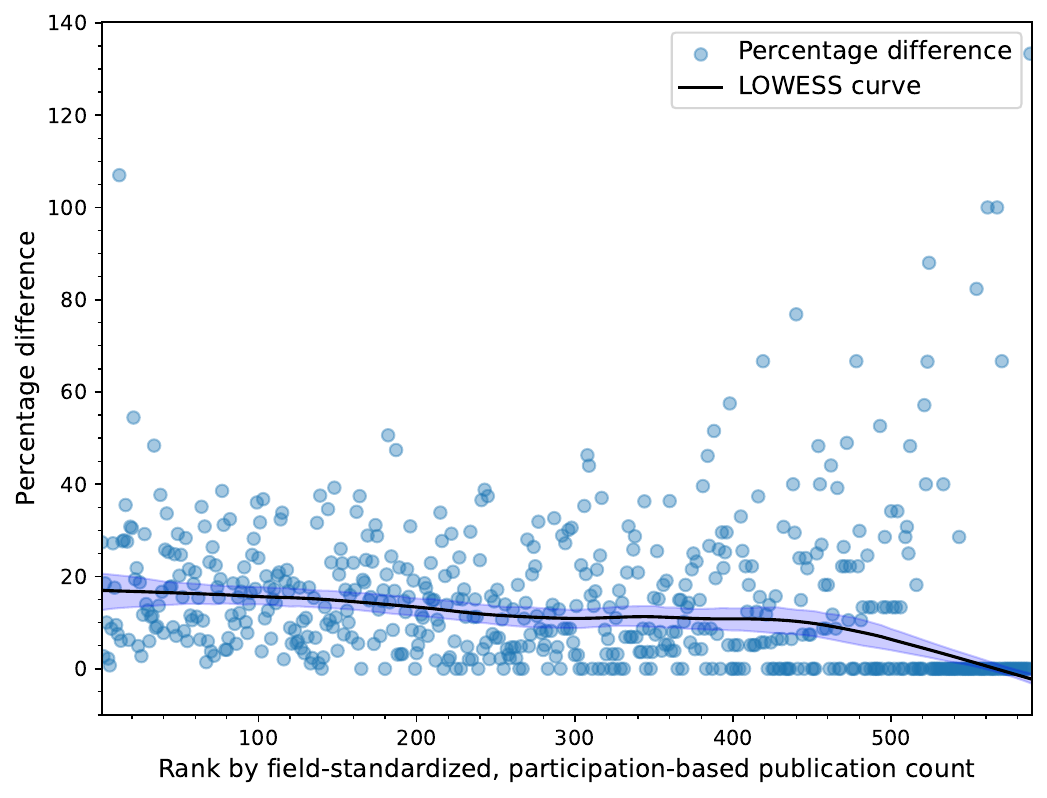}
		\caption{Marketing}
	\end{subfigure}
	\begin{subfigure}[t]{0.32\textwidth}
		\centering
		\includegraphics[width=\linewidth,keepaspectratio]{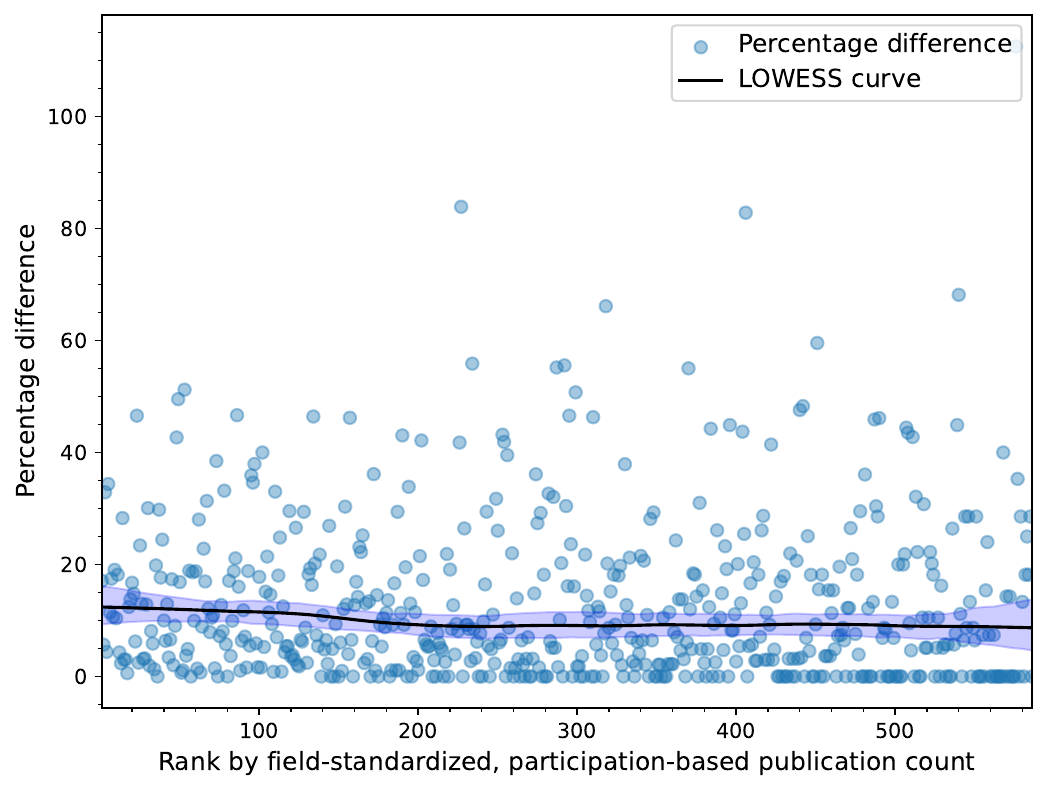}
		\caption{Mathematics}
	\end{subfigure}
	
	\caption{Percentage differences in publications by participation and by contribution in the combined and field datasets.}
	\label{fig:percentage_diff_pub}
\end{figure*}

\FloatBarrier
\clearpage

\section{$L^\prime$-index of Authors By Standardized Publication Participation Ranks in the Combined and Field Datasets}

\label{t_index_by_publication_participation}
\begin{figure*}[!htbp]
	\centering
	\begin{subfigure}[t]{0.32\textwidth}
		\centering
		\includegraphics[width=\linewidth,keepaspectratio]{plots/t_prime_index.pdf}
		\caption{Combined}
	\end{subfigure}
	\hfill
	\begin{subfigure}[t]{0.32\textwidth}
		\centering
		\includegraphics[width=\linewidth,keepaspectratio]{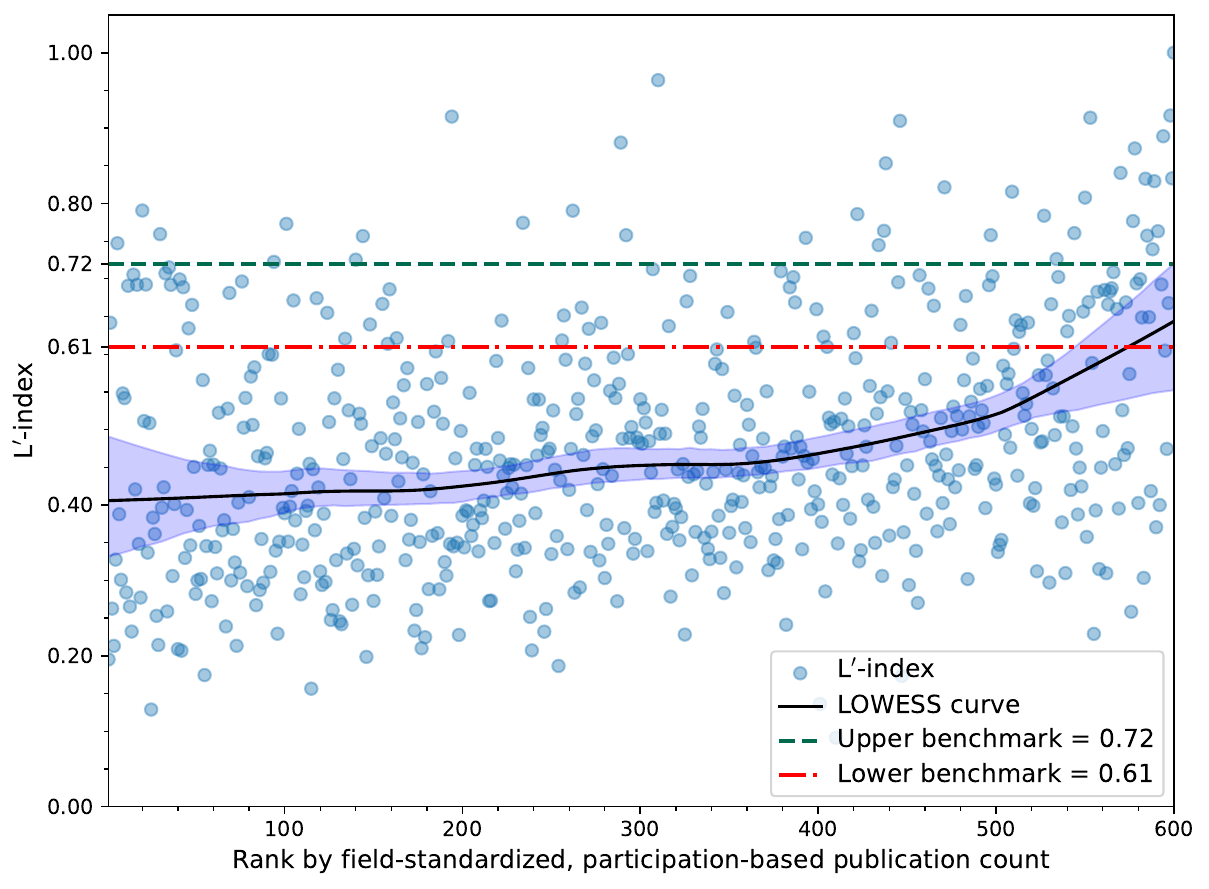}
		\caption{Biology}
	\end{subfigure}
	\hfill
	\begin{subfigure}[t]{0.32\textwidth}
		\centering
		\includegraphics[width=\linewidth,keepaspectratio]{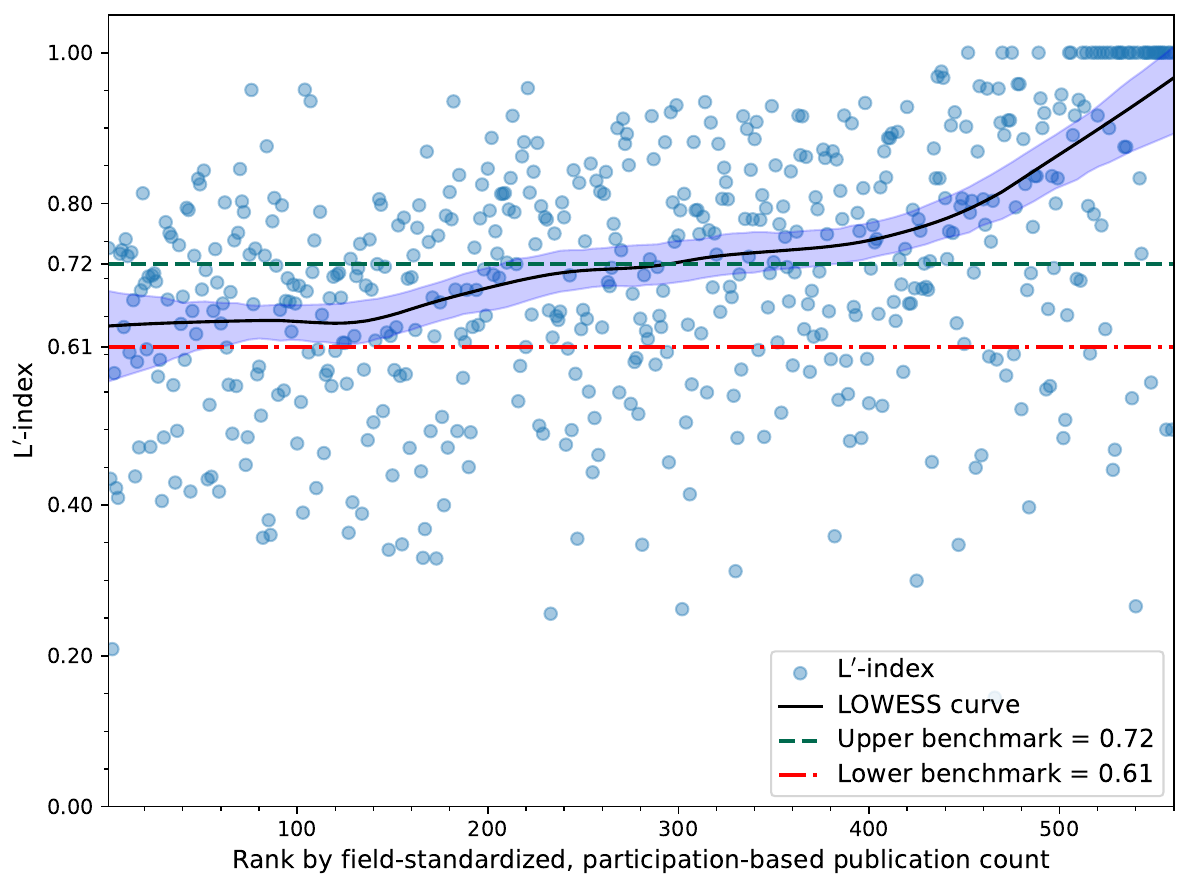}
		\caption{Electrical and Electronics Engineering}
	\end{subfigure}
	
	\begin{subfigure}[t]{0.32\textwidth}
		\centering
		\includegraphics[width=\linewidth,keepaspectratio]{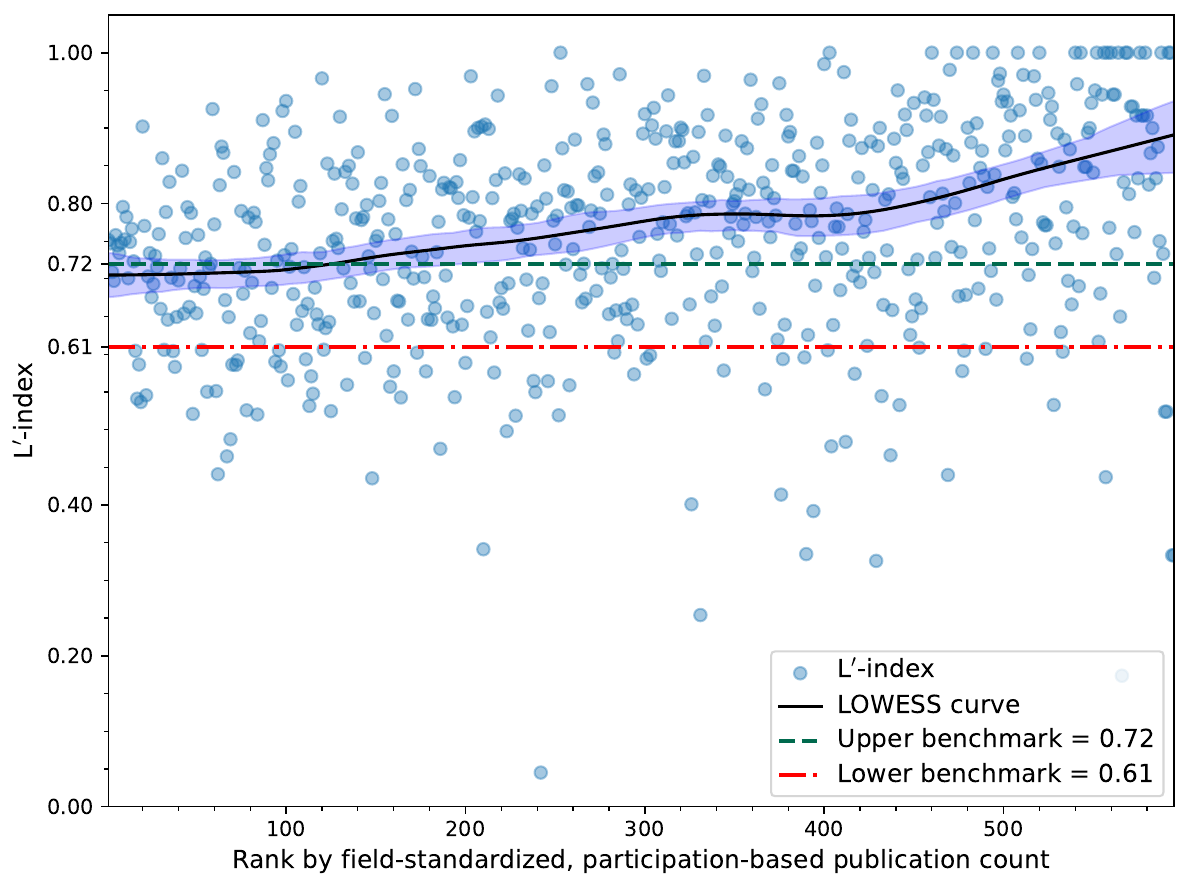}
		\caption{Computer Science and Engineering}
	\end{subfigure}
	\hfill
	\begin{subfigure}[t]{0.32\textwidth}
		\centering
		\includegraphics[width=\linewidth,keepaspectratio]{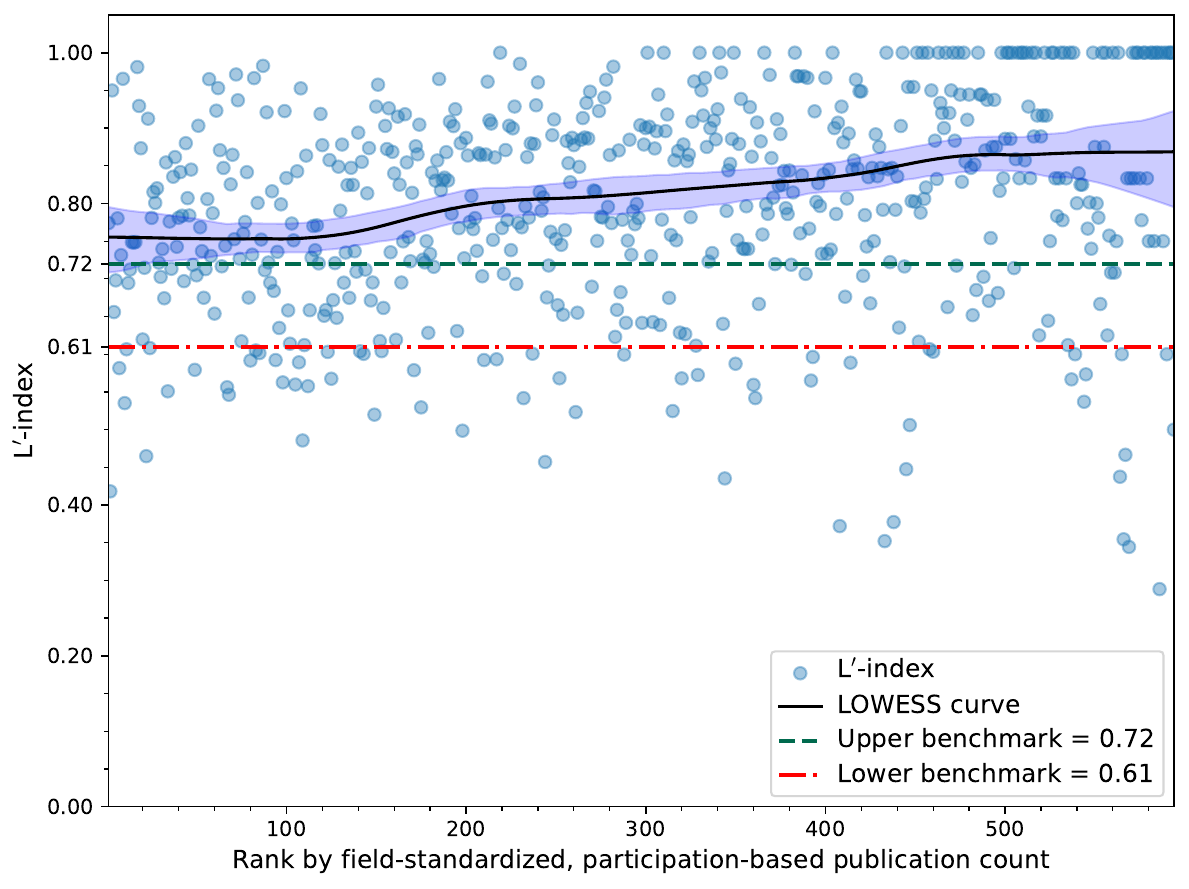}
		\caption{Psychology}
	\end{subfigure}
	\hfill
	\begin{subfigure}[t]{0.32\textwidth}
		\centering
		\includegraphics[width=\linewidth,keepaspectratio]{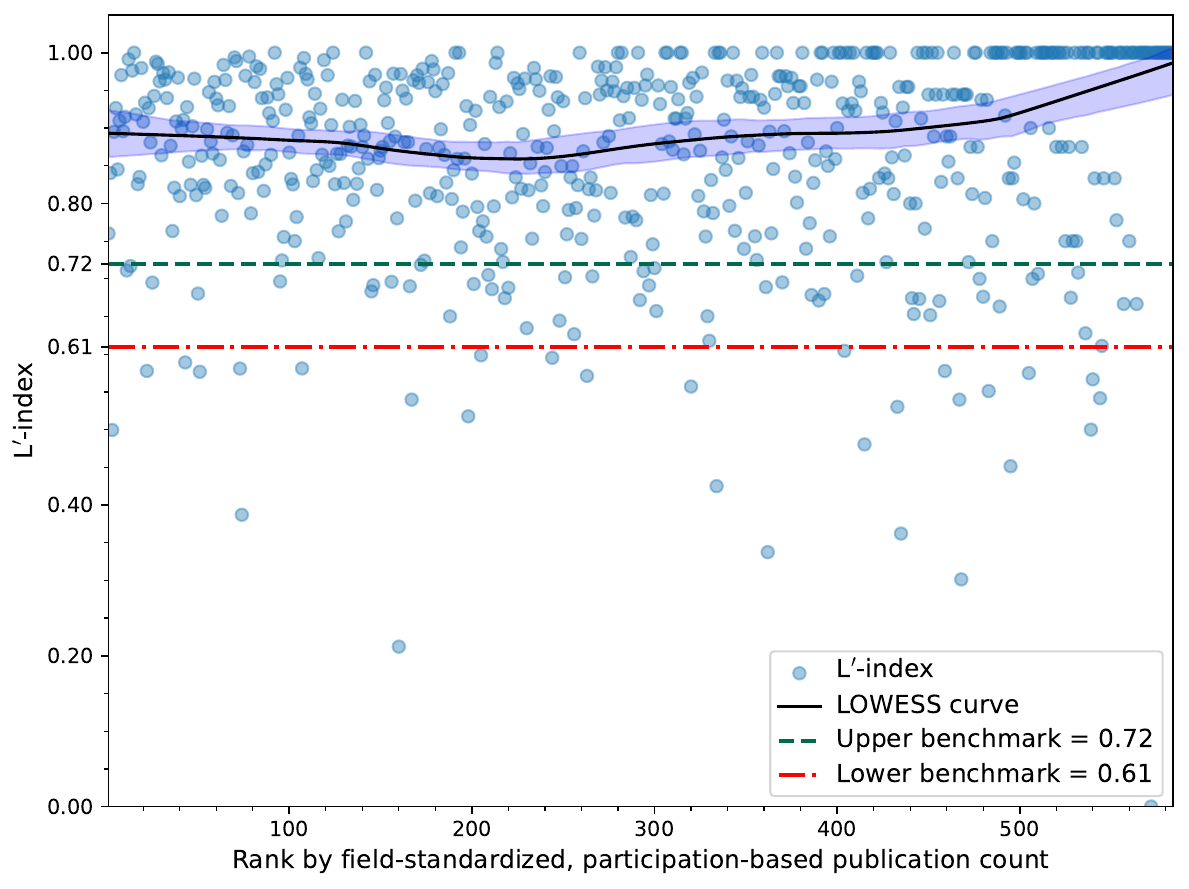}
		\caption{Sociology}
	\end{subfigure}
	
	\begin{subfigure}[t]{0.32\textwidth}
		\centering
		\includegraphics[width=\linewidth,keepaspectratio]{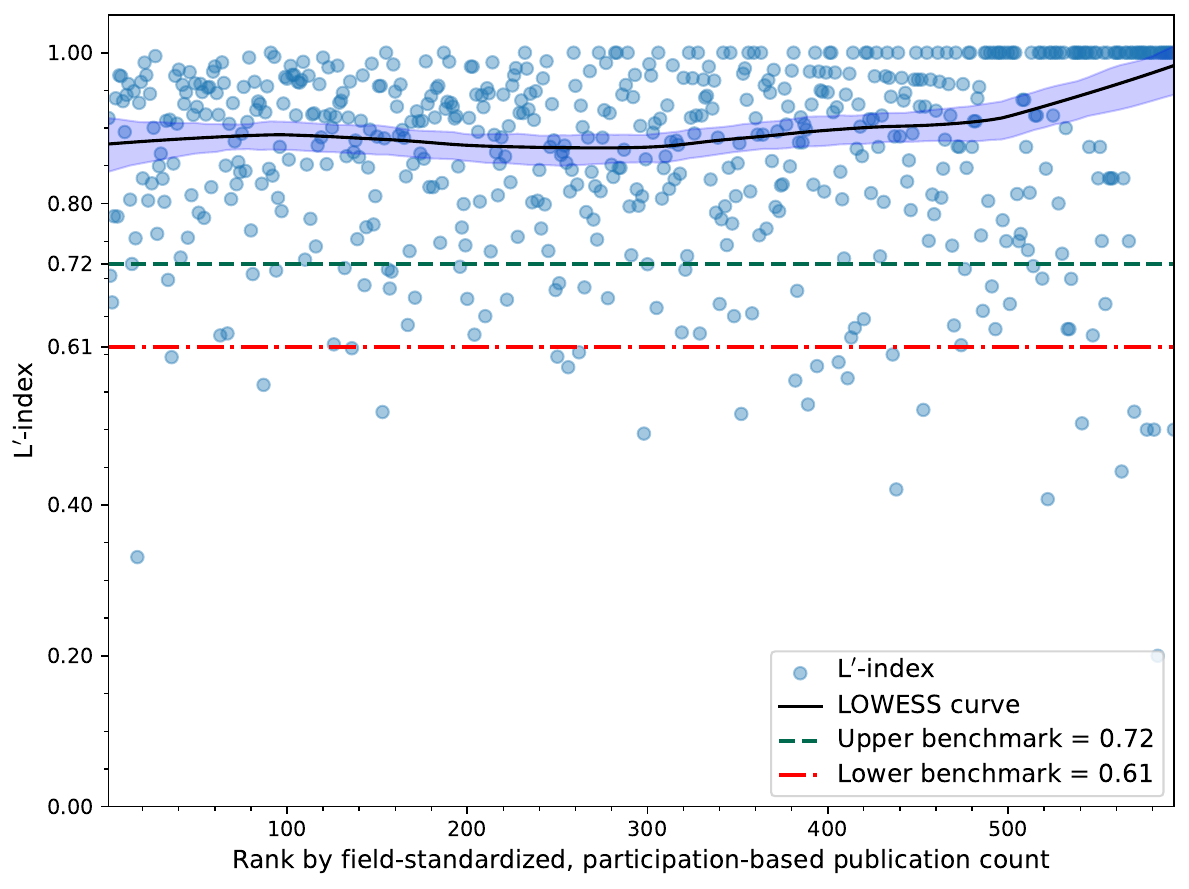}
		\caption{Economics}
	\end{subfigure}
	\hfill
	\begin{subfigure}[t]{0.32\textwidth}
		\centering
		\includegraphics[width=\linewidth,keepaspectratio]{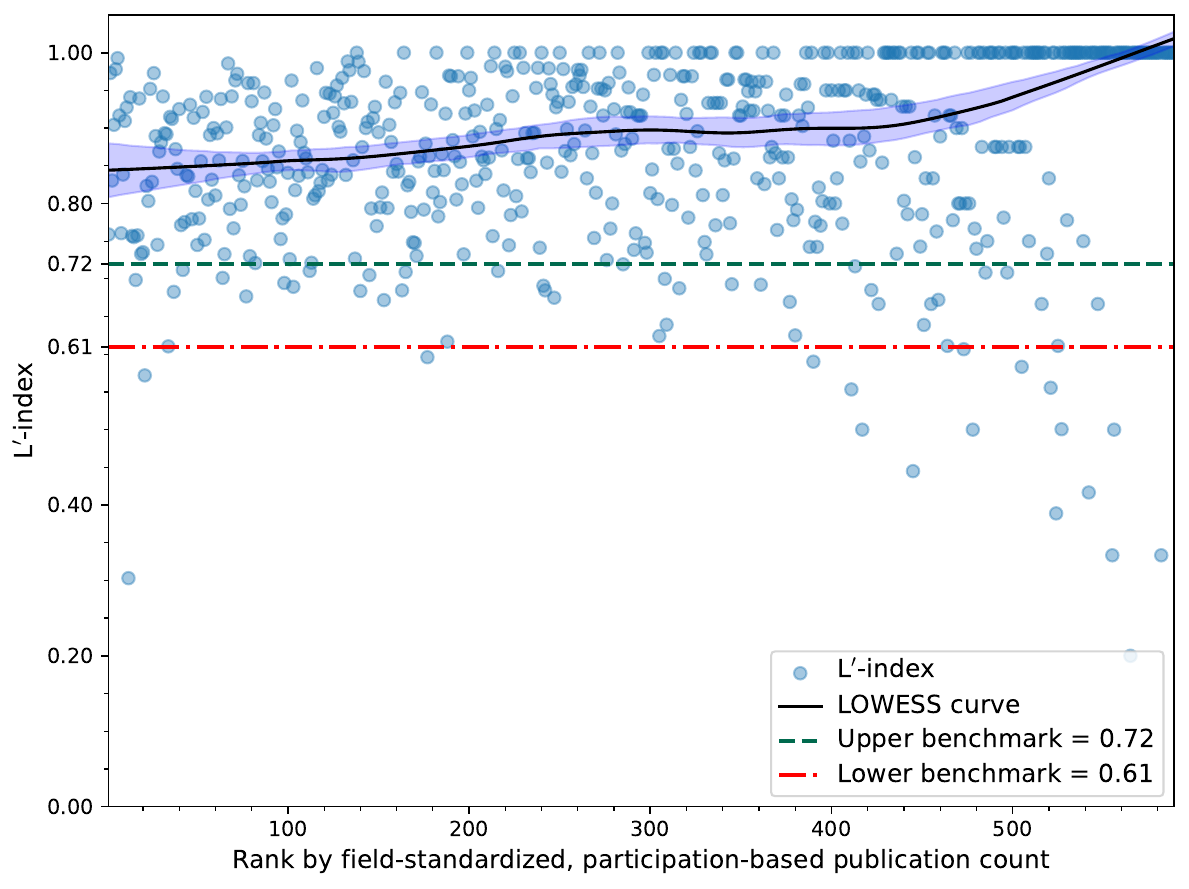}
		\caption{Marketing}
	\end{subfigure}
	\begin{subfigure}[t]{0.32\textwidth}
		\centering
		\includegraphics[width=\linewidth,keepaspectratio]{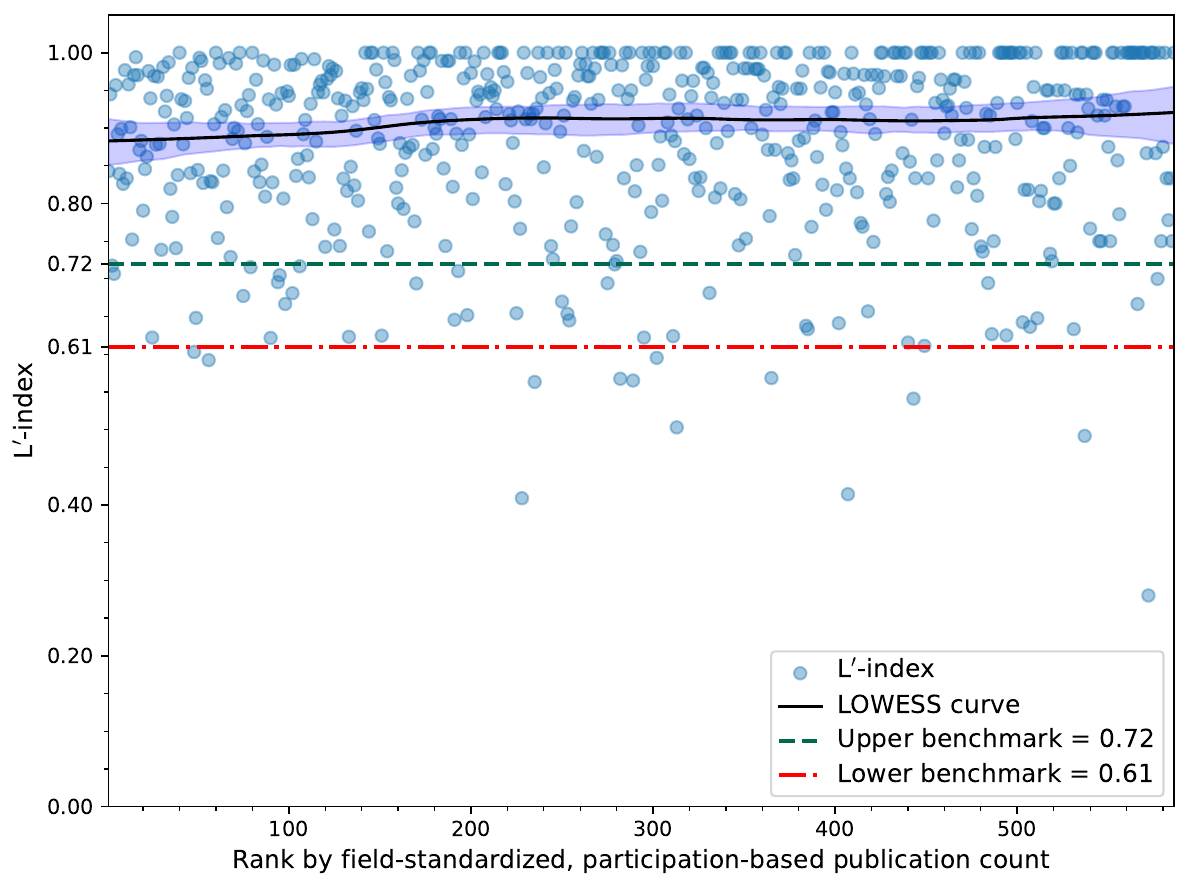}
		\caption{Mathematics}
	\end{subfigure}
	
	\caption{$L^\prime$-index of authors by standardized publication participation ranks in the combined and field datasets.}
	\label{fig:t_index_field}
\end{figure*}
\clearpage
\FloatBarrier

\section{Distributions of Citations in the Combined and Field Datasets}

\label{citation_distribution_per_publication}
\begin{figure*}[!htbp]
	\centering
	\begin{subfigure}[t]{0.32\textwidth}
		\centering
		\includegraphics[width=\linewidth,keepaspectratio]{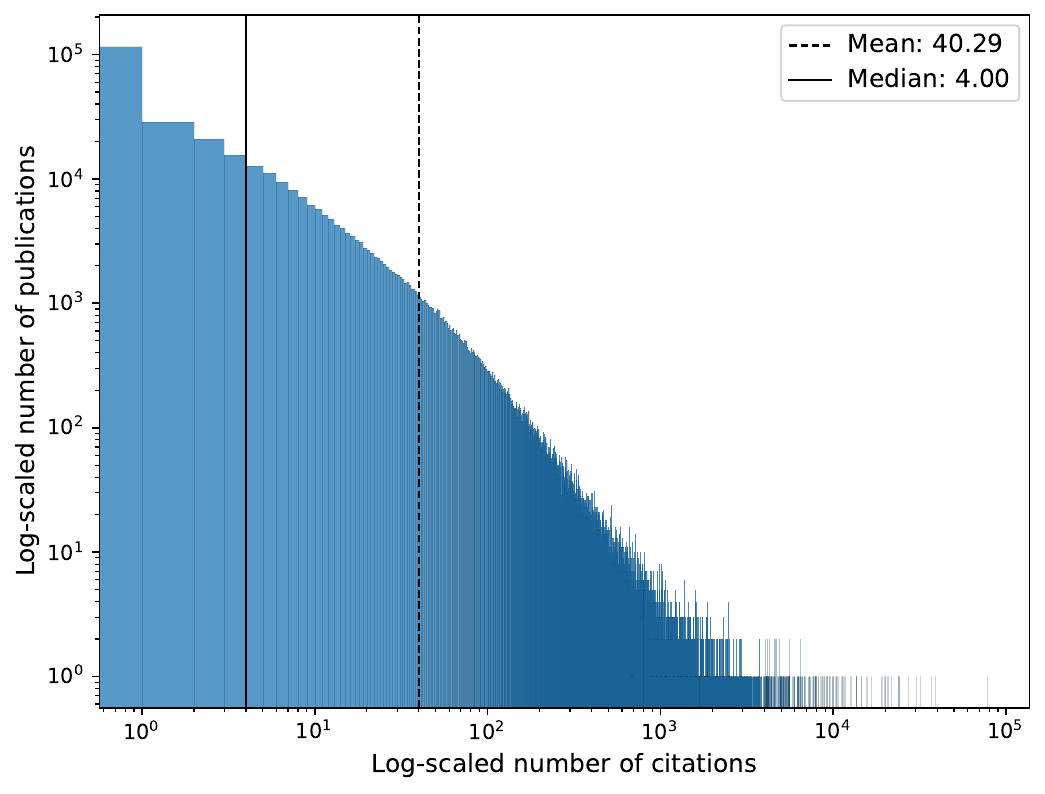}
		\caption{Combined}
	\end{subfigure}
	\hfill
	\begin{subfigure}[t]{0.32\textwidth}
		\centering
		\includegraphics[width=\linewidth,keepaspectratio]{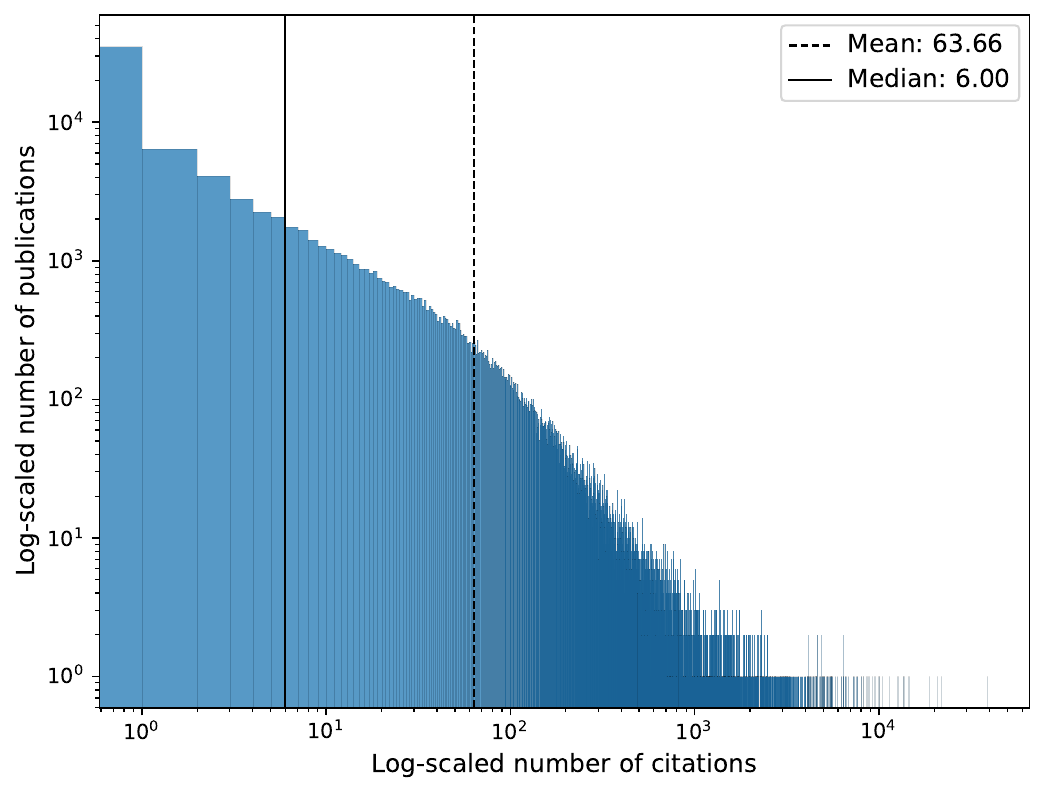}
		\caption{Biology}
	\end{subfigure}
	\hfill
	\begin{subfigure}[t]{0.32\textwidth}
		\centering
		\includegraphics[width=\linewidth,keepaspectratio]{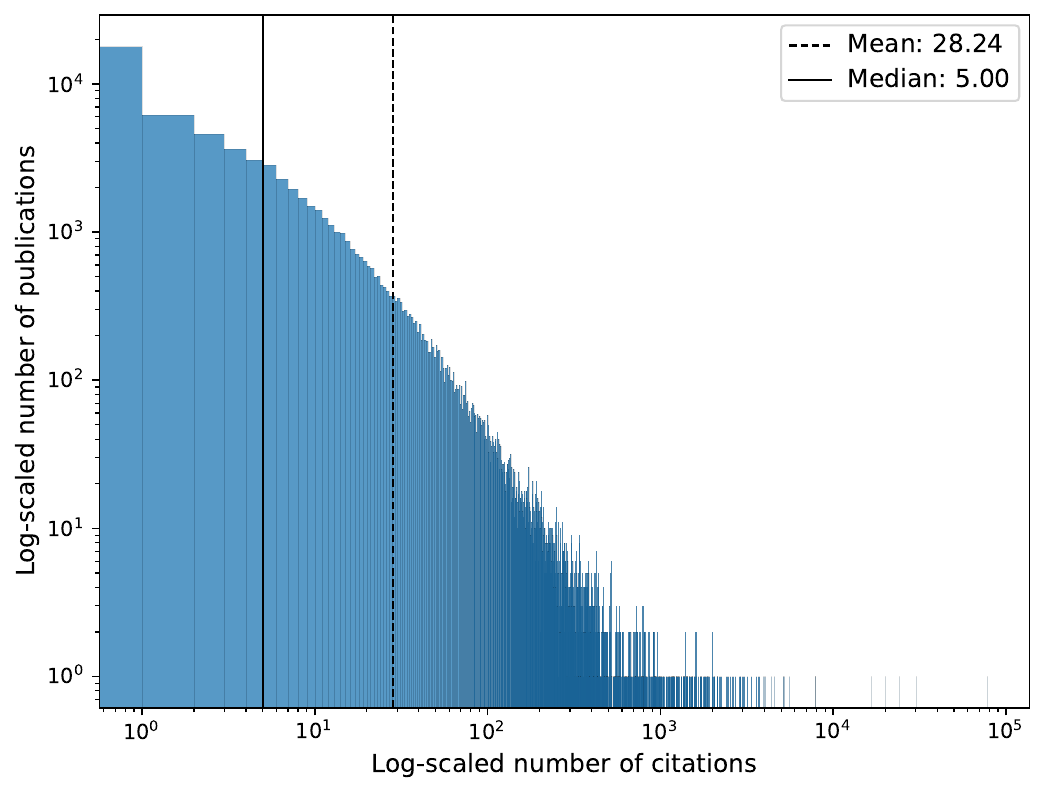}
		\caption{Electrical and Electronics Engineering}
	\end{subfigure}
	
	\begin{subfigure}[t]{0.32\textwidth}
		\centering
		\includegraphics[width=\linewidth,keepaspectratio]{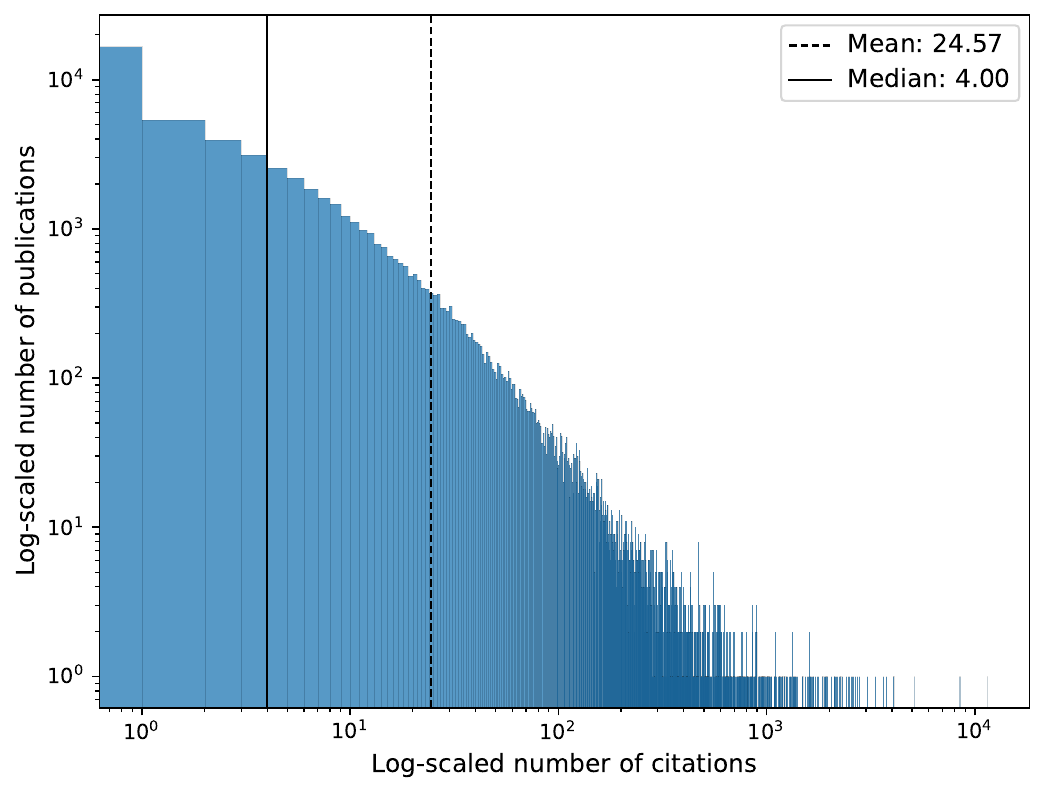}
		\caption{Computer Science and Engineering}
	\end{subfigure}
	\hfill
	\begin{subfigure}[t]{0.32\textwidth}
		\centering
		\includegraphics[width=\linewidth,keepaspectratio]{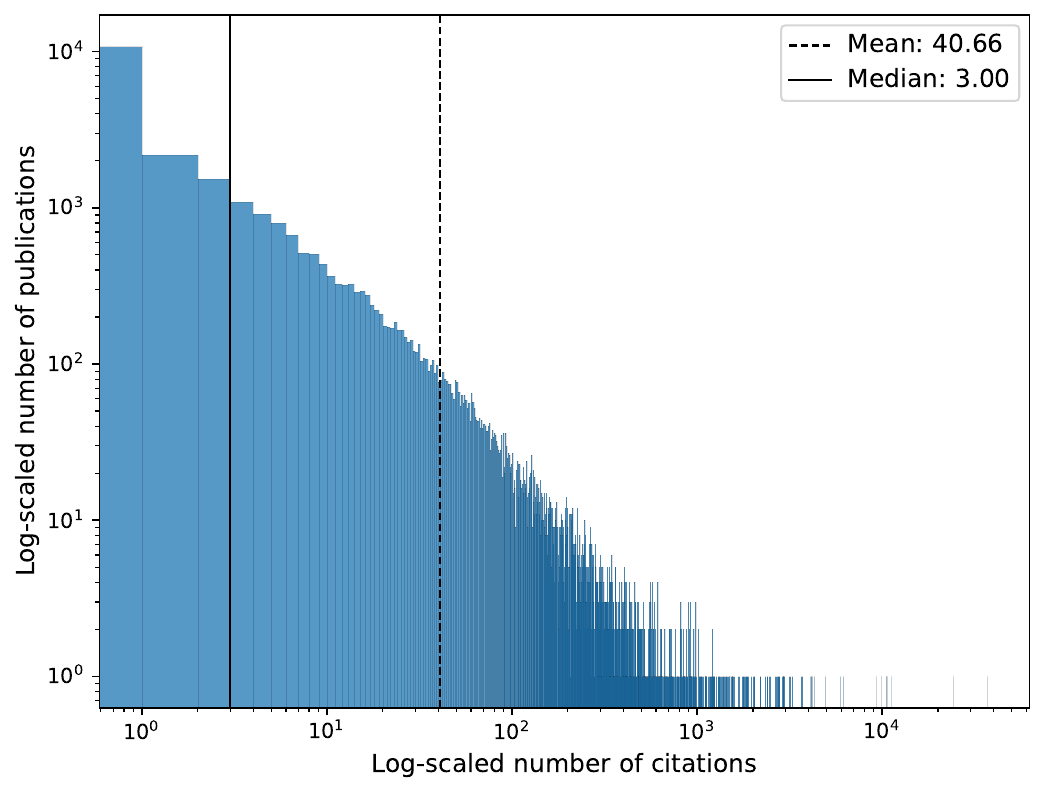}
		\caption{Psychology}
	\end{subfigure}
	\hfill
	\begin{subfigure}[t]{0.32\textwidth}
		\centering
		\includegraphics[width=\linewidth,keepaspectratio]{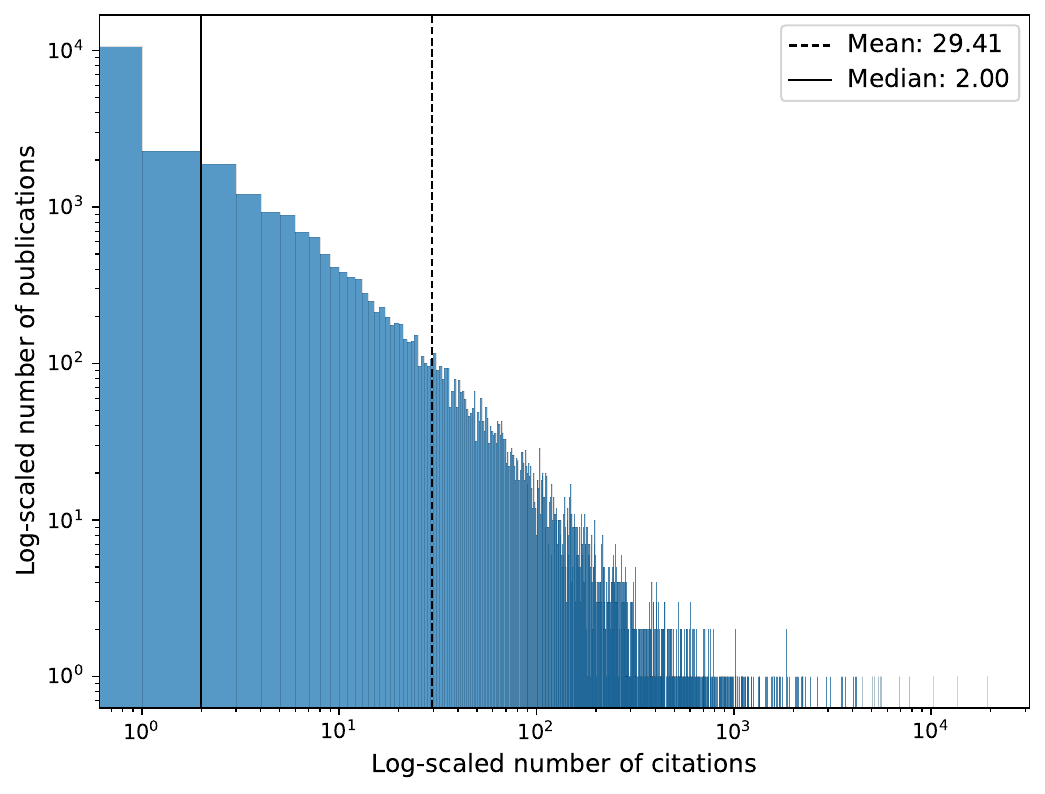}
		\caption{Sociology}
	\end{subfigure}
	
	\begin{subfigure}[t]{0.32\textwidth}
		\centering
		\includegraphics[width=\linewidth,keepaspectratio]{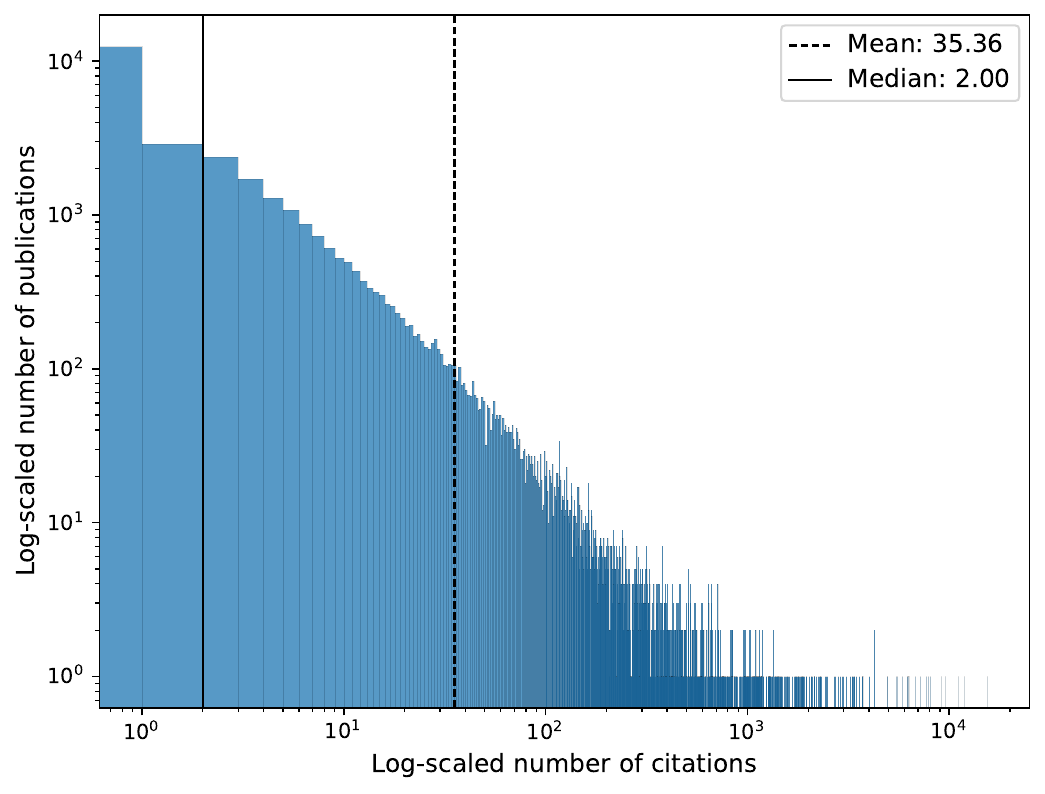}
		\caption{Economics}
	\end{subfigure}
	\hfill
	\begin{subfigure}[t]{0.32\textwidth}
		\centering
		\includegraphics[width=\linewidth,keepaspectratio]{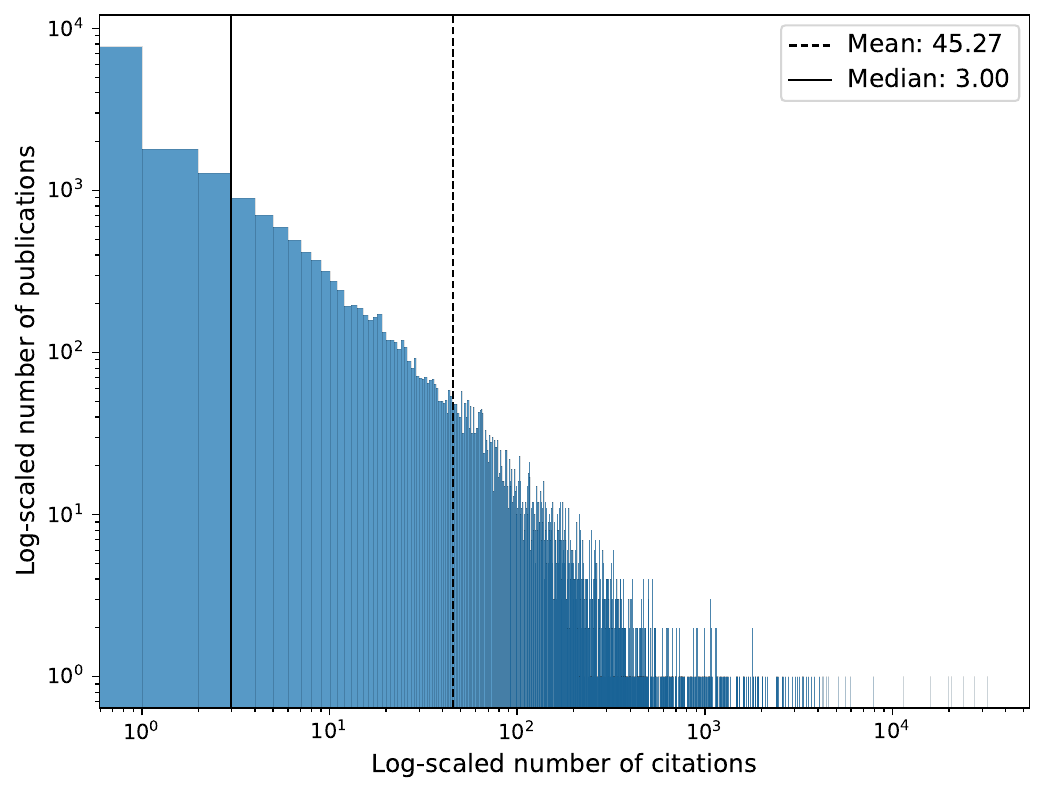}
		\caption{Marketing}
	\end{subfigure}
	\hfill
	\begin{subfigure}[t]{0.32\textwidth}
		\centering
		\includegraphics[width=\linewidth,keepaspectratio]{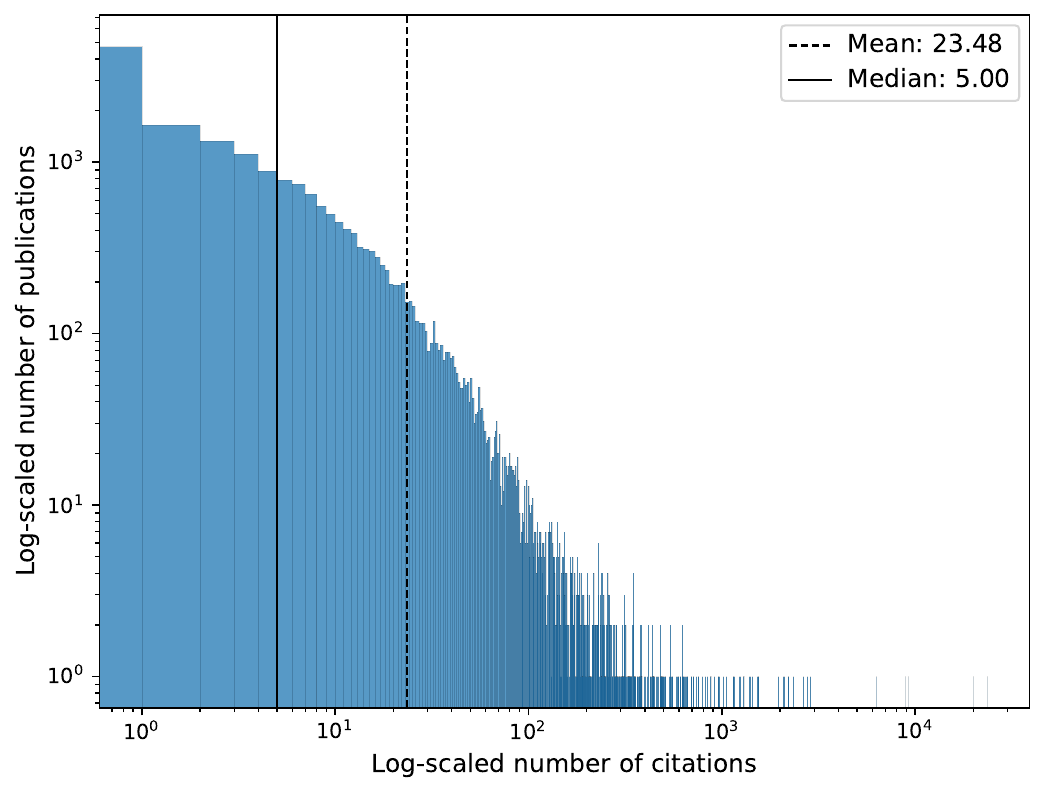}
		\caption{Mathematics}
	\end{subfigure}
	
	\caption{Log transformed distributions of number of citations in the combined and field datasets.}
	\label{fig:dist_citation}
\end{figure*}
\clearpage
\FloatBarrier

\section{Percentage Difference in Citations by Participation and by Contribution}

\label{percentage_difference_citation_appendix}
\begin{figure*}[!htbp]
	\centering
	\begin{subfigure}[t]{0.32\textwidth}
		\centering
		\includegraphics[width=\linewidth,keepaspectratio]{plots/percent_diff_cc.pdf}
		\caption{Combined}
	\end{subfigure}
	\hfill
	\begin{subfigure}[t]{0.32\textwidth}
		\centering
		\includegraphics[width=\linewidth,keepaspectratio]{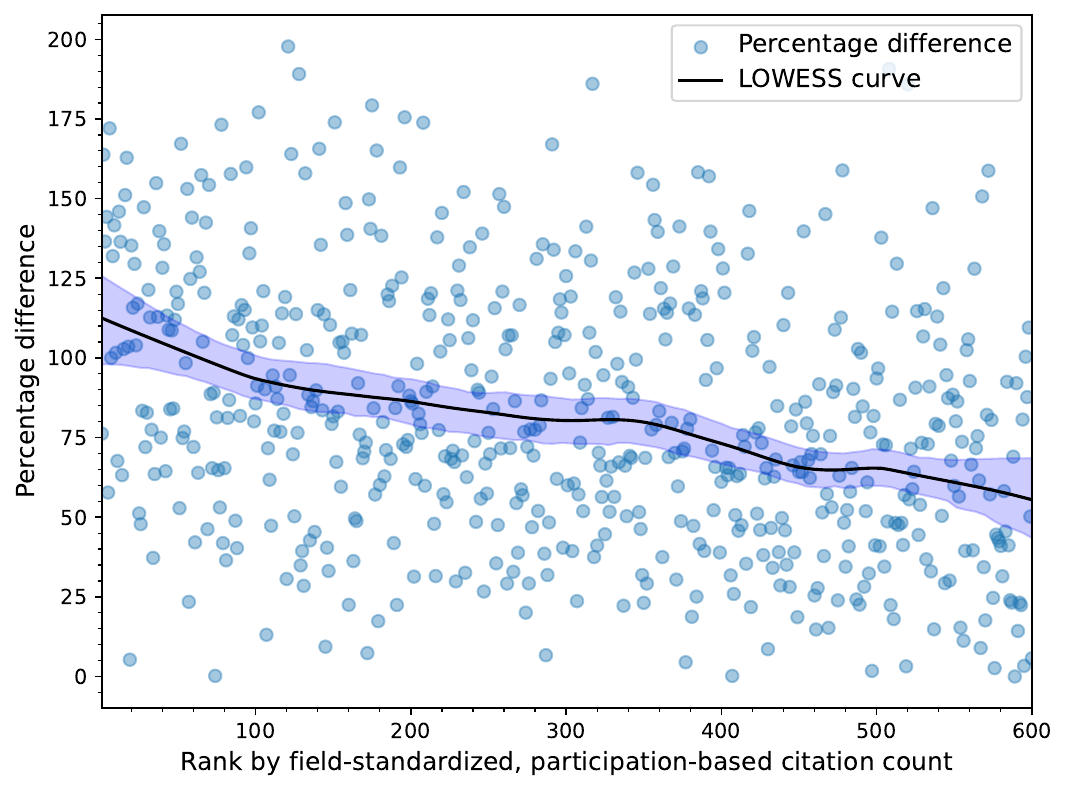}
		\caption{Biology}
	\end{subfigure}
	\hfill
	\begin{subfigure}[t]{0.32\textwidth}
		\centering
		\includegraphics[width=\linewidth,keepaspectratio]{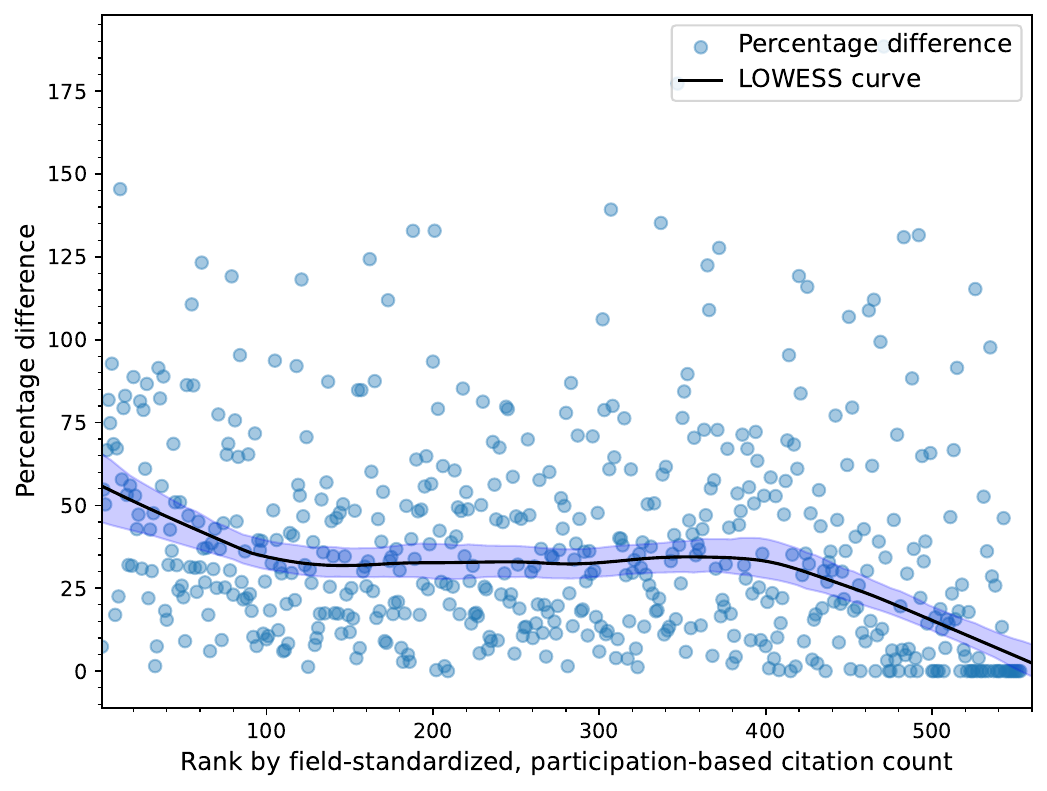}
		\caption{Electrical and Electronics Engineering}
	\end{subfigure}
	
	
	\begin{subfigure}[t]{0.32\textwidth}
		\centering
		\includegraphics[width=\linewidth,keepaspectratio]{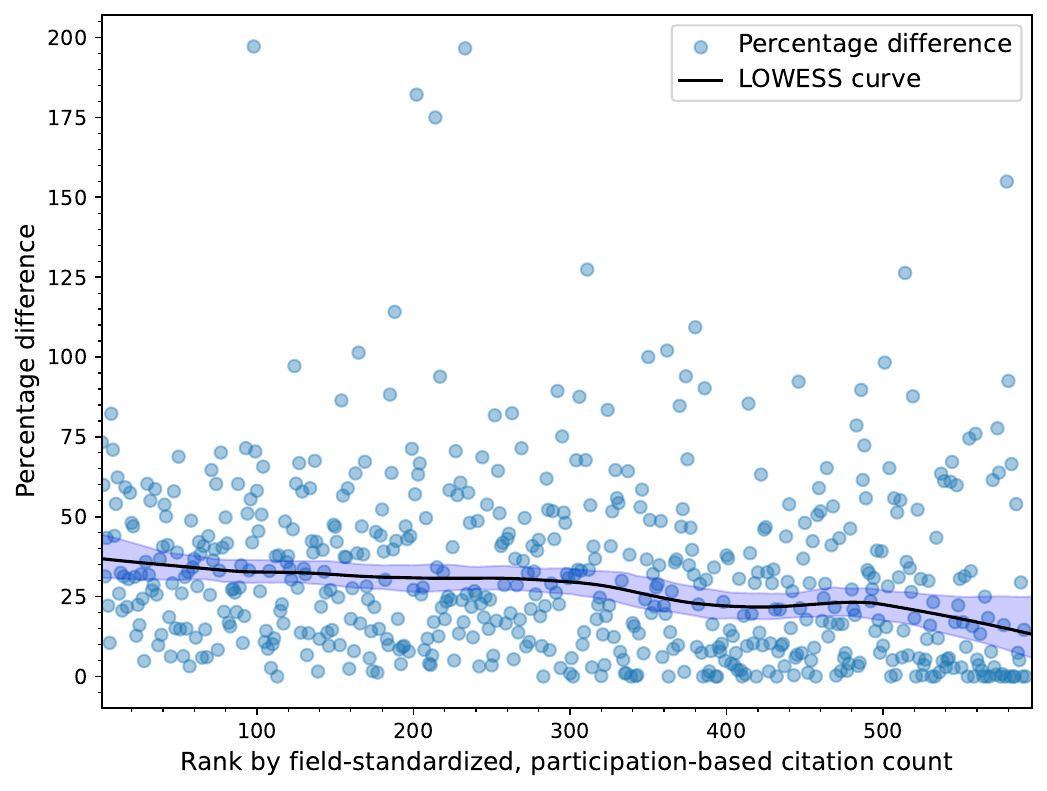}
		\caption{Computer Science and Engineering}
	\end{subfigure}
	\hfill
	\begin{subfigure}[t]{0.32\textwidth}
		\centering
		\includegraphics[width=\linewidth,keepaspectratio]{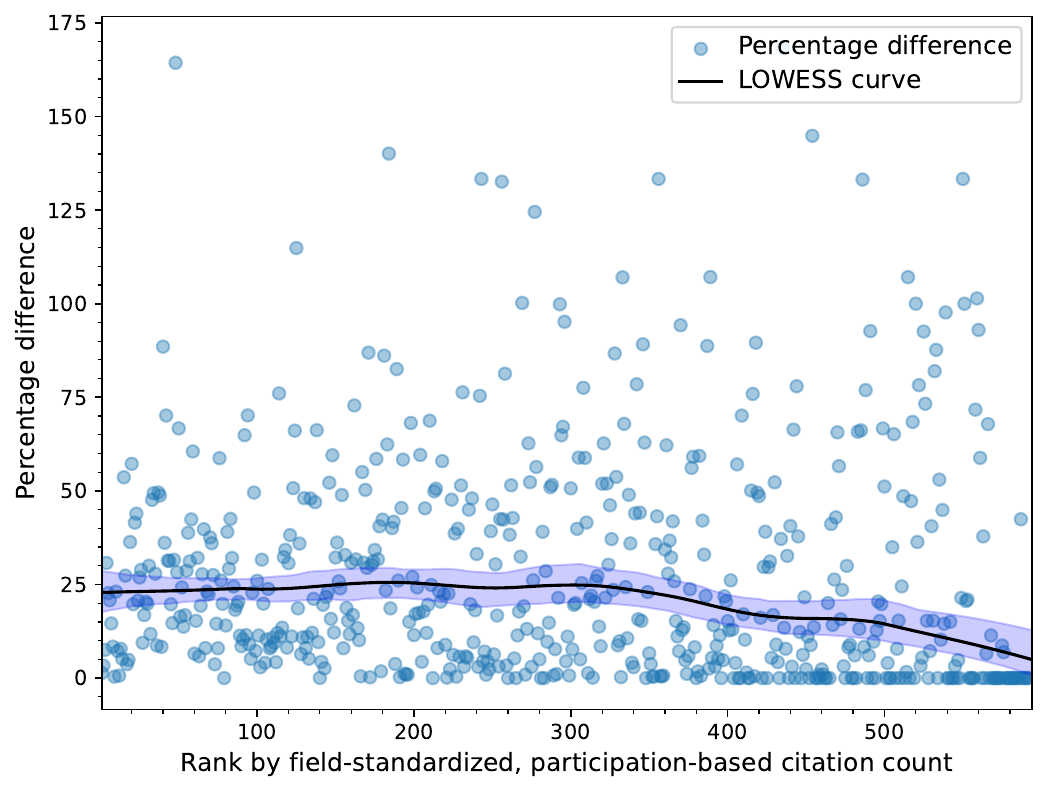}
		\caption{Psychology}
	\end{subfigure}
	\hfill
	\begin{subfigure}[t]{0.32\textwidth}
		\centering
		\includegraphics[width=\linewidth,keepaspectratio]{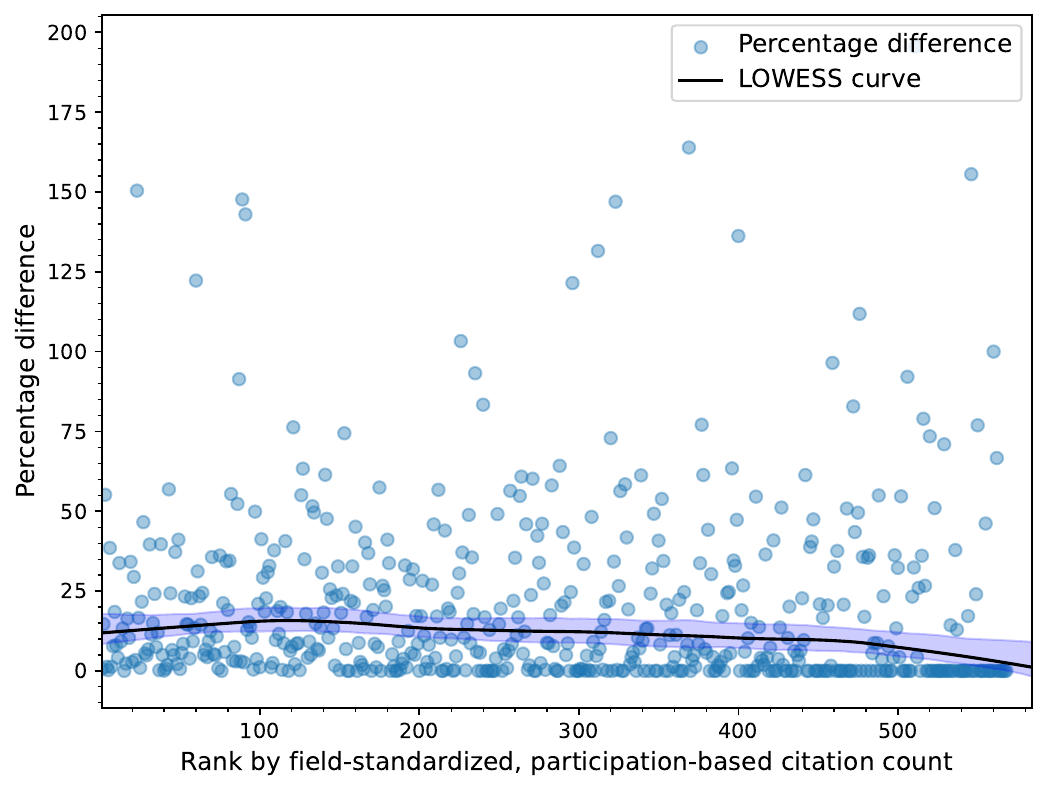}
		\caption{Sociology}
	\end{subfigure}
	
	\begin{subfigure}[t]{0.32\textwidth}
		\centering
		\includegraphics[width=\linewidth,keepaspectratio]{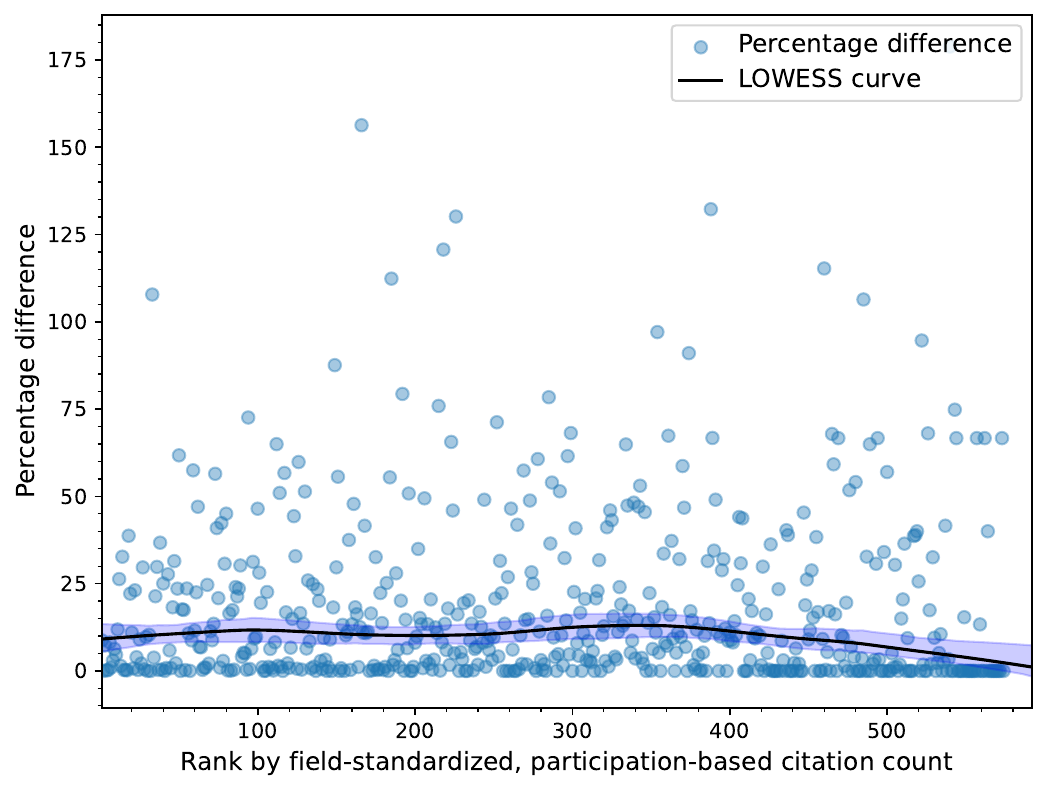}
		\caption{Economics}
	\end{subfigure}
	\hfill
	\begin{subfigure}[t]{0.32\textwidth}
		\centering
		\includegraphics[width=\linewidth,keepaspectratio]{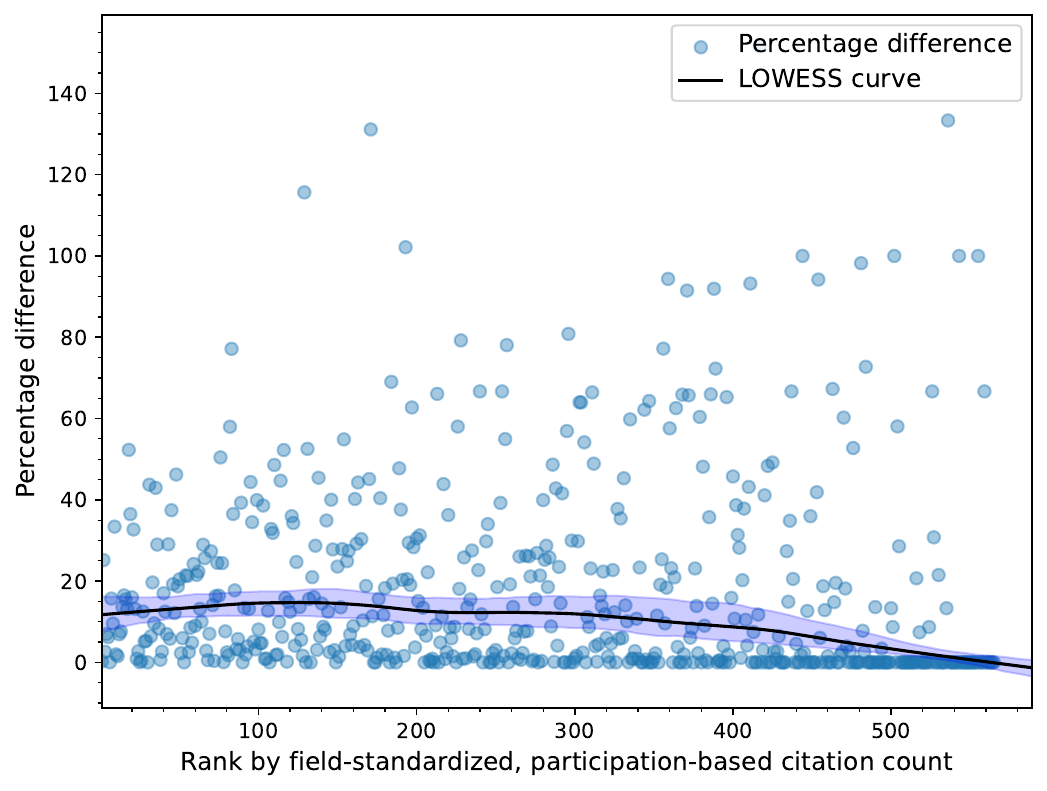}
		\caption{Marketing}
	\end{subfigure}
	\hfill
	\begin{subfigure}[t]{0.32\textwidth}
		\centering
		\includegraphics[width=\linewidth,keepaspectratio]{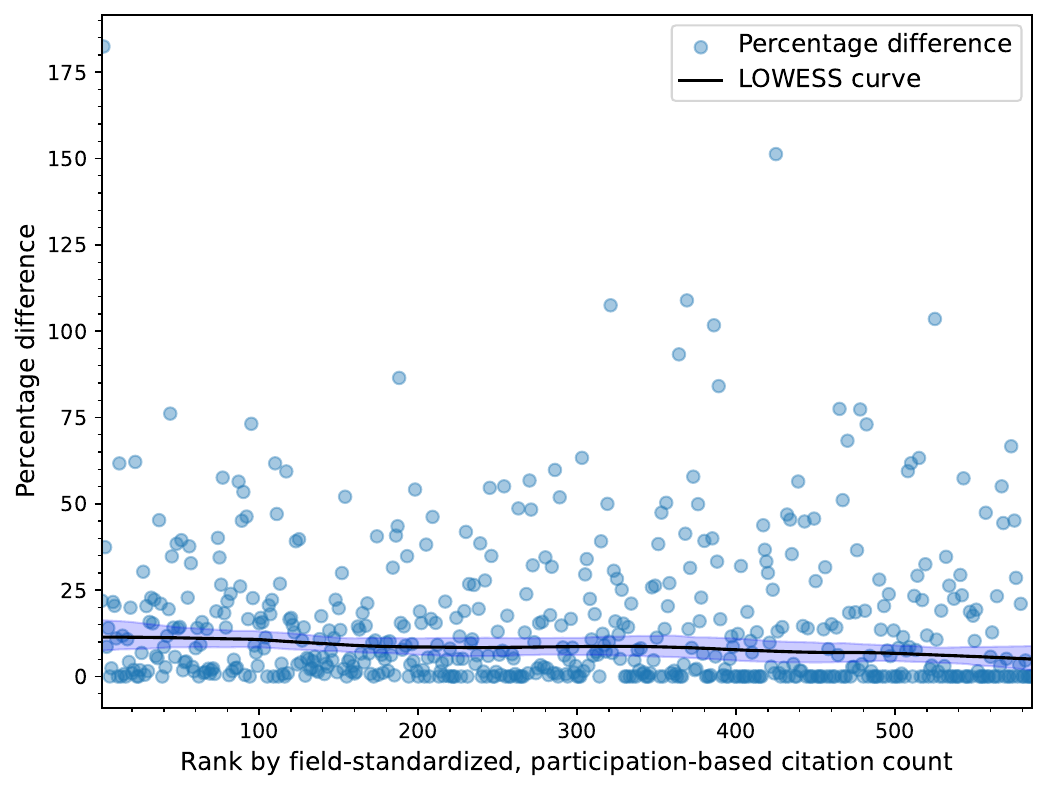}
		\caption{Mathematics}
	\end{subfigure}
	
	\caption{Percentage differences in citations by participation and by contribution in the combined and field datasets.}
	\label{fig:percentage_diff_citation}
\end{figure*}
\FloatBarrier
\clearpage

\section{ECDF Plot for Percentage Differences in Citation Counts by participation and by contribution}
\label{percentage_difference_citation_ecdf_appendix}
\begin{figure*}[h]
	\centering
	\includegraphics[width=0.8\textwidth]{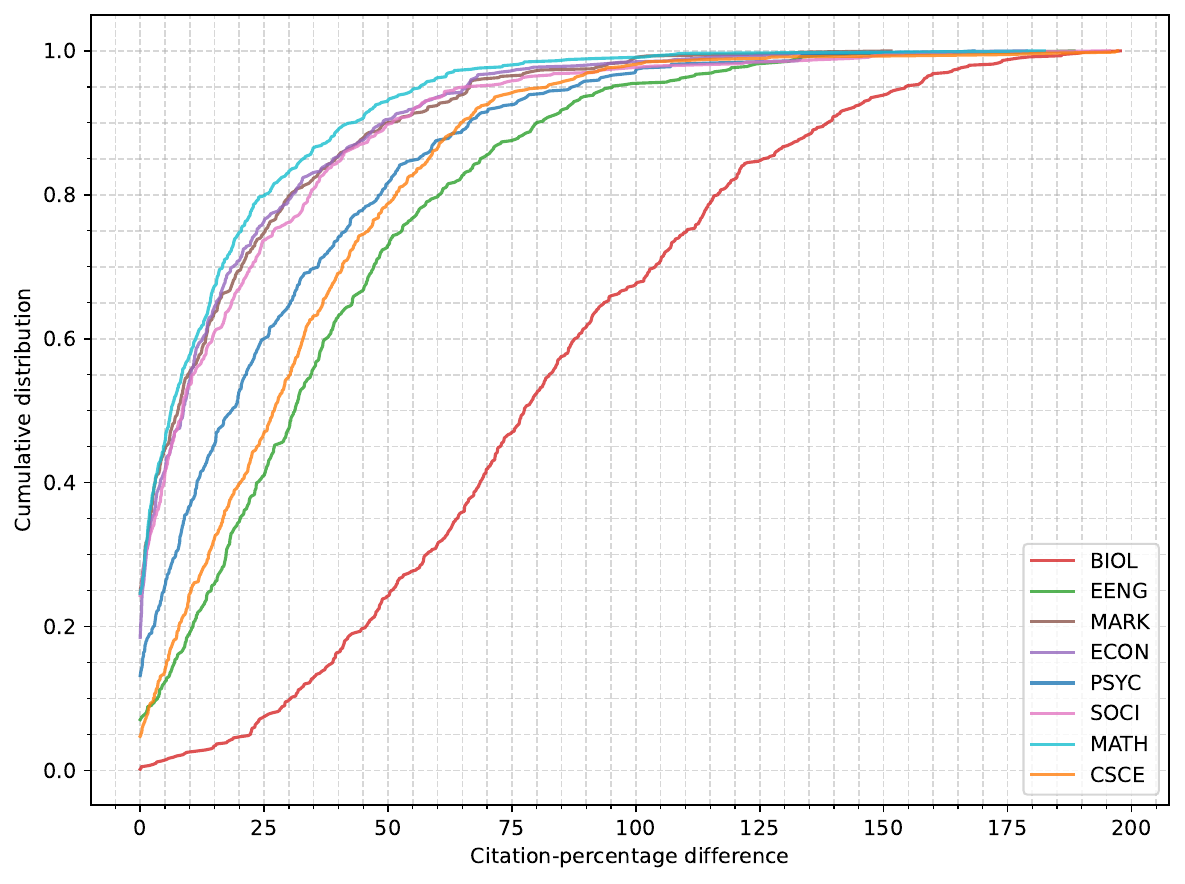} 
	\caption{The Empirical Cumulative Distribution Function (ECDF) for percentage differences in citation counts by participation and by contribution.}
	\label{fig:ecdf_citation}
\end{figure*}
\FloatBarrier
\clearpage

\section{Distributions of $h$-index in the Combined Field Datasets}

\label{h_index_distribution_by_participation}
\begin{figure*}[!htbp]
	\centering
	\begin{subfigure}[t]{0.32\textwidth}
		\centering
		\includegraphics[width=\linewidth,keepaspectratio]{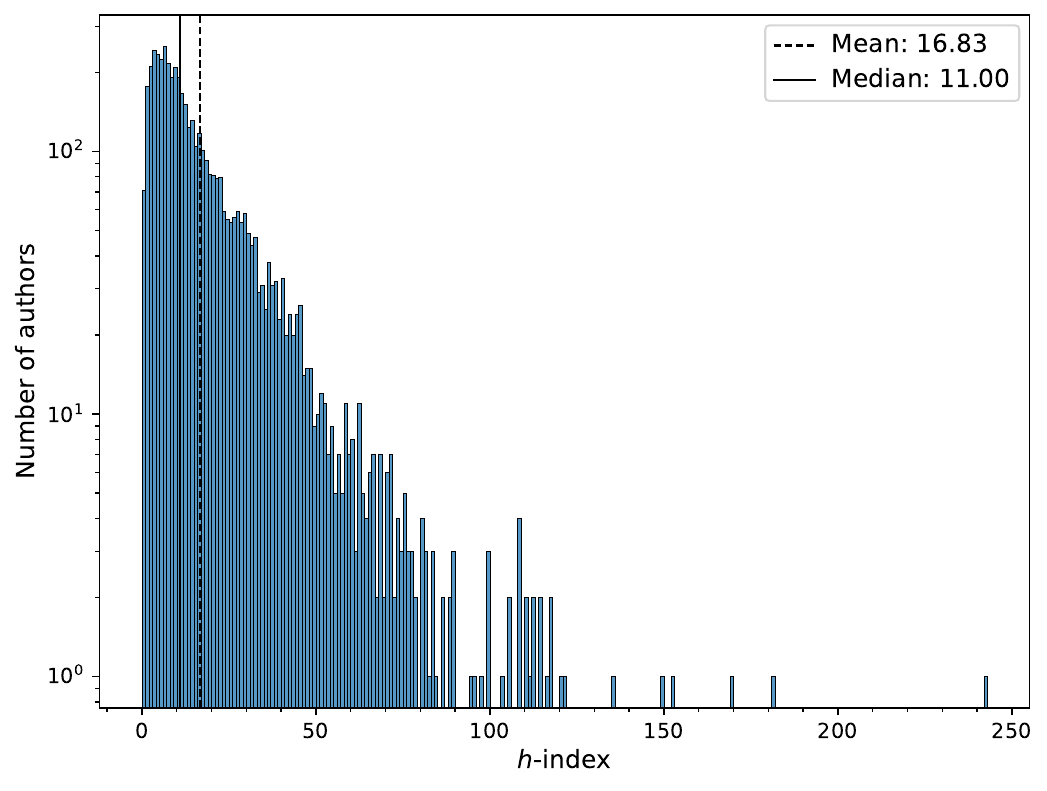}
		\caption{Combined}
	\end{subfigure}
	\hfill
	\begin{subfigure}[t]{0.32\textwidth}
		\centering
		\includegraphics[width=\linewidth,keepaspectratio]{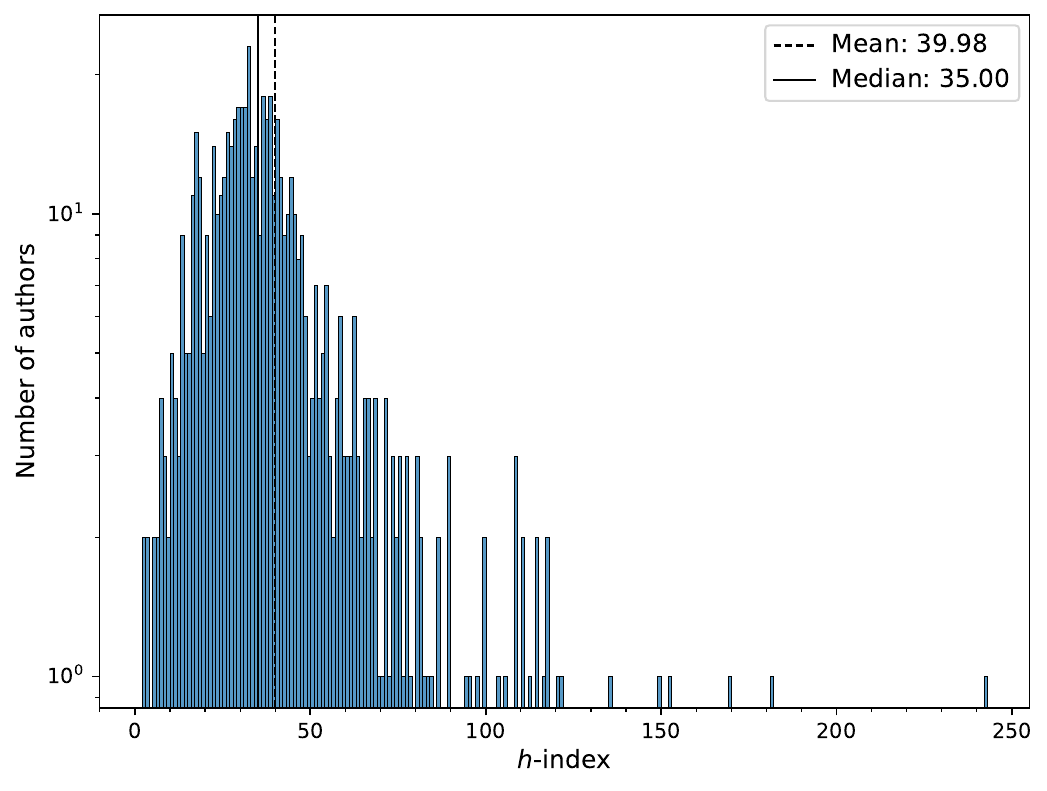}
		\caption{Biology}
	\end{subfigure}
	\hfill
	\begin{subfigure}[t]{0.32\textwidth}
		\centering
		\includegraphics[width=\linewidth,keepaspectratio]{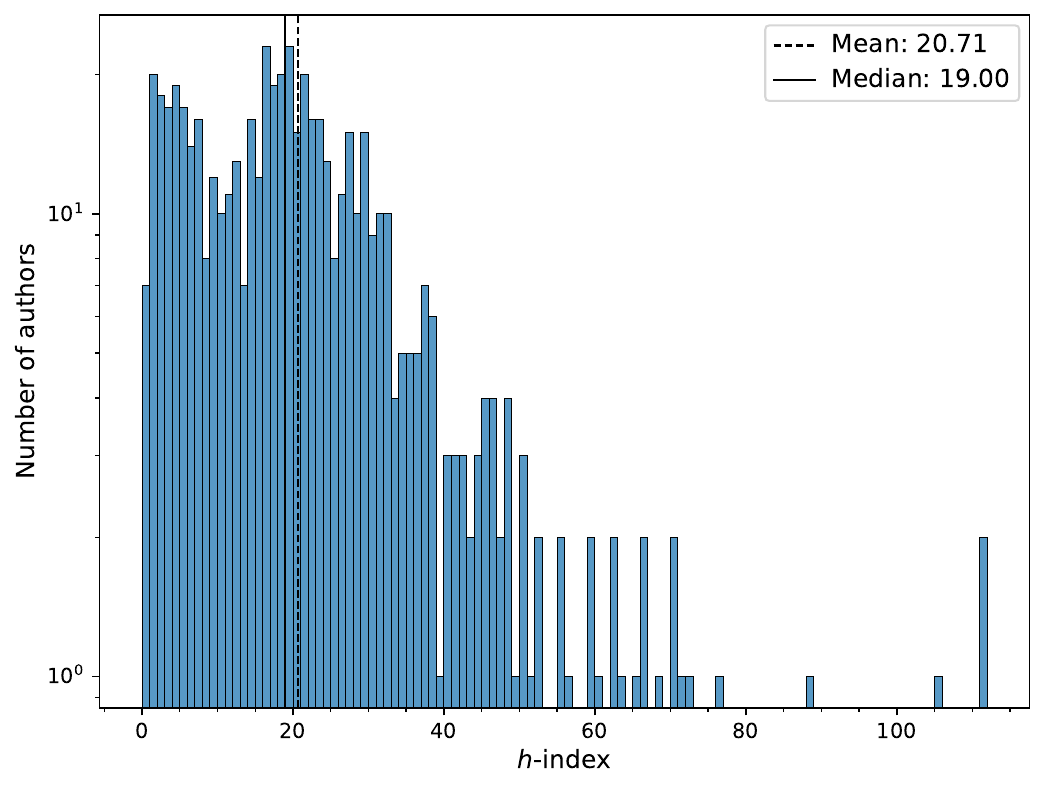}
		\caption{Electrical and Electronics Engineering}
	\end{subfigure}
	
	\begin{subfigure}[t]{0.32\textwidth}
		\centering
		\includegraphics[width=\linewidth,keepaspectratio]{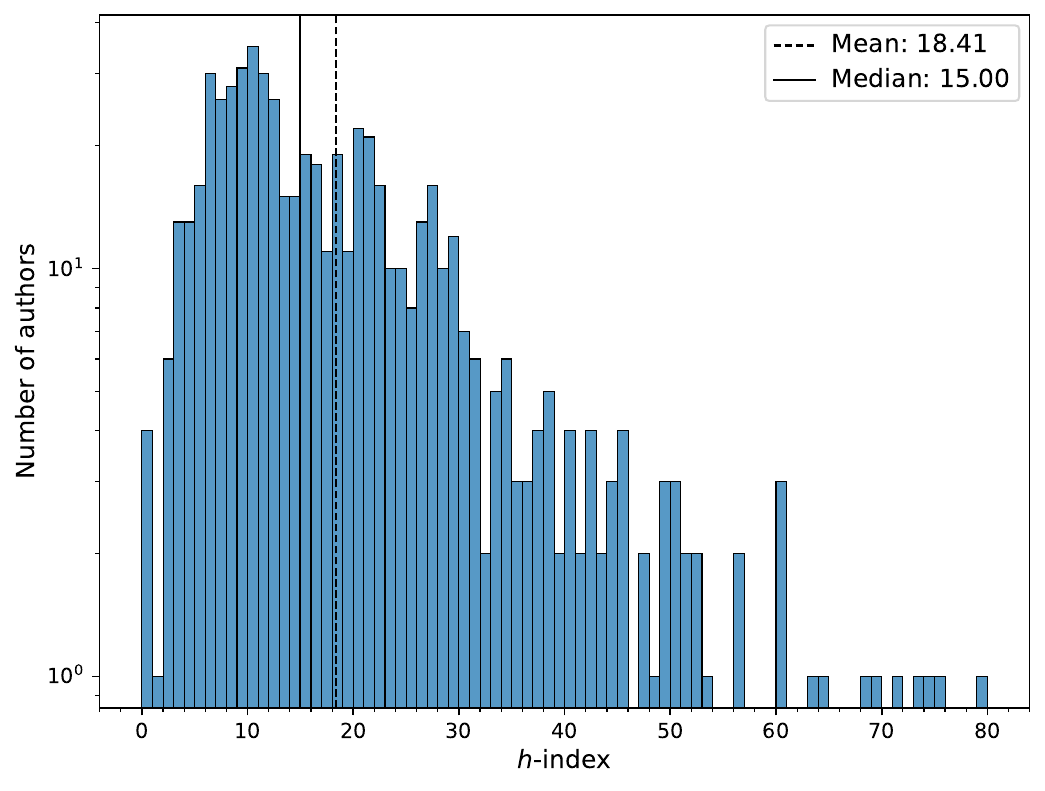}
		\caption{Computer Science and Engineering}
	\end{subfigure}
	\hfill
	\begin{subfigure}[t]{0.32\textwidth}
		\centering
		\includegraphics[width=\linewidth,keepaspectratio]{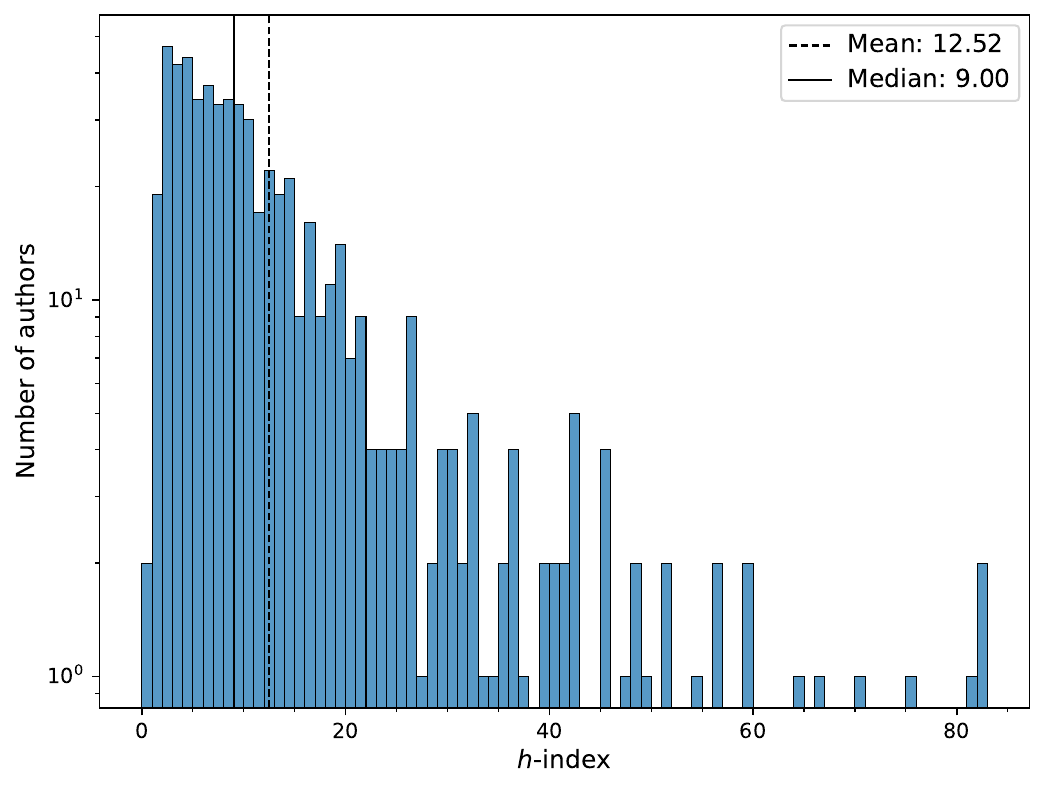}
		\caption{Psychology}
	\end{subfigure}
	\hfill
	\begin{subfigure}[t]{0.32\textwidth}
		\centering
		\includegraphics[width=\linewidth,keepaspectratio]{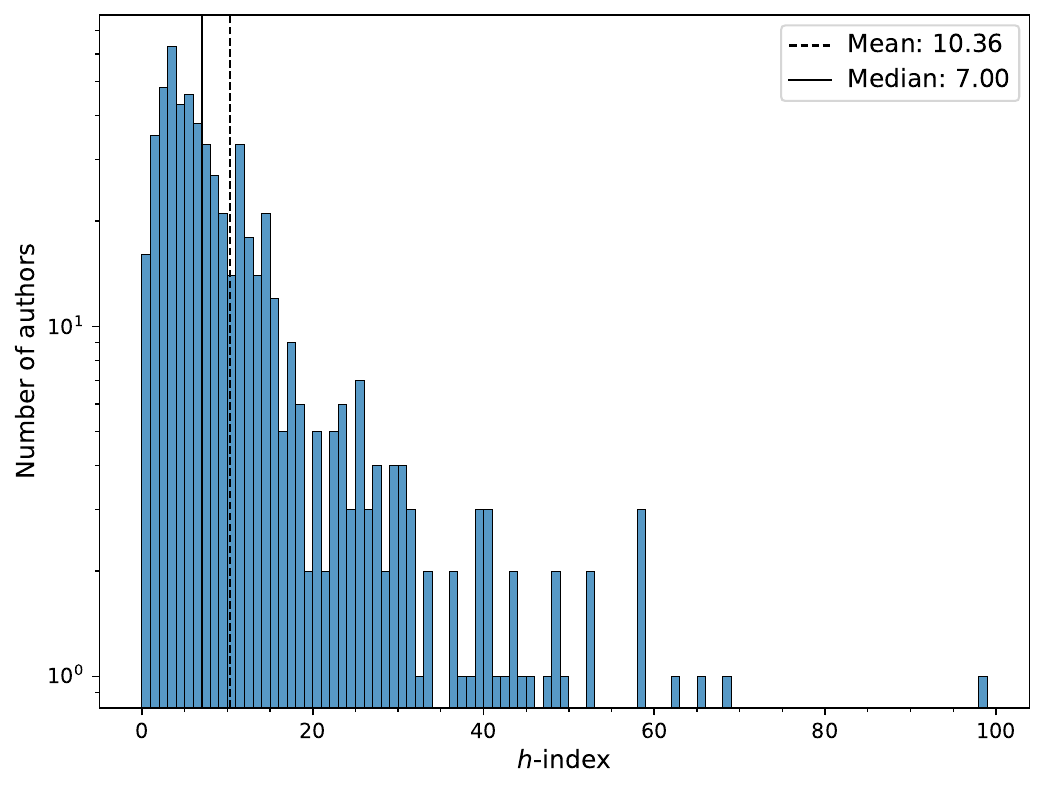}
		\caption{Sociology}
	\end{subfigure}
	
	\begin{subfigure}[t]{0.32\textwidth}
		\centering
		\includegraphics[width=\linewidth,keepaspectratio]{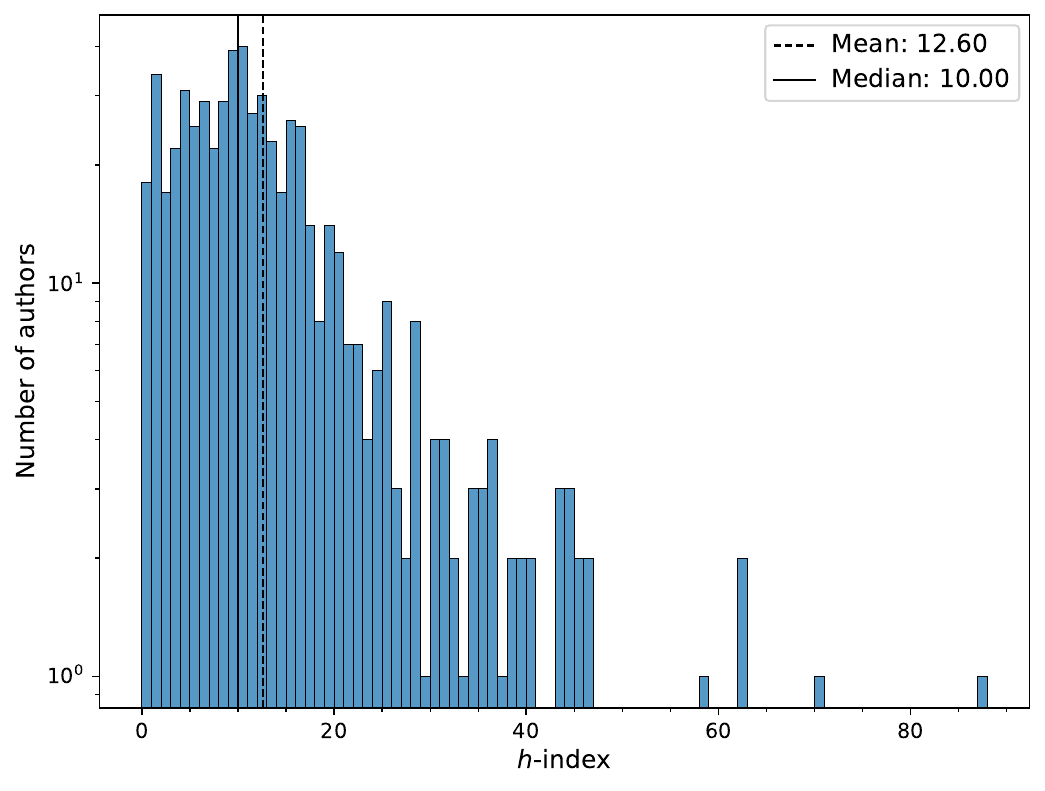}
		\caption{Economics}
	\end{subfigure}
	\hfill
	\begin{subfigure}[t]{0.32\textwidth}
		\centering
		\includegraphics[width=\linewidth,keepaspectratio]{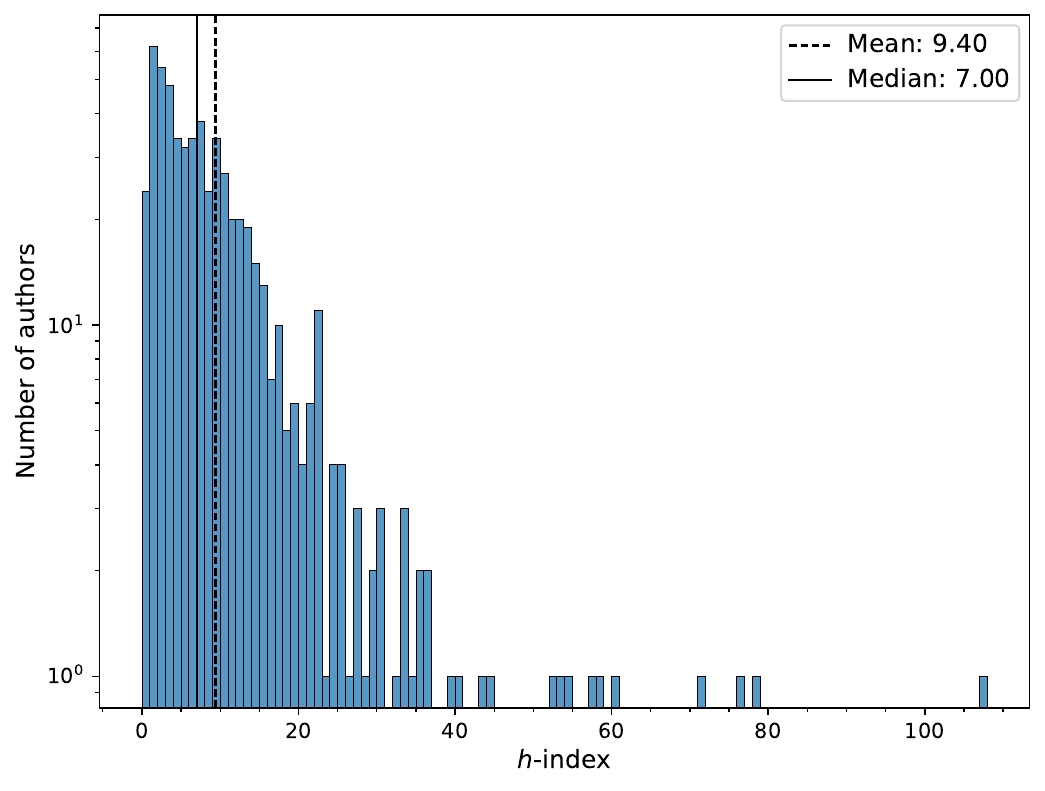}
		\caption{Marketing}
	\end{subfigure}
	\hfill
	\begin{subfigure}[t]{0.32\textwidth}
		\centering
		\includegraphics[width=\linewidth,keepaspectratio]{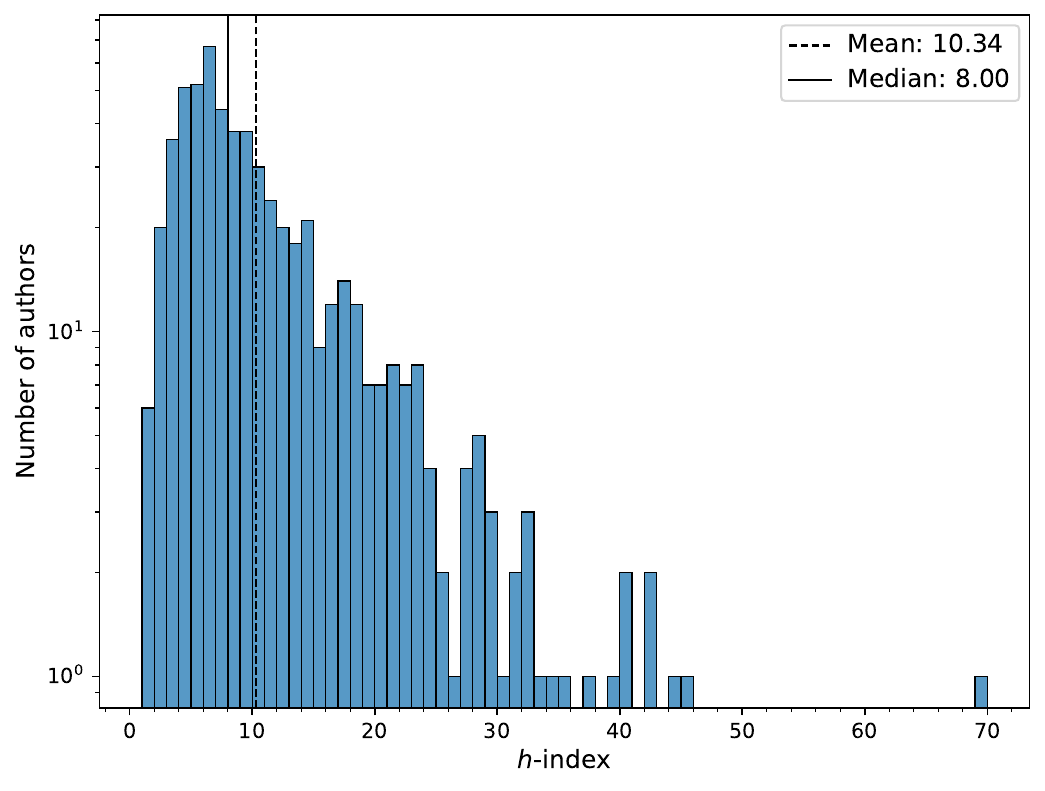}
		\caption{Mathematics}
	\end{subfigure}
	
	\caption{Log transformed distributions of $h$-index by participation in the combined and field datasets.}
	\label{fig:h_index_distribution_by_participation}
\end{figure*}
\clearpage
\FloatBarrier

\section{Percentage Differences in $h$-index by Participation and by Contribution}

\label{percentage_difference_h_index_appendix}
\begin{figure*}[!htbp]
	\centering
	\begin{subfigure}[t]{0.32\textwidth}
		\centering
		\includegraphics[width=\linewidth,keepaspectratio]{plots/percent_diff_hc.pdf}
		\caption{Combined}
	\end{subfigure}
	\hfill
	\begin{subfigure}[t]{0.32\textwidth}
		\centering
		\includegraphics[width=\linewidth,keepaspectratio]{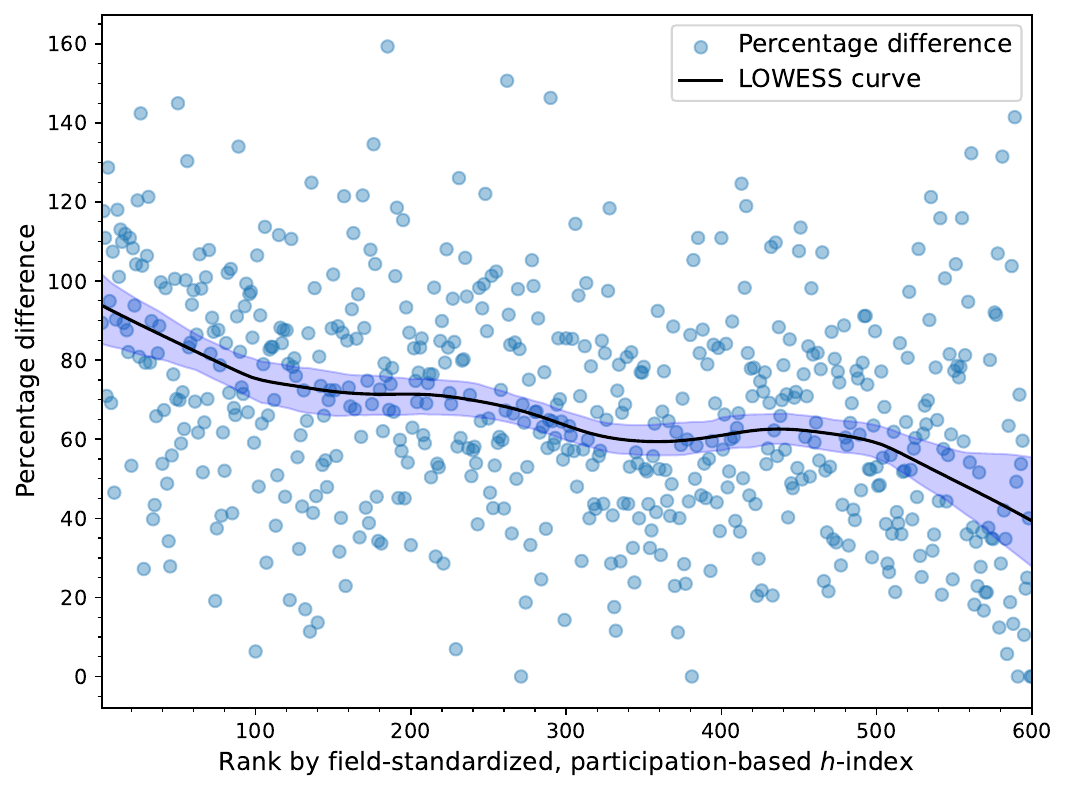}
		\caption{Biology}
	\end{subfigure}
	\hfill
	\begin{subfigure}[t]{0.32\textwidth}
		\centering
		\includegraphics[width=\linewidth,keepaspectratio]{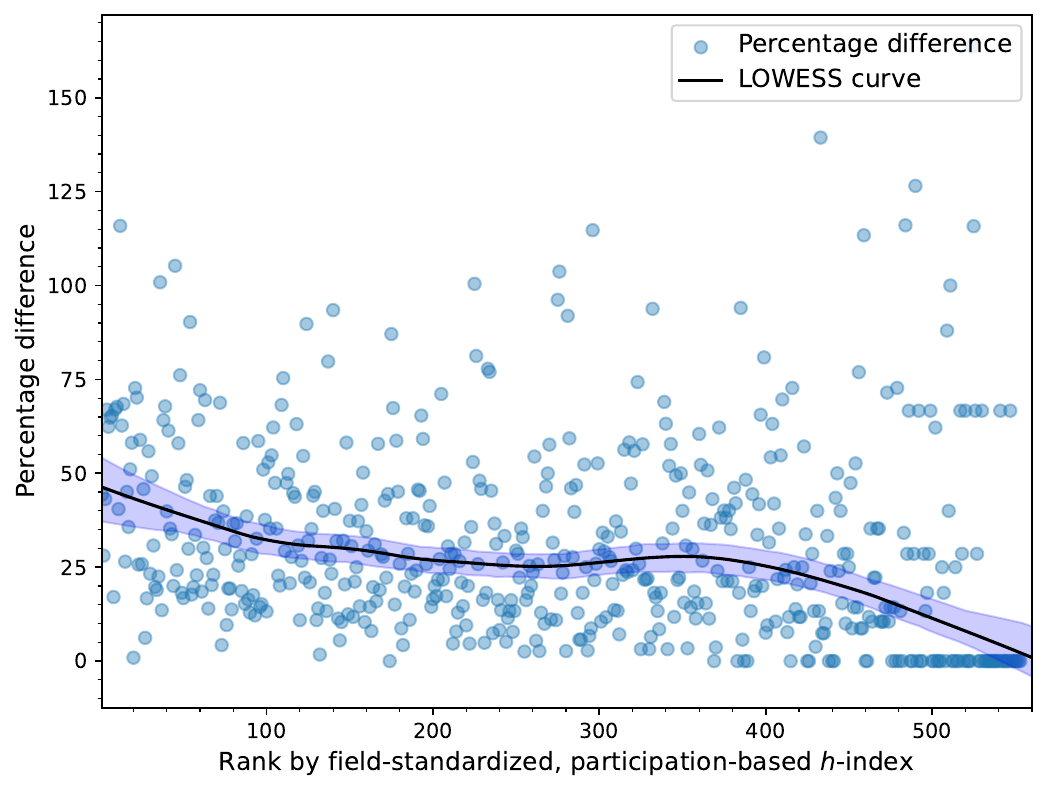}
		\caption{Electrical and Electronics Engineering}
	\end{subfigure}
	
	\begin{subfigure}[t]{0.32\textwidth}
		\centering
		\includegraphics[width=\linewidth,keepaspectratio]{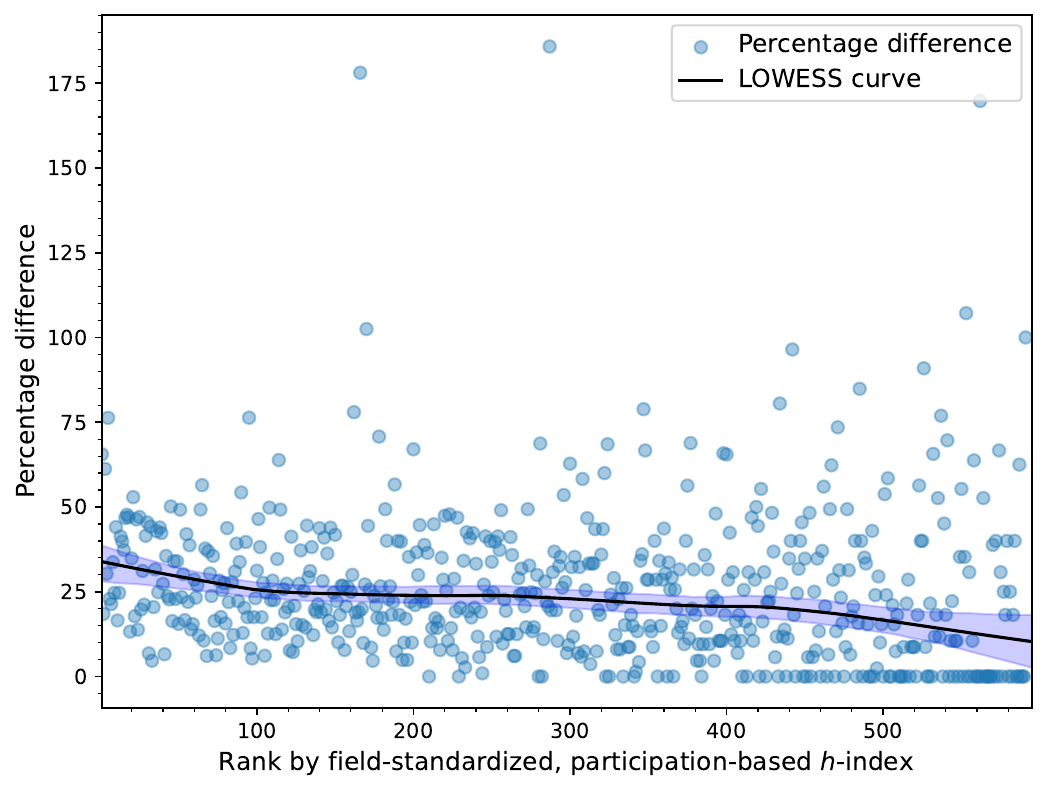}
		\caption{Computer Science and Engineering}
	\end{subfigure}
	\hfill
	\begin{subfigure}[t]{0.32\textwidth}
		\centering
		\includegraphics[width=\linewidth,keepaspectratio]{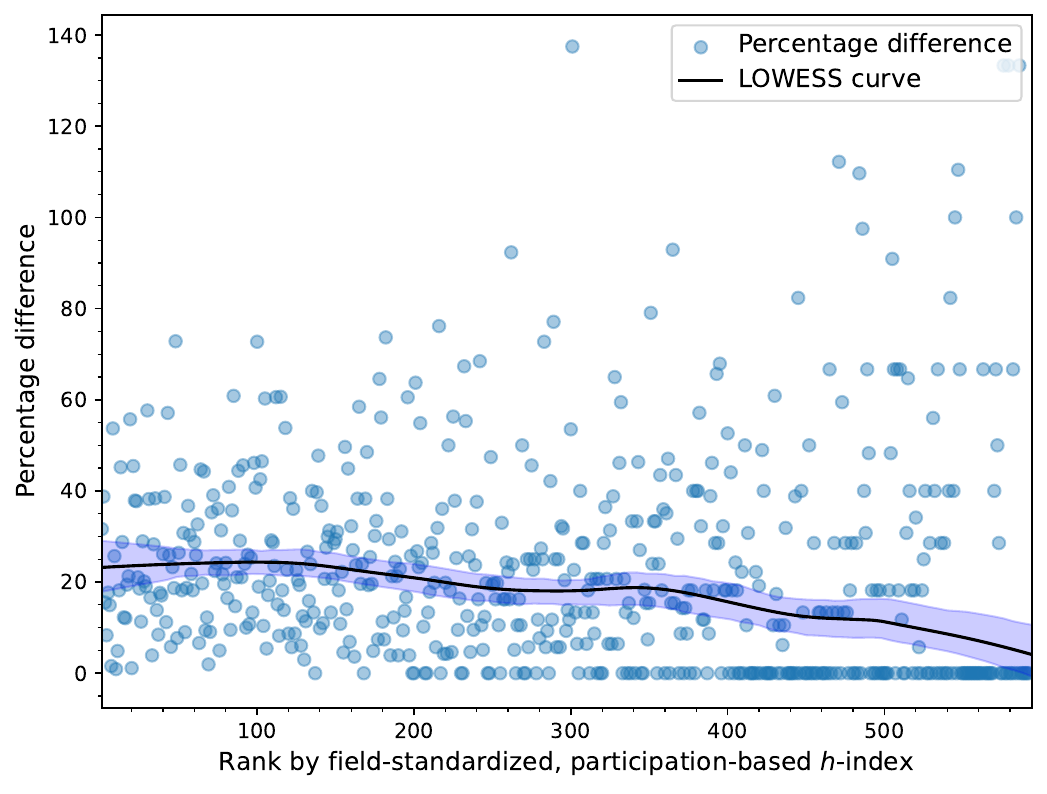}
		\caption{Psychology}
	\end{subfigure}
	\hfill
	\begin{subfigure}[t]{0.32\textwidth}
		\centering
		\includegraphics[width=\linewidth,keepaspectratio]{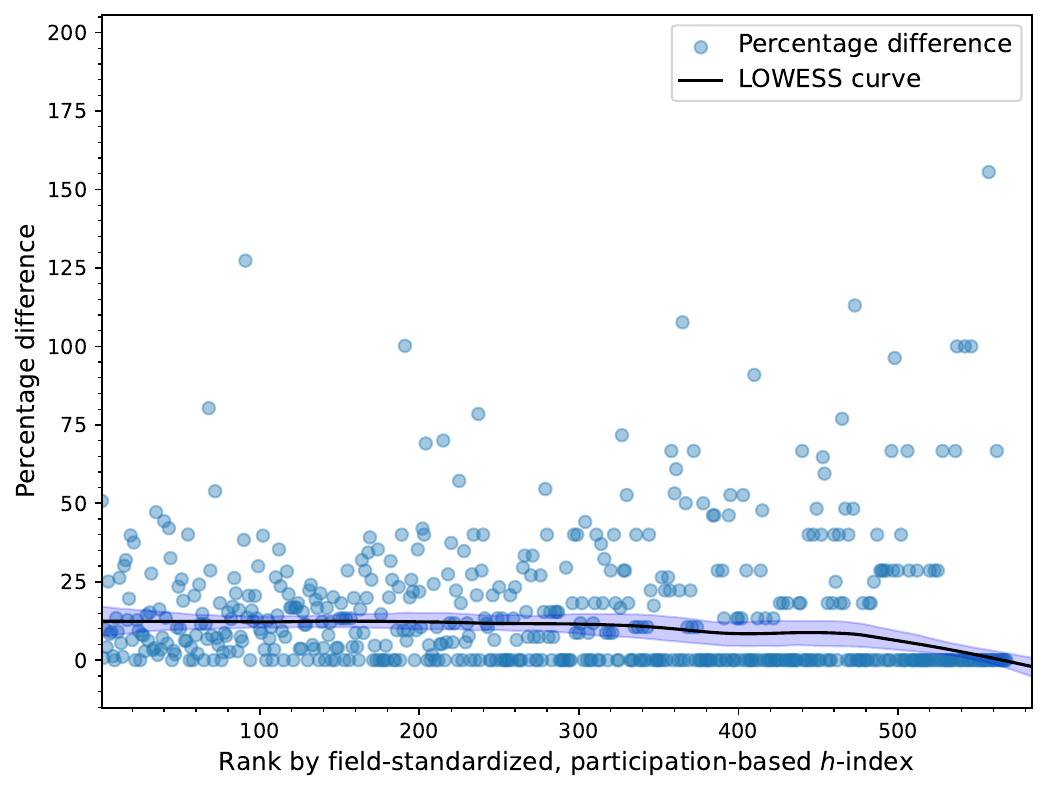}
		\caption{Sociology}
	\end{subfigure}
	
	
	\begin{subfigure}[t]{0.32\textwidth}
		\centering
		\includegraphics[width=\linewidth,keepaspectratio]{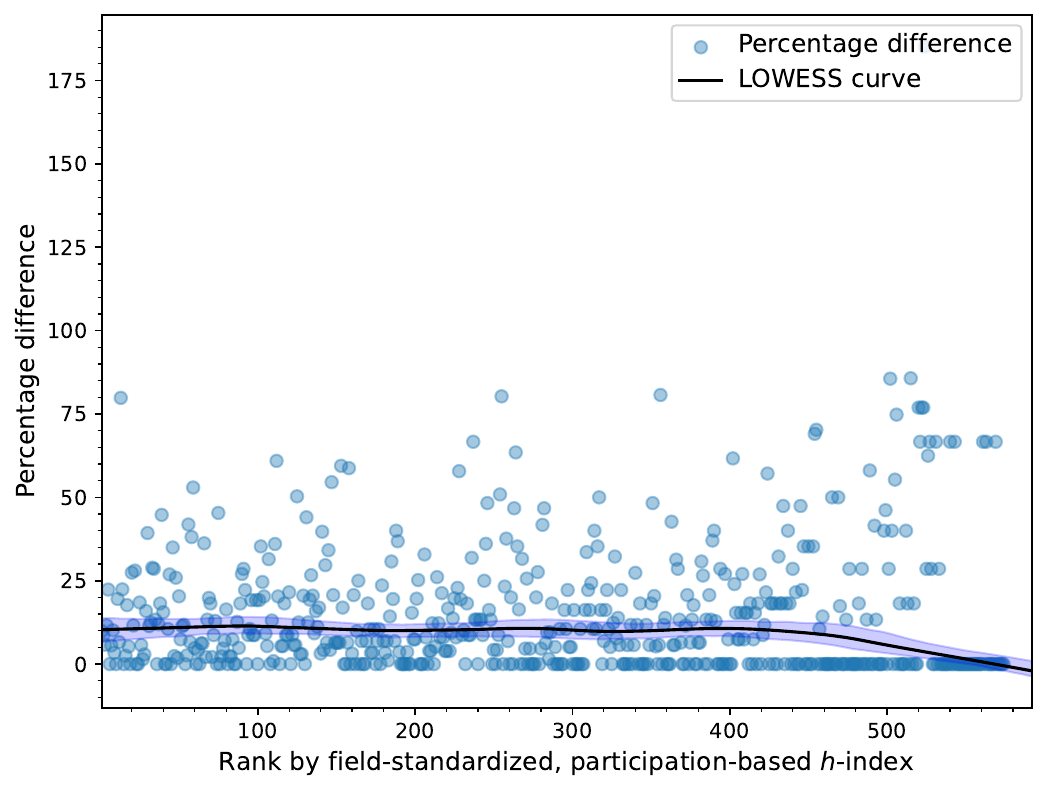}
		\caption{Economics}
	\end{subfigure}
	\hfill
	\begin{subfigure}[t]{0.32\textwidth}
		\centering
		\includegraphics[width=\linewidth,keepaspectratio]{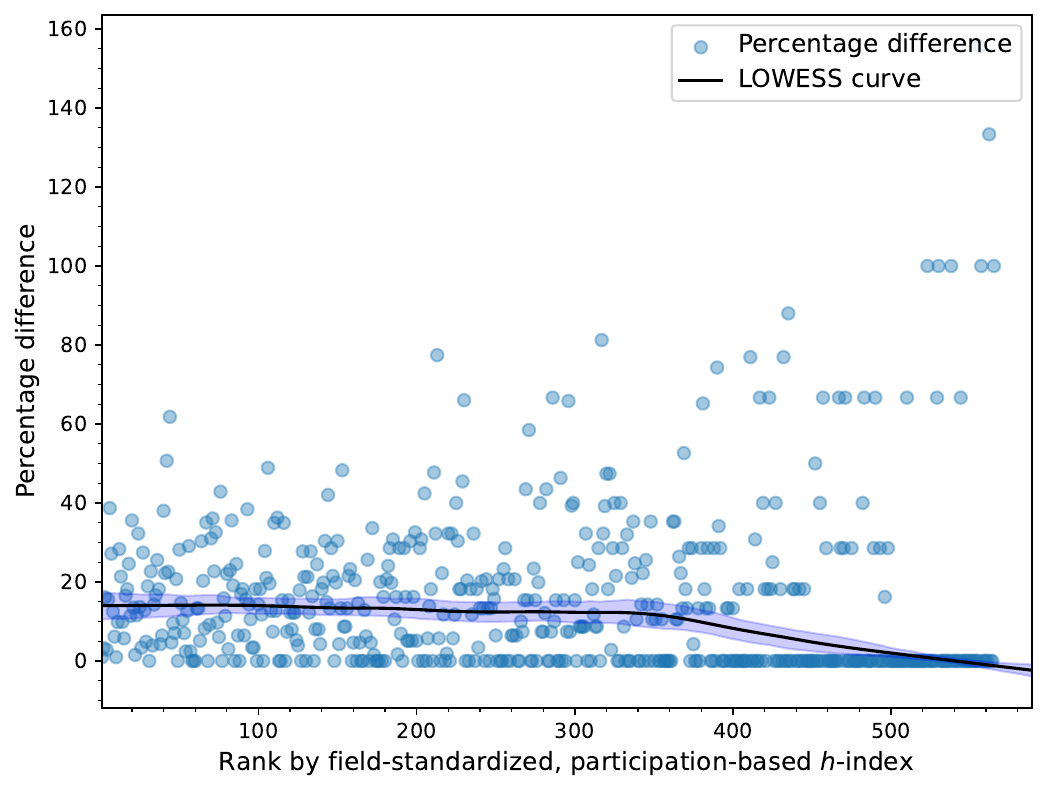}
		\caption{Marketing}
	\end{subfigure}
	\hfill
	\begin{subfigure}[t]{0.32\textwidth}
		\centering
		\includegraphics[width=\linewidth,keepaspectratio]{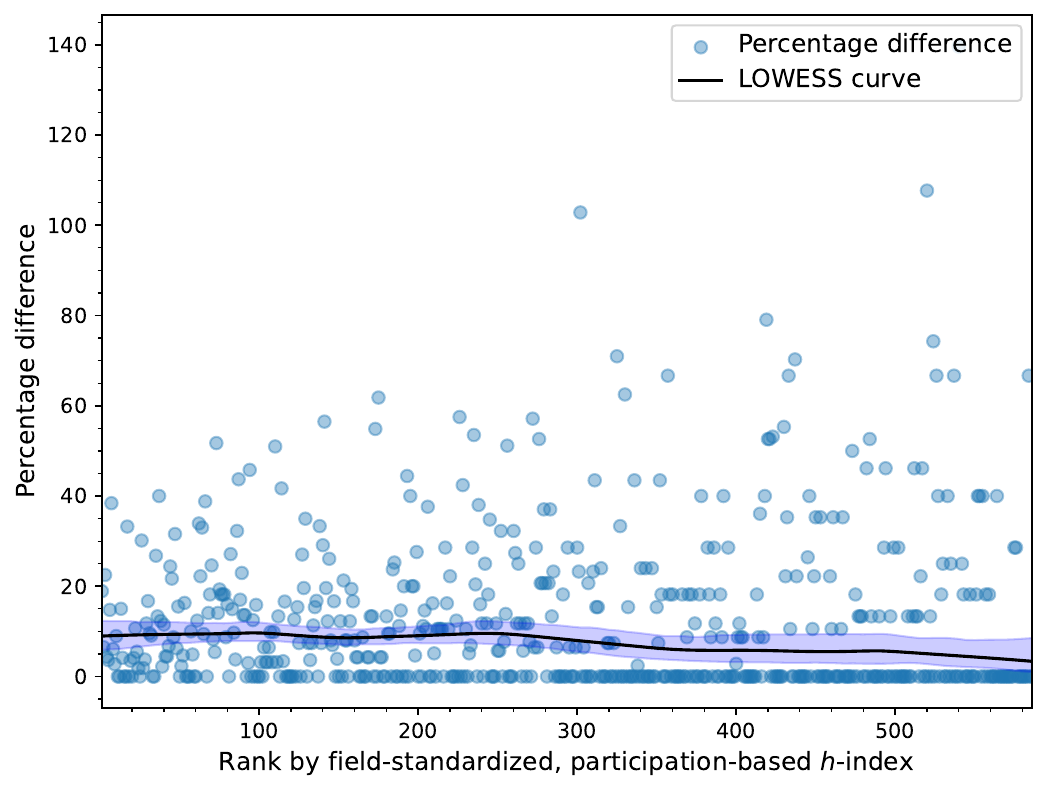}
		\caption{Mathematics}
	\end{subfigure}
	
	\caption{Percentage differences in $h$-index by participation and by contribution in the combined and field datasets.}
	\label{fig:percentage_diff_h_index}
\end{figure*}
\clearpage
\FloatBarrier

\section{ECDF Plot for Percentage Differences in $h$-index by participation and by contribution}
\label{percentage_difference_h_index_ecdf_appendix}
\begin{figure*}[h]
	\centering
	\includegraphics[width=0.8\textwidth]{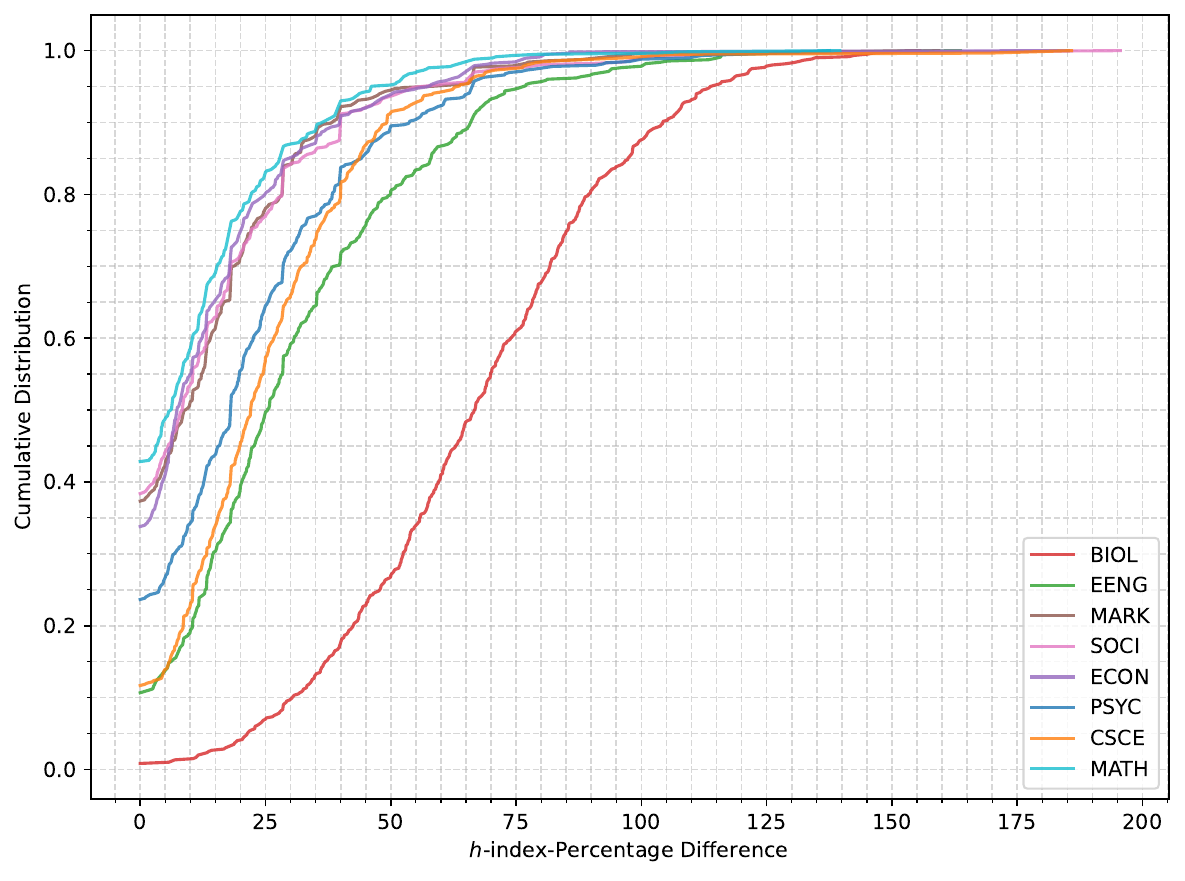}
	\caption{The Empirical Cumulative Distribution Function (ECDF) for percentage differences in $h$-index by participation and contribution.}
	\label{fig:ecdf_h_index}
\end{figure*}
\FloatBarrier
\clearpage

\section{ECDF Plot for Percentage Differences in Publication Counts by participation and by contribution}
\label{percentage_difference_pub_ecdf_appendix}
\begin{figure*}[h]
	\centering
	\includegraphics[width=0.8\textwidth]{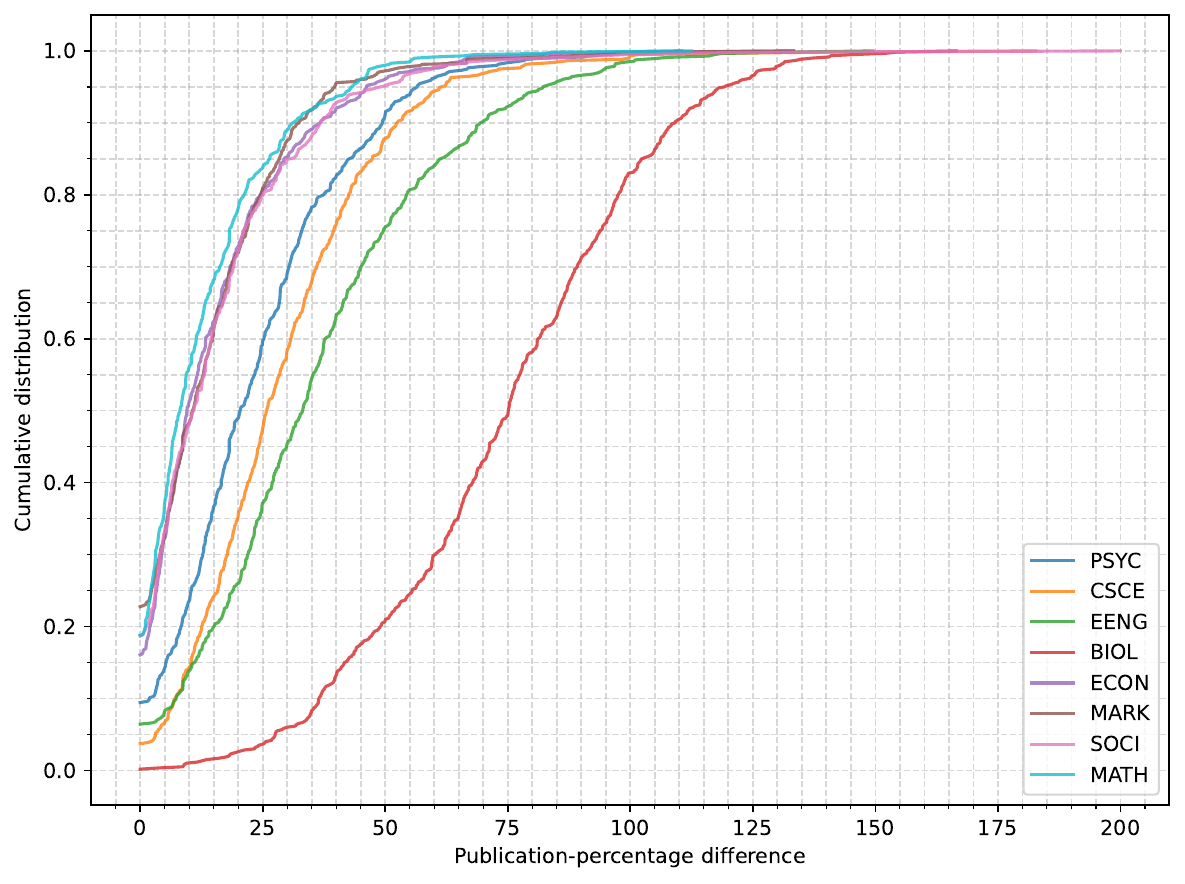}
	\caption{Empirical Cumulative Distribution Function (ECDF) for percentage difference in publication count by participation and contribution.}
	\label{fig:ecdf_publication}
\end{figure*}
\FloatBarrier
\clearpage
\end{document}